\documentclass[smallextended]{svjour3}
\smartqed

\usepackage{graphicx}
\usepackage{hyperref}
\usepackage{url}
\usepackage{epsf}
\usepackage{bm}
\usepackage{amsmath,amssymb}
\usepackage{color}

\def\d{\delta}

\newcommand{\na}{\nabla}

\def\a{\alpha}
\def\b{\beta}
\def\g{\gamma}
\def\t{\theta}

\def\be{\begin{equation}}
\def\ee{\end{equation}}
\def\ba{\begin{eqnarray}}
\def\ea{\end{eqnarray}}
\def\bea{\begin{eqnarray}}
\def\eea{\end{eqnarray}}
\newcommand{\dis}{\displaystyle}

\begin{document}

\title{Rotating Stars in Relativity}

\author{Vasileios Paschalidis         \and
        Nikolaos Stergioulas
}

\institute{Vasileios Paschalidis \at
              Theoretical Astrophysics Program, Departments of Astronomy and Physics, \\
              University of Arizona, Tucson, AZ 85721, USA \\
              Department of Physics, Princeton University \\
              Princeton, NJ 08544, USA \\
              \email{vpaschal@email.arizona.edu}           \\
              \url{http://physics.princeton.edu/~vp16/}
           \and
           Nikolaos Stergioulas \at
           Department of Physics, Aristotle University of Thessaloniki \\
           Thessaloniki, 54124, Greece \\
           \email{niksterg@auth.gr} \\
           \url{http://www.astro.auth.gr/~niksterg}
}


\date{Received: date / Accepted: date}
\maketitle


\begin{abstract}
  Rotating relativistic stars have been studied extensively in recent
  years, both theoretically and observationally, because of the
  information they might yield about the equation of state of matter
  at extremely high densities and because they are considered to be
  promising sources of gravitational waves. The latest theoretical
  understanding of rotating stars in relativity is reviewed in this
  updated article. The sections on equilibrium properties and on
  nonaxisymmetric oscillations and instabilities in $f$-modes and
  $r$-modes have been updated. Several new sections have been added on
  equilibria in modified theories of gravity, approximate universal
  relationships, the one-arm spiral instability, on analytic solutions
  for the exterior spacetime, rotating stars in LMXBs, rotating
  strange stars, and on rotating stars in numerical relativity
  including both hydrodynamic and magnetohydrodynamic studies of these
  objects.
  \keywords{Relativistic stars \and Rotation \and Stability \and
  Oscillations \and Magnetic fields \and Numerical relativity}
\end{abstract}


\newpage

The article has been substantially revised and updated. 
New Section~\ref{rotating-stars} and various 
Subsections (2.3\,--\,2.5, 2.7.5\,--\,2.7.7, 2.8, 2.10-2.11,4.5.7) 
have been added. The number of references (877) has more than doubled.


\section{Introduction}

Rotating relativistic stars are of fundamental interest in
physics. Their bulk properties constrain the proposed equations of
state for densities greater than the nuclear saturation
density. Accreted matter in their gravitational fields undergoes
high-frequency oscillations that could become a sensitive probe for
general relativistic effects. Temporal changes in the rotational
period of millisecond pulsars can also reveal a wealth of information
about important physical processes inside the stars or of cosmological
relevance. In addition, rotational instabilities can result in the
generation of copious amounts of gravitational waves the detection of
which would initiate a new field of observational asteroseismology of
relativistic stars. The latter is of particular importance because
with the first direct detections of gravitational waves by the LIGO
and VIRGO collaborations~\cite{LIGO_first_direct_GW,LIGOseconddirect}
the era of gravitational wave astronomy has arrived.

There exist several independent numerical codes for obtaining accurate
models of rotating neutron stars in full general relativity, including
two that are publicly available.  The uncertainty in the high-density
equation of state still allows numerically constructed maximum mass
models to differ by more than 50\% in mass, radius and angular
velocity, and by a larger factor in the moment of inertia. Given these
uncertainties, an absolute upper limit on the rotation of relativistic
stars can be obtained by imposing causality as the only requirement on
the equation of state.  It then follows that gravitationally bound
stars cannot rotate faster than 0.41~ms.

In rotating stars, nonaxisymmetric oscillations have been studied in
various approximations (the Newtonian limit and the post-Newtonian 
approximation, the slow rotation limit, the Cowling approximation, the 
spatial conformal flatness approximation) as an eigenvalue problem.
Normal modes in full general relativity have been obtained through 
numerical simulations only. Time evolutions of the
linearized equations have improved our understanding of the
spectrum of axial and hybrid modes in relativistic stars.

Nonaxisymmetric instabilities in rotating stars can be driven by the
emission of gravitational waves [Chandrasekhar-Friedman-Schutz (CFS)
  instability] or by viscosity.  Relativity strengthens the former,
but weakens the latter.  Nascent neutron stars can be subject to the
$l=2$ bar mode CFS instability, which would turn them into a strong
gravitational wave source. Axial fluid modes in rotating stars
($r$-modes) have received considerable attention since the discovery
that they are generically unstable to the emission of gravitational
waves. The $r$-mode instability could slow down newly-born
relativistic stars and limit their spin during accretion-induced
spin-up, which would explain the absence of millisecond pulsars with
rotational periods less than $\sim$~1.5~ms. Gravitational waves from
the $r$-mode instability could become detectable if the amplitude of
$r$-modes is sufficiently large, however, nonlinear effects seem to
set a small saturation amplitude on long timescales. Nevertheless, if
the signal persists for a long time, even a small amplitude could
become detectable. Highly differentially rotating neutron stars are
also subject to the development of a one-arm ($m=1$) instability, as
well as to the development of a dynamical bar-mode ($m=2$) instability
which both act as emitters of potentially detectable gravitational
waves.

Recent advances in numerical relativity have enabled the long-term
dynamical evolution of rotating stars. Several interesting phenomena,
such as dynamical instabilities, pulsation modes, and neutron star and
black hole formation in rotating collapse have now been studied in
full general relativity. The latest studies include realistic
equations of state and also magnetic fields.

The aim of this article is to present a summary of theoretical and
numerical methods that are used to describe the equilibrium properties
of rotating relativistic stars, their oscillations and dynamical
evolution. The focus is on the most recently available publications in
the field, in order to rapidly communicate new methods and results. At
the end of some sections, the reader is directed to papers that could
not be presented in detail here, or to other review articles.  As new
developments in the field occur, updated versions of this article will
appear. Another review on rotating relativistic stars has appeared by
Gourgoulhon~\cite{Gourgou2010arXiv1003.5015G} while monographs
appeared by Meinel et al.~\cite{Meinel2008rfe..book.....M}, and
Friedman and Stergioulas~\cite{FSbook}.  In several sections, our
Living Review article updates and extends previous
versions~\cite{Stergioulas98,Stergioulas2003} using also abridged
discussions of topics from~\cite{FSbook}.\vskip0.5cm

 {\it Notation and conventions.}  Throughout the article,
gravitational units, where $G = c = 1$ (these units are also referred
to as geometrized), will be adopted in writing the equations governing
stellar structure and dynamics, while numerical properties of stellar
models will be listed in cgs units, unless otherwise noted. We use the
conventions of Misner, Thorne, and Wheeler (1973) for the signature of
the spacetime metric $(-+++)$ and for signs of the curvature tensor
and its contractions.  Spacetime indices will be denoted by Greek
letters, $\alpha$, $\beta, \ldots$, while Latin $a, b, \ldots$
characters will be reserved to denote spatial indices. (Readers
familiar with abstract indices can regard indices early in the
alphabet as abstract, while indices $\mu, \nu, \lambda$ and $i,j,k$
will be concrete, taking values $\mu = 0,1,2,3$, $i=1,2,3$.)
Components of a vector $u^\alpha$ in an orthonormal frame, $\{ {\bf
  e}_{\hat 0}, . .  ., {\bf e}_{\hat 3}\} $, will be written as
$\{u^{\hat 0}, . . ., u^{\hat 3}\}$. Parentheses enclosing a set of
indices indicate symmetrization, while square brackets indicate
anti-symmetrization.

Numbers that rely on physical constants are based on the values
$c = 2.9979  \times10^{10}\ $cm$\ $s$^{-1}$, \enskip
$G = 6.670\times10^{-8} $g$^{-1} $cm$^{3} $s$^{-2}$, \enskip
$\hbar = 1.0545\times10^{-27} $g cm$^{2}$s$^{-1}$, \enskip
baryon mass $m_B = 1.659\times 10^{-24}$g, and
$M_\odot = 1.989\times10^{33}$g $ = 1.477$ km.

\newpage


\section{The Equilibrium Structure of Rotating Relativistic Stars}


\subsection{Assumptions}
\label{assumptions}

A relativistic star can have a complicated structure (such as a solid
crust, magnetic field, possible superfluid interior, possible quark
core, etc.). Depending on which phase in the lifetime of the star one
wants to study, a number of physical effects can be ignored, so that
the description becomes significantly simplified. In the following, we
will take the case of a {\it cold, uniformly rotating relativistic
  star} as a reference case and mention additional assumptions for
other cases, where necessary.

The matter can be modeled as a perfect fluid because observations
of pulsar glitches are consistent with departures from a perfect fluid
equilibrium (due to the presence of a solid crust) of order
$10^{-5}$ (see~\cite{FI92}). The temperature of a cold neutron star
has a negligible affect on its bulk properties and can be assumed to be 0~K,
because its thermal energy
($ \ll 1 \mathrm{\ MeV} \sim 10^{10} \mathrm{\ K} $) is much smaller than
Fermi energies of the interior ($> 60 \mathrm{\ MeV}$). One can then use a
{\it one-parameter}, \emph{barotropic} equation of state (EOS) to describe
the matter:
\begin{equation}
  \varepsilon = \varepsilon(P),
\end{equation}
where $\varepsilon$ is the energy density and $P$ is the pressure. At
birth, a neutron star is expected to be rotating differentially, but
as the neutron star cools, several mechanisms can act to enforce
uniform rotation. Kinematical shear viscosity is acting against
differential rotation on a timescale that has been estimated to
be~\cite{FI76,FI79,Cutler87}
\begin{equation}
  \tau \sim 18
  \left( \frac{\rho}{10^{15}\mathrm{\ g\ cm}^{-3}}\right)^{-5/4} 
  \left(\frac{T}{10^9\mathrm{\ K}}\right)^2
  \left(\frac{R}{10^6\mathrm{\ cm}}\right) \mathrm{\ yr},
\end{equation}
where $\rho$, $T$ and $R$ are the central density, temperature, and
radius of the star. It has also been suggested that convective and
turbulent motions may enforce uniform rotation on a timescale of the
order of days~\cite{Hegyi77}. Shapiro~\cite{Shapiro01}
suggested that magnetic braking of differential rotation by Alfv{\'e}n
waves could be the most effective damping mechanism, acting on short
timescales, possibly of the order of minutes.  
 
Within a short time after its formation, the temperature of a
neutron star becomes less than $10^{10} \mathrm{\ K}$ (due to neutrino
emission). When the temperature drops further, below roughly $10^{9} \mathrm{\ K}$, 
its outer core is expected to become superfluid 
(see~\cite{Me98} and references
therein). Rotation causes superfluid neutrons to form an array of
quantized vortices, with an intervortex spacing of
\begin{equation}
  d_{\mathrm{n}} \sim 3.4 \times 10^{-3} \Omega_2^{-1/2} \mathrm{\ cm},
\end{equation}
where $\Omega_2$ is the angular velocity of the star in
$10^2 \mathrm{\ s}^{-1}$. On scales much larger than the intervortex
spacing, e.g., of the order of centimeters or meters, the fluid 
motions can be averaged and the rotation can be considered to be
uniform~\cite{So87}. With such an assumption, the error in computing
the metric is of order
\begin{equation}
  \left( \frac{1 \mathrm{\ cm}}{R} \right)^2 \sim 10^{-12}, 
\end{equation}
assuming $R\sim 10 \mathrm{\ km}$ to be a typical neutron star radius. 

The above arguments show that the bulk properties of a cold, isolated
rotating relativistic star can be modeled accurately by a uniformly
rotating, one-parameter perfect fluid. Effects of differential
rotation and of finite temperature need only be considered during the
first year (or less) after the formation of a relativistic star. Furthermore,
magnetic fields, while important for high-energy phenomena in the 
magnetosphere and for the damping of differential rotation and oscillations, do
not alter the structure of the star, unless one assumes magnetic field
strengths significantly higher than typical observed values.


\subsection{Geometry of spacetime}

In general relativity, the spacetime geometry of a rotating star in
equilibrium can be described by a stationary and axisymmetric metric
$g_{\a\b}$ of the form
\begin{equation}
  ds^2 = -e^{2 \nu} dt^2 + e^{2 \psi} (d \phi - \omega dt)^2 + e^{2 \mu}
  (dr^2+r^2 d \theta^2),
  \label{e:metric}
\end{equation}
where $\nu$, $\psi$, $\omega$ and $\mu $ are four metric functions
that depend on the coordinates $r$ and $\theta$ only (see e.g.,
Bardeen and Wagoner~\cite{Bardeen71}). For a discussion and historical
overview of other coordinate choices for axisymmetric rotating spacetimes see \cite{Gourgou2010arXiv1003.5015G,FSbook}. 
In
the exterior vacuum, it is possible to reduce the number of metric
functions to three, but as long as one is interested in describing the
whole spacetime (including the source-region of nonzero pressure),
four different metric functions are required. It is convenient to
write $e^\psi$ in the form
\begin{equation}
  e^\psi=r \sin \theta B e^{-\nu},
\end{equation}
where $B$ is again a function of $r$ and $\theta$ only~\cite{Bardeen73}.

One arrives at the above form of the metric assuming that

\begin{enumerate}

\item The spacetime is {\it stationary} and {\it axisymmetric}:
There exist an asymptotically timelike symmetry vector $t^\alpha$
and a rotational symmetry vector $\phi^\alpha$.

The spacetime is said to be {\it strictly} stationary if $t^\alpha$ is
everywhere timelike. (Some rapidly rotating stellar models, as well as
black-hole spacetimes, have {\it ergospheres}, regions in which
$t^\alpha$ is spacelike.)

\item  The Killing vectors commute, 
\be
        [t, \phi]=0,
\label{eq:commute}\ee
and there is an isometry of the spacetime that simultaneously
reverses the direction of $t^\alpha$ and $\phi^\alpha$,
\be
 t^\alpha\rightarrow -t^\alpha,\quad \phi^\alpha \rightarrow - \phi^\alpha.
\label{eq:isometry} \ee
\item
The spacetime is asymptotically flat, i.e., $t_at^a=-1$,
$\phi_a\phi^a=+\infty $ and $t_a\phi^a=0$ at spatial
infinity
\item
The spacetime is circular (there are no meridional currents in the fluid).
\end{enumerate}

If the spacetime is strictly stationary, one does not need
(\ref{eq:commute}) as a separate assumption: A theorem by
Carter \cite{Carter70} shows that $[t, \phi]=0$.
The Frobenius theorem now implies the existence of scalars
$t$ and $\phi$ \cite{Kundt66,Carter69} for which there exists 
a family of 2-surfaces orthogonal to $t^\alpha$ and
$\phi^\alpha$, the surfaces of constant $t$ and $\phi$; and 
it is natural to choose as coordinates $x^0=t$ and $x^3=\phi $.
In the absence of meridional currents, the
2-surfaces orthogonal to $t^\a$ and $\phi^\a$ can be described by the
remaining two coordinates $x^1$ and $x^2$ ~\cite{Carter70}.
Requiring that these are Lie derived by $t^\alpha$
and $\phi^\alpha$, we have 
\begin{eqnarray}
t^\alpha &=&{\bm\partial}_t, \\
\phi^\alpha &=&{\bm\partial}_\phi.
\end{eqnarray}
With coordinates chosen in this way, the metric components are
independent of $t$ and $\phi$.

Because time reversal inverts the direction of rotation, the 
fluid is not invariant under $t\rightarrow -t$ alone, implying that 
$t^\alpha$ and $\phi^\alpha$ are
not orthogonal to each other. The lack of orthogonality is measured by the 
metric function $\omega$ that describes the {\it dragging of
inertial frames}.
  
In a fluid with meridional convective currents one loses both
time-reversal invariance and invariance under the simultaneous
inversion $t\rightarrow -t, \phi \rightarrow -\phi$, because the
inversion changes the direction of the circulation. In this case, the
spacetime metric will have additional off-diagonal components
\cite{GrougPhysRevD.48.2635,BirklPhysRevD.84.023003}.

A common choice for $x^1$ and $x^2$ are \emph{quasi-isotropic
  coordinates}, for which $g_{r\theta}=0$ and $g_{\theta \theta }=r^2
g_{rr}$ (in spherical polar coordinates), or $g_{\varpi z}=0$ and
$g_{zz}=r^2 g_{\varpi \varpi}$ (in cylindrical coordinates).  In the
nonrotating limit, the metric~(\ref{e:metric}) reduces to the metric
of a nonrotating relativistic star in \emph{isotropic coordinates}
(see~\cite{Weinberg72} for the definition of these coordinates). In
the slow rotation formalism by Hartle~\cite{H67}, a different form of
the metric is used, requiring $g_{\theta \theta}=g_{\phi \phi }/
\sin^2 \theta$, which corresponds to the choice of Schwarzschild
coordinates in the vacuum region.

The three metric functions $\nu$, $\psi$ and $\omega$ can be written as
invariant combinations of the two Killing vectors $t^\a$ and $\phi^\a$,
through the relations
\begin{eqnarray}
t_\alpha t^\alpha &=& g_{tt}= -e^{2\nu} + \omega^2 e^{2\psi}, \label{eq:killnorms1}\\
\phi_\alpha\phi^\alpha &=& g_{\phi \phi}= e^{2\psi}, \label{eq:killnorms2}\\
t_\alpha\phi^\alpha &=& g_{t\phi}= -\omega e^{2\psi}, \label{eq:killnorms3}
\end{eqnarray}
The corresponding components of the contravariant metric are 
\ba
g^{tt} &=& \nabla_\alpha t \nabla^\alpha t = -e^{-2\nu},\\
g^{\phi\phi}&=& \nabla_\alpha \phi \nabla^\alpha\phi 
        =  e^{-2\psi}-\omega^2 e^{-2\nu},\\
g^{t\phi}&=& \nabla_\alpha t \nabla^\alpha\phi = -\omega e^{-2\nu}.
\ea
The fourth metric function $\mu$ determines the conformal factor
$e^{2\mu}$ that characterizes the geometry of the orthogonal 2-surfaces.

There are two main effects that distinguish a rotating relativistic
star from its nonrotating counterpart: The shape of the star is
flattened by centrifugal forces (an effect that first appears at
second order in the rotation rate), and the local inertial frames are
dragged by the rotation of the source of the gravitational field.
While the former effect is also present in the Newtonian limit, the
latter is a purely relativistic effect. 

The study of the dragging of inertial frames in the spacetime of a
rotating star is assisted by the introduction of the local
Zero-Angular-Momentum-Observers
(ZAMO)~\cite{Bardeen70,Bardeen73}. These are observers whose
worldlines are normal to the $t=\mathrm{\ const.}$ hypersurfaces (also
called \emph{Eulerian} or \emph{normal} observers in the 3+1
formalism~\cite{ADM2008}). Then, the metric function $\omega$ is the
angular velocity $d\phi/dt$ of the local ZAMO with respect to an
observer at rest at infinity. Also, $e^{-\nu}$ is the time dilation
factor between the proper time of the local ZAMO and coordinate time
$t$ (proper time at infinity) along a radial coordinate line. The
metric function $\psi$ has a geometrical meaning: $e^\psi $ is the
\emph{proper circumferential radius} of a circle around the axis of
symmetry.

In rapidly rotating models, an \emph{ergosphere} can appear, where
$g_{tt}>0$ (as long as we are using the Killing coordinates described
above). In this region, the rotational frame-dragging is strong enough
to prohibit counter-rotating time-like or null geodesics to exist, and
particles can have negative energy with respect to a stationary
observer at infinity. Radiation fields (scalar, electromagnetic, or
gravitational) can become unstable in the
ergosphere~\cite{Friedman78}, but the associated growth time is
comparable to the age of the universe~\cite{Comins78}.

The lowest-order asymptotic behaviour of the metric functions $\nu$ and $\omega$ is
\begin{eqnarray}
    \nu &\sim& -{M \over r},
  \label{nuatr} \\
    \omega &\sim& {2J \over r^3},
\end{eqnarray}
where $M$ and $J$ are the total gravitational mass and angular momentum 
(see Section~\ref{RotEquil} for definitions). The asymptotic expansion of
the dragging potential $\omega$ shows that it decays rapidly far from
the star, so that its effect will be significant mainly in the
vicinity of the star.


\subsection{The rotating fluid}

When sources of non-isotropic stresses (such as a magnetic field or a
solid state of parts of the star), viscous stresses, and heat transport
are neglected in constructing an equilibrium model of a relativistic
star, then the matter can be modeled as a perfect fluid, described by
the stress-energy tensor
\begin{equation}
  T^{\a\b} = (\varepsilon+P)u^\a u^\b + P g^{\a\b},
\end{equation}
where $u^\a$ is the fluid's 4-velocity. In terms of the two Killing
vectors $t^\a$ and $\phi^\a$, the 4-velocity can be written as
\begin{equation}
  u^\a = \frac{e^{-\nu}}{\sqrt{1-v^2}} (t^\a + \Omega \phi^\a),
\end{equation}
where $v$ is the 3-velocity of the fluid with respect to a local ZAMO,
given by
\begin{equation}
  v = (\Omega-\omega)e^{\psi-\nu},
\end{equation}
and $\Omega\equiv u^\phi / u^t=d\phi / dt$ is the angular velocity of
the fluid in the \emph{coordinate frame}, which is equivalent to the
{\it angular velocity of the fluid as seen by an  observer at rest at
infinity}. Stationary configurations can be differentially rotating,
while uniform rotation ($\Omega= \mathrm{\ const.}$) is a special case (see
Section~\ref{RotEquil}).

The covariant components of the 4-velocity take
the form
\ba
u_t &=& -\frac{e^\nu}{\sqrt{1-v^2}}(1+e^{\psi-\nu}\omega v ), \qquad
u_{\phi} = \frac{e^\psi v}{\sqrt{1-v^2}}.
\label{ultphi} 
\ea
Notice that the components of the 4-velocity are proportional to the
{\it Lorentz factor} $W:=(1-v^2)^{-1/2}$.


\subsection{Equations of structure}

Having specified an equation of state of the form
$\varepsilon = \varepsilon (P)$, the structure of the star is
determined by solving four components of Einstein's gravitational
field equation
\begin{equation}
  R_{\a\b} = 8 \pi \left (T_{\a\b}- \frac{1}{2} g_{\a\b}T \right),
\end{equation}
(where $R_{\a\b}$ is the Ricci tensor and $T=T_\a{}^\a$) and the equation
of hydrostationary equilibrium. Setting $\zeta = \mu + \nu$, one
common choice \cite{BI76} 
for the components of the gravitational field equation are the three
equations of elliptic type
\begin{eqnarray}
  \bm \nabla \cdot (B \bm\nabla \nu) &=&  \frac{1}{2}
  r^2\sin^2\theta B^3e^{-4\nu} \bm\nabla\omega \cdot \bm\nabla\omega \nonumber \\ 
  && + 4 \pi B e^{2\zeta -2\nu}
  \left[\frac{(\varepsilon+P)(1+v^2)}{1-v^2} +2P \right], \\
  \bm\nabla\cdot (r^2\sin^2\theta B^3e^{-4\nu}\bm\nabla\omega) &=&
  -16 \pi r \sin \theta B^2e^{2\zeta -4 \nu}
  \frac{(\varepsilon +P)v}{1-v^2}, \\
  \bm\nabla\cdot (r \sin \theta \bm\nabla B) &=&
  16 \pi r \sin \theta Be^{2\zeta -2\nu}P,
\end{eqnarray}%
supplemented by a first order differential equation for $\zeta$
\begin{eqnarray}  
\frac1\varpi\zeta,_\varpi + \frac1B(B,_\varpi\zeta,_\varpi -B,_z\zeta,_z) 
 &= &\frac1{2\varpi^2B}(\varpi^2B,_\varpi),_\varpi -\frac1{2B}B,_{zz}+(\nu,_\varpi)^2
\nonumber\\
&&
-(\nu,_z)^2 -\frac14\varpi^2B^2e^{-4\nu}\left[(\omega,_\varpi)^2-(\omega,_z)^2\right]. 
\nonumber\\
&& 
\label{eq:zeta}\end{eqnarray}  
Above, $\bm\nabla$ is the 3-dimensional derivative
operator in a flat 3-space with spherical polar coordinates $r$,
$\theta$, $\phi$. The remaining nonzero
components of the gravitational field equation yield two more
elliptic equations and one first order partial differential equation,
which are consistent with the above set of four equations.

The equation of motion ({\it Euler equation}) follows from the
projection of the conservation of the stress-energy tensor orthogonal to
the 4-velocity $(\delta^\g{}_\b+u^\gamma u_\b)\nabla_\a T^{\a\b}=0$
\ba
\frac{\nabla_\a p}{(\epsilon+p)} &=& -u^\beta \nabla_\beta u_\a \nonumber \\ 
  &=& \nabla_\a \ln u^t 
 -u^tu_\phi\nabla_\a \Omega. \label{EulerEq1}
\ea
In the $r-\theta$ subspace, one can find the following equivalent forms
\ba
\frac{\nabla  p}{(\epsilon+p)}  
&=& -\frac1{1-v^2} \left(\nabla\nu
         -v^2\nabla\psi +e^{\psi-\nu} v \nabla\omega\right), \\
&=& \nabla \ln u^t - u^t u_\phi \nabla \Omega, 
\label{eq:EulerEq1a}\\
&=&  \nabla \ln u^t - \frac{l}{1-\Omega l} \nabla \Omega, \\
&=& -\nabla \ln (-u_t) + \frac{\Omega}{1-\Omega l} \nabla l, \\
&=& -\nabla \nu + \frac{1}{1-v^2} \left( v \nabla v - \frac{v^2\nabla \Omega}
{\Omega-\omega} \right), 
\label{EulerEq5}
\end{eqnarray}
where $l:=-u_\phi/u_t$ is conserved along fluid trajectories (since $h u_t$
and $h u_\phi$ are conserved, so is their ratio and $l$ is the {\it angular
momentum per unit energy}).

For barotropes, one can arrive at a first integral of the equations of motion
in the following way. Since 
$\epsilon=\epsilon(p)$, one can define a function
\be
H(p) := \int_0^p \frac{dp'}{\epsilon(p')+p'},
\ee
so that (\ref{EulerEq1}) becomes
\be
\nabla (H-\ln u^t) = -F \nabla  \Omega,
\label{PW0}
\ee
where we have set $F:=u^t u_\phi=l/(1-l\Omega)$. For {\it homentropic stars} 
(stars with a homogeneous entropy distribution) one obtains
$H=\ln h$ (where $h$ is the {\it specific enthalpy}) 
and the equation of hydrostationary equilibrium takes the form 
\be
\nabla \left(\ln \frac h{u^t} \right) = -F \nabla  \Omega.
\label{eq:hequil}
\ee

Taking the curl of (\ref{PW0}) one finds that either 
\be
\Omega={\rm constant},
\ee
({\it uniform rotation}), or 
\be
F=F(\Omega),
\label{Fomega}
\ee
in the case of {\it differential rotation}. In the latter case,
(\ref{PW0}) becomes
\be
H-\ln u^t+\int^\Omega_{\Omega_{\rm pole}} F(\Omega')d\Omega' =\nu|_{\rm pole},
\label{PW2}
\ee
where the lower limit, $\Omega_0$ is chosen as the value of $\Omega$
at the pole, where $H$ and $v$ vanish.  The above {\it global} first
integral of the hydrostationary equilibrium equations is useful in
constructing numerical models of rotating stars.

For a uniformly rotating star, (\ref{PW2}) can be written as
\be
H-\ln u^t = \nu|_{\rm pole},
\ee
which, in the case of a homentropic star, becomes
\be 
        \frac{h}{u^t} = {\cal E},
\label{calE}
\ee 
with ${\cal E}=\left.e^\nu\right.|_{\rm pole}$ constant over the star
($\cal E$ has the meaning of the {\it injection energy}~\cite{FSbook},
the increase in a star's mass when a unit mass of baryons is injected
at a point in the star).

In the Newtonian limit $\displaystyle e^\psi = \varpi +
O(\lambda^2), \quad e^\nu = 1+O(\lambda^2)$, so to Newtonian
order we have
\be 
u^t u_\phi = v\varpi = \varpi^2\Omega, 
\ee
and the functional dependence of $\Omega$ implied by
Eq.~(\ref{Fomega}) becomes the familiar requirement that, for a
barotropic equation of state, $\Omega$ be stratified on cylinders,
\be
\Omega = \Omega(\varpi),
\ee 
where $\varpi=r\sin(\theta)$. The Newtonian limit of the integral of
motion (\ref{PW2}) is
\be
   h_{\rm Newtonian} -\frac12 v^2 + \Phi = \mbox{constant} 
\label{eq:eulernewt},\ee  
where, in the Newtonian limit, $h_{\rm Newtonian}=h-1$ differs from
the relativistic definition by the rest mass per unit rest mass.


\subsection{Rotation law and equilibrium quantities}
\label{RotEquil}

A special case of rotation law is \emph{uniform rotation} (uniform
angular velocity in the coordinate frame), which minimizes the total
mass-energy of a configuration for a given baryon number and total
angular momentum~\cite{Boyer66,Hartle67}. In this case, the term
involving $F(\Omega)$ in~(\ref{PW2}) vanishes.

More generally, a simple, {\it one-parameter} 
choice of a differential-rotation law is
\begin{equation}
  F(\Omega)= A^2(\Omega_{\mathrm{c}}-\Omega) =
  \frac{(\Omega-\omega)r^2\sin^2\theta~e^{2(\beta-\nu)}}
  {1-(\Omega -\omega)^2r^2\sin^2\theta~e^{2(\beta-\nu)}},
\label{eq:diffrot}
\end{equation}
where $A$ is a constant~\cite{KEH89a,KEH89b}. When $A \to \infty$, the
above rotation law reduces to the uniform rotation case. In the
Newtonian limit and when $A \to 0$, the rotation law becomes a
so-called $j$-constant rotation law (with specific angular momentum $j$, angular momentum per unit mass, being constant in space), which satisfies the Rayleigh
criterion for local dynamical stability against axisymmetric
disturbances ($j$ should not decrease outwards, $dj/d\Omega<0$). The
same criterion is also satisfied in the relativistic case, but with
$j\rightarrow \tilde j=u_\phi(\varepsilon+P)/\rho$~\cite{KEH89b},
where $\rho$ is the fluid rest-mass density. It should be noted that
differentially rotating stars may also be subject to a shear
instability that tends to suppress differential
rotation~\cite{Zahn93}.

The above rotation law is a simple choice that has proven to be
computationally convenient. A new, {\it 3-parameter} generalization of
the above rotation law was recently proposed in \cite{Galeazzi2012} and
is defined by
\begin{equation}
  F(\Omega)= \frac{\frac{R_0^2}{\Omega_c^\alpha}\Omega(\Omega^\alpha-\Omega_c^\alpha)}
  {1-\frac{R_0^2}{\Omega_c^\alpha}\Omega^2(\Omega^\alpha-\Omega_c^\alpha)} 
\label{eq:diffrot2}
\end{equation}
where $\alpha,\ R_0$ and $\Omega_c$ are constants. The specific angular momentum
corresponding to this law is
\begin{equation}
 l= \frac{R_0^2}{\Omega_c^\alpha}\Omega(\Omega^\alpha-\Omega_c^\alpha)
\label{eq:diffrot2b}.
\end{equation}
The Newtonian limit for this law yields an angular frequency of
\begin{equation}
\Omega = \Omega_c\left[1+\left(\frac{\varpi}{R_0}\right)^2\right]^{\frac{1}{\alpha}},
\label{eq:diffrot2c}
\end{equation}
thus, for $\varpi \ll R_0$, $\Omega \sim \Omega_c$, whereas for
$\varpi \gg R_0$, $\Omega_c\sim\Omega(\varpi/R_0)^{2/\alpha}$. A more
recent {\it 4-parameter} family of rotation laws was proposed
in~\cite{Mach2015PhRvD} mainly for accretion tori (it has not yet been
applied to models of rotating neutron stars). It remains to be seen
how well the above laws can match the angular velocity profiles of
proto-neutron stars and remnants of binary neutron star mergers formed
in numerical simulations.

\begin{table}[htbp]
  \caption{Equilibrium properties.}
  \label{tab_equ}
  \renewcommand{\arraystretch}{1.3}
  \centering
    \begin{tabular}{l|l}
      \hline\noalign{\smallskip}
      circumferential radius & $R=e^\psi$ \\
      gravitational mass & $M=-2 \int(T_\a{}^\b-{1 \over 2} \delta_\a^\b T) t^\a \hat n^\b dV$ \\
      baryon mass  & $M_0 = \int \rho u_\b \hat n^\b dV$  \\
      internal energy & $U = \int u u_\b \hat n^\b dV $ \\
      proper mass & $M_{\mathrm{p}} = M_0 +U$ \\
      gravitational binding energy & $W=M-M_{\mathrm{p}}-T$ \\
      angular momentum & $J=\int T_{\a\b} \phi^\a \hat n^\b dV$ \\
      moment of inertia (uniform rotation) & $I=J / \Omega$ \\
      kinetic energy & $T={1 \over 2} J \Omega$ \\
      \noalign{\smallskip}\hline
    \end{tabular}
\end{table}

Equilibrium quantities for rotating stars, such as gravitational mass,
baryon mass, or angular momentum, for example, can be obtained as
integrals over the source of the gravitational field. A list of the
most important equilibrium quantities that can be computed for
axisymmetric models, along with the equations that define them, is
displayed in Table~\ref{tab_equ}. There, $\rho$ is the {\it rest-mass
  density}, $u=\varepsilon-\rho c^2$ is the {\it internal energy
  density}, $\hat n^a= \nabla_at/|\nabla_bt\nabla^bt|^{1/2}$ is the
{\it unit normal vector} to the $t= \mathrm{\ const.}$ spacelike
hypersurfaces, and $dV=\sqrt{|{}^3g|}~d^3x$ is the proper 3-volume
element (with ${}^3g$ being the determinant of the 3-metric of
spacelike hypersurfaces). It should be noted that the moment of
inertia cannot be computed directly as an integral quantity over the
source of the gravitational field. In addition, there exists no unique
generalization of the Newtonian definition of the moment of inertia in
general relativity and $I=J/\Omega$ is a common choice.


\subsection{Equations of state}


\subsubsection{Relativistic polytropes}

Because old neutron-stars have temperatures much
smaller than the Fermi energy of their constituent particles, one
can ignore entropy gradients and assume a uniform specific entropy 
$s$. The increase in 
pressure and density toward the star's center
are therefore adiabatic, if one neglects the slow change in
composition. That is, they are related by the first law of thermodynamics,
with $ds=0$,

\be 
d\epsilon = \frac{\epsilon+p}{\rho} d\rho, 
\label{eq:dedrho}\ee
with $p$ given in terms of $\rho$ by 
\be 
 \frac\rho p \frac{dp}{d\rho}  = \frac{\epsilon+p}{p}\frac{dp}{d \epsilon} 
= \Gamma_1.
\label{eq:dpdrho}
\ee 
Here $\Gamma_1$ is the {\it adiabatic index}, the fractional  
change in pressure per fractional change in comoving volume, at 
constant entropy and composition. In an ideal degenerate Fermi gas, 
in the nonrelativistic and 
ultrarelativistic regimes, $\Gamma_1$ has the constant 
values $5/3$ and $4/3$, respectively. 
Except in the outer crust, neutron-star matter is far from an ideal Fermi 
gas, but models often assume a constant effective adiabatic index, chosen to 
match an average stellar compressibility.  An equation of state of the 
form 
\be 
        p = K \rho^\Gamma,
\label{eq:gamma}
\ee 
with $K$ and $\Gamma$ constants, is called {\it polytropic};
$K$ and $\Gamma$ are the {\it polytropic constant} and {\it polytropic
exponent}, respectively.  
The corresponding relation between $\epsilon$ and $p$ follows from 
(\ref{eq:dedrho})
\be 
        \epsilon = \rho + \frac{p}{\Gamma-1}.
\label{eq:polytrope}\ee
The polytropic exponent $\Gamma$ is commonly 
replaced by a {\em polytropic index} $N$, given by   
\begin{equation}
  \Gamma=1+\frac{1}{N}.
\index{polytrope!polytropic index}\end{equation}

For the above polytropic EOS, the quantity $c^{(\Gamma-2)/(\Gamma-1)}
\sqrt{K^{1/(\Gamma-1)}/G}$ has units of length. In gravitational units
one can thus use $K^{N/2}$ as a fundamental length scale to
define dimensionless quantities. Equilibrium models are then
characterized by the polytropic index $N$ and their dimensionless
central energy density. Equilibrium properties can be scaled to
different dimensional values, using appropriate values for $K$.  For
$N<1.0$ ($N>1.0$) one obtains stiff (soft) models, while for $N\sim 0.5
- 1.0$, one obtains models whose masses and radii are roughly consistent 
with observed neutron-star masses and with the weak constraints on radius 
imposed by present observations and by candidate equations of state.

The definition (\ref{eq:gamma}), (\ref{eq:polytrope}) of the 
relativistic polytropic EOS was introduced by 
Tooper \cite{T65}, to allow a polytropic exponent $\Gamma$
that coincides with the adiabatic index of a relativistic fluid 
with constant entropy per baryon (a homentropic fluid). \index{adiabatic index} 
A different form, $p=K\epsilon^\Gamma$, previously also introduced by Tooper \cite{Tooper64}, 
does not satisfy Eq.~(\ref{eq:dedrho}) and therefore it is not consistent 
with the first law of thermodynamics for a fluid with uniform entropy.


\subsubsection{Hadronic equations of state}

Cold matter below the nuclear saturation density, $\rho_0 =
2.7\times10^{14}$ g/cm$^3$ (or $n_0 = 0.16$ fm$^{-3}$), is thought to
be well understood. A derivation of a sequence of equations of state
at increasing densities, beginning with the semi-empirical mass
formula for nuclei, can be found in \cite{ST83} (see also
\cite{HaenselBook2007}).  Another treatment, using experimental data
on neutron-rich nuclei was given in \cite{HP94}.  In a neutron star,
matter below nuclear density forms a crust, whose outer part is a
lattice of nuclei in a relativistic electron gas.  At $4\times
10^{11}$ g/cm$^3$, the electron Fermi energy is high enough to induce
{\em neutron drip}\index{neutron drip}: Above this density nucleons
begin leaving their nuclei to become free neutrons. The inner crust is
then a two-phase equilibrium of the lattice nuclei and electrons and a
gas of free neutrons.  The emergence of a free-neutron phase means
that the equation of state softens immediately above neutron drip:
Increasing the density leads to an increase in free neutrons and to a
correspondingly smaller increase in pressure.  Melting of the Coulomb
lattice, marking the transition from crust to a liquid core of
neutrons, protons and electrons occurs between 10$^{14}$ g/cm$^3$ and
$\rho_0$.

A review by Heiselberg and Pandharipande \cite{HP00} describes the
partly phenomenological construction of a primarily nonrelativistic
many-body theory that gives the equation of state at and slightly below
nuclear density.  Two-nucleon interactions are matched to
neutron-neutron scattering data and the experimentally determined
structure of the deuteron. Parameters of the three-nucleon interaction
are fixed by matching the observed energy levels of light nuclei.

Above nuclear density, however, the equation of state is still beset
by substantial uncertainties.  For a typical range of current
candidate equations of state, values of the pressure differ by more
than a factor of 5 at $2\rho_0 \sim 5\times10^{14}$ g/cm$^3$, and by
at least that much at higher densities \cite{Haensel03}.  Although
scattering experiments probe the interactions of nucleons (and quarks)
at distances small compared to the radius of a nucleon, the many-body
theory required to deduce the equation of state from fundamental
interactions is poorly understood.  Heavy ion collisions do produce
collections of nucleons at supranuclear densities, but here the
unknown extrapolation is from the high temperature of the experiment
to the low temperature of neutron-star matter.

\index{neutron star!maximum observed mass}\index{maximum mass}
Observations of neutron stars provide a few additional constraints, of
which, two are unambiguous and precise: The equation of state must
allow a mass at least as large as $1.97 M_{\odot}$, the largest
accurately determined mass of a neutron star. (The observation by
Antoniadis et al.~\cite{Antoniadis2013} is of a $2.01 \pm 0.04$
neutron star. There is also an observation by Demorest et
al. \cite{demorest10} of a $1.97 \pm0.04 M_\odot$ neutron star). The
equation of state must also allow a rotational period \index{neutron
  star!  maximum observed spin} at least as small as $1.4$ ms, the
period of the fastest confirmed millisecond pulsar \cite{Hessels2006}.
Observations of neutron star radii are much less precise, but a large number
of observations of type I X-ray bursts or transient X-ray binaries may allow 
for the reconstruction of the neutron star equation of state 
\cite{Ozel2009,Ozel2010,Steiner2010}.

The uncertainty in the equation of state above nuclear density is
dramatically seen in the array of competing alternatives for the nature
of matter in neutron star cores:  Cores that are dominantly neutron
matter may have sharply different equations of state, depending on the
presence or absence of pion or kaon condensates, of hyperons\index{hyperons}, and of
droplets of strange quark matter (described below).  The inner core of the most massive
neutron stars may be entirely strange quark matter. Other differences
in candidate equations of state arise from constructions based on
relativistic and on nonrelativistic many-body theory.  A classic
collection of early proposed EOSs was compiled by Arnett and Bowers
\cite{AB}, 
while reviews of many modern EOSs have been compiled by Haensel
\cite{Haensel03} and Lattimer and Prakash \cite{Lattimer2007}.
Detailed descriptions and tables of several modern EOSs, especially
EOSs with phase transitions, can be found in Glendenning
\cite{Glendenning97}; his treatment is particularly helpful in showing
how one constructs an equation of state from a relativistic field
theory. The Heiselberg-Pandharipande review \cite{HP00}, in contrast,
presents a more phenomenological construction of equations of state
that match experimental data. Detailed theoretical derivations of
equations of state are presented in the book by Haensel, Potekhin and
Yakovlev \cite{HaenselBook2007}. For recent reviews on nuclear EOSs
see~\cite{SagertHempel2010JPhG...37i4064S,Lattimer2012ARNPS..62..485L,FischerHempel2014EPJA...50...46F,LattimerPrakash2016Review,OertelHempel2016arXiv161003361O}.

Candidate EOSs are supplied in the form of an energy density vs.
pressure table and intermediate values are interpolated. This results
in some loss of accuracy because the usual interpolation methods do
not preserve thermodynamic consistency. Swesty \cite{S96} devised a
cubic Hermite interpolation scheme that does preserve thermodynamical
consistency and the scheme has been shown to indeed produce higher-accuracy 
neutron star models \cite{N97}.

\index{equation of state!pion and kaon condensates}
\index{pion condensate}\index{kaon condensate}
High density equations of state with pion condensation were proposed in
 \cite{Mi71,SS73} (see also \cite{K93}).  Beyond nuclear density,
the electron chemical potential could exceed the rest mass of $\pi^-$
(139 MeV) by a margin large enough to overcome a pion-neutron repulsion
and thus allow a condensate of zero-momentum pions.  
The critical density is thought to be 2$\rho_0$ 
or higher, but the uncertainty is greater than a factor of 2; and a
condensate with both $\pi^0$ and $\pi^-$ has also been suggested.
Because the s-wave kaon-neutron interaction is attractive, kaon
condensation may also occur, despite the higher kaon mass, a
possibility suggested in \cite{KN} (for
discussions with differing viewpoints see \cite{BB94,P95,HP00}).  
Pion and kaon condensates lead to significant softening of the equation
of state. 

As initially suggested in \cite{AS60}, when the Fermi energy of the
degenerate neutrons exceeds the mass of a $\Lambda$ or $\Sigma$, weak
interactions convert neutrons to these hyperons\index{hyperons}:
Examples are $2n\leftrightarrow p+\Sigma^-$, $n+p^+\rightarrow
p^++\Lambda$.  Reviews and further references can be found in
\cite{Glendenning97,BG97,P97}, and more recent work, spurred by the
$r$-mode instability (see Sec. \ref{s_axial}), is reported in
\cite{Lindblom01,Haensel02,Lackey2006}.  The critical density above
which hyperons appear is estimated at 2 or 3 times nuclear density.
Above that density, the presence of copious hyperons can significantly
soften the equation of state.  Because a softer core equation of state
can support less mass against collapse, the larger the observed
maximum mass, the less likely that neutron stars have cores with
hyperons (or with pion or kaon condensates).  In particular, a
measured mass of $1.97\pm0.04 M_\odot$ for the pulsar PSR J1614−2230
with a white dwarf companion \cite{demorest10} limits the equation of
state parameter space \cite{rlof09}, ruling out several candidate
equations of states with hyperons \cite{ozel10}.  Whether a hyperon
core is consistent with a mass this large remains an open question
\cite{stone10}.

A new hadron-quark hybrid equation of state was recently introduced by
Beni\'c et al.~\cite{BeniBlaschke2015A&A...577A..40B} (see
also~\cite{Bejger:2016emu} for potential observational signatures of
these objects). The quark matter description is based on a quantum
chromodynamics approach, while the hadronic matter is modeled by means
of a relativistic mean-field method with an excluded volume correction
at supranuclear densities to treat the finite size of the
nucleons. The excluded volume correction in conjunction with the quark
repulsive interactions, result in a first-order phase transition,
which leads to a new family of compact stars in a mass-radius
relationship plot whose masses can exceed the $2M_\odot$ limit that is
set by observations. These new stars are termed ``twin'' stars. The
twin star phenomenon was predicted a long time ago by
Gerlach~\cite{Gerlach1968PhRv..172.1325G} (see
also~\cite{Kampfer1981JPhA...14L.471K,Schertler2000NuPhA.677..463S,Glendenning2000A&A...353L...9G}).
Twin stars consist of a quark core with a shell made of hadrons and a
first-order phase transition at their interface. Recently, rotating
twin star solutions were constructed by Haensel et
al.~\cite{Haensel2016EPJA...52...59H}.


\subsubsection{Strange quark equations of state}
\label{s:strange_eos}

Before a density of about 6$\rho_0$ is reached, lattice QCD
calculations indicate a phase transition from quarks confined to
nucleons (or hyperons) to a collection of free quarks (and gluons).
Heavy ion collisions at CERN and RHIC show evidence of the formation
of such a quark-gluon plasma. A density for the phase transition
higher than that needed for strange quarks in hyperons is similarly
high enough to give a mixture of up, down and strange quarks in quark
matter, and the expected strangeness per unit baryon number is $\simeq
-1$.  If densities become high enough for a phase transition to quark
matter to occur, neutron-star cores may contain a transition region
with a mixed phase of quark droplets in neutron matter
\cite{Glendenning97}.

Bodmer \cite{Bodmer71} and later Witten \cite{Witten84} pointed out
that experimental data do not rule out the possibility that the ground
state of matter at zero pressure and large baryon number is not iron
but strange quark matter.  If this is the case, all ``neutron stars''
may be strange quark stars, a lower density version of the quark-gluon
plasma, again with roughly equal numbers of up, down and strange
quarks, together with electrons to give overall charge neutrality
\cite{Bodmer71,Farhi84}.  The first extensive study of strange quark
star properties is due to Witten \cite{Witten84} (but, see also
\cite{Ipser75,Brecher76}), while hybrid stars that have a mixed-phase
region of quark and hadronic matter, have also been studied
extensively (see, for example, Glendenning's review
\cite{Glendenning97}).

The strange quark matter equation of state can be represented by the
following linear relation between pressure and energy density
\begin{equation}
p=a(\epsilon-\epsilon_0),
\label{MIT}
\end{equation}
where $\epsilon_0$ is the energy density at the surface of a bare
strange star (neglecting a possible thin crust of normal matter).  The
MIT bag model of strange quark matter involves three parameters, the
bag constant, ${\cal B}=\epsilon_0/4$, the mass of the strange quark,
$m_s$, and the QCD coupling constant, $\alpha_c$.  The constant $a$ in
(\ref{MIT}) is equal to $1/3$ if one neglects the mass of the strange
quark, while it takes the value of $a=0.289$ for $m_s=250$ MeV. When
measured in units of ${\cal B}_{60}={\cal B}/(60~{\rm MeV~fm^{-3}})$,
the constant ${\cal B}$ is restricted to be in the range
\begin{equation}
0.9821<{\cal B}_{60}<1.525,
\end{equation}
assuming $m_s=0$.  \index{bag constant}
The lower limit is set by the requirement of
stability of neutrons with respect to a spontaneous fusion into
strangelets, while the upper limit is determined by the energy per
baryon of ${}^{56}$Fe at zero pressure (930.4 MeV).  For other values
of $m_s$ the above limits are modified somewhat (see also 
\cite{Dey98,Gondek00} for other attempts to describe deconfined 
strange quark matter).


\subsection{Numerical schemes}

All available methods for solving the system of equations describing
the equilibrium of rotating relativistic stars are numerical, as no
self-consistent solution for both the interior and exterior
spacetime in an algebraic closed form has been found. The first 
numerical solutions were obtained
by Wilson~\cite{W72} and by Bonazzola and Schneider~\cite{BS74}. 
In the following, we give a description of several available numerical 
methods and their various implementations (codes) and extensions.


\subsubsection{Hartle}

To order ${\cal O}(\Omega^2)$, the structure of a star changes only by
quadrupole terms and the equilibrium equations become a set of
ordinary differential equations. Hartle's~\cite{H67,HT68} method
computes rotating stars in this slow rotation approximation, and a
review of slowly rotating models has been compiled by
Datta~\cite{D88}. Weber et al.~\cite{WG91,WGW91} also implement
Hartle's formalism to explore the rotational properties of four new
EOSs.

Weber and Glendenning~\cite{WG92} improve on Hartle's formalism in
order to obtain a more accurate estimate of the angular velocity at
the mass-shedding limit, but their models still show large
discrepancies compared to corresponding models computed without the
slow rotation approximation~\cite{S94}. Thus, Hartle's formalism is
appropriate for typical pulsar (and most millisecond pulsar)
rotational periods, but it is not the method of choice for computing
models of rapidly rotating relativistic stars near the mass-shedding
limit. An extension of Hartle's scheme to 3rd order was presented by
Benhar et al. \cite{Benhar2005}.


\subsubsection{Butterworth and Ipser (BI)}

The BI scheme~\cite{BI76} solves the four field equations following a
Newton--Raphson-like linearization and iteration procedure. One starts
with a nonrotating model and increases the angular velocity in small
steps, treating a new rotating model as a linear perturbation of the
previously computed rotating model. Each linearized field equation is
discretized and the resulting linear system is solved. The four field
equations and the hydrostationary equilibrium equation are solved
separately and iteratively until convergence is achieved.

Space is truncated at a finite distance from the star and the boundary
conditions there are imposed by expanding the metric potentials in
powers of $1/r$. Angular derivatives are approximated by high-accuracy
formulae and models with density discontinuities are treated specially
at the surface. An equilibrium model is specified by fixing its rest
mass and angular velocity.

The original BI code was used to construct uniform density models and
polytropic models~\cite{BI76,B76}. Friedman et al.~\cite{FIP86,FIP89} (FIP) extend the BI code to obtain a large number of rapidly
rotating models based on a variety of realistic EOSs. Lattimer {\it et
  al.}~\cite{L90} used a code that was also based on the BI scheme to
construct rotating stars using ``exotic'' and schematic EOSs,
including pion or kaon condensation and strange quark matter.


\subsubsection{Komatsu, Eriguchi, and Hachisu (KEH)}

In the KEH scheme~\cite{KEH89a,KEH89b}, the same set of field
equations as in BI is used, but the three elliptic-type field
equations are converted into integral equations using appropriate
Green's functions. The boundary conditions at large distance from the
star are thus incorporated into the integral equations, but the region
of integration is truncated at a finite distance from the star. The
fourth field equation is an ordinary first order differential
equation. The field equations and the equation of hydrostationary
equilibrium are solved iteratively, fixing the maximum energy density
and the ratio of the polar radius to the equatorial radius, until
convergence is achieved. In~\cite{KEH89a,KEH89b,EHN94} the original
KEH code is used to construct uniformly and differentially rotating
stars for both polytropic and realistic EOSs.

Cook, Shapiro, and Teukolsky (CST) improve on the KEH scheme by
introducing a new radial variable that maps the semi-infinite region
$[0,\infty)$ to the closed region $[0,1]$. In this way, the region of
integration is not truncated and the model converges to a higher
accuracy. Details of the code are presented in~\cite{CST92} and
polytropic and realistic models are computed in~\cite{CST94a}
and~\cite{CST94b}.

Stergioulas and Friedman (SF) implement their own KEH code following
the CST scheme. They improve on the accuracy of the code by a special
treatment of the second order radial derivative that appears in the
source term of the first order differential equation for one of the
metric functions. This derivative was introducing a numerical error
of 1\,--\,2\% in the bulk properties of the most rapidly rotating
stars computed in the original implementation of the KEH scheme. The
SF code is presented in~\cite{SF95} and in~\cite{SPHD}. It is
available as a public domain code, named {\tt RNS}, and can be
downloaded from~\cite{RNS}.

A generalized KEH-type numerical code, suitable also for binary
compact objects, was presented by Ury\=u and Tsokaros
\cite{Uryu2012,Uryu2012b}.  The {\tt COCAL} code has been applied to
black hole models, and was recently extended to neutron star models,
either in isolation~\cite{UryuTsokaros2014,UryuTsokaros2016a} or in
binaries~\cite{TsokarosUryu2015}. The extended {\tt COCAL} code allows
for the generation of (quasi)equilribrium, magnetized, and rotating
axisymmetric neutron star models, as well as quasiequilibrium
corotational, irrotational, and spinning neutron star binaries. The
code can also build models of isolated, quasiequilibrium, triaxial
neutron stars~\cite{UryuTsokaros2016a,UryuTsokaros2016b} -- a
generalization of Jacobi ellipsoids in general relativity. Such
configurations were recently studied dynamically in Tsokaros et
al.~\cite{Tsokaros:2017ueb} and were found to be dynamically stable,
though their secular stability still remains an open question.


\subsubsection{Bonazzola \textit{et al.} (BGSM)}

In the BGSM scheme~\cite{BGSM93}, the field equations are derived in
the 3+1 formulation. All four chosen equations that describe the
gravitational field are of elliptic type. This avoids the problem with
the second order radial derivative in the source term of the ODE used
in BI and KEH. The equations are solved using a spectral method, i.e.,
all functions are expanded in terms of trigonometric functions in both
the angular and radial directions and a Fast Fourier Transform (FFT)
is used to obtain coefficients. Outside the star a redefined radial
variable is used, which maps infinity to a finite distance.

In~\cite{S94,SAL94} the code is used to construct a large number of
models based on recent EOSs. The accuracy of the computed models is
estimated using two general relativistic virial identities, valid for
general asymptotically flat spacetimes~\cite{GB94,BG94} (see
Section~\ref{s:virial}).

While the field equations used in the BI and KEH schemes assume a
perfect fluid, isotropic stress-energy tensor, the BGSM formulation
makes no assumption about the isotropy of $T_{ab}$. Thus, the BGSM
code can compute stars with a magnetic field, a solid crust, or a solid
interior, and it can also be used to construct rotating boson stars.


\subsubsection{Lorene/rotstar}

Bonazzola et al.~\cite{BGM98} have improved the BGSM spectral 
method by allowing for several domains of integration. One of the
domain boundaries is chosen to coincide with the surface of the star
and a regularization procedure is introduced for the divergent
derivatives at the surface (that appear in the density field when
stiff equations of state are used). This allows models to be computed
that are nearly free of Gibbs phenomena at the surface. The same
method is also suitable for constructing quasi-stationary models of
binary neutron stars. The new method has been used
in~\cite{Gourgoulhon99} for computing models of rapidly rotating
strange stars and it has also been used in 3D computations of the
onset of the viscosity-driven instability to bar-mode
formation~\cite{Gondek02}. 

The {\tt LORENE} library is available as public domain software 
\cite{LORENE}. It has also been used to construct equilibrium models
of rotating stars as initial data for a fully constraint evolution
scheme in the Dirac gauge and with maximal slicing \cite{LinNovak2006}.


\subsubsection{Ansorg \textit{et al.} (AKM)}

Another multi-domain spectral scheme was introduced in
\cite{Ansorg01,Ansorg03}. The scheme can use several domains inside
the star, one for each possible phase transition in the equation of
state. Surface-adapted
coordinates are used and approximated by a two-dimensional
Chebyshev-expansion.  Transition conditions are satisfied
at the boundary of each domain, and the field and fluid equations are
solved as a free boundary value problem by a full Newton-Raphson method,
starting from an initial guess. The field-equation components are simplified by
using a corotating reference frame. Applying this new method to the
computation of rapidly rotating homogeneous relativistic stars, Ansorg
et al. achieve near machine accuracy, when about 24 expansion 
coefficients are used (see Section \ref{s:compare}).   
For configurations near the mass-shedding limit the relative error
increases to about $10^{-5}$, even with 24 expansion coefficients, 
due to the low differentiability of the solution at the surface. 
The AKM code has been used in systematic studies of uniformly rotating 
homogeneous stars \cite{Schoebel03} and differentially rotating polytropes 
\cite{Ansorg2009}.  A detailed 
description of the numerical method and a review of the results is 
given in \cite{rfoe08}.  
 
A public domain library which implements spectral methods for solving
nonlinear systems of partial differential equations with a Newton-Rapshon
method was presented by Grandcl\'ement \cite{Grandclement2010,KADATH}.
The {\tt KADATH} library could be used to construct equilibrium models
of rotating relativistic stars in a similar manner as in 
\cite{Ansorg01,Ansorg03}.

\subsubsection{IWM-CFC approximation}

The spatial conformal flatness condition (IWM-CFC)
\cite{isenberg7808,wilsonCFC96} is an approximation, in which the
spatial part of the metric is assumed to be conformally
flat. Computationally, one has to solve one equation less than in full
GR, for isolated stars. The accuracy of this approximation has been
tested for uniformly rotating stars by Cook, Shapiro and Teukolsky
\cite{CST96} and it is satisfactory for many applications.
Nonaxisymmetric configurations in the IWM-CFC approximation were
obtained in \cite{Huang2008}. The accuracy of the IWM-CFC
approximation was also tested for initial data of strongly
differentially rotating neutron star models \cite{Iosif2013}.

The conformal flatness approach has been extended to avoid
non-uniqueness issues arising in the solution of the standard CFC
equations by Cordero-Carri{\'o}n et
al.~\cite{Cordero2009PhRvD..79b4017C}. This method has also been
termed the ``extended CFC'' approach~\cite{Bucciantini2011} and has
been applied to the construction of general relativistic
magnetodydrodynamic
equilibria~\cite{Pili2014MNRAS.439.3541P,Pili:2017yxd}.


\subsubsection{The virial identities}
\label{s:virial}

Equilibrium configurations in Newtonian gravity satisfy the well-known
virial relation (assuming a polytropic equation of state)
\begin{equation}
  2T+3(\Gamma -1)U+W=0.
  \label{virial}
\end{equation}
This can be used to check the accuracy of computed numerical
solutions. In general relativity, a different identity, valid for a
stationary and axisymmetric spacetime, was found in~\cite{Bonazzola73}.
More recently, two relativistic virial identities, valid for general
asymptotically flat spacetimes, have been derived by Bonazzola and
Gourgoulhon~\cite{GB94,BG94}. The 3-dimensional virial identity
(GRV3)~\cite{GB94} is the extension of the Newtonian virial
identity~(\ref{virial}) to general relativity. The 2-dimensional
(GRV2)~\cite{BG94} virial identity is the generalization of the
identity found in~\cite{Bonazzola73} (for axisymmetric spacetimes) to
general asymptotically flat spacetimes. In~\cite{BG94}, the Newtonian
limit of GRV2, in axisymmetry, is also derived. Previously, such a
Newtonian identity had only been known for spherical
configurations~\cite{Chandrasekhar39}.

The two virial identities are an important tool for checking the
accuracy of numerical models and have been repeatedly used by several
authors~ (see, e.g. \cite{BGSM93,S94,SAL94,N97,Ansorg01}).


\subsubsection{Direct comparison of numerical codes}
\label{s:compare}

The accuracy of the above numerical codes can be estimated, if one
constructs exactly the same models with different codes and compares
them directly. The first such comparison of rapidly rotating models
constructed with the FIP and SF codes is presented by Stergioulas and
Friedman in~\cite{SF95}. Rapidly rotating models constructed with
several EOSs agree to 0.1\,--\,1.2\% in the masses and radii and
to better than 2\% in any other quantity that was compared (angular
velocity and momentum, central values of metric functions, etc.). This
is a very satisfactory agreement, considering that the BI code was
using relatively few grid points, due to limitations of computing
power at the time of its implementation.

In~\cite{SF95}, it is also shown that a large discrepancy between
certain rapidly rotating models (constructed with the FIP and KEH
codes) that was reported by Eriguchi et al.~\cite{EHN94},
resulted from the fact that Eriguchi et al.\ and FIP used
different versions of a tabulated EOS.

Nozawa et al.~\cite{N97} have completed an extensive direct
comparison of the BGSM, SF, and the original KEH codes, using a large
number of models and equations of state. More than twenty different
quantities for each model are compared and the relative differences
range from $10^{-3}$ to $10^{-4}$ or better, for smooth equations of
state. The agreement is also excellent for soft polytropes. These
checks show that all three codes are correct and successfully compute
the desired models to an accuracy that depends on the number of grid
points used to represent the spacetime.

If one makes the extreme assumption of uniform density, the agreement
is at the level of $10^{-2}$. In the BGSM code this is due to the fact
that the spectral expansion in terms of trigonometric functions cannot
accurately represent functions with discontinuous first order
derivatives at the surface of the star. In the KEH and SF codes, the
three-point finite-difference formulae cannot accurately represent
derivatives across the discontinuous surface of the star.

The accuracy of the three codes is also estimated by the use of the
two virial identities. Overall, the BGSM and SF codes show a better
and more consistent agreement than the KEH code with BGSM or SF. This
is largely due to the fact that the KEH code does not integrate over
the whole spacetime but within a finite region around the star, which
introduces some error in the computed models.

A direct comparison of different codes is also presented by Ansorg et
al.~\cite{Ansorg01}. Their multi-domain spectral code is compared to
the BGSM, KEH, and SF codes for a particular uniform density model of
a rapidly rotating relativistic star. An extension of the detailed
comparison in~\cite{Ansorg01}, which includes results obtained by the
Lorene/rotstar code in~\cite{Gondek02} and by the SF code with higher
resolution than the resolution used in~\cite{N97}, is shown in
Table~\ref{Comparison}. The comparison confirms that the virial
identity GRV3 is a good indicator of the accuracy of each code. For
the particular model in Table~\ref{Comparison}, the AKM code achieves
nearly double-precision accuracy, while the Lorene/rotstar code has a
typical relative accuracy of $2 \times 10^{-4}$ to $7\times 10^{-6}$
in various quantities. The SF code at high resolution comes close to
the accuracy of the Lorene/rotstar code for this model. Lower accuracy
is obtained with the SF, BGSM, and KEH codes at the resolutions used
in~\cite{N97}.

The AKM code converges to machine accuracy when a large number of
about 24 expansion coefficients are used at a high computational cost.
With significantly fewer expansion coefficients (and comparable
computational cost to the SF code at high resolution) the achieved
accuracy is comparable to the accuracy of the Lorene/rotstar and SF
codes. Moreover, at the mass-shedding limit, the accuracy of the
AKM code reduces to about 5 digits (which is still highly accurate, of
course), even with 24 expansion coefficients, due to the nonanalytic
behaviour of the solution at the surface. Nevertheless, the AKM
method represents a great achievement, as it is the first method to
converge to machine accuracy when computing rapidly rotating stars in
general relativity.

\vspace{1 em}\noindent{\bf Going further:}~~
A review of spectral methods in numerical relativity can be found
in~\cite{Grandclement2009}. Pseudo-Newtonian models of axisymmetric, 
rotating relativistic stars are treated in 
\cite{Kim2009}, while a formulation for nonaxisymmetric, uniformly
rotating equilibrium configurations in the second post-Newtonian
approximation is presented in~\cite{AS96}. 
Slowly-rotating models of white dwarfs in general relativity are 
presented in \cite{Boshkayev2013}. The validity of the slow-rotation
approximation is examined in Berti et al. \cite{Berti2005}. 
A minimal-surface scheme due to Neugebauer and Herold was presented in
~\cite{NH92}.
The convergence properties iterative self-consistent-field methods when applied to stellar equilibria are investigated by Price, Markakis and Friedman in \cite{Price2009}. 
\begin{table}[htbp]
  \caption[Detailed comparison of the accuracy of different
      numerical codes in computing a rapidly rotating, uniform density
      model. The absolute value of the relative error in each quantity,
      compared to the AKM code, is shown for the numerical codes
      Lorene/rotstar, SF (at two resolutions), BGSM, and KEH (see
      text). The resolutions for the SF code are (angular $\times$
      radial) grid points.]{Detailed comparison of the accuracy of different
      numerical codes in computing a rapidly rotating, uniform density
      model. The absolute value of the relative error in each quantity,
      compared to the AKM code, is shown for the numerical codes
      Lorene/rotstar, SF (at two resolutions), BGSM, and KEH (see
      text). The resolutions for the SF code are (angular $\times$
      radial) grid points. See~\cite{N97} for the
      definition of the various equilibrium quantities.}
  \label{Comparison}
    \renewcommand{\arraystretch}{1.3}
    \centering
      \begin{tabular}{llccccc}
        \hline\noalign{\smallskip}
        ~ &
        \textbf{AKM} &
        \textbf{Lorene/} &
        \textbf{SF} &
        \textbf{SF} &
        \textbf{BGSM} &
        \textbf{KEH} \\ 
        ~ &
        ~ &
        \textbf{rotstar} &
        {($260 \times 400$)} &
        {($70 \times 200$)} &
        ~ &
        ~ \\
        \noalign{\smallskip}\hline\noalign{\smallskip}
        $\bar{p}_{\mathrm{c}}$ & 1.0 \\
        $r_{\mathrm{p}}/r_{\mathrm{e}}$ & 0.7 & & & & $1 \times 10^{-3}$ & \\
        $\bar{\Omega}$ & 1.41170848318 & $9 \times 10^{-6}$ &
        $3 \times 10^{-4}$ & $ 3 \times 10^{-3}$ & $1 \times 10^{-2}$ &
        $1 \times 10^{-2}$ \\
        $\bar{M}$ & 0.135798178809 & $2 \times 10^{-4}$ &
        $2 \times 10^{-5}$ & $2 \times 10^{-3}$ & $9 \times 10^{-3}$ &
        $2 \times 10^{-2}$ \\
        $\bar{M}_0$ & 0.186338658186 & $2 \times 10^{-4}$ &
        $2 \times 10^{-4}$ & $3 \times 10^{-3}$ & $1 \times 10^{-2}$ &
        $2 \times 10^{-3}$ \\
        $\bar{R}_{\mathrm{circ}}$ & 0.345476187602 & $5 \times 10^{-5}$ &
        $3 \times 10^{-5}$ & $5 \times 10^{-4}$ & $3 \times 10^{-3}$ &
        $1 \times 10^{-3}$ \\
        $\bar{J}$ & 0.0140585992949 & $2 \times 10^{-5}$ &
        $4 \times 10^{-4}$ & $5 \times 10^{-4}$ & $2 \times 10^{-2}$ &
        $2 \times 10^{-2}$ \\
        $Z_{\mathrm{p}}$ & 1.70735395213 & $1 \times 10^{-5}$ &
        $4 \times 10^{-5}$ & $1 \times 10^{-4}$ & $2 \times 10^{-2}$ &
        $6 \times 10^{-2}$ \\
        $Z_{\mathrm{eq}}^{\mathrm{f}}$ & $-$0.162534082217 & $2 \times 10^{-4}$ &
        $2 \times 10^{-3}$ & $2 \times 10^{-2}$ & $4 \times 10^{-2}$ &
        $2 \times 10^{-2}$ \\
        $Z_{\mathrm{eq}}^{\mathrm{b}}$ & 11.3539142587 & $7 \times 10^{-6}$ &
        $7 \times 10^{-5}$ & $1 \times 10^{-3}$ & $8 \times 10^{-2}$ &
        $2 \times 10^{-1}$ \\
        \noalign{\smallskip}\hline\noalign{\smallskip}
        $|\mathrm{GRV3}|$ & $4 \times 10^{-13}$ &
        $3 \times 10^{-6}$ & $3 \times 10^{-5}$ & $1 \times 10^{-3}$ &
        $4 \times 10^{-3}$ & $1 \times 10^{-1}$ \\
        \noalign{\smallskip}\hline
      \end{tabular}
\end{table}


\subsection{Analytic approximations to the exterior spacetime}

The exterior metric of a rapidly rotating neutron star differs
considerably from the Kerr metric. The two metrics agree only to
lowest order in the rotational velocity~\cite{Hartle69}. At higher
order, the multipole moments of the gravitational field created by a
rapidly rotating compact star are different from the multipole moments
of the Kerr field. There have been many attempts in the past to find
analytic solutions to the Einstein equations in the stationary,
axisymmetric case, that could describe a rapidly rotating neutron
star.


In the vacuum region surrounding a stationary and axisymmetric star,
the spacetime only depends on three metric functions (while four
metric functions are needed for the interior). The most general form
of the metric was given by Papapetrou \cite{Papapetrou1953} 
\be
\label{Manko}
ds^2=-f(dt-\omega d\phi)^2+f^{-1}\left\{
e^{2\gamma} (d\tilde\varpi^2+d\tilde z^2)+\tilde\varpi^2d\phi^2
\right\}.
\ee
Here $f$, $\omega$ and $\gamma$ are functions of the
quasi-cylindrical Weyl-Lewis-Papapetrou coordinates $(\tilde
\varpi,~\tilde z)$.  
Starting from this metric, one can write the
vacuum Einstein-Maxwell equations as two equations for two complex
potentials ${\cal E}$ and $\Phi$, 
\index{exact vacuum solutions! Ernst potentials}
following a procedure due to Ernst
\cite{Ernst1968a,Ernst1968b}.  
Once the potentials are known, the metric can be reconstructed.
Sibgatullin (1991) devised a powerful procedure for reducing the
solution of the Ernst equations to simple integral
equations. The exact solutions are generated as a series expansion, in terms
of the physical multipole moments of the spacetime, by choosing the values 
of the Ernst potentials on the symmetry axis.


An interesting exact vacuum solution, given in a closed, algebraic
form, was found by Manko et al. \cite{Manko00,Manko00b}. For
non-magnetized sources of zero net charge, it reduces to a 3-parameter
solution, involving the gravitational mass, $M$, the specific angular
momentum, $a=J/M$, and a third parameter, $b$, related to the
quadrupole moment of the source.  The Ernst potential $\cal E$ on the
symmetry axis is
\begin{equation} 
 e(z)={(z-M-ia)(z+ib)+d-\delta-ab \over
  (z+M-ia)(z+ib)+d-\delta-ab},
\end{equation}
where
\ba
\delta&=&{-M^2b^2\over M^2-(a-b)^2},\\
d&=&{1\over 4}\left[M^2-(a-b)^2\right].  
\ea 
Since $a$ and $b$ are independent parameters, setting $a$
equal to zero does not automatically imply a vanishing quadrupole moment. 
Instead, the nonrotating solution ($a=0$) has a quadrupole
moment equal to
\begin{equation}
Q(a=0)=-{M \over 4} {\left( M^2+b^2 \right)^2 \over \left( M^2-b^2 \right)},
\end{equation}
and there exists no real value of the parameter $b$ for
which the quadrupole moment vanishes for a nonrotating star.  
Hence, the  3-parameter solution by Manko et al. does not reduce
continuously to the Schwarzschild solution as the rotation
vanishes and is not suitable for describing slowly rotating
stars. 

For rapidly rotating models, when the quadrupole 
deformation induced by rotation roughly exceeds the minimum nonvanishing oblate
quadrupole deformation of the solution in the absence of rotation, 
the  3-parameter solution by Manko et al. is still relevant.  
A matching of the vacuum exterior solution to 
numerically-constructed interior solutions of rapidly rotating
stars (by identifying three multipole moments) was presented by
Berti \& Stergioulas in \cite{Berti2004}.  
For a wide range of candidate EOSs, the critical rotation rate
$\Omega_{\rm crit}/\Omega_{\rm K}$ above which the Manko et al. 3-parameter solution is
relevant, ranges from $\sim 0.4$ to $\sim 0.7$ for sequences of models with
$M=1.4M_\odot$, with the lower ratio corresponding to
the stiffest EOS. For the maximum-mass sequence the ratio is $\sim
0.9$, nearly independent of the EOS.  
In \cite{Manko00} the 
quadrupole moment was also used
for matching the exact vacuum solution to numerical interior solutions, but
only along a different solution branch which is not a good approximation to 
rotating stars. 

A more versatile exact exterior vacuum solution found by Manko,
Mart\'in \& Ruiz \cite{Manko1995} involves (in the case of vanishing
charge and magnetic field) four parameters, which can be directly
related to the four lowest-order multipole moments of a source (mass,
angular momentum, quadrupole moment and current octupole moment).  The
advantage of the above solution is that its four parameters are
introduced linearly in the first moment it appears.  For this reason,
one can always match the exact solution to a numerical solution by
identifying the four lowest-order multipole moments. Therefore, the
4-parameter Manko, Mart\'in \& Ruiz \cite{Manko1995} solution is
relevant for studying rotating relativistic stars at any rotation
rate.  Pappas~\cite{Pappas2009} compared the two Manko et
al. solutions to numerical solutions of rapidly rotating relativistic
stars, finding good agreement. In Pappas and Apostolatos
\cite{Pappas2012b} a more detailed comparison is shown, using a
corrected expression for the numerical computation of the quadrupole
moment. Manko and Ruiz~\cite{Manko2016PhRvD..93j4051M} express the
Manko et al. 4-parameter solution explicitly in terms of only three
potentials, and compare the multipole structure of the solution with
physically realistic numerical models of Berti and
Stergioulas~\cite{Berti2004MNRAS.350.1416B}.

Another exact exterior solution (that is related to the 4-parameter
Manko et al. solution) was presented by Pach\'on, Rueda and
Sanabria-G\'omez \cite{Pachon2006} and was applied to relativistic
precession and oscillation frequencies of test particles around
rotating compact stars. Furthermore, an exact vacuum solution
(constructed via B\"acklund transformations), that can be matched to
numerically constructed solutions with an arbitrary number of
constants, was presented by Teichm\"uller, Fr\"ob and Maucher
\cite{Teichmueller2011}, who found very good agreement with numerical
solutions even for a small number of parameters.

A very recent analytic solution for the exterior spacetime is provided
by Pappas~\cite{Pappas2016}.  The metric is constructed by adopting
the Ernst formulation, it is written as an expansion in
Weyl-Papapetrou coordinates and has 3 free parameters -- multipole
moments of the NS. The metric compares favourably with numerically
computed general relativistic neutron star spacetimes. An extension of
the approximate metric to scalar-tensor theories with massless fields
is also provided.

\index{exact vacuum solutions|)}
\index{approximation methods|)}


\subsection{Properties of equilibrium models}


\subsubsection{Bulk properties and sequences of equilibrium models}

Neutron star models constructed with various realistic EOSs have
considerably different bulk properties, due to the large uncertainties
in the equation of state at high densities. Very compressible (soft)
EOSs produce models with small maximum mass, small radius, and large
rotation rate. On the other hand, less compressible (stiff) EOSs
produce models with a large maximum mass, large radius, and low
rotation rate. The sensitivity of the maximum mass to the
compressibility of the neutron-star core is responsible for the
strongest astrophysical constraint on the equation of state of cold
matter above nuclear density.  With the mass measurement of
$1.97\pm0.04 M_\odot$ for PSR J1614−2230 \cite{demorest10} and of
$2.01\pm 0.04$ for PSR J0348+0432 \cite{Antoniadis2013}, several
candidate EOSs that yielded models with maximum masses of nonrotating
stars below this limit are ruled out, but the remaining range of
candidate EOSs is still large, yielding compact objects with
substantially different properties.


Not all properties of the maximum mass models between proposed EOSs
differ considerably, at least not within groups of similar EOSs. For
example, most realistic hadronic EOSs predict a maximum mass model
with a ratio of rotational to gravitational energy $T/|W|$ of
$0.11 \pm 0.02$, a dimensionless angular momentum $cJ/GM^2$ of
$0.64 \pm 0.06$, and an eccentricity of
$0.66 \pm 0.04$~\cite{FI92}. Hence, within the set of realistic
hadronic EOSs, some properties are directly related to the stiffness
of the EOS while other properties are rather insensitive to
stiffness. On the other hand, if one considers strange quark EOSs,
then for the maximum mass model, $T/|W|$ can become more than 60\% 
larger than for hadronic EOSs.

\begin{figure*}[t]
  \center
   \includegraphics[width=0.75\textwidth]{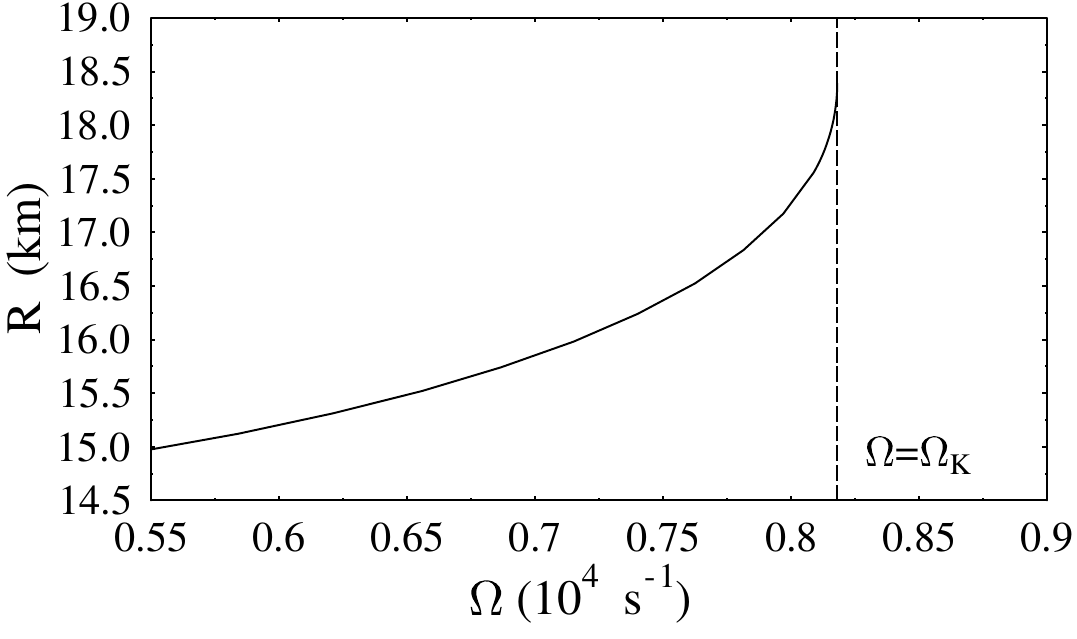}
  \caption{ The radius $R$ of a uniformly rotating star increases
    sharply as the Kepler (mass-shedding) limit ($\Omega=\Omega_K$) is
    approached. The particular sequence of models shown here has a
    constant central energy density of $\epsilon_c = 1.21 \times
    10^{15}$ g cm$^{-3}$ and was constructed with EOS L.  (Image
    reproduced with permission from \cite{SF95}, copyright by AAS.) }
    \label{effectradius}
\end{figure*}

\begin{figure*}[t]
  \center
  \includegraphics[width=0.75\textwidth]{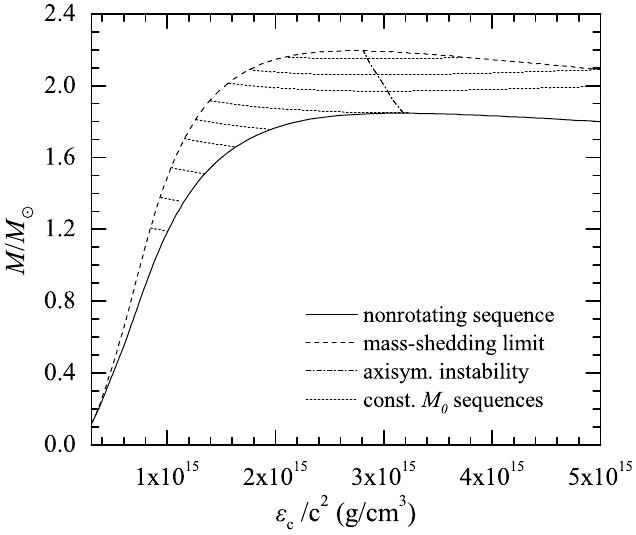}
  \caption{ Representative sequences of rotating stars with fixed
    baryon mass, for EOS WFF3 \cite{Wiringa1988}.  Above a rest mass
    of $M_0=2.17 M_\odot$ only supramassive stars exist, which reach
    the axisymmetric instability limit when spun down. The onset of
    axisymmetric instability approximately coincides with the minima
    of the constant rest mass sequences. Image reproduced with
    permission from~\cite{FSbook}.}
 \label{fig:Mvsec-M0}
\end{figure*}

Compared to nonrotating stars, the effect of rotation is to increase
the equatorial radius of the star and also to increase the mass that
can be sustained at a given central energy density. As a result, the
mass of the maximum-mass rigidly rotating model is roughly
15\,--\,20\% higher than the mass of the maximum mass nonrotating
model~\cite{Morrison2004ApJ}, for typical realistic hadronic EOSs. The
corresponding increase in radius is 30\,--\,40\%.
Fig. \ref{effectradius} shows an example of a sequence of uniformly
rotating equilibrium models with fixed central energy
density\footnote{Following the standard convention, we report
  numerical values of $\epsilon_c$ as $\epsilon_c/c^2$.}, constructed
with EOS L \cite{PS1975,PS1976}.  Near the Kepler (mass-shedding)
limit ($\Omega=\Omega_K$), the radius increases sharply.  This leads
to the appearance of a cusp in the equatorial plane. The effect of
rotation in increasing the mass and radius becomes more pronounced in
the case of strange quark EOSs (see Section \ref{s:strange}).

For a given zero-temperature EOS, the uniformly rotating equilibrium
models form a two-dimensional surface in the three-dimensional space
of central energy density, gravitational mass, and angular
momentum~\cite{SF95}. The surface is limited by the nonrotating models
and by the models rotating at the mass-shedding (Kepler) limit.  Cook
et al.~\cite{CST92,CST94a,CST94b} have shown that the model with
maximum angular velocity does not coincide with the maximum mass
model, but is generally very close to it in central density and
mass. Stergioulas and Friedman~\cite{SF95} showed that the maximum
angular velocity and maximum baryon mass equilibrium models are also
distinct. The distinction becomes significant in the case where the
EOS has a large phase transition near the central density of the
maximum mass model; otherwise the models of maximum mass, baryon mass,
angular velocity, and angular momentum can be considered to coincide
for most purposes.

In the two-dimensional parameter space of uniformly rotating models
one can construct different one-dimensional sequences, depending on
which quantity is held fixed. Examples are sequences of constant
central energy density, constant angular momentum or constant rest
mass.  Fig.  \ref{fig:Mvsec-M0} displays a representative sample of
fixed rest mass sequences for EOS WFF3 \cite{Wiringa1988} in a mass
versus central energy density graph, where the sequence of nonrotating
models and the sequence of models at the mass-shedding limit are also
shown\footnote{Notice that, although this particular EOS does not
  satisfy the current observational constraint of a 2$M_\odot$
  maximum-mass non-rotating model, the qualitative features of all
  sequences of models discussed here are generic for practically all
  EOSs.}. The rest mass of the maximum-mass nonrotating model is $2.17
M_\odot$.  Below this value, all fixed rest mass sequences have a
nonrotating member. Along such a sequence, the gravitational mass
increases somewhat, since it also includes the rotational kinetic
energy. Above $M_0=2.17 M_\odot$ none of the fixed rest mass sequences
have a nonrotating member.  Instead, the sequences terminate at the
{\it axisymmetric instability limit} (see
Sec. \ref{axisyminstability}). The onset of the instability occurs
just prior to the minimum of each fixed rest mass sequence, and models
to the right of the instability line are unstable.

Models with $M_0>2.17 M_\odot$ have masses larger than the
maximum-mass nonrotating model and are called {\it supramassive}
\cite{CST92}. A millisecond pulsar spun up by accretion can become
supramassive, in which case it would subsequently spin down along a
sequence with approximately fixed rest mass, finally reaching the
axisymmetric instability limit and collapsing to a black hole.  Some
relativistic stars could also be born supramassive or become so as the
result of a binary merger; here, however, the star would be initially
differentially rotating, and collapse would be triggered by a
combination of spin-down and by viscosity (or magnetic-field braking)
driving the star to uniform rotation. The {\it maximum mass of
  differentially rotating} supramassive neutron stars can be
significantly larger than in the case of uniform rotation
\cite{Lyford03} and typically 50\% or more than the TOV
limit~\cite{Morrison2004ApJ}. 

A supramassive relativistic star approaching the axisymmetric
instability will actually {\it spin up} before collapse, even though
it loses angular momentum \cite{CST92,CST94a,CST94b}. This
potentially observable effect is independent of the equation of state
and it is more pronounced for rapidly rotating massive
stars. Similarly, stars can be spun up by loss of angular momentum near
the mass-shedding limit, if the equation of state is extremely stiff
or extremely soft.



\subsubsection{Multipole moments}
\label{sec:multipolemoments}
The deformed shape of a rapidly rotating star creates a non-spherical
distortion in the spacetime metric, and in the exterior vacuum region
the metric is determined by a set of multipole moments, which arise at
successively higher powers of $r^{-1}$.  As in electromagnetism, the
asymptotic spacetime is characterized by two sets of multipoles, mass
multipoles and current multipoles, analogs of the electromagnetic
charge multipoles and current multipoles.

The dependence of metric components on the choice of coordinates leads
to the complication that in coordinate choices natural for a rotating
star (including the quasi-isotropic coordinates) the asymptotic form
of the metric includes information about the coordinates as well as
about the multipole structure of the geometry.  Because the metric
potentials $\nu$, $\omega$ and $\psi$ are scalars constructed locally
from the metric and the symmetry vectors $t^\alpha$ and $\phi^\alpha$,
as in Eqs. (\ref{eq:killnorms1} - \ref{eq:killnorms3}), their
definition is in this sense coordinate-independent. But, the
functional forms, $\nu(r,\theta)$, $\omega(r,\theta)$,
$\psi(r,\theta)$, depend on $r$ and $\theta$ and one must disentangle
the physical mass and current moments from the coordinate
contributions.

Up to $O(r^{-3})$, the only contributing multipoles are the monopole
and quadrupole mass moments and the $l=1$ current moment.  Two
approaches to asymptotic multipoles of stationary systems, developed
by Thorne \cite{Thorne80} and by Geroch \cite{geroch70} and Hansen
\cite{hansen74} yield identical definitions for $l\leq 2$, while
higher multipoles differ only in the normalization
chosen. Ryan~\cite{Ryan1995PhRvD..52.5707R} and Laarakkers and
Poisson~\cite{LP97} provide coordinate invariant definitions of
multiple moments.

In the nonrotating limit, the quasi-isotropic metric 
(\ref{e:metric}) takes the isotropic form 
\be
 ds^2 = -\left(\frac{1-M/2r}{1+M/2r}\right)^2 dt^2 
        + \left(1+\frac M{2r}\right)^4(dr^2+ r^2\sin^2\theta d\phi^2 + r^2d\theta^2),
\label{eq:isotropic_schwarzschild}
\ee 
with asymptotic form  
\bea
 ds^2 
 &=& -\left[1-\frac{2M}r+2\frac{M^2}{r^2}-\frac14\frac{M^3}{r^3}+O(r^{-5})\right] dt^2 \nonumber\\
         && + \left[1+\frac{2M}r+\frac32\frac{M^2}{r^2}+O(r^{-3})\right]
                (dr^2+r^2\sin^2\theta d\phi^2 + r^2d\theta^2), \phantom{xxx}
\eea
Thus, the metric potentials $\nu$, $\mu$ and $\psi$ have asymptotic behavior 
\bea 
 \nu  &=&-\frac Mr -\frac1{12}\frac{M^3}{r^3} +O(r^{-5}), \\
 \mu  &=& \frac Mr - \frac14\frac {M^2}{r^2}+\frac1{12}\frac{M^3}{r^3} +O(r^{-4}),\\ 
 \psi  &=& \log(r\sin\theta) +\mu. 
\label{eq:asymp_schwarz} 
\eea  

For a rotating star, the asymptotic metric differs from the
nonrotating form already at $O(r^{-2})$. Through $O(r^{-3})$ there are
three corrections due to rotation: (i) the frame dragging potential
$\dis\omega\sim \frac{2J}{r^3}$; (ii) a quadrupole correction to the
diagonal metric coefficients at $O(r^{-3})$ associated with the mass
quadrupole moment $Q$ of the rotating star; and (iii) {\it
  coordinate-dependent} monopole and quadrupole corrections to the
diagonal metric coefficients (reflecting the asymptotic shape of the
$r$- and $\theta$- surfaces) which can be described by a dimensionless
parameter $a$.

For convenience, one can define a dimensionless qudrupole moment
parameter $q := Q/M^3$. Then, Friedman and Stergioulas \cite{FSbook}
show that the asymptotic form of the metric is given in terms of the
parameters $M$, $J$, $q$ and $a$ by:
\bea
\nu &=& -\frac Mr -\frac1{12}\frac{M^3}{r^3} +\left(a - 4aP_2 - qP_2\right)\frac{M^3}{r^3} 
           +O(r^{-4}),
\label{eq:asymp_nu}\\
\mu &=& \frac Mr - \frac14\frac {M^2}{r^2}+\frac1{12}\frac{M^3}{r^3} -(a-4a P_2)\frac {M^2}{r^2}
        -(a-4a P_2 - qP_2)\frac{M^3}{r^3} \nonumber\\
               && + O(r^{-4}),\\
\psi &=& \log(r\sin\theta) +\mu + O(r^{-4}),
\label{eq:asymp_psi}\\
\omega &=& \frac{2J}{r^3} + O(r^{-4}),
\label{eq:asymp_omega}
\eea 
where $P_2$ is the Legendre polynomial $P_2(\cos\theta)$. The
coefficient of $-P_2/r^3$ in the expansion of the metric potential $\nu$
is thus $Q+4aM^3$, from which the quadrupole moment $Q$ can be
extracted, if the parameter $a$ has been determined from the
coefficient of $P_2/r^2$ in the expansion of the metric potential
$\mu$. Notice that sometimes the coeffient of $-P_2/r^3$ in the
expansion of $\nu$ is identified with $Q$ (instead of $Q+4aM^3$),
which can lead to a deviation of up to about $20\%$ in the numerical
values of the quadrupole moment. Pappas and Apostolatos
\cite{Pappas2012} have independently verified the correctness of the
identification in \cite{FSbook} and also provide the correct
identification of the current-octupole moment.

Laarakkers and Poisson~\cite{LP97} found that along a sequence of
fixed gravitational mass $M$, the quadrupole moment $Q$ scales
quadratically with the angular momentum, as
\begin{equation}
  Q = -a_2 \frac{J^2}{Mc^2} = -a_2 \chi^2 M^3,
\label{Qexp}
\end{equation}
where $a_2$ is a dimensionless coefficient that depends on the
equation of state, and $\chi :=J/M^2$.  In \cite{LP97}, the coefficient
$a_2$ varied between $a \sim 2$ for very soft EOSs and $a \sim 8$ for
very stiff EOSs, for sequences of $M=1.4\,M_{\odot}$, but these values
were computed with the erroneous identification of $Q$ discussed
above.  Pappas and Apostolatos \cite{Pappas2012} verify the simple
form of (\ref{Qexp}) and provide corrected values for the parameter
$a_2$ as well as similar relations for other multipole moments. Pappas
and Apostolatos~\cite{Pappas2014} and Yagi et al.~\cite{YKPYA2014}
have further found that in addition to $Q$, the spin octupole $S_3$
and mass hexadecapole $M_4$ also have scaling relationships for
realistic equations of state as follows
\begin{eqnarray}
 S_3 & = & -\beta_3 \chi^3 M^4,  \\
 M_4 & = & \gamma_4 \chi^4 M^5,
\end{eqnarray}
where $\beta_3$ and $\gamma_4$ are dimensionless constants.


\subsubsection{Mass-shedding limit and the empirical formula}

Mass-shedding occurs when the angular velocity of the star reaches the
angular velocity of a particle in a circular Keplerian orbit at the
equator, i.e., when
\begin{equation}
  \Omega = \Omega_{\mathrm{K}},
\end{equation}
where
\begin{equation}
  \Omega_{\mathrm{K}}=\frac{\omega'}{2\psi'} +
  e^{\nu-\psi}\left[c^2\frac{\nu'}{\psi'}+
  \left(\frac{\omega'}{2\psi'}e^{\psi-\nu}\right)^2\right]^{1/2}
  \!\!\!+\omega,
\end{equation}
(a prime indicates radial differentiation). In differentially rotating
stars, even a small amount of differential rotation can significantly
increase the angular velocity required for mass-shedding. Thus, a
newly-born, hot, differentially rotating neutron star or a massive,
compact object formed in a binary neutron star merger could be
sustained (temporarily) in equilibrium by differential rotation, even
if a uniformly rotating configuration with the same rest mass does not
exist.

In the Newtonian limit, one can use the Roche model to derive 
the maximum angular velocity for uniformly rotating polytropic 
stars, finding $\Omega_K \simeq (2/3)^{3/2} (GM/R^3)^{1/2}$ 
(see \cite{ST83}). An
identical result is obtained in the relativistic Roche model of 
Shapiro and Teukolsky \cite{Shapiro1983}.
 For relativistic stars, the empirical formula
\cite{HZ89,FIP89,Fr89,Haensel95}
\begin{equation}
\Omega_K = 0.67 \sqrt{\frac{G M^{\rm max}_{\rm sph}}{(R^{\rm max}_{\rm sph})^3}},
                 \label{eq:empirical}
\end{equation}
gives the maximum angular velocity in terms of the mass and radius of
the maximum mass {\em nonrotating} (spherical) model with an accuracy of $5 \% -7
\%$, without actually having to construct rotating models. Expressed
in terms of the minimum period $P_{\rm min}=2\pi/\Omega_K$, 
the empirical formula reads
\begin{equation}
P_{\rm min} \simeq 0.82 \left( \frac{M_\odot}{M_{\rm sph}^{\rm max}} \right )^{1/2}
\left ( \frac{R_{\rm sph}^{\rm max}}{10{\rm km}} \right )^{3/2} {\rm ms}.
\end{equation}

The empirical formula results from universal proportionality relations
that exist between the mass and radius of the maximum mass rotating
model and those of the maximum mass nonrotating model for the same
EOS. Lasota et al. \cite{LHA96} found that, for most EOSs, the
numerical coefficient in the empirical formula is an almost linear function of
the parameter
\begin{equation}
\chi_s = \frac{2GM^{\rm max}_{\rm sph}}{R^{\rm max}_{\rm sph} c^2}.
\end{equation}
The Lasota et al. empirical formula
\begin{equation}
\Omega_K = (0.468+0.378 \chi_s) \sqrt{\frac{G M^{\rm max}_{\rm sph}}
            {(R^{\rm max}_{\rm sph})^3}},
                 \label{eq:lasota}
\end{equation}
reproduces the exact values with
a relative error of only $1.5 \%$. The corresponding formula for $P_{\min}$
is
\begin{equation}
P_{\rm min} \simeq \frac{0.187}{(\chi_s)^{3/2}(1+0.808\chi_s)}
\left (\frac{M_\odot}{M_{\rm sph}^{\rm max}} \right ) {\rm ms}.
\end{equation}
The above empirical relations are specifically constructed for the
most rapidly rotating model for a given EOS. 

Lattimer and Prakash \cite{Lattimer2004} suggest the following empirical relation
\begin{equation}
P_{\rm min} \simeq 0.96 \left( \frac{M_\odot}{M} \right )^{1/2}
\left ( \frac{R_{\rm sph}}{10{\rm km}} \right )^{3/2} {\rm ms},
\end{equation}
for any neutron star model with mass $M$ and radius $R_{\rm sph}$ of the nonrotating
model with same mass, as long as its mass is not close to the
maximum mass allowed by the EOS. Haensel et al. \cite{Haensel2009} refine 
the above formula, giving a factor of 0.93 for hadronic EOSs and 0.87 for
strange stars. A corresponding empirical relation between the radius at
maximal rotation and the radius of a nonrotating configuration of same
mass also exists.

Using the above relation, one can set an approximate
constraint on the radius of a nonrotating star with mass $M$, given the minimum
observed rotational period of pulsars.

\subsubsection{Upper limits on mass and rotation: theory vs. observation}

\noindent{\sl Maximum mass.} \index{neutron stars!maximum mass}\index{mass!maximum}
Candidate EOSs for high density matter
predict vastly different maximum masses for nonrotating models. One of 
the stiffest proposed EOSs (EOS L) has a nonrotating maximum mass of 
3.3$M_\odot$. Some core-collapse simulations suggest a bi-modal mass 
distribution of the remnant, with peaks at about $1.3M_\odot$ and 
$1.7M_\odot$ \cite{Timmes96}. 

Observationally, the masses of a large number of compact objects have
been determined, but, in most cases, the observational error bars are
still large. A recent review of masses and spins of neutron stars as
determined by observations was presented by Miller \&
Miller~\cite{MillerMiller2015PhR}. The heaviest neutron stars with the
most accurately determined masses ever observed are PSR J1614−2230,
with $M=1.97\pm0.04 M_\odot$ \cite{demorest10} and PSR J0348+0432,
with $2.01\pm 0.04$ \cite{Antoniadis2013}, and there are indications
for even higher masses (see \cite{HaenselBook2007} for a detailed
account). Masses of compact objects have been measured in different
types of binary systems: double neutron star binaries, neutron
star-white dwarf binaries, X-ray binaries and binaries composed of a
compact object around a main sequence star. For most double neutron
star binaries, masses have already been determined with good precision
and are restricted to a narrow range of about $1.2-1.4 M_\odot$
\cite{Thorsett1999}.  This narrow range of relatively small masses is
probably associated with an upper mass limit on iron cores, which in
turn is related to the stability of the core of each progenitor
star.\index{binary neutron stars} Masses determined for compact stars
in X-ray binaries still have large error bars, but are consistently
higher than $1.4M_\odot$, which is probably the result of
mass-accretion. A similar finding seems to apply to white
dwarf--neutron star binaries (see~\cite{Paschalidis:2009zz} and
references therein).

\vskip0.4cm

\noindent{\sl Minimum period.}
\index{neutron stars!maximum spin} 
When magnetic-field effects are ignored, conservation of angular
momentum can yield very rapidly rotating neutron stars at birth.
Simulations of the rotational core collapse of evolved rotating
progenitors \cite{Heger00,Fryer00} have demonstrated that rotational
core collapse could result in the creation of neutron stars with
rotational periods of the order of 1 ms (and similar initial rotation
periods have been estimated for neutron stars created in the
accretion-induced collapse of a white dwarf \cite{LL01}). However, 
 magnetic fields may complicate this picture. Spruit \&
Phinney \cite{Spruit98} have presented a model in which a strong
internal magnetic field couples the angular velocity between core and
surface during most evolutionary phases. The core rotation decouples
from the rotation of the surface only after central carbon depletion
takes place. Neutron stars born in this way would have very small
initial rotation rates, even smaller than the ones that have been
observed in pulsars associated with supernova remnants. In this model,
an additional mechanism is required to spin up the neutron star to
observed periods. On the other hand, Livio \& Pringle \cite{Livio98}
argue for a much weaker rotational coupling between core and surface
by a magnetic field, allowing for the production of more rapidly
rotating neutron stars than in \cite{Spruit98}. In \cite{Heger03} 
 intermediate initial rotation rates were obtained.  
Clearly, more detailed studies of the role of magnetic fields
are needed to resolve this important question.

Independently of their initial rotation rate, compact stars in binary
systems are spun up by accretion, reaching high rotation rates. In
principle, accretion could drive a compact star to its mass-shedding
limit. For a wide range of candidates for the neutron-star EOS, the
mass-shedding limit sets a minimum period of about 0.5 - 0.9 ms
\cite{Fr95}.  However, there are a number of different processes that
could limit the maximum spin to lower values. In one model, the
minimum rotational period of pulsars could be set by the occurrence of
the $r$-mode instability in accreting neutron stars in LMXBs
\cite{Bildsten98,Andersson00}, during which gravitational waves carry
away angular momentum. Other models are based on the standard
magnetospheric model for accretion-induced spin-up \cite{White97}, or
on the idea that the spin-up torque is balanced by gravitational
radiation produced by an accretion-induced quadrupole deformation of
the deep crust \cite{Bildsten98,Ushomirsky00}, by deformations induced
by a very strong toroidal field \cite{Cutler2002} or by magnetically
confined ``mountains'' \cite{Melatos2005,Vigelius2008}. With the
maximum observed pulsar spin frequency at 716Hz \cite{Hessels2006} and
a few more pulsars at somewhat lower rotation rates
\cite{Chakrabarty2008}, it is likely that one of the above mechanisms
ultimately dominates over the accretion-induced spin-up, setting an
upper limit that may be somewhat dependent on the final mass, the
magnetic field or the spin-up history of the star.
This is consistent with the absence of sub-millisecond pulsars in pulsar 
surveys that were in principle sensitive down to a few tenths of a
millisecond \cite{DA96,Damico00,Crawford00,Edwards01}. 

\vskip0.4cm

\noindent{\sl EOS constraints.}  One can systematize the observational
constraints on the neutron-star EOS by introducing a parameterized EOS
above nuclear density with a set of parameters large enough to
encompass the wide range of candidate EOSs and small enough that the
number of parameters is smaller than the number of relevant
observations.  Read et al. \cite{rlof09}. found that one can match a
representative set of EOSs to within about 3\% rms accuracy with a
4-parameter EOS based on piecewise polytropes.

Using spectral modeling to simultaneously estimate the radius and mass
of a set of neutron stars in transient low-mass x-ray binaries, \"Ozel
et al. \cite{Ozel2010} and Steiner et al. \cite{Steiner2010} find more
stringent constraints.  They also adopt piecewise-polytropic
parametrizations to find the more restricted region of the EOS space.
Future gravitational-wave observations of inspiraling neutron-star
binaries
\cite{FlaHin2008PhRvD..77b1502F,readetal09,Markakis2009,Markakis2010,Duez2010CQGra..27k4106D,Bernuzzi2012PhRvD..86d4030B,Damour2012PhRvD..85l3007D}
and of oscillating, post-merger remnants
\cite{2005PhRvD..71h4021S,Bauswein2012,Bauswein2012PhRvL.108a1101B,Bauswein2014PhysRevD.90.023002,Bauswein2016EPJA...52...56B,Clark2016CQGra}
may yield comparable or more accurate constraints without the
model-dependence of the current electromagnetic studies.

The existence of $2.0M_\odot$ neutron stars in conjunction with
nuclear physics place constraints on the neutron star EOS. For
example, Hebeler et al.~\cite{Hebeler2013ApJ} use microscopic
calculations of neutron matter based on nuclear interactions derived
from chiral effective field theory to constrain the equation of state
of neutron-rich matter at sub- and supranuclear densities, arriving at
a range of $9.7-13.9$ km for the radius of nonrotating neutron stars,
which is somewhat smaller than the range that a large sample of
various proposed EOSs allow (the authors use a piecewise polytropic
approach to derive the constraints). The corresponding range of
compactness is $0.149-0.213$.

A review of efforts to observationally constrain the EOS is given by
Lattimer~\cite{Lattimer2011}. For recent reviews and the most
up-to-date constraints on the neutron star radii and masses from
electromagnetic observations see
Lattimer~\cite{Lattimer2012ARNPS..62..485L} and \"Ozel and
Freire~\cite{Ozel2016ARA&A..54..401O} and references therein.  Future
observations with missions such as NICER~\cite{NICER} and the proposed
LOFT~\cite{LOFT} have the potential to determine the neutron star
radius with $ \sim5-10\%$ uncertainty, which will be useful in placing
stringent (albeit model dependent) constraints on the EOS (see
Psaltis, {\"O}zel and Chakrabarty~\cite{Psaltis2014ApJ...787..136P}).

\subsubsection{Maximum mass set by causality}
\label{Sec:maxmass}
\index{maximum mass}\index{neutron stars!maximum mass}  
\index{rotating stars!maximum mass}

If one is interested in obtaining an upper limit on the mass, 
independent of the current uncertainty in the high-density part
of the EOS for compact stars, one can construct a schematic EOS 
that satisfies only a {\it minimal set of physical constraints} and
which yields a model of absolute maximum mass. The minimal set 
of constraints are

\begin{itemize} 
\item[(0)] {\it A relativistic star is described as a self-gravitating,
uniformly rotating perfect fluid with a one-parameter EOS}, an assumption
that is satisfied to high accuracy by cold neutron stars.

\item[(1)] {\it Matter at high densities satisfies the causality constraint} $c_s\equiv\sqrt{dp/d\epsilon} < 1$, where $c_{\rm s}$ is the sound speed.
Relativistic fluids are governed by hyperbolic equations
whose characteristics lie inside the light cone (consistent  
with the requirement of causality) only if $c_{\rm s}< 1$ 
\cite{GL}.  

\item[(2)] {\it The EOS is known at low densities.} One assumes that 
the EOS describing the crust of cold relativistic stars is accurately known 
up to a matching energy density $\epsilon_m$.
\end{itemize}
 
\index{spherical stars!maximum mass}
For {\it nonrotating stars}, Rhoades and Ruffini \cite{Rhoades1974}
showed that the EOS that
satisfies the above constraints and yields the maximum mass
consists of a high density region at the
causal limit, $dp/d \epsilon=1$ (as stiff as possible), that matches 
directly to the assumed low density EOS at $\epsilon=\epsilon_m$
\begin{eqnarray}
p(\epsilon)&=&{\left\{
\begin{array}{ll} 
p_{\rm crust}(\epsilon)  & \epsilon < \epsilon_m, \\ & \\
p_m+ \epsilon-\epsilon_m & \epsilon > \epsilon_m,
\end{array}
\right.}
\label{eq:maxmasseos}
\end{eqnarray}
where $p_m = p_{\rm crust}(\epsilon_m)$. For this {\it maximum mass EOS} and a 
{\it specific value} of the 
matching density, they computed a maximum mass of $3.2 M_{\odot}$. 
More generally, $M_{\rm max}$
depends on $\epsilon_m$ as \cite{HS77,H78}
\begin{equation}
     M_{\rm max} = 4.8 \ \left ( \frac{2 \times 10^{14} {\rm g/cm}^3} 
                   { \epsilon_m/c^2} \right )^{1/2} M_{\odot}.
\end{equation}

In the case of {\it uniformly rotating stars}, one obtains the 
following limit on the mass, when matching to the FPS EOS at
low densities 
\begin{equation}
     M^{\rm rot}_{\rm max} = 6.1 \ \left ( \frac{2 \times 10^{14} {\rm g/cm}^3} 
                   { \epsilon_m/c^2} \right )^{1/2} M_{\odot},
\label{eq:maxmass}
\end{equation} 
(see \cite{FI87,KSF97}).

\subsubsection{Minimum period set by causality}
 
A rigorous limit on the minimum period of uniformly rotating,
gravitationally bound stars, allowed by causality, has been obtained in \cite{KSF97} 
(hereafter KSF), extending previous results by 
Glendenning \cite{G92}. The same three minimal constraints (0), (1) and (2)
of Section \ref{Sec:maxmass},
as in the case of the maximum mass allowed by causality, yield the
minimum period. However, the {\it minimum period EOS} is different
from the maximum mass EOS (\ref{eq:maxmasseos}). KSF 
found that just the two constraints (0), (1) 
(without matching to a known low-density part) suffice to yield a simpler,
{\it absolute minimum period EOS} and an absolute lower bound on 
the minimum period. 

\vskip0.4cm

\begin{figure*}[t]
  \center
  \includegraphics[width=0.75\textwidth]{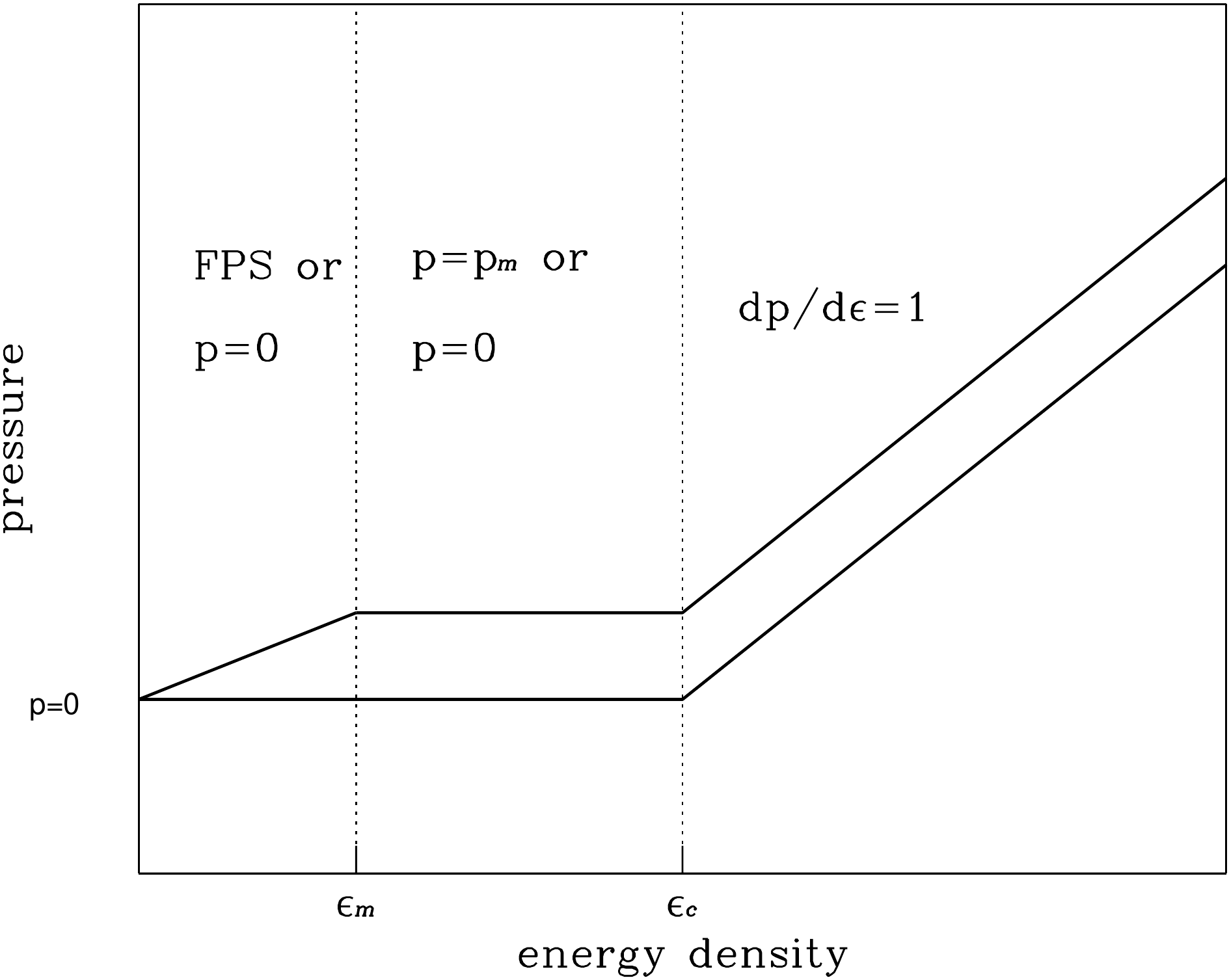}
 \caption{Schematic representations of the minimum-period EOSs
   (\protect\ref{eq:minimaleos1}) and (\protect\ref{eq:minimaleos}).
   For the minimum-period EOS (\protect\ref{eq:minimaleos1}) the
   pressure vanishes for $\epsilon <\epsilon_C$.  The minimum-period
   EOS (\protect\ref{eq:minimaleos}) matches the FPS EOS to a constant
   pressure region at an energy density $\epsilon_m$.  For $\epsilon >
   \epsilon_C$ both EOSs are at the causal limit with
   $dp/d\epsilon=1$. (Image reproduced with permission from
   \cite{KSF97}, copyright by AAS.)}
  \label{Fig:KSF97-f1}
\end{figure*}  

\noindent{\sl Absolute minimum period, without matching to low-density EOS.}
Considering only assumptions (0) and (1), so that the EOS is
constrained only by causality, the minimum period EOS is simply 
\begin{eqnarray}
p(\epsilon)&=&{\left\{
\begin{array}{ll} 
0  & \epsilon \leq \epsilon_C, \\ & \\
\epsilon-\epsilon_C & \epsilon \geq \epsilon_C,
\end{array}
\right.}
\label{eq:minimaleos1}
\end{eqnarray}
describing a star entirely at the causal limit $dp/d\epsilon=1$, with surface 
energy density $\epsilon_C$. This is not too surprising. A
soft EOS yields stellar models with dense central cores and thus 
small rotational periods. Soft EOSs, however, cannot support 
massive stars. This
suggests that the model with minimum period arises from an EOS which is
{\it maximally stiff} ($dp/d\epsilon =1$) at high density, allowing stiff cores to support against
collapse, but {\it maximally soft} at low density ($dp/d\epsilon =0$), allowing small radii and thus fast 
rotation, in agreement with (\ref{eq:minimaleos1}).
The minimum period EOS is depicted in Fig. \ref{Fig:KSF97-f1} and 
yields an absolute lower bound on the period of uniformly rotating stars obeying
the causality constraint, independent of any specific
knowledge about the EOS for the matter composing the star. Choosing 
different values for $\epsilon_C$, one constructs EOSs with
different $M_{\rm sph}^{\rm max}$. All properties of stars constructed with EOS 
(\ref{eq:minimaleos1}) scale
according to their dimensions in gravitational units and thus, 
the following relations hold between different
maximally rotating stars computed from minimum-period EOSs with
different $\epsilon_C$: 
\begin{eqnarray}
P_{\rm min}&\propto &M_{\rm sph}^{\rm max} \propto R_{\rm sph}^{\rm max}, \\
\epsilon_{\rm sph}^{\rm max} &\propto& \frac{1}{\bigl ( M_{\rm sph}^{\rm max} \bigr )^2}, \\
M_{\rm rot}^{\rm max} &\propto& M_{\rm sph}^{\rm max}, \\
R_{\rm rot}^{\rm max} &\propto& R_{\rm sph}^{\rm max}, \\
\epsilon_{\rm rot}^{\rm max} &\propto& \epsilon_{\rm sph}^{\rm max}. 
\label{eq:relations}
\end{eqnarray}

A fit to the numerical results, yields the following
relation for the absolute minimum period
\begin{equation}
\frac{P_{\min}}{\rm ms} = 0.196 \ \left( 
   \frac{M_{\rm sph}^{\rm max}}{{ M}_\odot} \right ).
   \label{pmin3}
\end{equation}  
Thus, for $M_{\rm sph}^{\rm max}=2{M}_\odot$ the absolute minimum 
period is 0.39ms.

\vskip0.4cm

\noindent{\it Minimum period when low-density EOS is known.}
Assuming all three constraints (0), (1) and (2) of Section \ref{Sec:maxmass} 
(so that the EOS
matches to a known EOS at low density), the minimum-period EOS is 
\begin{eqnarray}
p(\epsilon)&=&{\left\{
\begin{array}{ll} 
p_{\rm crust}(\epsilon) & \epsilon\leq \epsilon_m, \\ & \\
p_m  & \epsilon_m\leq \epsilon \leq \epsilon_C, \\ & \\
p_m + \epsilon-\epsilon_C & \epsilon \geq \epsilon_C.
\end{array}
\right.}
\label{eq:minimaleos}
\end{eqnarray}
\begin{figure*}[t]
  \center
  \includegraphics[width=0.75\textwidth]{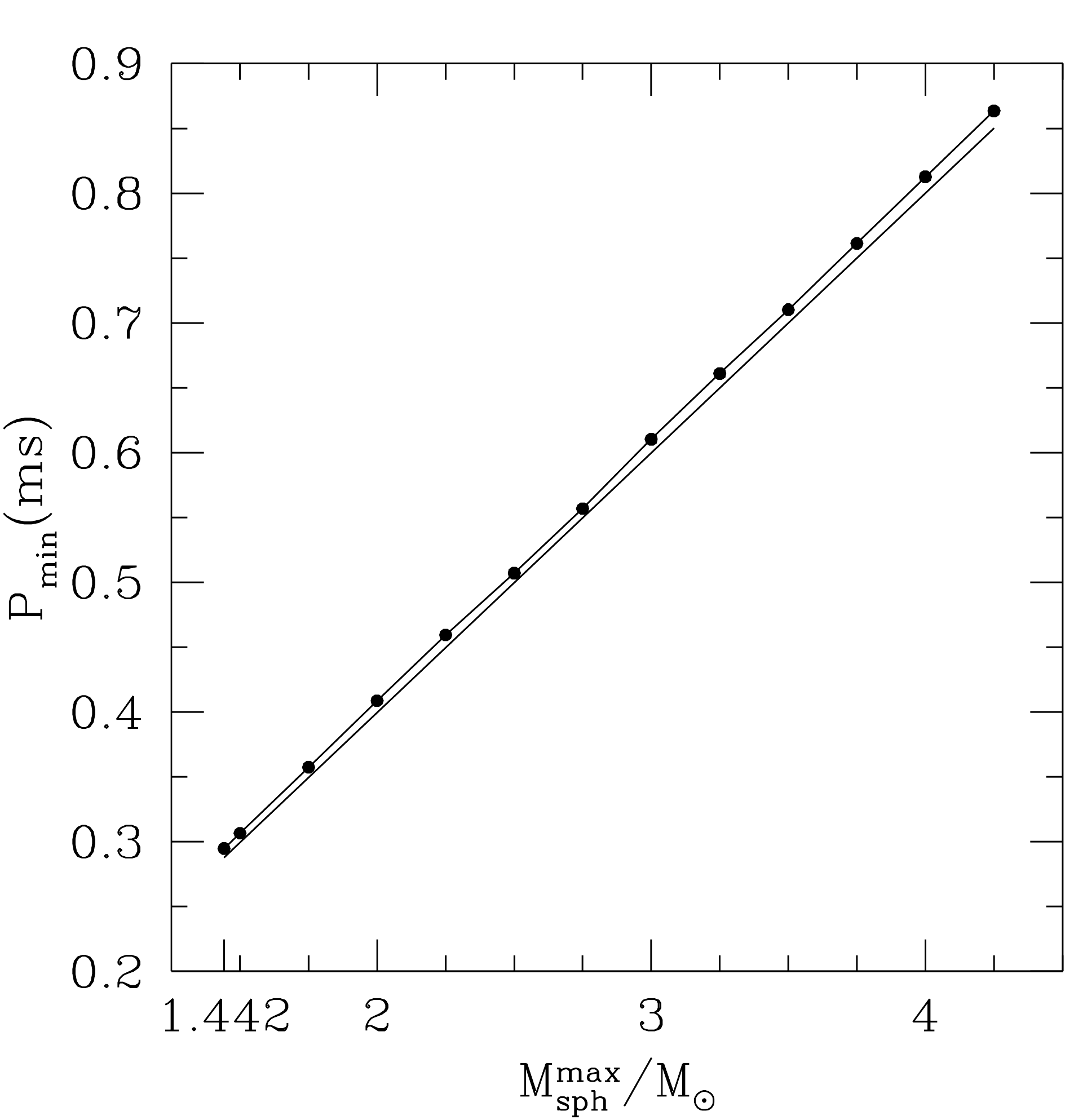}
    \caption{Minimum period $P_{min}$ allowed by causality for
      uniformly rotating, relativistic stars as a function of the mass
      $M_{sph}^{max}$ of the maximum mass nonrotating model. Lower
      curve: constructed using the absolute minimum-period EOS
      (\protect\ref{eq:minimaleos1}), which does not match at low
      densities to a known EOS.  Upper curve: constructed using the
      minimum-period EOS (\protect\ref{eq:minimaleos}), which matches
      at low densities to the FPS EOS. Due to the causality
      constraint, the region below the curves is inaccessible to
      stars.  (Image reproduced with permission from \cite{KSF97},
      copyright by AAS.)}
\label{Fig:Pmin}
\end{figure*}

Between $\epsilon_m$ and $\epsilon_C$ the minimum period EOS has a
constant pressure region (a first order phase transition) and is
maximally soft, while above $\epsilon_{C}$ the EOS is maximally stiff,
see Fig.  \ref{Fig:KSF97-f1}.  For a matching number density of
$n_m=0.1\>{ fm}^{-3}$ to the FPS EOS, the minimum period allowed by
causality is shown as a function of $M^{\rm max}_{\rm sph}$ in
Fig. \ref{Fig:Pmin}.  A quite accurate linear fit of the numerical results is
\begin{equation}
\frac{P_{\min}}{ ms} = 0.295 + 0.203 \ \left( 
   \frac{M_{\rm sph}^{\rm max}}{{ M}_\odot}-1.442 \right ).   
\label{pmin2}
\end{equation}
Thus, if $M_{\rm sph}^{\rm max}=2 M_\odot$, the minimum period is 
$P_{\rm min}=0.41$ms. This result is rather insensitive to $n_m$, 
for $n_m <0.2 fm^{-3}$, but starts to depend significantly on $n_m$
for larger matching densities. 

Comparing (\ref{pmin2}) to (\ref{pmin3}) it is evident that the 
currently trusted part of the nuclear EOS
plays a negligible role in determining the minimum period due to causality.
In addition, since matching to a known low-density EOS {\it raises} $P_{\min}$,
(\ref{pmin3}) represents an {\it absolute minimum period}.

\subsubsection{Moment of inertia and ellipticity}
\index{moment of inertia}\index{rotating stars!moment of inertia}

  The scalar moment of inertia of a neutron star, defined
as the ratio $I=J/\Omega$, has been computed for polytropes 
and for a wide variety of candidate 
equations of state (see, for example, \cite{Stergioulas99,CST94b,CST94a,FIP86}).  
For a given equation of state the maximum value of the moment of inertia 
typically exceeds its maximum value for a spherical star by a factor 
of $1.5 - 1.6$.  For spherical models, Bejger et al.~\cite{bejger05} obtain an
empirical formula for the maximum value of $I$ 
for a given EOS in terms of the maximum mass for that EOS and the 
radius of that maximum-mass configuration,     
\ba
  I_{\rm max,\Omega=0} &\approx& 0.97\times 10^{45}
\left(\frac{M_{\rm max}}{M_\odot} \right) \left(\frac{R_{M_{\rm max}}}{10\,\rm km}\right)^2 
{\rm g\,cm^2}.
\ea
   Neutron-star moments of inertia can in principle be measured by 
observing the periastron advance of a binary pulsar \cite{damour88}.
Because the mass of each star can be found to high 
accuracy, this would allow a simultaneous measurement of two properties of 
a single neutron star \cite{Morrison2004b,lattimerschutz04,bejger05,rlof09}. 
 
The departure of the shape of a rotating neutron star from axisymmetry
can be expressed in terms of its ellipticity
$\varepsilon$, defined in a Newtonian context by  
\be
   \varepsilon := \frac{I_{xx}-I_{yy}}I = \sqrt{\frac{8\pi}{15}} \frac{Q_{22}}I,
\ee
\index{ellipticity!textbf}%
where $I=I_{zz}$ is the moment of inertia about the star's rotation axis
and the $m=2$ part of a neutron star's quadrupole moment is given by 
\be
   Q_{22} := {\rm Re}\int \rho Y_{22} r^2 dV,
\ee
where $Y_{22}$ is the $l=2,m=2$ spherical harmonic.
   
Following Ushomirsky et al. \cite{Ushomirsky00}, Owen \cite{owen05}
finds for the maximum value of a neutron star's ellipticity the
expression \index{neutron star!ellipticity}
\begin{eqnarray}
\varepsilon_{\rm max} &=& 3.3\times 10^{-7} \frac{\sigma_{\rm max}}{10^{-2}}
                \left( \frac{1.4 M_\odot}M\right)^{2.2}
                \left( \frac R{10\,\rm km}\right)^{4.26} \nonumber\\
               &&\times \left[1+0.7\left( \frac M{1.4 M_\odot}\right)
                \left( \frac {10\,\rm km}R\right)\right]^{-1},
\end{eqnarray}
where $\sigma_{\rm max}$ is the breaking strain of the crust, with 
an estimated value of order $10^{-2}$ for crusts below $10^8K$ \cite{CH10}.

\subsubsection{Rotating strange quark stars}
\label{s:strange}
\index{strange quark stars}\index{rotating stars!strange quark stars}
\index{quark stars}

Most rotational properties of strange quark stars differ considerably
from the properties of rotating stars constructed with hadronic EOSs.
First models of rapidly rotating strange quark stars were computed by
Friedman et al. \cite{FIP89} and by Lattimer et al.
\cite{L90}. Nonrotating strange stars obey relations that scale with the
constant ${\cal B}$ in the MIT bag-model of the strange quark matter EOS. In 
\cite{Gourgoulhon99},  scaling relations for the model with
maximum rotation rate were also found.  The maximum angular velocity scales as
\begin{equation}
\Omega_{\rm max}=9.92 \times 10^3 \sqrt{{\cal B}_{60}} \ {\rm s}^{-1},
\end{equation}
while the allowed range of ${\cal B}$ implies an allowed range of $0.513 \ 
{\rm ms}<P_{\rm min}<0.640 \ {\rm ms}$.  The empirical formula
(\ref{eq:empirical}) also holds for rotating strange stars with an
accuracy of better than $2\% $. 
Rotation increases the mass  and 
radius of the maximum mass model by 44\% and 54\%, correspondingly, 
significantly more than for hadronic EOSs.

Accreting strange stars in LMXBs will follow different evolutionary
paths in a mass vs.  central energy density diagram than accreting
hadronic stars \cite{Zdunik01b}. When (and if) strange stars reach the
mass-shedding limit, the ISCO still exists \cite{Stergioulas99} (while
it disappears for most hadronic EOSs). In \cite{Stergioulas99} 
it was shown that the radius and location of the
ISCO for the sequence of mass-shedding models also scales as
${\cal B}^{-1/2}$, while the angular velocity of particles in circular orbit
at the ISCO scales as ${\cal B}^{1/2}$.  Additional scalings with the
constant $a$ in the strange quark EOS~\eqref{MIT} (that were proposed in
\cite{L90}) were found to hold within an accuracy of better than $\sim
9\%$ for the mass-shedding sequence:
\begin{equation}
M \propto a^{1/2}, \ \ \ \ R \propto a^{1/4}, \ \ \ \ \Omega \propto a^{-1/8}.
\end{equation}
In addition, it was found that models at the mass-shedding limit can have $T/|W|$
as large as $0.28$ for $M=1.34 \ M_\odot$.

If strange stars have a solid normal crust, then the density at the
bottom of the crust is the neutron drip density $\epsilon_{\rm ND}\simeq 
4.1 \times 10^{11} \ {\rm g} \ {\rm cm}^{-3}$, as neutrons are absorbed by
strange quark matter. A strong electric field separates the nuclei of
the crust from the quark plasma. In general, the mass of the crust
that a strange star can support is very small, of the order of
$10^{-5}M_\odot$. Rapid rotation increases by a few times the mass of the
crust and the thickness at the equator becomes much larger than the
thickness at the poles \cite{Zdunik01}. The mass $M_{\rm crust}$ and thickness
$t_{\rm crust}$ of the crust can be expanded in powers of the spin
frequency $\nu_3=\nu/(10^3 \ {\rm Hz})$ as 
\begin{eqnarray}
M_{\rm crust} &=& M_{\rm crust,0}(1+0.24 \nu_3^2+0.16 \nu_3^8), \\
t_{\rm crust} &=& t_{\rm crust,0} (1+0.4 \nu_3^2+0.3\nu_3^6),
\end{eqnarray}
where a subscript ``0'' denotes nonrotating values \cite{Zdunik01}. 
For $\nu\leq 500$ Hz,
the above expansion agrees well with a quadratic expansion derived
previously in \cite{Glendenning92}. The presence of the crust reduces
the maximum angular momentum and ratio of $T/|W|$ by about 20\%, 
compared to corresponding bare strange star models.


\subsubsection{Rotating magnetized neutron stars}

The presence of a magnetic field has been ignored in the models of
rapidly rotating relativistic stars that were considered in the
previous sections. The reason is that the inferred surface dipole
magnetic field strength of pulsars ranges between $10^8$ G and $2
\times 10^{13}$ G. These values of the magnetic field strength imply a
magnetic field energy density that is too small compared to the energy
density of the fluid, to significantly affect the structure of a
neutron star. However, there exists another class of compact objects
with much stronger magnetic fields than normal pulsars -- {\it
  magnetars}, that could have global fields up to the order of
$10^{15}$ G \cite{Duncan1992}, possibly born initially with high spin
(but quickly spinning down to rotational periods of a few seconds). In
addition, even though moderate magnetic field strengths do not alter
the bulk properties of neutron stars, they may have an effect on the
damping or growth rate of various perturbations of an equilibrium
star, affecting its stability.  For these reasons, a fully
relativistic description of magnetized neutron stars is
necessary. However, for fields less than $10^{15}$ G a {\it passive}
description, where one ignores the influence of the magnetic field on
the equilibrium properties of the fluid and the spacetime is
sufficient for most practical purposes.

The equations of electromagnetism and magnetohydrodynamics (MHD) in
general relativity have been discussed in a number of works; see, for
example
{\cite{Lichnerowicz1967,MTW,Bekenstein1978,Anile1989,Gourgoulhon2011}
} and references therein. The electromagnetic (E/M) field is described
by a vector potential $A_\alpha$, from which one constructs the
antisymmetric Faraday tensor $ F_{\a\b}= \na_\a A_\b- \na_\b A_\a, $
satisfying Maxwell's equations \index{Maxwell's equations}
\begin{eqnarray}
\na_\b{}^*F^{\a\b}&=&0, \label{Max1}\\
\na_\b F^{\a\b}&=&4\pi J^\a, \label{Max2}
\end{eqnarray}
where ${}^*F_{\a\b}{}=\frac{1}{2}\epsilon_{\a\b\g\d}F^{\g\d}$, with
$\epsilon_{\a\b\g\d}$ the totally antisymmetric Levi-Civita tensor. In
(\ref{Max2}), $J^\a$ is the 4-current creating the E/M field and the
Faraday tensor can be decomposed in terms of an electric 4-vector
$E_\a=F_{\a\b}u^\b$ and a magnetic 4-vector $B_\a={}^*F_{\b\a}u^\b$
which are measured by an observer comoving with the plasma and satisfy
$E_\a u^\a=B_\a u^\a=0$.

The stress-energy tensor of the E/M field is 
\be
T_{\a\b}^{\rm (em)}=\frac{1}{4\pi} \left(F_{\a\g}F_\b{}^\g-\frac{1}{4}F^{\g\d}F_{\g\d}
g_{\a\b} \right),
\ee
and the conservation of the total stress-energy tensor leads to 
the Euler equation in magnetohydrodynamics  
\be
(\epsilon +p)u^\beta\nabla_\beta u^\alpha
        = -q^{\alpha\beta}\nabla_\beta p + q^\alpha{}_\delta F^\delta{}_\g J^\g,
        \label{MHDEuler}
\ee
where $q_{\alpha\beta}:=g_{\alpha\beta}+u_\alpha u_\beta$. In the
ideal MHD approximation, where the conductivity ($\sigma$) is assumed
to be $\sigma \rightarrow \infty$, the MHD Euler equation takes the
form
\be
\left(\epsilon +p+\frac{B_\g B^\g}{4\pi} \right) 
u^\b \na_\b u_\a = - q_\a{}^\b \left[ \na_\b\left(p+\frac{B_\g B^\g}{8\pi} \right)
- \frac1{4\pi}\na_\g(B_\b B^\g) \right]. \label{Euler-iMHD}
\ee 
In general, a magnetized compact star will possess a magnetic field
with both poloidal and toroidal components. Then its velocity field
may include {\it non-circular} flows that give rise to the toroidal component. 
In such case, the spacetime metric  will
include additional non-vanishing components. The general formalism describing
such a spacetime has been presented by \cite{Gourgoulhon1993}, but
no numerical solutions of equilibrium models have been constructed, so far.
Instead, one can  look for special cases, where the 
velocity field is circular or {\it assume} that it is approximately so. 

If the current is purely toroidal, i.e. of the form $(J_t, 0,0, J_\phi)$, then a
theorem by Carter   \cite{carterlh} allows  for equilibrium solutions
with circular velocity flows  and a purely poloidal magnetic field, of the 
form $(0,B_r, B_\theta, 0)$. 
In ideal MHD, a purely  toroidal magnetic field, $(B_t,0,0,B_\phi)$, is also
allowed, generated by a current of the form $(0, J_r,J_\theta, 0)$ \cite{Oron2002}.
 
For {\it purely poloidal} magnetic fields, rotating stars must be uniformly rotating   
in order to be in a stationary equilibrium and the Euler equation becomes
\be
\nabla(H-\ln u^t) -\frac{1}{\epsilon+p}(j^\phi-\Omega j^t) \nabla A_\phi=0,
\ee
where $j^\alpha$ is the conduction current (the component of $J^\alpha$ normal
to the fluid 4-velocity). The hydrostationary equilibrium equation has a 
first integral in three different cases. These are a) $(j^\phi-\Omega j^t)$=0, 
b) $(\epsilon+p)^{-1}(j^\phi-\Omega j^t)$=const., and 
c) $(\epsilon+p)^{-1}(j^\phi-\Omega j^t)=f(A_\phi)$.
The first case corresponds to a vanishing Lorentz force and
has been considered in \cite{Bekenstein1979,Oron2002} (force-free field).
The second case is difficult to realize, but has been considered 
as an approximation in, e.g., \cite{Colaiuda2008}.  The third case is more
general and was first considered in \cite{BGSM93,BBGN95}.  
After making a choice for the
current and for the total charge, the system consisting of the Einstein equations, the 
hydrostationary equilibrium equation and Maxwell's equations can
be solved for the spacetime metric, the hydrodynamical variables and the 
vector-potential components $A_t$ and $A_\phi$, from which the magnetic and
electric fields in various observer frames are obtained. 

\begin{figure*}[htbp]
  \center
  \includegraphics[width=0.75\textwidth]{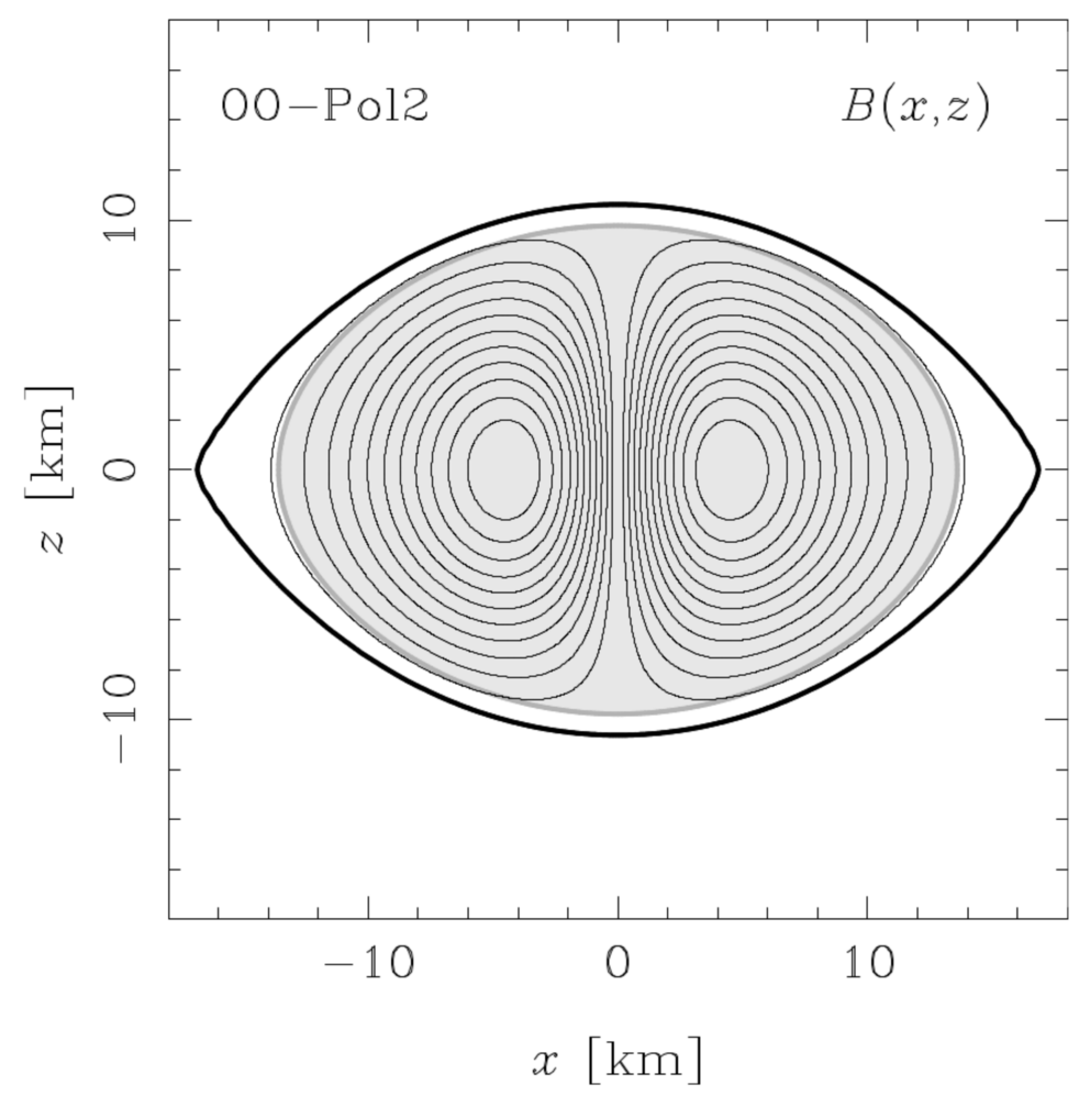}
  \caption{Isocontours of magnetic field strength in the meridional
    plane, for a rapidly rotating model with a purely toroidal
    magnetic field.  (Figure from~\cite{Frieben2012}, copyright by MNRAS.)}
  \label{fig:Frieben2012}
\end{figure*}

For a {\it purely toroidal magnetic field}, the only non-vanishing
component of the Faraday tensor is $F_{r\theta}$. Then, the ideal MHD
condition does not lead to a restriction on the angular velocity of
the star.  For uniformly rotating stars, the Euler equation becomes
\cite{Kiuchi2008b,Gourgoulhon2011} \be \nabla ( H -\ln u^t) +
\frac{1}{4\pi(\epsilon+p)g_2} \sqrt{\frac{g_2}{g_1}} F_{r\theta}
\nabla \left( \sqrt{\frac{g_2}{g_1}} F_{r\theta} \right)=0, \ee where
$g_1 = g_{rr} g_{\t\t}-(g_{r\t})^2$, $g_2 = -g_{tt}
g_{\phi\phi}+(g_{t\phi})^2$, which implies the existence of solutions
for which $\sqrt{\frac{g_2}{g_1}} F_{r\theta}$ is a function of
$(\epsilon+p)g_2$ (see \cite{Kiuchi2008b,Kiuchi2009} for
representative numerical solutions). A detailed study of rapidly
rotating equilibrium models with purely toroidal fields (in uniform
rotation) was recently presented by Frieben and Rezzolla
\cite{Frieben2012} and Fig. \ref{fig:Frieben2012} shows the
isocontours of magnetic field strength in the meridional plane, for a
representative case.

Gourgoulhon et al. \cite{Gourgoulhon2011} find a general form of 
stationary axisymmetric magnetic fields, including non-circular equilibria.  

Equilibria with purely toroidal or purely poloidal magnetic fields are
unstable in nonrotating stars (and likely unstable in rotating stars), see,
e.g. \cite{Wright1973,tayler73,Markey1973,Lasky2011,Ciolfi2012,Lasky2012}.
Some mixed poloidal/toroidal configurations seem more promising
for stability (see, for example, Duez et al. \cite{duez10}, who infer
stability of mixed equilibria in a Newtonian context from numerical
evolutions).

A poloidal magnetic field in a differentially rotating star will be
wound up, leading to the appearance of a toroidal component. This has
several consequences, such as magnetic braking of the differential
rotation, amplification of the magnetic field through dynamo action
and the development of the magnetorotational instability see
Sec.~\ref{Diffrotstars}.

\subsection{Approximate Universal relationships}

Yagi and Yunes recently discovered a set of universal relationships
that relate the moment of inertia, the tidal love number and the
(spin-induced) quadrupole moment for slowly rotating neutron stars and
quark stars~\cite{Yagi2013Sci,YagiYun2013} (for another review of the
$I$-Love-$Q$ relations see also~\cite{YagiYunReview2016} where
applications of these relations are also presented). The word
``universal'' in this context means within the framework of a
particular theory of gravitation, but independent of the equation of
state, provided the equation of state belongs to the class of
\emph{cold, realistic} equations of state, i.e., those that for the
most part agree below the nuclear saturation density where our
knowledge of nuclear physics is robust. More specifically, the
universal relationships were established numerically between properly
defined non-dimensional versions of the moment of inertia, the tidal
Love number and the quadrupole moment. In particular, if $M$ is the
gravitational mass of the star, Yagi and Yunes introduced the
following dimensionless quantities: $\bar I \equiv I/M^3$, $\bar Q
\equiv -Q/(M^3\chi^2)$, where $\chi\equiv J/M^2$ is the dimensionless
NS spin parameter, and $\bar \lambda^{\rm (tid)}\equiv \lambda^{\rm
  (tid)}/M^5$. Here $\lambda^{\rm (tid)}$ is the tidal Love number,
which determines the magnitude of the quadrupole moment tensor,
$Q_{ij}$, induced on the star by an external quadrupole tidal tensor
field $\mathcal{E}_{ij}$ through the relation $Q_{ij}=- \lambda^{\rm
  (tid)}\mathcal{E}_{ij}$. The universal relations can be expressed
through the following fitting formulae~\cite{Yagi2013Sci} (see
also~\cite{LattimerLim2013})
\be
\ln y_i = a_i+b_i\ln x_i+c_i(\ln x_i)^2 + d_i(\ln x_i)^3+e_i(\ln x_i)^4,
\ee
where $y_i$ and $x_i$ are a pair of two variables from the trio $\bar
I$, $\bar \lambda^{\rm (tid)}$ and $\bar Q$, and the values of the
coefficients $a_i,b_i,c_i,d_i,e_i$ are given in
Table~\ref{tab_IloveQ}.

\begin{table}[htbp]
  \caption{Coefficients for the fitting formulae of the neutron star
    and quark star $I$-Love, $I-Q$ and Love-$Q$ relations.}
  \label{tab_IloveQ}
  \renewcommand{\arraystretch}{1.3}
  \centering
    \begin{tabular}{ccccccc}
      \hline\noalign{\smallskip}
      $y_i$ & $x_i$ & $a_i$ & $b_i$ & $c_i$ & $d_i$ & $e_i$ \\
      \noalign{\smallskip}\hline\noalign{\smallskip}
      $\bar I$ & $\bar \lambda^{\rm (tid)}$ & $1.47$ & 0.0817 & 0.0149 & $2.87\times 10^{-4}$ & $-3.64 \times 10^{-5}$\\
      $\bar I$ & $\bar Q$ & $1.35$ & 0.697 & -0.143 & $9.94\times 10^{-2}$ & $-1.24 \times 10^{-2}$\\
      $\bar Q$ & $\bar \lambda^{\rm (tid)}$ & $0.94$ & 0.0936 & 0.0474 & $-4.21\times 10^{-3}$ & $1.23 \times 10^{-4}$\\
      \noalign{\smallskip}\hline
    \end{tabular}
\end{table}

As pointed out in~\cite{YagiYun2013} these relations could have been
anticipated because in the Newtonian limit $\bar I \propto C^{-2}$,
$\bar Q \propto C^{-1}$ and $\bar \lambda^{\rm (tid)} \propto C^{-5}$,
indicating the existence of one-parameter relation between the trio
$\bar I$, $\bar \lambda^{\rm (tid)}$, $\bar Q$. Here, $C$ is the
compactness of the star. The advantage of the existence of such
universal relations is that in principle the measurement of one of the
$I$-Love-$Q$ parameters determines the other two, and one can use these
relations to lift quadrupole moment and spin-spin degeneracies that
arise in parameter estimation from future gravitational wave
observations of compact binaries involving neutron
stars~\cite{Yagi2013Sci,YagiYun2013}. These relations could also help
constrain modified theories of gravity~\cite{Yagi2013Sci,YagiYun2013}
(but see below).

Shortly after the discovery of these relations several works attempted
to test the limits of the universality of these relations. Maselli et
al.~\cite{MaselliCardoso2013} relaxed the small tidal deformation
approximation assumed in~\cite{Yagi2013Sci,YagiYun2013} and derived
universal relations for the different phases during a neutron star
inspiral, concluding that these relations do not deviate significantly
from those reported in~\cite{Yagi2013Sci,YagiYun2013}. On the other
hand, Haskell et al.~\cite{Haskell2014}, considered neutron star
quadrupole deformations that are induced by the presence of a magnetic
field. They built self-consistent magnetized equilibria with the {\tt
  LORENE} libraries and concluded that the $I$-Love-$Q$ universal
relations break down for slowly rotating neutron stars (spin periods
$> 10$s), and for polar magnetic field strengths $B_p >
10^{12}$G. Doneva et al.~\cite{Doneva2014a} considered
self-consistent, equilibrium models of spinning neutron stars beyond
the slow-rotation approximation adopted
in~\cite{Yagi2013Sci,YagiYun2013}. They use the {\tt RNS} code to
built rapidly rotating stars, and find that with increasing rotation
rate, the $\bar I-\bar Q$ relation departs significantly from its
slow-rotation limit deviating up to 40\% for neutron stars and up to
75\% for quark stars. Moreover, they find that the deviation is EOS
dependent and for a broad set of hadronic and strange matter EOS the
spread due to rotation is comparable to the spread due to the EOS, if
one considers sequences with fixed rotational frequency. For a
restricted set of EOSs, that do not include models with extremely
small or large radii, they were still able to find relations that are
roughly EOS-independent at fixed rotational frequencies. However,
Pappas and Apostolatos~\cite{Pappas2014} using the {\tt RNS} code,
showed that even for rapidly rotating neutron stars universality is
again recovered, if instead of the $\bar I$-$\bar Q$ and angular
frequency parameters, one focuses on the 3 dimensional parameter space
spanned by the dimensionless spin angular momentum $\chi$, the
dimensionless mass quadrupole $\bar Q$ and the dimensionless spin
octupole moment $\beta_2\equiv -s_3/\chi^3$, where $s_3 \equiv
-S_3/M^4$, and where, again, $S_3$ is the spin octupole moment of the
Hansen-Geroch
moments~\cite{Geroch1970Moments,Hansen1974Moments}. Moreover, Pappas
and Apostolatos~\cite{Pappas2014} show that if one considers the
parameter space $(\chi,\bar I,\bar Q)$, then the $I-Q$ EOS
universality is recovered, in the sense that for each value $\chi$
there exists a unique universal $\bar I$-$\bar Q$ relation.

It should be pointed out~\cite{Yagi2013Sci,YagiYun2013,Pappas2014}
that these ``universal'' relations hold not among the moments
themselves, but among the rescaled, dimensionless moments, where the
mass scale is factored out. Thus, the introduction of a scale will
lift the apparent degeneracy among different EOSs.

Given the existence of such universal relations relating moments of
neutron and quark stars, a fundamental question then arises: what is
the origin of the universality? Yagi et al.~\cite{YagiStein2014}
performed a thorough study to answer this question and concluded that
universality arises as an emergent approximate symmetry in that
relativistic stars have an approximate self-similarity in their
isodensity contours, which leads to the universal behavior observed in
their exterior multipole moments. Work by Chan et al.~\cite{Chan2015}
has explored the origin of the I-Love relation through a
post-Minkowskian expansion for the moment of inertia and the tidal
deformability of incompressible stars.

Another way deviations from universality can take place are in
protoneutron stars for which a cold, nuclear EOS in not
applicable. Martinon et al.~\cite{Martinon2014} find that the
$I$-Love-$Q$ relations do not apply following one second after the
birth of a protoneutron star, but that they are satisfied as soon as
the entropy gradients are smoothed out typically within a few seconds.
See also Marques et al.~\cite{Marques:2017zju} where a new finite
temperature hyperonic equation of state is constructed and finds a
similar conclusion as Martinon et al.~\cite{Martinon2014} regarding
thermal effects.

Pani and Berti~\cite{Pani2014} have extended the Hartle-Thorne
formalism for slowly rotating stars to the case of scalar tensor
theories of gravity and explored the validity of the $I$-Love-$Q$
relations in scalar-tensor theories of gravity focusing on theories
exhibiting the phenomenon of spontaneous
scalarization~\cite{DamourEsposito1993,DamourEsposito1996}.  Pani and
Berti find that $I$-Love-$Q$ relations exist in scalar-tensor gravity
and interestingly also for spontaneously scalarized stars. Most
remarkably, the relations in scalar-tensor theories coincide with
their general relativity counterparts to within less than a few
percent. This result implies that the $I$-Love-$Q$ relations may not
be used to distinguish between general relativity and scalar-tensor
theories. We note that a similar conclusion was drawn by Sham et
al.~\cite{ShamLin2014ApJ} in the context of Eddington-inspired
Born-Infeld gravity where the $I$-Love-$Q$ relations were found to be
indistinguishable than those of GR -- an anticipated
result~\cite{Pani2014}. More recently Sakstein et
al.~\cite{Sakstein:2016oel} have found the the $I-C$ ($C$ for
compactness) in beyond Hordenski theories are clearly distinct from
those in GR.

The effects of anisotropic pressure have been explored by Yagi and
Yunes in~\cite{YagiYun2015}. They find that anisotropy breaks the
universality, but that the $I$-Love-$Q$ relations remain approximately
universal to within 10\%. Finally, Yagi and Yunes \cite{YagiYun2016}
considered anisotropic pressure to build slowly rotating, very high compactness
stars that approach the black hole compactness limit, in order to
answer the question of how the approximate $I$-Love-$Q$ relations
become exact in the BH limit. While the adopted methodology provides
some hints into how the BH limit is approached, an interesting, and
perhaps, definitive way to probe this is to consider unstable rotating
neutron stars and perform dynamical simulations of neutron star
collapse to black hole with full GR simulations.

In addition to the $I$-Love-$Q$ relations, Pappas and
Apostolatos~\cite{Pappas2014} find that for realistic equations of
state there exists a universal relation between $\alpha_2$ and
$\beta_3$, i.e., $\beta_3=\beta_3(\alpha_2)$, while Yagi et
al.~\cite{YKPYA2014} discover a similar universal relation between
$\gamma_4$ and $\alpha_2$, i.e., $\gamma_4=\gamma_4(\alpha_2)$. These
new approximate universal relations provide a type of ``no-hair''
relations among the multipole moments for neutron stars and quark
stars. Motivated by these studies, Manko and
Ruiz~\cite{Manko:2016ncb}, show that there exists an infinite
hierarchy of universal relations for neutron star multipole moments,
assuming that neutron star exterior field can be described by four
arbitrary parameters as in~\cite{Manko1995}.

\subsection{Rapidly Rotating equilibrium configurations in modified theories of gravitation}

With the arrival of ``multimessenger'' astronomy, gravitational wave
and electromagnetic signatures of compact objects will soon offer a
unique probe to test the limits of general relativity.  Neutron stars
are an ideal astrophysical laboratory for testing gravity in the
strong field regime, because of their high compactness and because of
the coupling of possible extra mediator fields with the matter.

Testing for deviations from general relativity would preferably
require a generalized framework that parametrizes such deviations in
an agnostic way as in the spirit of the parameterized post-Newtonian
approach~\cite{CliffWilllrr-2014-4} (which systematically models
post-Newtonian deviations from GR), or in the spirit of the
parametrized post-Einsteinian
approach~\cite{YunesPret2009,CornishSampson2011PhRvD}, which
parametrizes a class of deviations from general relativistic waveforms
within a certain regime. In the absence of such a complete
parameterized framework, the existence of alternative theories of
gravity are welcome not only as a means for testing for such
deviations, but also for gaining a better understanding of how to
develop a generalized, theory-agnostic framework of deviations from
general relativity. Motivated by these ideas and by observations that can be interpreted as an
accelerated expansion of the Universe~\cite{Riess1998AJ} a number of
extended theories of gravitation have been proposed as alternatives to a
cosmological constant in order to explain dark energy~(see
e.g.~\cite{Tsujikawa2010LNP,Paschalidis2011CQGra,Bloomfield2013JCAP,deTham2014LRR,Joyce2016a,Koyama2016RPPh}
for reviews and multiple aspects of such theories). The ``infrared''
predictions of modified gravity theories have been investigated
extensively, and recently their strong-field predictions have
attracted considerable attention (see e.g. Berti et
al.~\cite{BertiBarCar2015CQGra} for a review). Studies of spherically
symmetric and slowly rotating neutron stars in modified gravity are
reviewed in~\cite{BertiBarCar2015CQGra}, thus we focus here on the
bulk properties of equilibrium, rapidly rotating neutron stars in
modified theories of gravity.

Doneva et al.~\cite{Doneva2013} presented a study of rapidly rotating
neutron stars in scalar-tensor theories of gravity, by extending the
{\tt RNS} code to treat these theories in the Einstein frame, while
computing physical quantities in the Jordan frame. The Jordan frame
action considered in~\cite{Doneva2013} is given by
\be
S = \frac{1}{16\pi}\int d^4 x\sqrt{-\tilde g}[F(\Phi)\tilde R-Z(\Phi)g^{\mu\nu}\partial_\mu \Phi\partial_\nu \Phi- 2 U(\Phi)]+ S_m(\Psi_m;\tilde g_{\mu\nu}),
\ee
where $\tilde g_{\mu\nu}$ is the Jordan frame metric, $\tilde R$ the
Ricci scalar accociated with $\tilde g_{\mu\nu}$, $\Phi$ the scalar
field, $U(\Phi)$ the potential and $S_m$ denotes the matter action and
$\Psi_m$ denotes the matter fields. The functions $U(\Phi)$, $F(\Phi)$
and $Z(\Phi)$ control the dynamics of the scalar field. However,
requiring that the gravitons carry positive energy implies $Z(\Phi) >
0$, and non-negativity of the scalar field kinetic energy requires
$2F(\Phi)Z(\Phi)+3(dF/d\Phi)^2 \geq 0$. Note that the matter action
does not involve $\Phi$. Via a conformal transformation 
\be
g_{\mu\nu} = F(\Phi) \tilde g_{\mu\nu}
\ee
and a scalar field redefinition via 
\be
\left(\frac{d\phi}{d\Phi}\right)^2 = \frac{3}{4}\left(\frac{d\ln F(\Phi)}{d\Phi}\right)^2+ \frac{Z(\Phi)}{2F(\Phi)}
\ee
and letting
\be
\mathcal{A}(\phi)=1/\sqrt{F(\Phi)}, 2V(\phi)=U(\Phi)/F(\Phi)^2,
\ee
one recovers the Einstein frame action
\be
S = \frac{1}{16\pi}\int d^4 x\sqrt{- g}[ R-2g^{\mu\nu}\partial_\mu \phi\partial_\nu \phi- 4 V(\phi)]+ S_m(\Psi_m;A(\phi)^2 g_{\mu\nu}) \label{Ein_frame},
\ee
where $R$ is the Ricci scalar associated with the Einstein frame
metric $g_{\mu\nu}$. Variation of this action with respect to
$g_{\mu\nu}$ and $\phi$ yields the equations of motion for the metric
and for the scalar field, which are then cast in a convenient form and
coupled to the equilibrium fluid equations that make it
straightforward to extend the general relativistic equations solved by
the {\tt RNS} code. In their study, Doneva et al.~\cite{Doneva2013}
set $V(\phi)=0$ and consider two choices for the function
$\mathcal{A}(\phi)$, namely $\ln\mathcal{A}(\phi)=k_0\phi$ and
$\ln\mathcal{A}(\phi)=\beta \phi^2/2$, while setting
$\lim_{r\rightarrow \infty} \phi = 0$ and focusing on rigidly rotating
equilibria. The former choice for $\mathcal{A}(\phi)$ is equivalent to
Brans-Dicke theory, but the latter choice while it is
indistinguishable from general relativity in the weak field regime,
leads to the emergence of new phenomenology, such as a bifurcation due
to non-uniqueness of
solutions~\cite{DamourEsposito1993,DamourEsposito1996}. Observations
currently constrain $k_0$ and $\beta$ to values $k_0 < 4\times
10^{-3}$ and $\beta \gtrsim
-4.5$~\cite{CliffWilllrr-2014-4,FreireWex2012MNRAS,Antoniadis2013,Shibata:2013pra}. However,
as pointed out in~\cite{Popchev,RamazanoPreto2016PhRvD}, a massive
scalar field naturally circumvents these observational bounds if the
Compton wavelength of the scalar field is small compared to the binary
orbital separation. The equation of state adopted in~\cite{Doneva2013}
is a polytrope $P=k\rho^{1+1/n}$, with $n=0.7463$ and $k=1186$ in
units where $G=c=M_{\odot}=1$.
%
  \begin{figure*}[t]
  \center
      \includegraphics[width=0.46\textwidth]{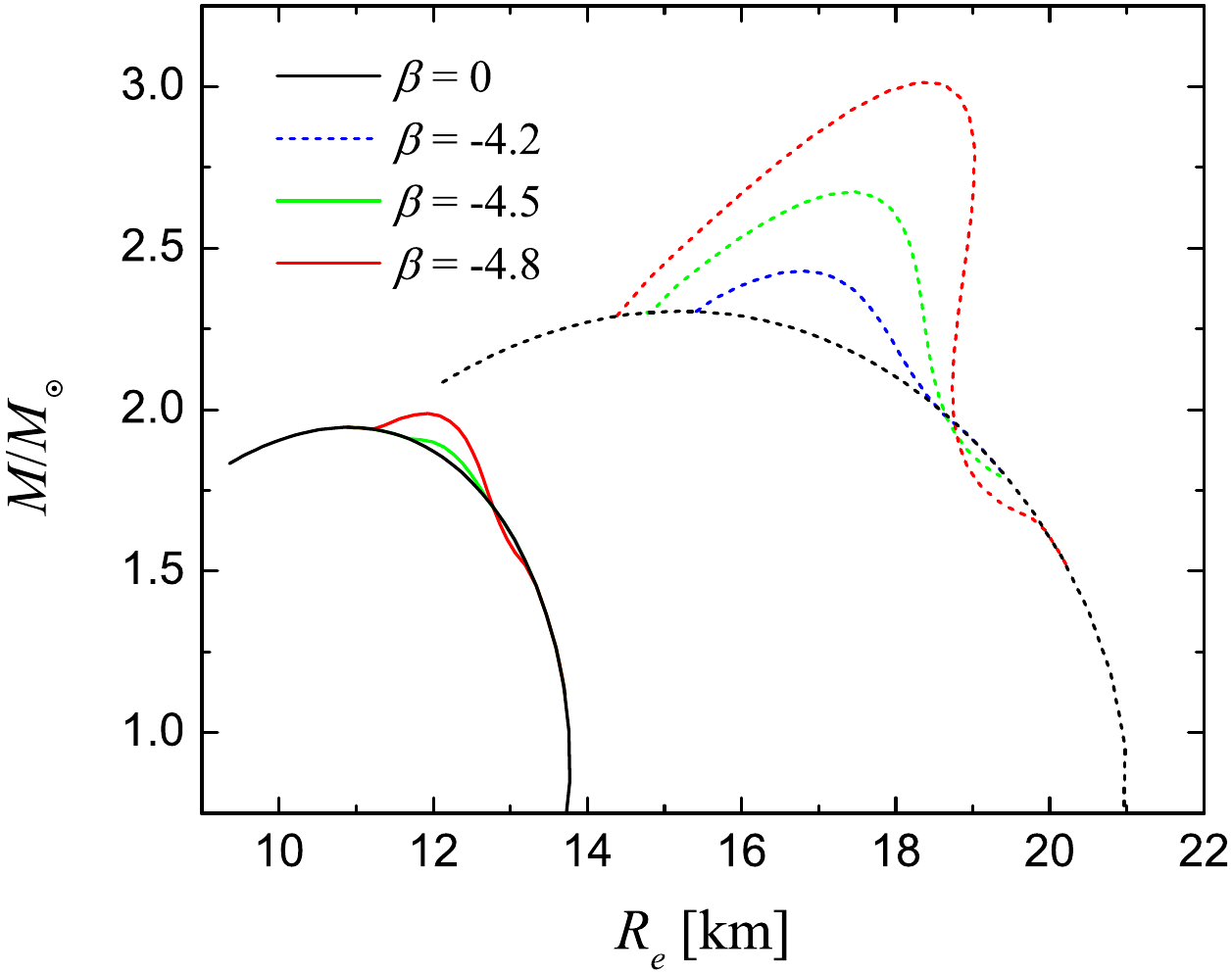}
      \includegraphics[width=0.45\textwidth]{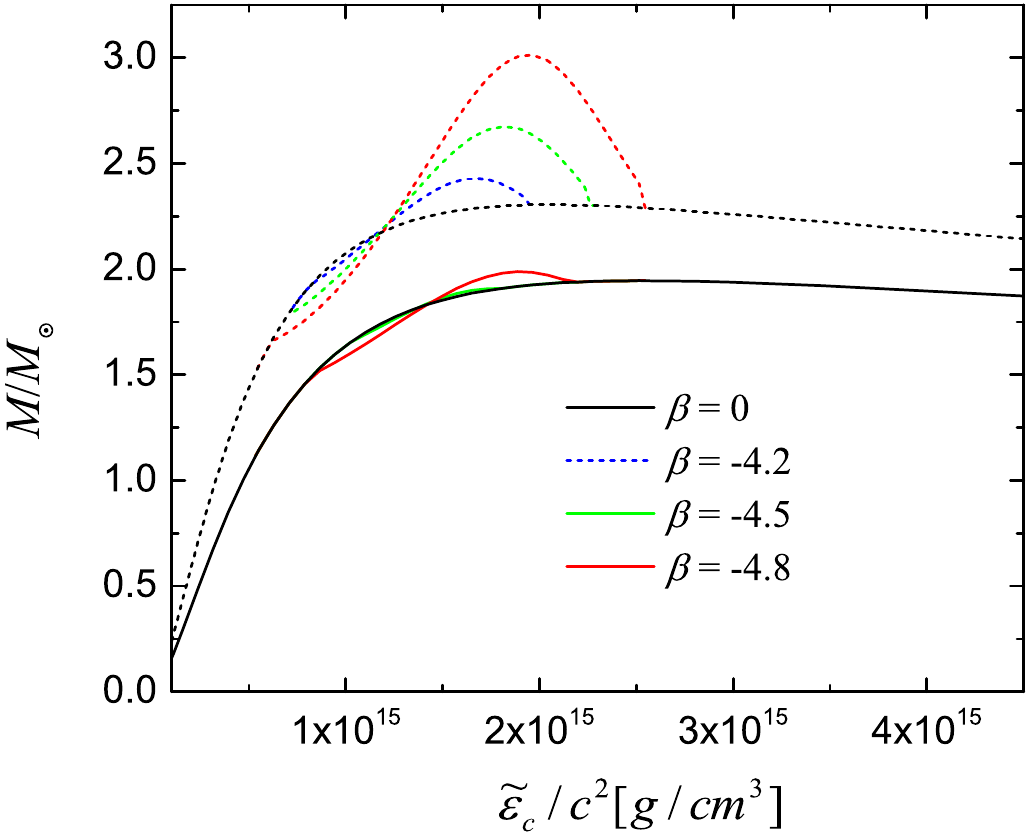} 
      \includegraphics[width=0.46\textwidth]{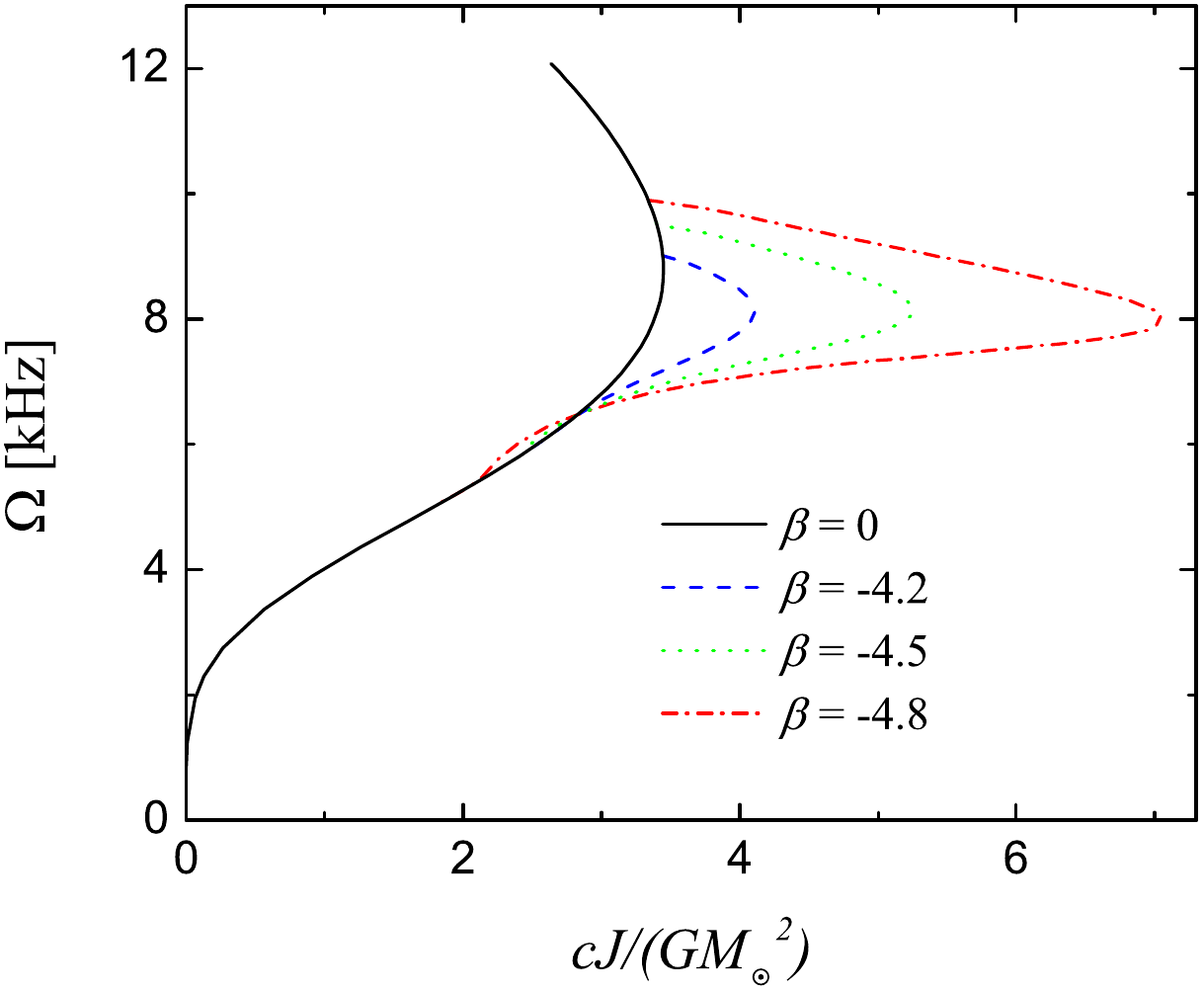}
      \includegraphics[width=0.45\textwidth]{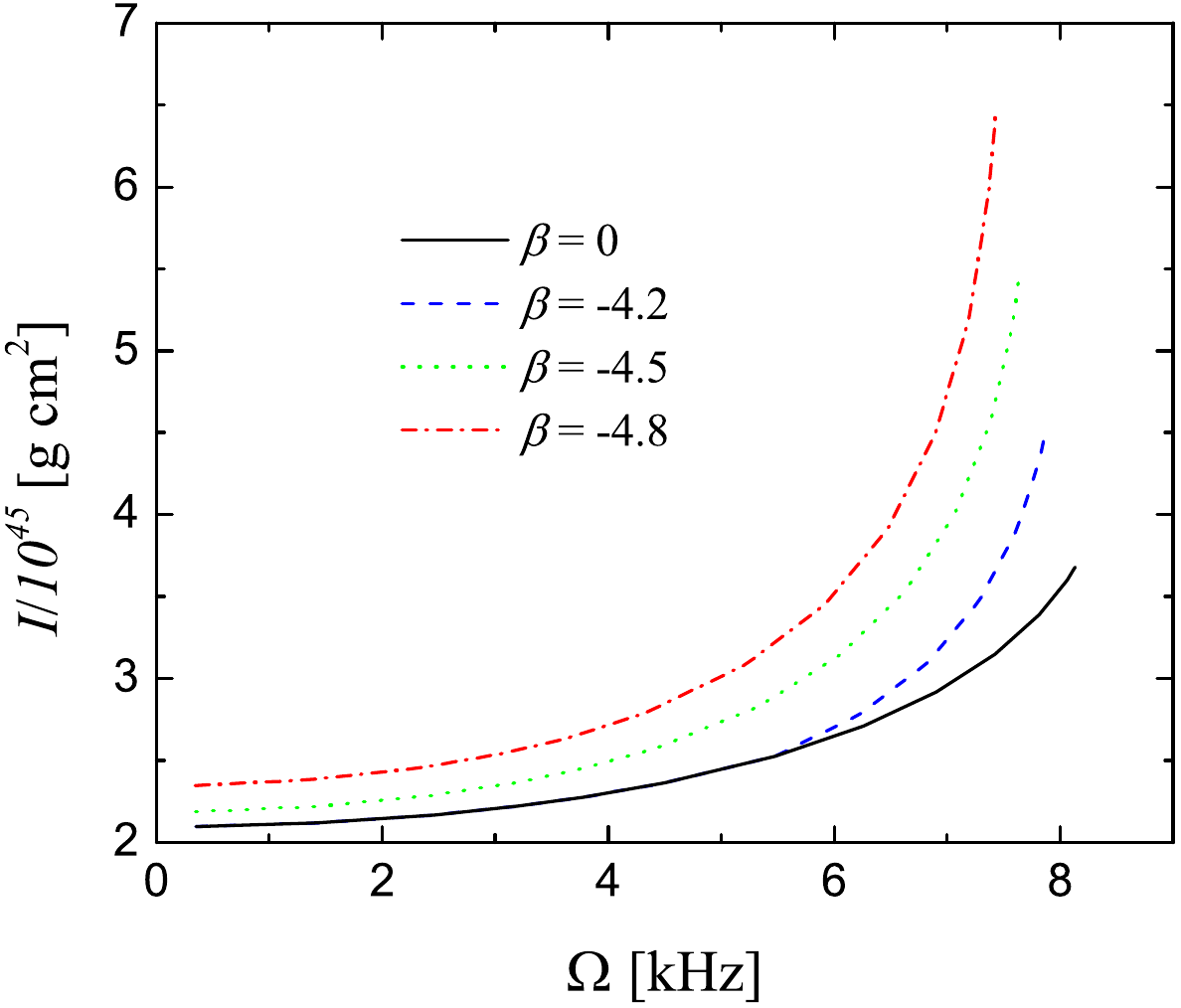} 
    \caption{From left to right, top to bottom the plots correspond to
      mass vs. radius, mass vs central energy density, angular
      frequency vs dimensionless angular momentum, and moment of
      inertia vs. angular frequency. For the plots in the top row, solid lines
      correspond to non-rotating stars, while dotted lines correspond
      to the mass-shedding sequence. The angular frequency vs.
      dimensionless angular momentum has only the mass-shedding
      sequence, while the moment of inertia plot corresponds to models
      for which the central energy density is fixed at $\sim
      \epsilon_c/c^2 = 1.5 \times 10^{15} \rm g/cm^3$. (Image
      reproduced with permission from \cite{Doneva2013}, copyright by
      APS.)  }
    \label{spont_scalar_rapid_r}
\end{figure*}

For $\ln\mathcal{A}(\phi)=k_0\phi$ with the largest allowed value $k_0
= 4\times 10^{-3}$, Doneva et al.~\cite{Doneva2013} find that even for
stars rotating at the mass shedding limit, their the total mass,
radius and angular momentum are practically indistinguishable from
their counterparts in general relativity. However, for
$\ln\mathcal{A}(\phi)=\beta \phi^2/2$, while all general relativity
solutions are also solutions of the scalar tensor theory with
$\phi=0$, for certain values of $\beta$ and a certain range of neutron
star densities new solutions emerge with non-trivial scalar field
values that are also energetically
favored~\cite{DamourEsposito1993,DamourEsposito1996}. This phenomenon
is known as spontaneous scalarization and for the equation of state
adopted in~\cite{Doneva2013}, Harada has argued that the phenomenon
occurs only for $\beta \lesssim -4.35$~\cite{Harada1998PhRvD}. One of
the important findings in~\cite{Doneva2013} is that rapid rotation
extends the range of $\beta$ values for which spontaneous
scalarization can take place, and in particular that along the
mass-shedding limit the bound becomes $\beta < − 3.9$. In addition, it
is found that rapid rotation changes significantly several bulk
properties from their GR counterparts. Examples of such properties
include the mass, radius, angular momentum, and moment of inertia as
can be seen in Fig.~\ref{spont_scalar_rapid_r}. Of all bulk quantities
those affected the most by the scalar field are the angular momentum
and the moment of inertia of the star, which can differ up to a factor
of two from their corresponding values in general relativity. It is
also worth noting that the deviation of the bulk properties from their
GR values, increases further if one considers smaller values of
$\beta$, that are still in agreement with the observations. Based on
the sensitivity of the moment of inertia (even at slow rotation
rates), Doneva et al. suggested that the moment of inertia could be an
astrophysical probe of theories exhibiting spontaneous scalarization.

In a subsequent paper~\cite{Doneva2014b}, Doneva et al. extended the
equilibrium solutions of rapidly rotating compact stars for the
spontaneous scalarization model $\ln\mathcal{A}(\phi)=\beta\phi^2/2$
with $V(\phi)=0$, for tabulated equations of state.  For the cases
when scalarization occurs, they find results similar to those reported
in~\cite{Doneva2013}. In addition, they compute orbital and epicyclic
frequencies for particles orbiting these neutron star models and find
considerable differences of these frequencies between the scalar
tensor theory and general relativity for the maximum-mass rotating
models (but not so for models with spin frequency of $\sim 700$Hz or
less, with the exception of very stiff equations of state).

The $I-Q$ relation for rapidly rotating stars in the model
$\ln\mathcal{A}(\phi)=\beta\phi^2/2$ was considered by Doneva et al.
in~\cite{DonevaYaza2014PhRvD}. The authors find that the $I-Q$ relation
is nearly EOS independent for scalarized rapidly rotating stars, and
that the spread of the relationship for higher rotation rates
increases compared to general relativity. They also find that smaller
negative values of $\beta$ lead to larger deviations from the general
relativitivstic $I-Q$ relation, but the deviations (at most 5\% for
$\beta=-4.5$) are less than the anticipated accuracy of future
observations. These results provide, yet, another example where the
$I-Q$ relation may not be able to provide strong constraints on
deviations from general relativity. We note that similar conclusions
hold for rapidly rotating stars in Einstein-Gauss-Bonnet-dilaton
gravity~\cite{Kleihaus2014PhRvD}.

In a recent paper, Doneva and Yazadjiev~\cite{DonevaYazad2016} studied
rapidly rotating stars for the model
$\ln\mathcal{A}(\phi)=\beta\phi^2/2$, but this time extending it to
the case of a massive scalar field by adding a potential
$V(\phi)=m_\phi^2 \phi^2/2$. In this case, In this case, the scalar
field is short-range and observations practically leave the value of
$\beta$ unconstrained. However, for the spontaneous scalarization of
the neutron star one must have $10^{-16}eV \lesssim m_\phi \lesssim
10^{-9} eV$. Adopting the value $\beta = -6$, Doneva and Yazadjiev
find that the $I-Q$ relation remains universal, but they deviate
substantially (up to $\sim 20\%$) from those in general
relativity. Thus, the $I-Q$ relation could be used to infer deviations
from general relativity.

Another modified gravity theory that has been considered in the
context of rapidly rotating stars is a particular model of $f(R)$
gravity~\cite{Sotiriou2010RvMP,DeFelice2010LRR} with $f(R)=R+a R^2$
sometimes referred to as $R^2$ gravity. It can be shown that the
Einstein frame action of this particular model of $f(R)$ gravity can
be cast in the form~\eqref{Ein_frame} with $\ln
A(\phi)=-\phi/\sqrt{3}$, but with a non-zero potential
$V(\phi)=(1-\exp(-2\phi/\sqrt{3}))^2/16a$~\cite{Yazadjiev2015}. Motivated
by the results found for static and slowly rotating stars in $R^2$
gravity~\cite{Yazadjiev2014,Staykov2014}, Yazadjiev et al. modified
the {\tt RNS} code to allow for the construction of rapidly rotating
neutron star models in $R^2$ gravity in~\cite{Yazadjiev2015}. Adopting
different equations of state, they find that rapid rotation enhances
the discrepancy in global quantities such as mass, radius, and angular
momentum between $R^2$-gravity and general relativistic stars. Also,
the differences become larger as the coupling constant $a$ increases.
Generically, the $R^2$-gravity maximum neutron star mass is larger than
the corresponding limit in general relativity. Yazadjiev et
al. adopted $a/M_{\odot}^2\in [0,10^4]$, which is within the Gravity
Probe B constraint $a\lesssim 5\times 10^{5}\rm km^2$, but much larger
than the E\"ot-Wash experiment constraint $a\lesssim 10^{-16}\rm
km^2$~\cite{NafJetzer2010PhRvD}. However, the bound from Gravity Probe
B is still relevant because the chameleon nature of $f(R)$ gravity can
give rise to different effective values at different
scales~\cite{NafJetzer2010PhRvD}. For the mass-shedding sequences and with
$a=10^4 M_\odot^2$, they find that for the equations of state
considered, the maximum fractional differences between general
relativity and $R^2$-gravity in maximum mass and maximum moment of
inertia are 16.6\% and 65.6\%, respectively.

Armed with the $R^2$-gravity code, Doneva et al. studied the
universality of the $I-Q$ relation~\cite{DonevaYazad2015PhRvD}. They
find that $R^2$ gravity exhibits an EOS-independent $I-Q$ relation,
but that the differences with the Einstein gravity can be as large as
$\sim 20\%$ for $a=10^4M_\odot^2$, similar to the deviations found in
\cite{DonevaYazad2016} for a scalar-tensor model
$\ln\mathcal{A}(\phi)=\beta\phi^2/2$ and a massive scalar field. Thus,
while it would be difficult to use the $I-Q$ relation in order to
single out a specific extended theory of gravity, this relation could
potentially be used to infer deviations from general relativity and to
exclude some theories of extended gravity.

In addition to theories mentioned that can be cast in the usual form of
scalar-tensor theories of gravity, the Einstein-dilaton-Gauss-Bonet
(EdGB) theory is another example that has received attention in the
context of rapidly rotating neutron stars. EdGB is inspired by
heterotic string theory~\cite{GrossSloan1987,MetsaevTseytlin1987}, and
the effective action is given by
\be S = \frac{1}{16\pi}\int d^4 x\sqrt{- g}[
  R-\frac{1}{2}g^{\mu\nu}\partial_\mu \Phi\partial_\nu \Phi+
  \alpha e^{-\beta\Phi}R_{GB}^2]+ S_m(\Psi_m; g_{\mu\nu}), 
\ee
where $\Phi$ is the dilaton field, $\gamma$ is a coupling constant,
$\alpha$ is a positive coefficient and
$R_{GB}^2=R_{\mu\nu\rho\sigma}R^{\mu\nu\rho\sigma}-4R_{\mu\nu}R^{\mu\nu}+R^2$
is the Gauss-Bonet term. The equations of motion for this theory are given by (see e.g.~\cite{Kleihaus2016PhRvD})
\be
\Box\Phi = \alpha\gamma e^{-\beta\Phi}R_{GB}^2
\ee
\bea
G_{\mu\nu} & = & 8\pi T_{\mu\nu}+\frac{1}{2}\left[\nabla_\mu\Phi\nabla_\nu\Phi-\frac{1}{2}\nabla_\rho\Phi\nabla^\rho\Phi\right] \nonumber \\
         &   & -\alpha e^{\beta\Phi}\left[H_{\mu\nu}+4(\beta^2\nabla^\rho\Phi\nabla^\sigma\Phi-\beta\nabla^\rho\nabla^\sigma\Phi)P_{\mu\rho\nu\sigma}\right],
\eea
where
\bea
H_{\mu\nu} & = & 2\left[RR_{\mu\nu}-2R_{\mu\rho}R^{\rho}{}_{\nu}-2R_{\mu\rho\nu\sigma}R^{\rho\sigma}+R_{\mu\rho\sigma\lambda}R_{\nu}{}^{\rho\sigma\lambda}\right] \\&&-\frac{1}{2}g_{\mu\nu}R^2_{GB}, \nonumber\\
P_{\mu\nu\rho\sigma} & = & R_{\mu\nu\rho\sigma}+2g_{\mu[\sigma} R_{\rho]\nu}+2g_{\nu[\rho} R_{\sigma]\mu}+Rg_{\mu[\rho}g_{\sigma]\nu},
\eea
are second-order partial differential equations because of the
particular form of the Gauss-Bonet term.  In this theory black hole
solutions exist only for up to a maximum value of
$|\alpha|$~\cite{Kanti1996PhRvD}, hence rotating neutron star
solutions are interesting to find only in this regime.  Pani et
al.~\cite{PBCR2011PhRvD} build models of slowly rotating compact stars
in this theory and find that only the product $\alpha\beta$ matters
for the structure of compact stars in EdGB theory, whereas  the larger
the value of this product the smaller the maximum neutron star mass
that can be supported in this theory. They also find that stellar
solutions do not exist for arbitrarily large values of $\alpha\beta$
(this was already known about the existence of black hole
solutions~\cite{Kanti1996PhRvD} in this theory). As a result, the
maximum observed mass could be used to place constraints on
$\alpha\beta$.

\begin{figure*}[t]
    \center
      \includegraphics[width=0.46\textwidth]{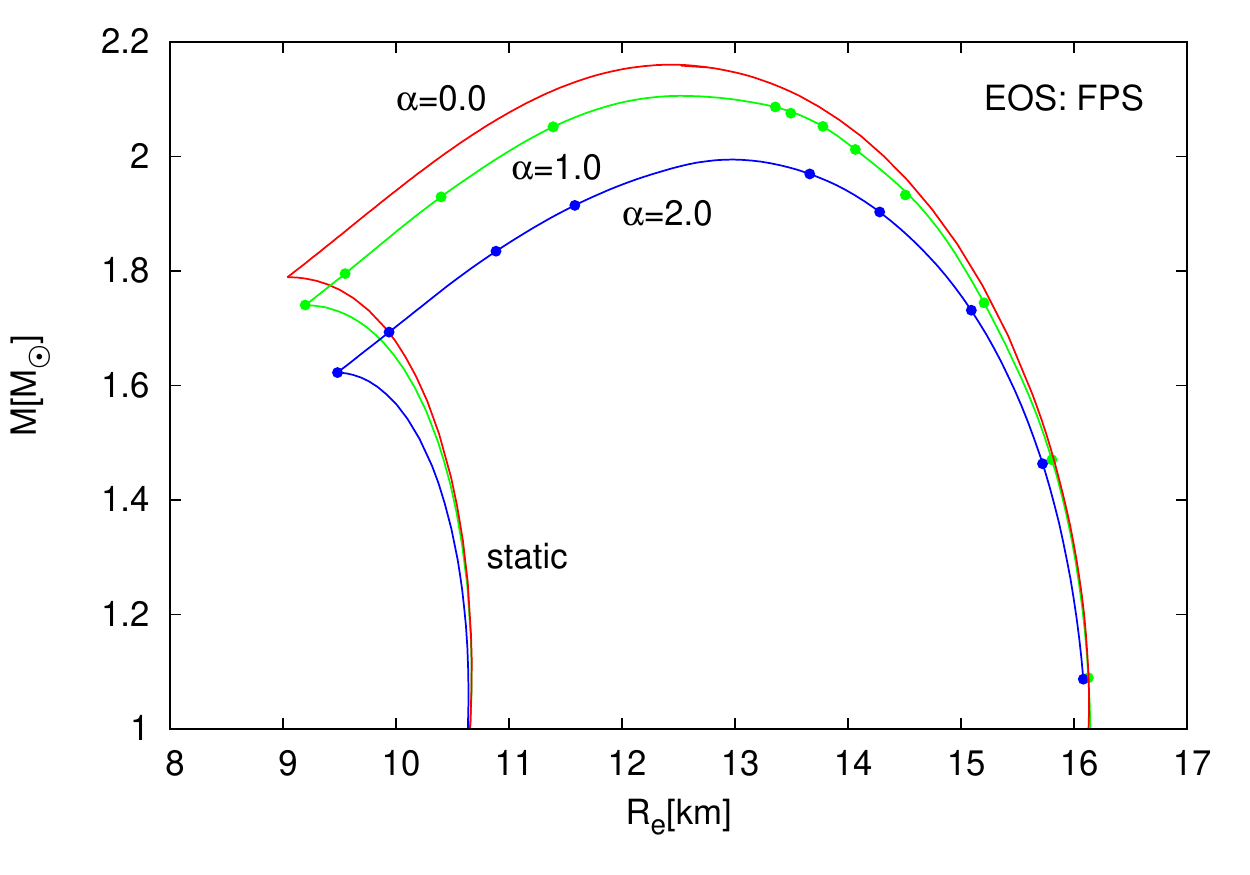}
      \includegraphics[width=0.46\textwidth]{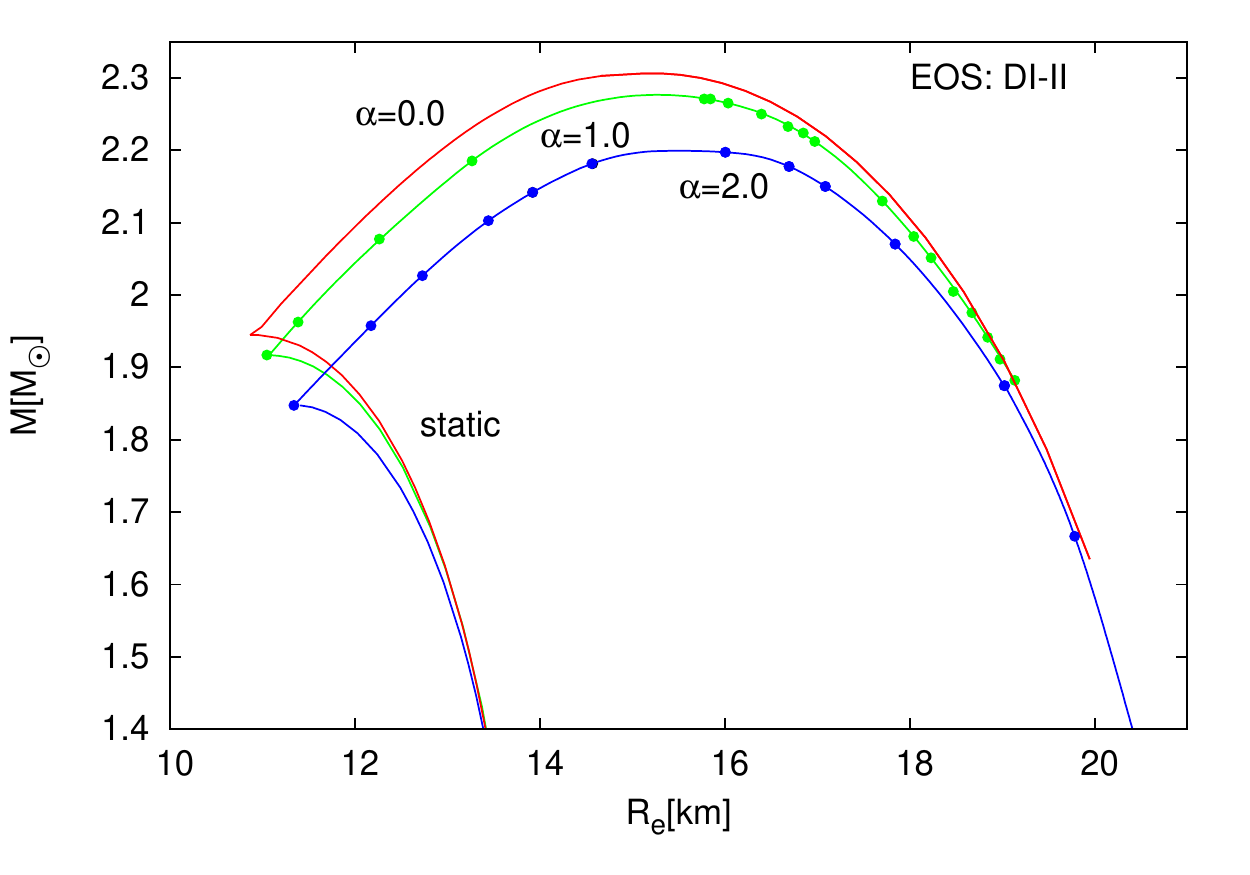} 
    \caption{The physical regime on the mass-circumferential equatorial
      radius plane for neutron star solutions in EdGB theory for the
      FPS (left) and DI-II (right) EOSs. The left boundary in each
      panel designates the static sequence and the upper and right
      boundary the mass-shedding sequence. The values of the coupling
      constant $\alpha=0,1,2$ are in units of $M_\odot^2$.  (Image
      reproduced with permission from \cite{Kleihaus2016PhRvD},
      copyright by APS.)  }
    \label{EdGB_rapid_r}
  \end{figure*}

Kleihaus et al.~\cite{Kleihaus2016PhRvD} develop a code for building
rapidly rotating neutron stars in EdGB theory. The authors consider
two different equations of state, FPS and DI-II~\cite{DI1985ApJ}. They
confirm the results of Pani et al.~\cite{PBCR2011PhRvD} and in
addition find that rotation enhances the effects of deviations from GR
(see Fig.~\ref{EdGB_rapid_r}).  Furthermore, the authors find that the
quadrupole moment depends on the value of the EdGB coupling constant
and that the dependence is enhanced for larger value of the angular
velocity. Finally, Kleihaus et al. discover that the GR $I-Q$ relation
extends to EdGB theory with weak dependence on the value of the
coupling parameter $\alpha$ when the NS dimensionless spin is
0.4. Therefore, EdGB theory provides yet another example where the
$I-Q$ relations cannot be utilized to constrain deviations from GR.
%

\subsection{Differentially rotating neutron stars} 
\label{Diffrotstars}

The non-uniformity of rotation in the early stages of the life of a
compact object (or right after the merger of a binary system) opens
another dimension in the allowed parameter space of equilibrium
models. The simplest description appropriate for neutron stars is the
1-parameter law (\ref{eq:diffrot}), introduced in
\cite{KEH89a,KEH89b,EHN94}. Relativistic models of differentially
rotating stars were constructed numerically by Baumgarte, Shapiro and
Shibata~\cite{Baumgarte00}, where it was pointed out that these
configurations can support more mass than uniformly rotating
stars. The authors coined the term ``hypermassive'' neutron stars for
these compact objects whose mass exceeds the supramassive limit.

\begin{figure*}[htbp]
  \center
  \includegraphics[width=0.7\textwidth]{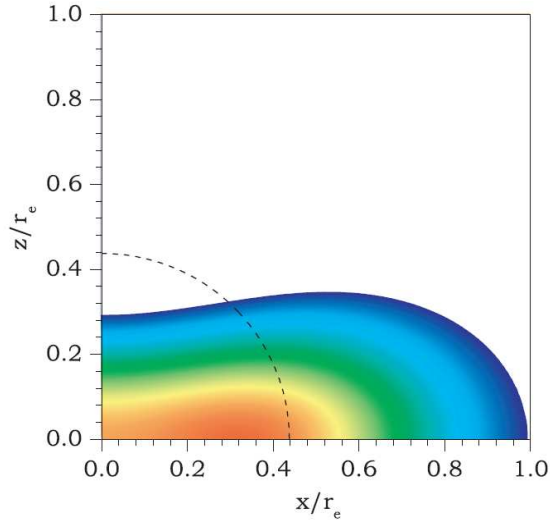}
  \caption{Density stratification for model A11 of Table \ref{tabeq},
    displaying an off-center density maximum. In comparison, the shape
    of the non-rotating star of same rest mass is shown, scaled by the
    equatorial radius of the rotating model (dashed line).  (Image
    reproduced with permission from~\cite{SAF}, copyright by MNRAS.)}
  \label{fig:SAF}
\end{figure*}

Examples of equilibrium sequences of differentially rotating
polytropic models, using the above rotation law, were constructed by
Stergioulas, Apostolatos and Font \cite{SAF}. Table \ref{tabeq} shows
the detailed properties of a fixed rest mass sequence (A), in which
the central density decreases as the star rotates more rapidly and of
a sequence of fixed central density (B), in which the mass increases
significantly with increasing rotation. $\Omega_c$ and $\Omega_e$ are
the values of the angular velocity at the center and at the equator,
respectively while $r_p$ and $r_e$ is the coordinate radius at the
pole and at the equator (other quantities shown are defined as in
Table \ref{tab_equ}). While most models along these sequences are
quasi-spherical (meaning that the maximum density appears at the
center), the fastest rotating members are quasi-toroidal, with an
off-center maximum density. An example is shown in Fig. \ref{fig:SAF}.

\begin{table}
\caption{Properties of two sequences of differentially rotating 
  equilibrium models. Sequence A is 
  a sequence of fixed rest mass with $M_0=1.506 M_\odot$  
  and sequence B is
  a sequence of fixed central rest mass density $\rho_c=1.28 \times
  10^{-3}$ with $A/r_e=1$.  All models are relativistic polytropes
  with $N=1$, $K=100$ and rotation law parameter $A/r_e=1$. 
  The definitions of the various quantities
  are given in the main text. All quantities are in
  dimensionless units with $c=G=M_\odot=1$. (Table adapted from 
  Stergioulas et al.~\cite{SAF}.)}
\begin{tabular}{*{9}{c}}
\hline\noalign{\smallskip}
model  &  $\varepsilon_c$  &     $M$  &  $R$  &  $r_e$  &  $r_p/r_e$  &  
$\Omega_c$  &  $\Omega_e$  &  $T/|W|$   \\
\noalign{\smallskip}\hline\noalign{\smallskip}
       &  $(\times 10^{-3})$ &       &       &   &   &  $(\times 10^{-2})$  &  
$(\times 10^{-2})$   &    \\
\hline
A0  &  1.444  &  1.400  &   9.59  &  8.13  &  1.0    &  0.0    &  0.0    &   0.0    
\\
A1  &  1.300  &  1.405  &  10.01  &  8.54  &  0.930  &  2.019  &  0.759  &   0.018  
\\
A2  &  1.187  &  1.408  &  10.40  &  8.92  &  0.875  &  2.580  &  0.977  &   0.033  
\\
A3  &  1.074  &  1.410  &  10.84  &  9.35  &  0.820  &  2.944  &  1.125  &   0.049  
\\
A4  &  0.961  &  1.413  &  11.37  &  9.87  &  0.762  &  3.192  &  1.232  &   0.066  
\\
A5  &  0.848  &  1.418  &  12.01  & 10.49  &  0.703  &  3.340  &  1.303  &   0.086  
\\
A6  &  0.735  &  1.422  &  12.78  & 11.25  &  0.643  &  3.383  &  1.336  &  0.107  
\\
A7  &  0.622  &  1.427  &  13.75  & 12.21  &  0.579  &  3.339  &  1.337  &  0.131  
\\
A8  &  0.509  &  1.433  &  15.01  & 13.45  &  0.513  &  3.197  &  1.300  &  0.158  
\\
A9 &  0.396  &  1.439  &  16.70  & 15.13  &  0.444  &  2.953  &  1.223  &   0.189  
\\
A10 &  0.283  &  1.447  &  19.03  & 17.44  &  0.370  &  2.604  &  1.101  &  0.223  
\\
A11 &  0.170  &  1.456  &  21.92  & 20.30  &  0.294  &  2.184  &  0.944  &  0.260  
\\
\hline
B0  &  1.444  &  1.400  &   9.59  &  8.13  &  1.0    &  0.0    &  0.0    &   0.0    
\\
B1  &   1.444  & 1.437   &   9.75  &  8.24  &  0.950  &  1.801  &  0.666  &  0.013   
\\
B2  &   1.444  & 1.478   &   9.92  &  8.36  &  0.900  &  2.574  &  0.944  &  0.026   
\\
B3  &   1.444  & 1.525   &  10.11  &  8.49  &  0.849  &  3.189  &  1.160  &  0.040   
\\
B4  &  1.444   & 1.578   &  10.31  &  8.63  &  0.800  &  3.728  &  1.342  &  0.055   
\\
B5  &   1.444  & 1.640   &  10.53  &  8.77  &  0.750  &  4.227  &  1.504  &  0.071   
\\
B6  &   1.444  & 1.713   &  10.76  &  8.91  &  0.700  &  4.707  &  1.651  &  0.087   
\\
B7  &   1.444  & 1.798   &  11.01  &  9.05  &  0.650  &  5.185  &  1.789  & 0.105   
\\
B8  &   1.444  & 1.899   &  11.26  &  9.17  &  0.600  &  5.683  &  1.921  & 0.124   
\\
B9  &   1.444  & 2.020   &  11.50  &  9.26  &  0.550  &  6.232  &  2.052  & 0.144   
\\
B10 &   1.444  & 2.167   &  11.71  &  9.27  &  0.500  &  6.889  &  2.192  & 0.165   
\\
B11 &   1.444  & 2.341   &  11.80  &  9.13  &  0.450  &  7.770  &  2.357  & 0.187   
\\
B12 &   1.444  & 2.532   &  11.64  &  8.72  &  0.400  &  9.118  &  2.584  & 0.207   
\\
\noalign{\smallskip}\hline
\label{tabeq}
\end{tabular}
\end{table}

\begin{figure*}[htbp]
  \center
  \includegraphics[width=0.7\textwidth]{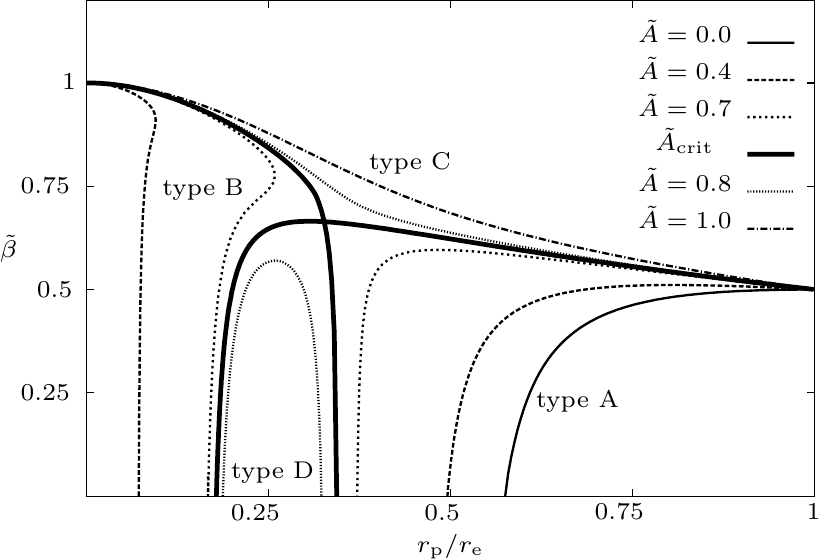}
  \caption{Different types of sequences of differentially rotating
    equilibrium models (see text for a detailed description). Here
    $\tilde A = \hat A^{-1}$.  (Image reproduced with permission
    from~\cite{Ansorg2009}, copyright by MNRAS.)}
  \label{fig:Ansorg2009}
\end{figure*}

Ansorg, Gondek-Rosi\'nska and Villain \cite{Ansorg2009} found 4
different types of differentially rotating models (which they label as
type A, B, C and D) for the same 1-parameter law (\ref{eq:diffrot})
which exists in parts of the allowed parameter space. For a
sufficiently weak degree of differential rotation, sequences with
increasing rotation terminate at the mass-shedding limit, but for
moderate and strong rates of differential rotation equilibrium
sequences can exhibit a continuous transition to a regime of toroidal
fluid bodies. Figure \ref{fig:Ansorg2009} displays sequences of $N=1$
polytropes with various values of the parameter $\tilde A=\hat
A^{-1}=r_e/A$ and a fixed central density. In the vertical axis, the
parameter $\tilde \beta$ is related to the shape of the surface of the
star and ranges between 0 (when rotation is limited by mass shedding
at the equator) and 1 (when the radius on the polar axis becomes 0,
indicating the transition to a toroidal configuration).  As the axis
ratio $r_p/r_e$ is varied, type A sequences start at a nonrotating
model and terminate at the mass-shedding limit. Type B sequences have
no nonrotating member, but connect models at the mass-shedding limit
to toroidal configurations.  Type C sequences connect the nonrotating
limit to toroidal configurations, while type D sequences connect
models at the mass-shedding limit to other models at the mass-shedding
limit. It will be interesting to study the stability properties of
models of these different types. One should keep in mind that these
different types arise for the simple 1-parameter law
(\ref{eq:diffrot}) and a more complicated picture may arise for
multi-parameter rotation laws.  More recently, Studszi{\'n}ska et
al. \cite{Studzinska2016} thoroughly explored the parameter space for
the rotation law~\eqref{eq:diffrot} and determined how the maximum
mass depends on the stiffness, on the degree of differential rotation
and on the maximum density, taking into account all types of solutions
that were shown to exist in \cite{Ansorg2009}.

A well know fact about differentially rotating neutron stars is that
they can support more mass than the supramassive limit - the maximum
mass when allowing for maximal uniform rotation. Neutron stars with
mass larger than the supramassive limit are known as hypermassive
neutron stars. Equilibrium sequences of differentially rotating models
with polytropic equations of state, and using the same differential
rotation law have been constructed by Lyford et al.~\cite{Lyford03}.
There, the focus was on the effects of differential rotation on the
maximum mass configuration. The authors find that differential
rotation can support about 50\% more mass than the TOV limit mass, as
opposed to uniform rotation that typically increases the TOV limit by
about 20\%. In a subsequent paper, Morrison et
al.~\cite{Morrison2004ApJ} extended this result to realistic equations
of state. However, recent calculations by Gondek-Rosi\'nska et
al.~\cite{Gondek2016} focussing on $n=1$ polytropes, discover that the
maximum mass depends not only on the degree of differential rotation,
but also on the type of solution identified in~\cite{Ansorg2009},
i.e., A, B, C or D. The authors find that different classes have
different maximum mass limits and even for moderate degrees of
differential rotation $\hat A^{-1} \sim 1$, the maximum rest-mass
configuration can be significantly higher than 2.0 times the TOV
limit. Although, masses greater than two times the TOV limit can never
be achieved in hypermassive neutron stars formed following a binary
neutron star merger, it would be interesting to investigated the
dynamical stability of the maximum mass configurations constructed
in~\cite{Gondek2016}.

\subsection{Proto-neutron stars} 
\label{protoneutronstars}

Following the gravitational collapse of a massive stellar core, a
proto-neutron star (PNS) is born with an initially large temperature
of order 50 MeV and a correspondingly large radius of up to 100 km. If
the PNS is slowly rotating, one can study its evolution assuming
spherical symmetry (see, for example
\cite{Burrows1986,Burrows1995,Bombaci1995,Keil1995,Keil1996,P97,Pons1999,Pons2000,Prakash2001,Strobel99}).
Up to a time of about 100 ms after core bounce, the PNS is lepton rich
and consists of an unshocked core at densities $n>0.1
\mathrm{fm}^{-3}$, with entropy per baryon $s\sim 1$, surrounded by a
transition region and a low-density but high-entropy, shocked envelope
with $s \sim 4-10$, which extends to large radii. The lepton number is
roughly $Y_l\sim 0.4$ and neutrinos in the core and in the transition
region are trapped (the PNS is opaque to neutrinos), while at
densities less than $n\sim 6\times 10^{-4} \ \mathrm{fm}^{-3}$ the
outer envelope becomes transparent to neutrinos.  Within about 0.5s,
the outer envelope cools and contracts with the entropy per baryon
becoming roughly $s\sim 2$ throughout the star (the lepton number in
the outer envelope drops to $Y_l\sim 0.3$). Further cooling results in
a fully deleptonized, hot neutron star at several tens of seconds
after core bounce, with a roughly constant entropy per baryon of
$s\sim 1-2$. After several minutes, when the neutron star has cooled
to $T<1 \mathrm{MeV}$, the thermal effects are negligibly small in the
bulk of the star and a zero-temperature EOS can be used to describe
its main properties.

The structure of hot PNSs is described by finite-temperature EOSs,
such as those presented in
\cite{LS91,Sugahara1994NuPhA.579..557S,Toki1995NuPhA.588..357T,Lala1997PhRvC..55..540L,Strobel1999b,Pons1999,Shen1998,Hempel2010NuPhA.837..210H,Typel2010PhRvC..81a5803T,Fattoyev2010PhRvC..82e5803F,Shen2011PhRvC..83c5802S,Shen2011PhRvC..83f5808S,Hempel2012ApJ...748...70H,Steiner2013ApJ...774...17S}
(see also~\cite{Oertel2016arXiv161003361O} for a review). These
candidate EOSs differ in several respects (for example in the thermal
pressure at high densities).  The sample of cold EOSs that has been
extended, so far, is quite limited and does not correspond to the wide
range of possibilities allowed by current observational
constraints. Therefore, PNS models that have been constructed only
cover a small region of the allowed parameter space.  Understanding
the detailed evolution of a PNS is significant, as the star could
undergo transformations that could be associated with direct or
indirect observational evidence, such as the delayed collapse of a
hypermassive PNS (see \cite{BB94,BST96,Baumgarte00}).
 
If the PNS is born rapidly rotating, its evolution will sensitively
depend on the rotation rate and other factors, such as the development
of the magnetorotational instability (MRI). Some partial understanding
has emerged by studying quasi-equilibrium sequences of rotating models
\cite{HOE94,Strobel99,Sumiyoshi99,Villain2004}.  Exact equilibria
can be found in the case that the model is considered to be
barotropic, where all thermodynamical quantities (energy density,
pressure, entropy, temperature) depend only on the baryon number
density.  Special cases, such as homentropic or isothermal stars have
also been considered.  In \cite{Villain2004} a barotropic EOS was
constructed by rescaling temperature, entropy and lepton number
profiles that were obtained from detailed, one-dimensional simulations
of PNS evolutions, while the rotational properties of the models were
taken from two-dimensional core-collapse simulations.

The main conclusion from the studies of sequences of quasi-equilibrium
models is that PNSs that are born with moderate rotation, will
contract and spin up during the cooling phase (see
e.g. \cite{GHZ98,Strobel99}).  This could lead to a PNS rotating with
large enough rotation rate for secular or dynamical instabilities to
become interesting.  It is not clear, however, whether the
quasi-stationary approximation is valid when the stars reach the
mass-shedding limit, as, upon further thermal contraction, the outer
envelope could actually be shed from the star, resulting in an
equatorial stellar wind. It should be noted here that a small amount
of differential rotation significantly affects the mass-shedding
limit, allowing more massive stars to exist than uniform rotation
allows.

Studies of PNSs are being extended to include additional effects, such
as entropy and lepton-driven convective instabilities and
hydromagnetic instabilities
\cite{Epstein1979,Livio1980,Burrows1986,Burrows1992,Miralles2000,Miralles2002,Miralles2004,Dessart2006,Lasky2012},
meridional flows \cite{Eriguchi1991}, local and mean-field magnetic
dynamos \cite{Thompson1993,Xu2001,Bonanno2003,Reinhardt2005,Naso2008},
magnetic braking and viscous damping of differential rotation
\cite{Shapiro01,LS04,Duez2004,Thompson2005,Duez2006b}, and the MRI and
Tayler instabilities
\cite{Akiyama2003,Kotake2004,Thompson2005,Ardeljan2005,Masada2006,Shibata2006,CerdaDuran2007,Masada2007,Stephens2007,BK2008,Spruit2008,Kiuchi2008,Obergaulinger2008,Siegel2013,Guilet2016arXiv161008532G}.
These effects will be important for the evolution of both PNSs formed
after core collapse, as well as for hypermassive or supramassive
neutron stars possibly formed after a binary neutron star merger.
\newpage


\section{Rotating relativistic stars in LMXBs}


\subsection{Particle orbits and kHz quasi-periodic oscillations}

X-ray observations of accreting sources in LMXBs have revealed a rich
phenomenology that is waiting to be interpreted correctly and could
lead to significant advances in our understanding of compact objects
(see~\cite{Lamb98,VanDerKlis00,Psaltis00}). The most important
feature of these sources is the observation (in most cases) of twin
kHz quasi-periodic oscillations (QPOs)
(see~\cite{vanderKlis20062675,Abramowiczlrr} for reviews on QPOs). The
high frequency of these variabilities and their quasi-periodic nature
are evidence that they are produced in high-velocity flows near the
surface of the compact star. To date, there exist a large number of
different theoretical models that attempt to explain the origin of
these oscillations. No consensus has been reached, yet, but once a
credible explanation is found, it will lead to important constraints
on the properties of the compact object that is the source of the
gravitational field in which the kHz oscillations take place. The
compact stars in LMXBs are spun up by accretion, so that many of them
may be rotating rapidly; therefore, the correct inclusion of
rotational effects in the theoretical models for kHz QPOs is
important. Under simplifying assumptions for the angular momentum and
mass evolution during accretion, one can use accurate rapidly rotating
relativistic models to follow the possible evolutionary tracks of
compact stars in LMXBs~\cite{Cook94,Zdunik01b}.

In most theoretical models, one or both kHz QPO frequencies are
associated with the orbital motion of inhomogeneities or blobs in a
thin accretion disk. In the actual calculations, the frequencies are
computed in the approximation of an orbiting test particle, neglecting
pressure terms. For most equations of state, stars that are massive
enough possess an ISCO, and the orbital frequency at the ISCO has been
proposed to be one of the two observed frequencies. To first order in
the rotation rate, the orbital frequency at the prograde ISCO is given
by (see Klu{\'z}niak, Michelson, and Wagoner~\cite{Kluzniak90})
\begin{equation}
  f_{\mathrm{ISCO}} \simeq
  2210 \, (1+0.75\chi) \left(\frac{1\,M_{\odot}}{M} \right) \mathrm{\ Hz},
\end{equation}
where $\chi:=J/M^2$. At larger rotation rates, higher order contributions
of $\chi$ as well as contributions from the quadrupole moment $Q$ become
important and an approximate expression has been derived by Shibata
and Sasaki~\cite{Shibata98}, which, when written as above and
truncated to the lowest order contribution of $Q$ and to
${\cal O}(\chi^2)$, becomes
\begin{equation}
  f_{\mathrm{ISCO}} \simeq 2210 \, (1+0.75\chi+0.78\chi^2-0.23Q_2)
  \left(\frac{1\,M_{\odot}}{M} \right) \mathrm{\ Hz},
\end{equation}
where $Q_2=-Q/M^3$.

Notice that, while rotation increases the orbital frequency at the
ISCO, the quadrupole moment has the opposite effect, which can become
important for rapidly rotating models. Numerical evaluations of
$f_{\mathrm{ISCO}}$ for rapidly rotating stars have been used
in~\cite{Miller98} to arrive at constraints on the properties of the
accreting compact object.

In other models, orbits of particles that are eccentric and slightly
tilted with respect to the equatorial plane are involved (the
relativistic precession model). For eccentric orbits, the periastron
advances with a frequency $\nu_{\mathrm{pa}}$ that is the difference
between the Keplerian frequency of azimuthal motion $\nu_{\mathrm{K}}$
and the radial epicyclic frequency $\nu_{\mathrm{r}}$. On the other
hand, particles in slightly tilted orbits fail to return to the
initial displacement $\psi$ from the equatorial plane, after a full
revolution around the star. This introduces a nodal precession
frequency $\nu_{\mathrm{pa}}$, which is the difference between
$\nu_{\mathrm{K}}$ and the frequency of the motion out of the orbital
plane (meridional frequency) $\nu_{\psi}$. Explicit expressions for
the above frequencies, in the gravitational field of a rapidly
rotating neutron star, have been derived recently by
Markovi{\'c}~\cite{Markovic00}, while in~\cite{Markovic00b} highly
eccentric orbits are considered. Morsink and Stella~\cite{Morsink99}
compute the nodal precession frequency for a wide range of neutron
star masses and equations of state and (in a post-Newtonian analysis)
separate the precession caused by the Lense--Thirring (frame-dragging)
effect from the precession caused by the quadrupole moment of the
star. The nodal and periastron precession of inclined orbits have also
been studied using an approximate analytic solution for the exterior
gravitational field of rapidly rotating
stars~\cite{Sibgatullin02}. These precession frequencies are
relativistic effects and have been used in several models to explain
the kHz QPO
frequencies~\cite{Stella99,Psaltis00a,Abramowicz01,Kluzniak02,Amsterdamski02}.

It is worth mentioning that it has recently been found that an ISCO
also exists in Newtonian gravity, for models of rapidly rotating
low-mass strange stars. The instability in the circular orbits is
produced by the large oblateness of the
star~\cite{Kluzniak01,Zdunik01c,Amsterdamski02} (see
also~\cite{Torok2014A&A...564L...5T} for a more recent
study). Epicyclic frequencies for Maclaurin spheroids in Newtonian
gravity have also been computed by Kluzniak and
Rosi{\'n}ska~\cite{Kluzniak2013}.

Epicyclic frequencies for rapidly rotating strange stars have been
computed by Gondek-Rosi\'nska et al.~\cite{Gondek2014PhRvD..89j4001G}
(2014) adopting the MIT bag model for the equation of state of quark
matter. They find that the orbits around rapidly rotating strange
quark stars are mostly affected by the stellar oblateness, rather than
by the effects of general relativity.

For reviews on applications of current QPO models and what one can
learn about the properties of NSs
see~\cite{Bhattacharyya2010AdSpR..45..949B,Torok2010ApJ,Pappas2012MNRAS.422.2581P,MillerLamb2016EPJA...52...63M,OzelFreire2016ARA&A..54..401O}.

\vspace{1 em}\noindent{\bf Going further:}~~ Observations of some LMXBs
finding that the difference in the frequencies of the peak QPOs is
equal to half the spin frequency of the star raise some questions
regarding the validity of the popular beat-frequency
model~\cite{Miller1998ApJ} (but
see~\cite{Lamb2003astro.ph..8179L}). Motivated by this tension,
another model for QPOs is suggested by Li and
Narayan~\cite{LiNar2004ApJ...601..414L} in which it is argued that a
strong magnetic field may truncate the inner parts of the disk and at
the interface between the accretion disk and the magnetosphere
surrounding the accreting star that gas becomes Rayleigh-Taylor and,
possible also, Kelvin-Helmholtz unstable, leading to nonaxisymmetric
structures which result in the high-frequency QPOs that can explain
observations. For other studies considering the impact of magnetic
fields see
also~\cite{KluznianRappaport2007,Kulkarni2008MNRAS.386..673K,TsangLai2009MNRAS.396..589T,Lovelave2009ApJ...701..225L,Bakala2010CQGra..27d5001B,Bakala2012CQGra..29f5012B,FuLai2012MNRAS.423..831F,RomUst2012MNRAS.421...63R,Long2012NewA...17..232L}
and references therein. Another model that can explain the
observations where the difference in the frequencies of the twin peaks
is equal to half the spin frequency of the star is the so-called
non-linear resonance model~\cite{Kluzniak2004ApJ} (see
also~\cite{LeeAbra2004ApJ,Horak2009A&A,Urbanec2010A&A}). For a recent
work investigating the compatibility of realistic neutron star
equations of state with several QPO modes see~\cite{Torok:2016tff}.
Constants of motion in stationary, axisymmetric spacetimes have been investigated recently in \cite{Markakis2014}.


\subsection{Angular momentum conservation during burst oscillations}

Some sources in LMXBs show signatures of type I X-ray bursts, which
are thermonuclear bursts on the surface of the compact
star~\cite{Lewin95}. Such bursts show nearly-coherent oscillations in
the range 270\,--\,620~Hz
(see~\cite{VanDerKlis00,Strohmayer01,Strohmayer2006csxs.book..113S,Watts2012ARA&A..50..609W}
for reviews). One interpretation of the burst oscillations is that
they are the result of rotational modulation of surface asymmetries
during the burst. In such a case, the oscillation frequency should be
nearly equal to the spin frequency of the star. This model currently
has difficulties in explaining some observed properties, such as the
oscillations seen in the tail of the burst, the frequency increase
during the burst, and the need for two anti-podal hot spots in some
sources that ignite at the same time. Alternative models also exist
(see, e.g., \cite{Psaltis00}).

Changes in the oscillation
frequency by a few Hz during bursts have been associated with expansion and contraction of the
burning shell. Cumming et al.~\cite{Cumming01} compute the
expected spin changes in general relativity taking into account rapid
rotation. Assuming that the angular momentum per unit mass is
conserved, the change in angular velocity with radius is given by
\begin{equation}
  \frac{d\ln \Omega}{d\ln r} =
  - 2 \left[ \left( 1-\frac{v^2}{2}-\frac{R}{2}
  \frac{\partial \nu}{\partial r} \right)
  \left(1-\frac{\omega}{\Omega}\right) -
  \frac{R}{2\Omega}\frac{\partial \omega}{\partial r} \right],
\end{equation}
where $R$ is the equatorial radius of the star and all quantities are
evaluated at the equator. The slow rotation limit of the above result
was derived previously by Abramowicz et al.~\cite{Abramowicz01b}. The
fractional change in angular velocity can then be
estimated as
\begin{equation}
  \frac{\Delta \Omega}{\Omega} =
  \frac{d\ln \Omega}{d\ln r}\left( \frac{\Delta r}{R} \right),
\end{equation}
where $\Delta r$ is the coordinate expansion of the burning shell, a
quantity that depends on the shell's composition. Cumming {\it et
  al.}\ find that in the expansion phase the expected spin down is a factor of two to three
times smaller than observed values, if the atmosphere rotates
rigidly. More detailed modeling is needed to fully explain the origin
and properties of burst oscillations
(see~\cite{Watts2012ARA&A..50..609W} for a recent review on
theoretical models of thermonuclear bursts).

A very interesting topic is modeling the expected X-ray spectrum of an
accretion disk in the gravitational field of a rapidly rotating
neutron star or of the ``hot spot'' on its surface as it could lead to
observational constraints on the source of the gravitational
field. See, e.g.,
\cite{Thampan98,Sibgatullin98,Sibgatullin00,Bhattacharyya02,Bhattacharyya02b},
where work initiated by Kluzniak and Wilson~\cite{Kluzniak91} in the
slow rotation limit is extended to rapidly rotating relativistic
stars.

Following an earlier work which uses approximate
spacetimes~\cite{Cadeau2005}, light curves from ray-tracing on
spacetimes corresponding to realistic models of rapidly rotating
neutron stars (generated with the {\tt RNS} code) are obtained by
Cadeau et al.~\cite{Cadeau2007} assuming that the X-ray photons arise
from a hot spot on the NS. There it was shown that the dominant effect
due to rotation comes from the stellar oblateness, and that
approximating a rapidly rotating star as a sphere results in large
errors if one is trying to fit for the radius and mass. However, for
cases with stellar spin frequencies less than $\sim 300$ Hz rapidly
rotating spacetime models are not necessary and only the stellar
oblateness has to be taken into consideration.  As a result, Morsink
et al.~\cite{Morsink2007} develop the Oblate Schwarzschild (OS) model
in which photons emerge from a hot spot in the NS oblate surface, and
they reach the observer following the geodesics of a corresponding
Schwarzschild spacetime, while doppler effects due to rotation are
taken into consideration as in the standard model of Miller and
Lamb~\cite{MillerLamb1998ApJ...499L..37M}. Morsink et al. demonstrate
that the OS model suffices to describe the effects due to the NS
rotation. An approximate analytic model for pulse profiles taking into
account gravitational light bending, doppler effect, anisotropic
emission and time delays is presented by Poutanen and
Beloborodov~\cite{PoutBel2006MNRAS.373..836P}. Another simple model
adopting the Hartle-Thorne approximation for generating pulse profiles
from rotating neutron stars is developed by Psaltis and
{\"O}zel~\cite{PsaltisOzel2014ApJ...792...87P}. For related studies
see
also~\cite{LeeStro2005MNRAS.361..659L,Baub2012ApJ...753..175B,Chan2013ApJ...777...13C,MillerLamb2015ApJ...808...31M}
and references therein.

\vspace{1 em}\noindent{\bf Going further:} A number of theoretical
works whose aim to model atomic lines in NS atmospheres in order to
infer the NS properties from the atomic line redshift see
e.g.~\cite{Ozel2003ApJ,Bildsten2003ApJ,Chang2005ApJ,Chang2006,Bhattacharyya2006ApJ,Baubock2013ApJ,Ozel2013RPPh,Baubock2013,Heinke2013JPhCS.432a2001H,Baubock2015ApJ}.

\newpage


\section{Oscillations and Stability}

The study of oscillations of relativistic stars is motivated by the
prospect of detecting such oscillations in electromagnetic or
gravitational wave signals. In the same way that helioseismology is
providing us with information about the interior of the Sun, the
observational identification of oscillation frequencies of
relativistic stars could constrain the high-density equation of
state~\cite{Ko97}. The oscillations could be excited after a core
collapse or during the final stages of a neutron star binary
merger. Rapidly rotating relativistic stars can become unstable to the
emission of gravitational waves.

When the displacement due to the oscillations of an equilibrium star
are small compared to its radius, it will suffice to approximate them
as linear perturbations. Such perturbations can be described in two
equivalent ways. In the Lagrangian approach, one studies the changes
in a given fluid element as it oscillates about its equilibrium
position. In the Eulerian approach, one studies the change in fluid
variables at a fixed point in space. Both approaches have their
strengths and weaknesses.

In the Newtonian limit, the Lagrangian approach has been used to
develop variational principles~\cite{LBO67,FS78}, but the Eulerian
approach proved to be more suitable for numerical computations of mode
frequencies and eigenfunctions~\cite{IM85,M85,IL90,IL91a,IL91b}.
Clement~\cite{C81} used the Lagrangian approach to obtain axisymmetric
normal modes of rotating stars, while nonaxisymmetric solutions were
obtained in the Lagrangian approach by Imamura {\it et
  al.}~\cite{IFD85} and in the Eulerian approach by Managan~\cite{M85}
and Ipser and Lindblom~\cite{IL90}. While a lot has been learned from
Newtonian studies, in the following we will focus on the relativistic
treatment of oscillations of rotating stars.


\subsection{Quasi-normal modes of oscillation}

A general linear perturbation of the energy density in a static and
spherically symmetric relativistic star can be written as a sum of
quasi-normal modes that are characterized by the indices $(l, m)$ of
the spherical harmonic functions $Y_l^m$ and have angular and
time dependence of the form
\begin{equation}
  \delta \varepsilon \sim f(r) P_l^m(\cos \theta)
  e^{i(m\phi+\omega_{\mathrm{i}} t)},
\end{equation}
where $\delta$ indicates the Eulerian perturbation of a quantity,
$\omega_{\mathrm{i}}$ is the angular frequency of the mode as measured by a
distant inertial observer, $f(r)$ represents the radial dependence of
the perturbation, and $P_l^m(\cos \theta)$ are the associated Legendre
polynomials. Normal modes of nonrotating stars are degenerate in $m$
and it suffices to study the axisymmetric $(m=0)$ case.

The Eulerian perturbation in the fluid 4-velocity $\delta u^a$ can be
expressed in terms of vector harmonics, while the metric perturbation
$\delta g_{ab}$ can be expressed in terms of spherical, vector, and tensor
harmonics. These are either of ``polar'' or ``axial'' parity. Here,
parity is defined to be the change in sign under a combination of
reflection in the equatorial plane and rotation by $\pi$. A polar
perturbation has parity $(-1)^l$, while an axial perturbation has
parity $(-1)^{l+1}$. Because of the spherical background, the polar
and axial perturbations of a nonrotating star are completely
decoupled.

A normal mode solution satisfies the perturbed gravitational field
equation,
\begin{equation}
  \delta(G^{ab}-8 \pi T^{ab})=0,
  \label{dGab}
\end{equation}
and the perturbation of the conservation of the stress-energy tensor,
\begin{equation}
  \delta(\nabla_aT^{ab})=0,
\end{equation}
with suitable boundary conditions at the center of the star and at
infinity. The latter equation is decomposed into an equation for the
perturbation in the energy density $\delta \varepsilon$ and into
equations for the three spatial components of the perturbation in the
4-velocity $\delta u^a$. As linear perturbations have a gauge freedom,
at most six components of the perturbed field equation~(\ref{dGab})
need to be considered.

For a given pair $(l, m)$, a solution exists for any value of the
frequency $\omega_{\mathrm{i}}$, consisting of a mixture of ingoing and
outgoing wave parts. Outgoing quasi-normal modes are defined by the
discrete set of eigenfrequencies for which there are no incoming waves
at infinity. These are the modes that will be excited in various
astrophysical situations.

The main modes of pulsation that are known to exist in relativistic
stars have been classified as follows ($f_0$ and $\tau_0$ are typical
frequencies and damping times of the most important modes in the
nonrotating limit):

\begin{enumerate}
\item \emph{Polar fluid modes} are slowly damped modes analogous to the
  Newtonian fluid pulsations:
  \begin{itemize} 
  \item $f$(undamental)-modes: surface modes due to the interface
    between the star and its surroundings ($f_0 \sim 2 \mathrm{\ kHz}$,
    $\tau_0<1 \mathrm{\ s}$),
  \item $p$(ressure)-modes: nearly radial ($f_0 > 4 \mathrm{\ kHz}$, 
    $\tau_0 > 1 \mathrm{\ s}$),
  \item $g$(ravity)-modes: nearly tangential, degenerate at zero frequency in nonrotating isentropic stars; they have nonzero frequencies in     stars that are non-isentropic or that have a composition
    gradient or a first order phase transition ($f_0<500 \mathrm{\ Hz}$,
    $\tau_0 > 5 \mathrm{\ s}$). 
  \end{itemize}
\item \emph{Axial and hybrid fluid modes}:
  \begin{itemize}
  \item \emph{inertial} modes: degenerate at zero frequency in
    nonrotating stars. In a rotating star, some inertial modes are
    generically unstable to the CFS instability; they have frequencies
    from zero to kHz and growth times inversely proportional to a high
    power of the star's angular velocity. Hybrid inertial modes have
    both axial and polar parts even in the limit of no rotation.
  \item $r$(otation)-modes: a special case of inertial modes that
    reduce to the classical axial $r$-modes in the Newtonian limit.
    Generically unstable to the CFS instability with growth times as
    short as a few seconds at high rotation rates.
  \end{itemize}  
\item \emph{Polar and axial spacetime modes}:
  \begin{itemize}
  \item $w$(ave)-modes: Analogous to the quasi-normal modes of a
    black hole (very weak coupling to the fluid). High frequency,
    strongly damped modes ($f_0>6 \mathrm{\ kHz}$,
    $\tau_0 \sim 0.1 \mathrm{\ ms}$).
  \end{itemize}
\end{enumerate}

For a more detailed description of various types of oscillation modes,
see~\cite{KB97,KLH96,MacD88,CAR86,KScha98}.


\subsection{Effect of rotation on quasi-normal modes}

In a continuous sequence of rotating stars that includes a nonrotating
member, a quasi-normal mode of index $l$ is defined as the mode which,
in the nonrotating limit, reduces to the quasi-normal mode of the same
index $l$. Rotation has several effects on the modes of a corresponding
nonrotating star:
\begin{enumerate}
\item The degeneracy in the index $m$ is removed and a nonrotating
  mode of index $l$ is split into $2l+1$ different $(l, m)$ modes.
\item Prograde ($m<0$) modes are now different from retrograde ($m>0$)
  modes.
\item A rotating ``polar'' $l$-mode consists of a sum of purely polar
  and purely axial terms~\cite{SPHD}, e.g., for $l=m$,
  \begin{equation}
    P_l^{\mathrm{rot}} \sim \sum_{l'=0}^\infty(P_{l+2l'} +A_{l+2l' \pm 1}),
  \end{equation}
  that is, rotation couples a polar $l$-term to an axial $l \pm 1$
  term (the coupling to the $l+1$ term is, however, strongly favoured
  over the coupling to the $l-1$ term~\cite{CF91}). Similarly, for a
  rotating ``axial'' mode with $l=m$,
  \begin{equation}
    A_l^{\mathrm{rot}} \sim \sum_{l'=0}^\infty(A_{l+2l'} +P_{l+2l' \pm 1}).
  \end{equation}
\item Frequencies and damping times are shifted. In general,
  frequencies (in the inertial frame) of prograde modes increase,
  while those of retrograde modes decrease with increasing rate of
  rotation.
\item In rapidly rotating stars, \emph{apparent intersections} between
  higher order modes of different $l$ can occur. In such cases, the
  shape of the eigenfunction is used in the mode classification.
\end{enumerate}

In rotating stars, quasi-normal modes of oscillation have been
studied only in the slow rotation limit, in the post-Newtonian, and in
the Cowling approximations. The solution of the fully relativistic
perturbation equations for a rapidly rotating star is still a very
challenging task and only recently have they been solved for
zero-frequency (neutral) modes~\cite{SPHD,SF97}. First frequencies of
quasi-radial modes have now been obtained through 3D numerical time
evolutions of the nonlinear equations~\cite{Font02}.

\vspace{1 em}\noindent{\bf Going further:}~~ The equations that
describe oscillations of the solid crust of a rapidly rotating
relativistic star are derived by Priou in~\cite{Pr92}. The effects of
superfluid hydrodynamics on the oscillations of neutron stars have
been investigated by several authors, see, e.g.,
\cite{LM94,Comer99,Andersson01b,Andersson02,Andersson2004MNRAS.355..918A,Prix2004MNRAS.348..625P,Sidery2008MNRAS.385..335S,Samuelsson2009CQGra..26o5016S,Passamonti2009MNRAS.396..951P,Andersson2011MNRAS.416..118A,2012MNRAS.419..638P,Passamonti2016MNRAS.455.1489P}
and references therein.


\subsection{Axisymmetric perturbations}


\subsubsection{Secular and dynamical axisymmetric instability}
\label{axisyminstability}

Along a sequence of nonrotating relativistic stars with increasing
central energy density, there is always a model for which the mass
becomes maximum. The maximum-mass turning point marks the onset of an
instability in the fundamental radial pulsation mode of the star.

Applying the turning point theorem provided by Sorkin~\cite{So82},
Friedman, Ipser, and Sorkin~\cite{FIS88} provide a sufficient
condition for a secular axisymmetric instability of rotating stars,
when the mass becomes maximum along a sequence of constant angular
momentum. An equivalent criterion (implied in~\cite{FIS88}) is
provided by Cook et al.~\cite{CST92}: The secular axisymmetric
instability sets in when the angular momentum becomes minimum along a
sequence of constant rest mass. The instability first develops on a
secular timescale that is set by the time required for viscosity to
redistribute the star's angular momentum. This timescale is long
compared to the dynamical timescale and comparable to the spin-up time
following a pulsar glitch. Eventually, the star encounters the onset
of dynamical instability and collapses to a black hole
(see~\cite{Shibata00c} for recent numerical simulations). Thus, the
onset of the secular instability to axisymmetric perturbations
separates stable neutron stars from neutron stars that will undergo
collapse to a black hole. More recently, Takami et
al.~\cite{Takami2011MNRAS} investigated the dynamical stability of
rotating stars, computing numerically the neutral point of the fundamental, quasi-radial $F$-mode
frequency, which signals the onset of the dynamical
stability. In their simulations, they found that the $F$-mode
frequency can go through zero before a star reaches turning point.
Prabhu, Schiffrin and Wald~\cite{Prabhu:2016pei} investigate the
axisymmetric stability of rotating relativistic stars through a
variational principle and show that the sign of the canonical energy
gives a necessary and sufficient condition for dynamical instability.
In addition, they determine a lower bound for exponential growth.

Goussard et al.~\cite{GHZ96} extend the stability criterion to
hot proto-neutron stars with nonzero total entropy. In this case, the
loss of stability is marked by the configuration with minimum angular
momentum along a sequence of both constant rest mass and total
entropy. In the nonrotating limit, Gondek et al.~\cite{GHZ97}
compute frequencies and eigenfunctions of radial pulsations of hot
proto-neutron stars and verify that the secular instability sets in at
the maximum mass turning point, as is the case for cold neutron stars.


\subsubsection{Axisymmetric pulsation modes}

\begin{figure*}[htbp]
  \center
  \includegraphics[width=0.7\textwidth]{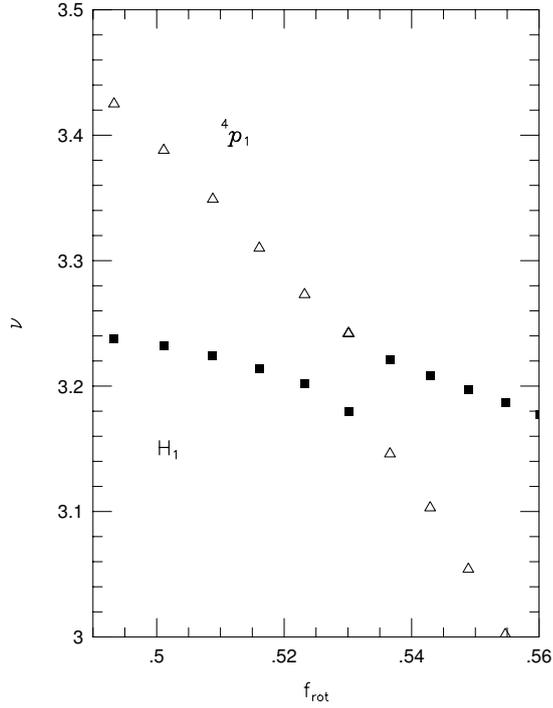}
  \caption{Apparent intersection (due to avoided crossing) of the
    axisymmetric first quasi-radial overtone ($H_1$) and the first
    overtone of the $l=4$ $p$-mode (in the Cowling
    approximation). Frequencies are normalized by
    $\sqrt{\rho_{\mathrm{c}}/4\pi}$, where $\rho_{\mathrm{c}}$ is the
    central energy density of the star. The rotational frequency
    $f_{\mathrm{rot}} $ at the mass-shedding limit is $0.597$ (in the
    above units). Along continuous sequences of computed frequencies,
    mode eigenfunctions are exchanged at the avoided
    crossing. Defining quasi-normal mode sequences by the shape of
    their eigenfunction, the $H_1$ sequence (filled boxes) appears to
    intersect with the ${}^4p_1$ sequence (triangle), but each
    sequence shows a discontinuity, when the region of apparent
    intersection is well resolved. In the notation ${}^l{\rm mode}_n$,
    the superscript indicates the $l$ of the perturbation, while the
    subscript indicates the harmonic overtone. (Figure reproduced with
    permission from~\cite{Yoshida00}, copyright by MNRAS.)}
  \label{fig_apparent}
\end{figure*}

Axisymmetric ($m=0$) pulsations in rotating relativistic stars could
be excited in a number of different astrophysical scenarios, such as
during core collapse, in star quakes induced by the secular spin-down
of a pulsar or during a large phase transition, or in the merger of two
relativistic stars in a binary system, among others. Due to rotational
couplings, the eigenfunction of any axisymmetric mode will involve a
sum of various spherical harmonics $Y_l^0$, so that even the
quasi-radial modes (with lowest order $l=0$ contribution) would, in
principle, radiate gravitational waves.

Quasi-radial modes in rotating relativistic stars have been studied by
Hartle and Friedman~\cite{Hartle75} and by Datta {\it et
  al.}~\cite{Datta98b} in the slow rotation approximation. Yoshida and
Eriguchi~\cite{Yoshida00} study quasi-radial modes of rapidly rotating
stars in the relativistic Cowling approximation and find that apparent
intersections between quasi-radial and other axisymmetric modes can
appear near the mass-shedding limit (see Figure~\ref{fig_apparent}).
These apparent intersections are due to \emph{avoided crossings}
between mode sequences, which are also known to occur for axisymmetric
modes of rotating Newtonian stars. Along a continuous sequence of
computed mode frequencies an avoided crossing occurs when another
sequence is encountered. In the region of the avoided crossing, the
eigenfunctions of the two modes become of mixed character. Away from
the avoided crossing and along the continuous sequences of computed
mode frequencies, the eigenfunctions are exchanged. However, each
``quasi-normal mode'' is characterized by the shape of its
eigenfunction and thus, the sequences of computed frequencies that
belong to particular quasi-normal modes are discontinuous at avoided
crossings (see Figure~\ref{fig_apparent} for more details). The
discontinuities can be found in numerical calculations, when
quasi-normal mode sequences are well resolved in the region of avoided
crossings. Otherwise, quasi-normal mode sequences will appear as
intersecting.

\begin{figure*}[htbp]
  \center
  \includegraphics[width=0.7\textwidth]{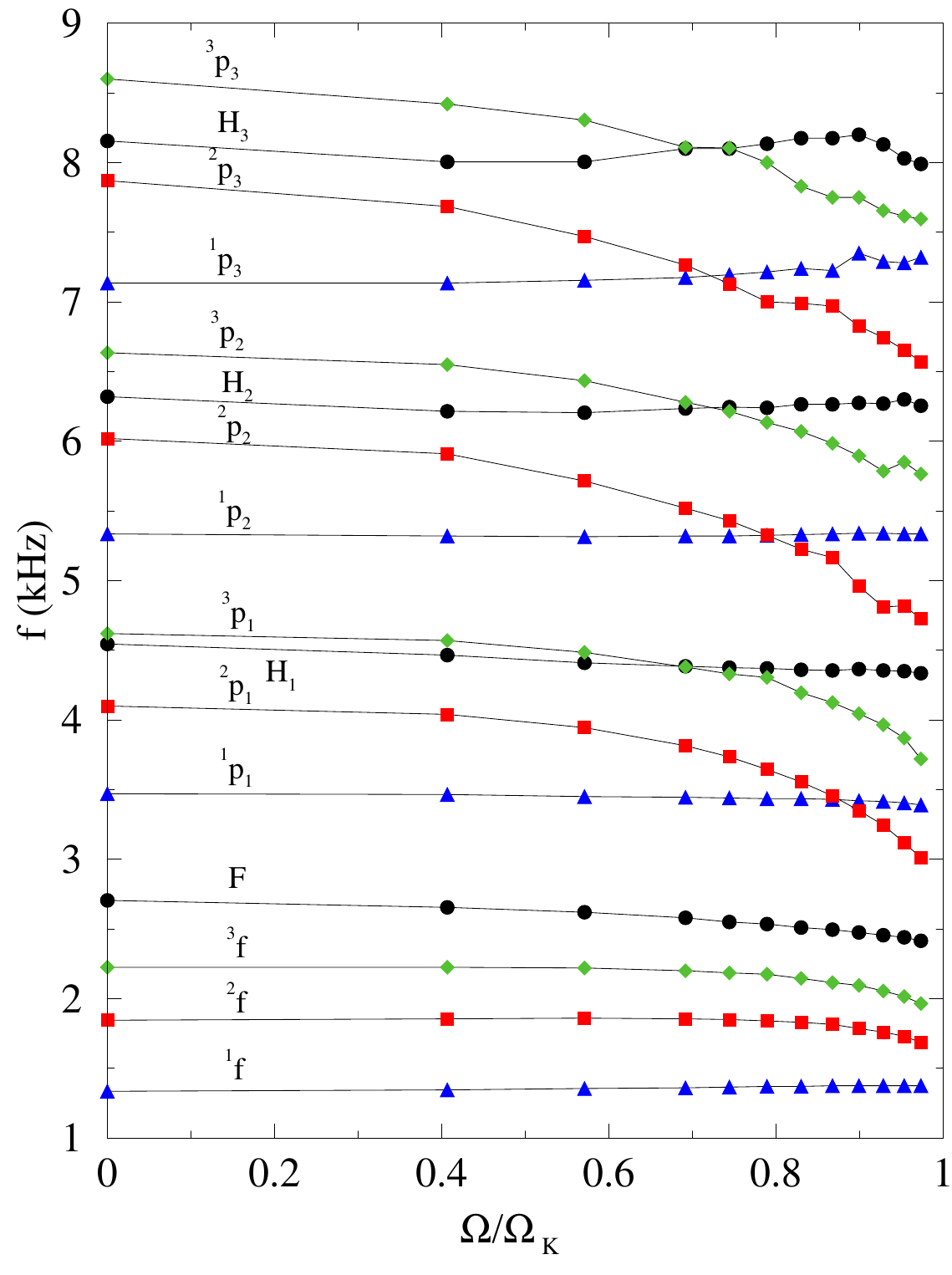}
  \caption{Frequencies of several axisymmetric modes along a
    sequence of rapidly rotating relativistic polytropes of $N=1.0$,
    in the Cowling approximation. On the horizontal axis, the angular
    velocity of each model is scaled to the angular velocity of the
    model at the mass-shedding limit. Lower order modes are weakly
    affected by rapid rotation, while higher order modes show apparent
    mode intersections. (Image reproduced with permission from~\cite{Font01}, copyright by MNRAS.)}
  \label{fig:axisym}
\end{figure*}

Several axisymmetric modes have recently been computed for rapidly
rotating relativistic stars in the Cowling approximation, using
time evolutions of the nonlinear hydrodynamical
equations~\cite{Font01} (see~\cite{Font00} for a description of the 2D
numerical evolution scheme). As in~\cite{Yoshida00}, Font {\it et
  al.}~\cite{Font01} find that apparent mode intersections are common
for various higher order axisymmetric modes (see
Figure~\ref{fig:axisym}). Axisymmetric inertial modes also appear in
the numerical evolutions.

The first fully relativistic frequencies of quasi-radial modes for
rapidly rotating stars (without assuming the Cowling approximation)
have been obtained recently, again through nonlinear time
evolutions~\cite{Font02} (see Section~\ref{pulsrot}).

\vspace{1 em}\noindent{\bf Going further:}~~
The stabilization, by an external gravitational field, of a
relativistic star that is marginally stable to axisymmetric
perturbations is discussed in~\cite{Th97}.


\subsection{Nonaxisymmetric perturbations}


\subsubsection{Nonrotating limit}

Thorne, Campolattaro, and Price, in a series of
papers~\cite{TC67,PT69,Th69}, initiated the computation of nonradial
modes by formulating the problem in the Regge--Wheeler (RW)
gauge~\cite{RW} and numerically computing nonradial modes for a number
of neutron star models. A variational method for obtaining
eigenfrequencies and eigenfunctions has been constructed by Detweiler
and Ipser~\cite{DI73}. Lindblom and Detweiler~\cite{LD83} explicitly
reduced the system of equations to four first order ordinary
differential equations and obtained more accurate eigenfrequencies and
damping times for a larger set of neutron star models. They later
realized that their system of equations is sometimes singular inside
the star and obtained an improved set of equations, which is free of
this singularity~\cite{DL85}.

Chandrasekhar and Ferrari~\cite{CF91} expressed the nonradial
pulsation problem in terms of a fifth order system in a diagonal
gauge, which is formally independent of fluid variables. Thus, they
reformulate the problem in a way analogous to the scattering of
gravitational waves off a black hole. Ipser and Price~\cite{IP91} show
that in the RW gauge, nonradial pulsations can be described by a
system of two second order differential equations, which can also be
independent of fluid variables. In addition, they find that the
diagonal gauge of Chandrasekhar and Ferrari has a remaining gauge
freedom which, when removed, also leads to a fourth order system of
equations~\cite{PI91}.

In order to locate purely outgoing wave modes, one has to be able to
distinguish the outgoing wave part from the ingoing wave part at
infinity. This is typically achieved using analytic approximations of
the solution at infinity.

$W$-modes pose a more challenging numerical problem because they are
strongly damped and the techniques used for $f$- and $p$-modes fail to
distinguish the outgoing wave part. The first accurate numerical solutions
were obtained by Kokkotas and Schutz~\cite{Kokkotas92}, followed
by Leins, Nollert, and Soffel~\cite{Leins93}. Andersson, Kokkotas, and
Schutz~\cite{AKS95} successfully combine a redefinition of variables
with a complex-coordinate integration method, obtaining highly
accurate complex frequencies for $w$-modes. In this method, the
ingoing and outgoing solutions are separated by numerically
calculating their analytic continuations to a place in the
complex-coordinate plane, where they have comparable amplitudes. Since
this approach is purely numerical, it could prove to be suitable for
the computation of quasi-normal modes in rotating stars, where
analytic solutions at infinity are not available.

The non-availability of asymptotic solutions at infinity in the case
of rotating stars is one of the major difficulties for computing
outgoing modes in rapidly rotating relativistic stars. A method that
may help to overcome this problem, at least to an acceptable
approximation, has been found by Lindblom, Mendell, and
Ipser~\cite{LMI97}. The authors obtain approximate near-zone boundary
conditions for the outgoing modes that replace the outgoing wave
condition at infinity and that enable one to compute the
eigenfrequencies with very satisfactory accuracy. First, the pulsation
equations of polar modes in the Regge--Wheeler gauge are reformulated
as a set of two second order radial equations for two potentials --
one corresponding to fluid perturbations and the other to the
perturbations of the spacetime. The equation for the spacetime
perturbation reduces to a scalar wave equation at infinity and to
Laplace's equation for zero-frequency solutions. From these, an
approximate boundary condition for outgoing modes is constructed and
imposed in the near zone of the star (in fact, on its surface) instead
of at infinity. For polytropic models, the near-zone boundary
condition yields $f$-mode eigenfrequencies with real parts accurate to
0.01\,--\,0.1\% and imaginary parts with accuracy at the 10\,--\,20\%
level, for the most relativistic stars. If the near zone boundary
condition can be applied to the oscillations of rapidly rotating
stars, the resulting frequencies and damping times should have
comparable accuracy.


\subsubsection{Slow rotation approximation}

The slow rotation approximation is useful for obtaining a first
estimate of the effect of rotation on the pulsations of relativistic
stars. To lowest order in rotation, a polar $l$-mode of an initially
nonrotating star couples to an axial $l \pm 1$ mode in the presence of
rotation. Conversely, an axial $l$-mode couples to a polar $l \pm 1$
mode as was first discussed by Chandrasekhar and Ferrari~\cite{CF91}.

The equations of nonaxisymmetric perturbations in the slow rotation
limit are derived in a diagonal gauge by Chandrasekhar and
Ferrari~\cite{CF91}, and in the Regge--Wheeler gauge by
Kojima~\cite{Koj92,Koj93}, where the complex frequencies $\sigma =
\sigma_R + i \sigma_I$ for the $l=m$ modes of various polytropes are
computed. For counterrotating modes, both $\sigma_R$ and $\sigma_I$
decrease, tending to zero, as the rotation rate increases (when
$\sigma$ passes through zero, the star becomes unstable to the CFS
instability). Extrapolating $\sigma_R$ and $\sigma_I$ to higher
rotation rates, Kojima finds a large discrepancy between the points
where $\sigma_R$ and $\sigma_I$ go through zero. This shows that the
slow rotation formalism cannot accurately determine the onset of the
CFS instability of polar modes in rapidly rotating neutron stars.

In~\cite{Koj93b}, it is shown that, for slowly rotating stars, the
coupling between polar and axial modes affects the frequency of $f$-
and $p$-modes only to second order in rotation, so that, in the slow
rotation approximation, to ${\cal O}( \Omega)$, the coupling can be
neglected when computing frequencies. This result was already known
from the original work of Hartle and
Thorne~\cite{HT1972ApJ...176..177H}, where it was noted that a
reversal of the direction of rotation cannot change the shape of the
mode or its frequency.

The linear perturbation equations in the slow rotation approximation
have been derived in a new gauge by Ruoff, Stavridis, and
Kokkotas~\cite{Ruoff02b}. Using the Arnowitt-Deser-Misner (ADM)
formalism~\cite{ADM2008}, a first order hyperbolic evolution system is
obtained, which is suitable for numerical integration without further
manipulations (as was required in the Regge--Wheeler gauge). In this
gauge (which is related to a gauge introduced for nonrotating stars
in~\cite{Battiston71}), the symmetry between the polar and axial
equations becomes directly apparent.

The case of relativistic inertial modes is different, as these modes
have both axial and polar parts at order ${\cal O}(\Omega)$, and the
presence of continuous bands in the spectrum (at this order in the
rotation rate) has led to a series of detailed investigations of the
properties of these modes (see~\cite{Kokkotas02} for a review).
Ruoff, Stavridis, and Kokkotas~\cite{Ruoff02} finally show that the
inclusion of both polar and axial parts in the computation of
relativistic $r$-modes, at order ${\cal O}(\Omega)$, allows for
discrete modes to be computed, in agreement with
post-Newtonian~\cite{Lockitch01} and nonlinear,
rapid-rotation~\cite{Stergioulas01} calculations.


\subsubsection{Post-Newtonian approximation}

A step toward the solution of the perturbation equations in full
general relativity has been taken by Cutler and
Lindblom~\cite{Cu91,CL92,Li95}, who obtain frequencies for the $l=m$
$f$-modes in rotating stars in the first post-Newtonian (1-PN)
approximation. The perturbation equations are derived in the
post-Newtonian formalism (see~\cite{Blanchet03}), i.e., the equations
are separated into equations of consistent order in $1/c$.

Cutler and Lindblom show that in this scheme, the perturbation of the
1-PN correction of the four-velocity of the fluid can be obtained
analytically in terms of other variables; this is similar to the
perturbation in the three-velocity in the Newtonian Ipser--Managan
scheme. The perturbation in the 1-PN corrections are obtained by
solving an eigenvalue problem, which consists of three second order
equations, with the 1-PN correction to the eigenfrequency of a mode
$\Delta \omega$ as the eigenvalue.

Cutler and Lindblom obtain a formula that yields $\Delta \omega$ if
one knows the 1-PN stationary solution and the solution to the
Newtonian perturbation equations. Thus, the frequency of a mode in the
1-PN approximation can be obtained without actually solving the 1-PN
perturbation equations numerically. The 1-PN code was checked in the
nonrotating limit and it was found to reproduce the exact general
relativistic frequencies for stars with $M/R=0.2$, obeying an $N=1$
polytropic EOS, with an accuracy of 3\,--\,8\%.

Along a sequence of rotating stars, the frequency of a mode is
commonly described by the ratio of the frequency of the mode in the
comoving frame to the frequency of the mode in the nonrotating limit.
For an $N=1$ polytrope and for $M/R=0.2$, this frequency ratio is
reduced by as much as 12\% (for the fundamental $l=m$ modes) in the
1-PN approximation compared to its Newtonian counterpart -- keeping
the gravitational mass fixed -- which is representative of the effect
that general relativity has on the frequency of quasi-normal modes in
rotating stars.


\subsubsection{Cowling approximation}

In several situations, the frequency of pulsations in relativistic
stars can be estimated even if one completely neglects the
perturbation in the gravitational field, i.e., if one sets
$\delta g_{ab}=0$ in the perturbation equations~\cite{MVS83}. In this
approximation, the pulsations are described only by the perturbation
in the fluid variables, and the scheme works quite well for $f$, $p$,
and $r$-modes~\cite{LS90}. A different version of the Cowling
approximation, in which $\delta g_{tr}$ is kept nonzero in the
perturbation equations, has been suggested to be more suitable for
$g$-modes~\cite{Fi88}, since these modes could have large fluid
velocities, even though the variation in the gravitational field is
weak.

Yoshida and Kojima~\cite{YK97} examine the accuracy of the
relativistic Cowling approximation in slowly rotating stars. The
first order correction to the frequency of a mode depends only on the
eigenfrequency and eigenfunctions of the mode in the absence of
rotation and on the angular velocity of the star. The eigenfrequencies
of $f$, $p_1$, and $p_2$ modes for slowly rotating stars with $M/R$
between 0.05 and 0.2 are computed (assuming polytropic EOSs with $N=1$
and $N=1.5$) and compared to their counterparts in the slow rotation
approximation.

For the $l=2$ $f$-mode, the relative error in the eigenfrequency
because of the Cowling approximation is 30\% for weakly
relativistic stars ($M/R=0.05$) and about 15\% for stars with
$M/R=0.2$; the error decreases for higher $l$-modes. For the $p_1$
and $p_2$ modes the relative error is similar in magnitude but it is
smaller for less relativistic stars. Also, for $p$-modes, the Cowling
approximation becomes more accurate for increasing radial mode number.

\begin{figure*}[htbp]
  \center
  \includegraphics[width=0.95\textwidth]{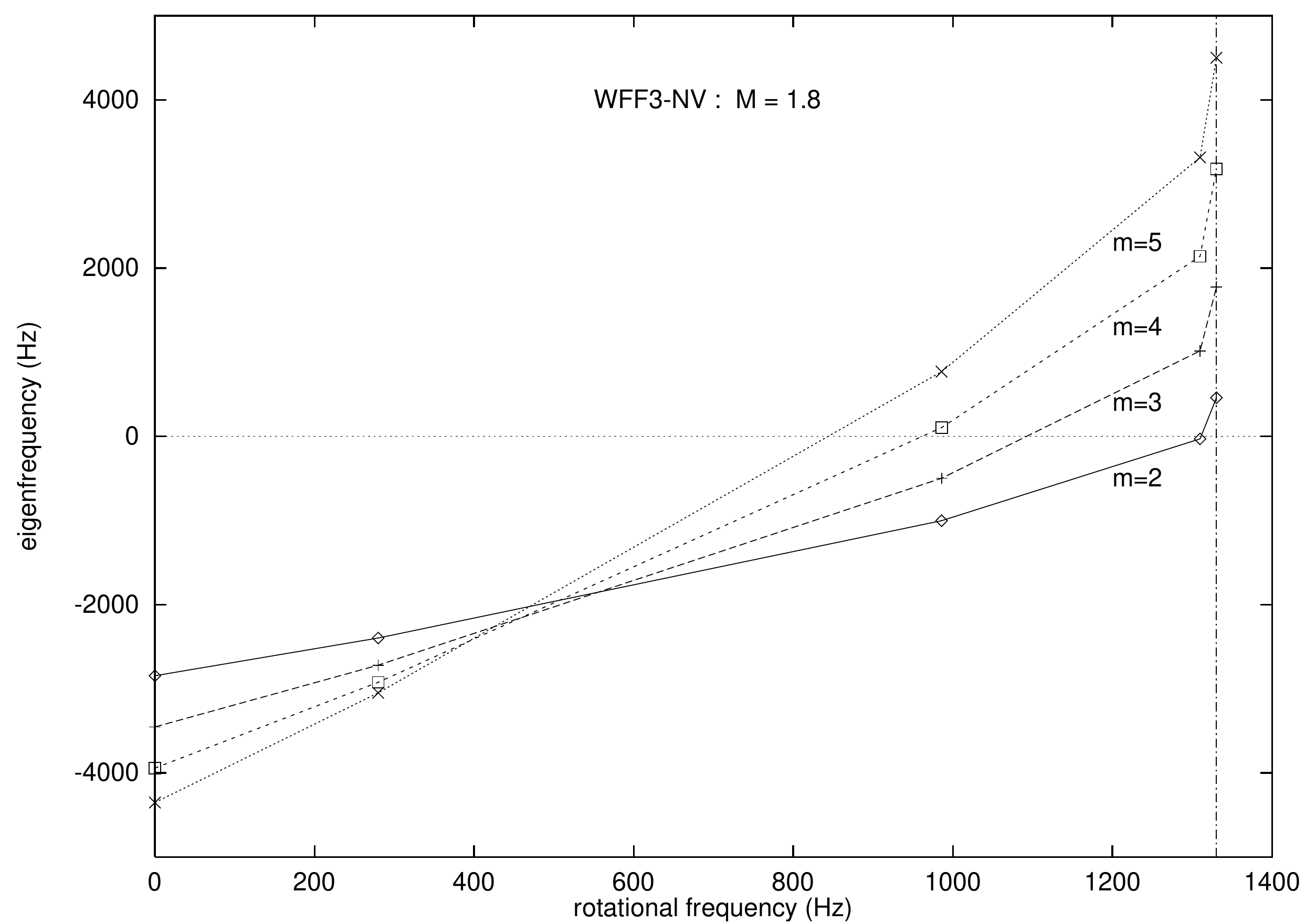}
  \caption{Eigenfrequencies (in the Cowling approximation) of
    $f$-modes along a $M = 1.8\,M_{\odot}$ sequence of models,
    constructed with the WFF3-NV EOS. The vertical line corresponds to
    the frequency of rotation of the model at the mass-shedding limit
    of the sequence. (Image reproduced with permission
    from~\cite{Yoshida99}, copyright by Ap. J.)}
  \label{fig:cowling}
\end{figure*}

As an application, Yoshida and Eriguchi~\cite{YE97,Yoshida99} use the
Cowling approximation to estimate the onset of the $f$-mode CFS
instability in rapidly rotating relativistic stars and to compute
frequencies of $f$-modes for several realistic equations of state (see
Figure~\ref{fig:cowling}).

The effects of rotation on the frequencies of the quasi-normal modes
in the case of a proto-neutron star are studied by Ferrari et
al. in~\cite{FerrariGua2004MNRAS.350..763F}, where the growth time of
unstable g modes is estimated based on a post-Newtonian formula.

Perturbations in Newtonian axisymmetric background configurations but
accounting for the superfluid hydrodynamics are studied by Passamonti
et al.~\cite{Passamonti2009MNRAS.396..951P}. Passamonti et
al.~\cite{Passamonti2009MNRAS.394..730P} study $g$-modes for stratified
rapidly rotating neutrons stars also in a Newtonian framework. In a
follow-up work, Gaertig and Kokkotas~\cite{Gaertig2009PhRvD..80f4026G}
extend the latter work in general relativity finding good
qualitative agreement with the Newtonian results.

Yoshida et al.~\cite{Yoshida2005} adopt the Cowling approximation to
study $r$-mode oscillations of rapidly and rigidly rotating,
barotropic, relativistic stars. Their formulation and method is the
general relativistic extension of the Yoshida and Eriguchi
method~\cite{YE97}, which amounts to the solution of a second-order,
time-indepedent partial differential equation for the eigenvalue
problem. Using the method they obtain the frequencies of the $r$-mode
oscillations as a function of $T/|W|$, and find that the normalized
oscillation frequencies $\sigma/\Omega$ (where $\Omega$ is the stellar
rotation frequency) scale almost linearly with $T/|W|$ and decrease as
$T/|W|$ increases.

Gaertig and Kokkotas~\cite{Gaertig2008PhRvD..78f4063G,Gaertig:2010kc}
adopt the Cowling approximation to study $m=\pm 2$ nonaxisymmetric
oscillations and instabilities of rapidly rotating general
relativistic, polytropic stars using a time-dependent approach for the
first time. They introduce a formulation for the linearized equations
of motion for a perfect fluid appropriate for a rapidly rotating star
in a comoving frame using surface fitted coordinates. The equations of
state they adopt have $(\Gamma,K)=(2,100)$ (labeled as the EOS BU),
$(\Gamma,K)=(2.46,0.00936)$ (labeled as the EOS A), and
$(\Gamma,K)=(2.34,0.0195)$ (labeled as the EOS II). The values for $K$
are in geometrized units with $M_\odot=1$. From the BU EOS they adopt
a neutron star model with $M=1.4M_\odot$ and circumferential radius
$R=14.15$ km. For the A (II) EOS they consider a model with
$M=1.61M_\odot$ ($M=1.91M_\odot$) and $R=9.51$ km ($R=11.68$ km).

For $l=2$ polar perturbations, Gaertig and Kokkotas find the
anticipated splitting of counter-rotating $m=2$ and corotating $m=-2$
f modes as the star is set to rotate at higher rates. They find that
even for rapidly rotating stars the higher the compactness of the star
the higher the $f$-mode frequency. They also discover an
equation-of-state independent fit for the $f$-mode frequencies
($\sigma_0$) in the corotating frame as a function of the rotation
frequency as follows
\be
\label{eq:asteroseismology}
\frac{\sigma}{\sigma_0} = 1.0 + C_{lm}^{(1)}\left(\frac{\Omega}{\Omega_K}\right) + C_{lm}^{(2)}\left(\frac{\Omega}{\Omega_K}\right)^2,
\ee 
where $\sigma_0$ is the $f$-mode frequency for a non-rotating star and
$\Omega_K$ the mass-shedding limit rotation frequency. From the numerical
calculations they find $C_{2-2}^{(1)} = -0.27$, $C_{2-2}^{(2)} =
-0.34$ and $C_{22}^{(1)} = 0.47$, $C_{22}^{(2)} = -0.51$. To transform
these frequencies to the stationary frame one must use $\sigma_{\rm
  stat} = \sigma_{\rm corot} -m\Omega$. Through this equation the
authors compute the stationary frame frequencies and are able to
determine the critical rotation rate at which $\sigma_{\rm stat}=0$
for $m=2$. For rotation rates higher than the critical one the
$f$-mode is retrograde in the corotating frame, and prograde in the
stationary frame, hence the mode becomes unstable to the
Chandrasekhar-Friedman-Schutz instability~\cite{C70,FS78} (see next
section). The authors are able to also study nonaxisymmetric axial
$l=2,m=2$ perturbations, i.e., $r$-modes, finding results that are in
excellent agreement with earlier results from nonlinear
general-relativistic simulations by Stergioulas and
Font~\cite{Stergioulas01}.

In follow up work Kr\"uger et al.~\cite{Krueger09} adopt the Cowling
approximation generalizing the approach of Gaertig and
Kokkotas~\cite{Gaertig2008PhRvD..78f4063G} to investigate
nonaxisymmetric oscillations of rapidly and differentially rotating
relativistic stars. Adopting polytropic equations of state, the
authors find that for nonaxisymmetric $f$-modes the higher the degree
of differential rotation the neutral point is reached at a lower value
of $T/|W|$, hence differential rotation favors the development of the
Chandrasekhar-Friedman-Schutz instability. Kr\"uger et al. also study
$r$-mode oscillations for which they find a discrete spectrum only, in
contrast to some earlier
studies~\cite{Koj97,Passamonti2008PhRvD..77b4029P} that found evidence
for a continuum spectrum (see also discussion in the next section).

\begin{figure*}[t]
    \center
      \includegraphics[width=0.485\textwidth]{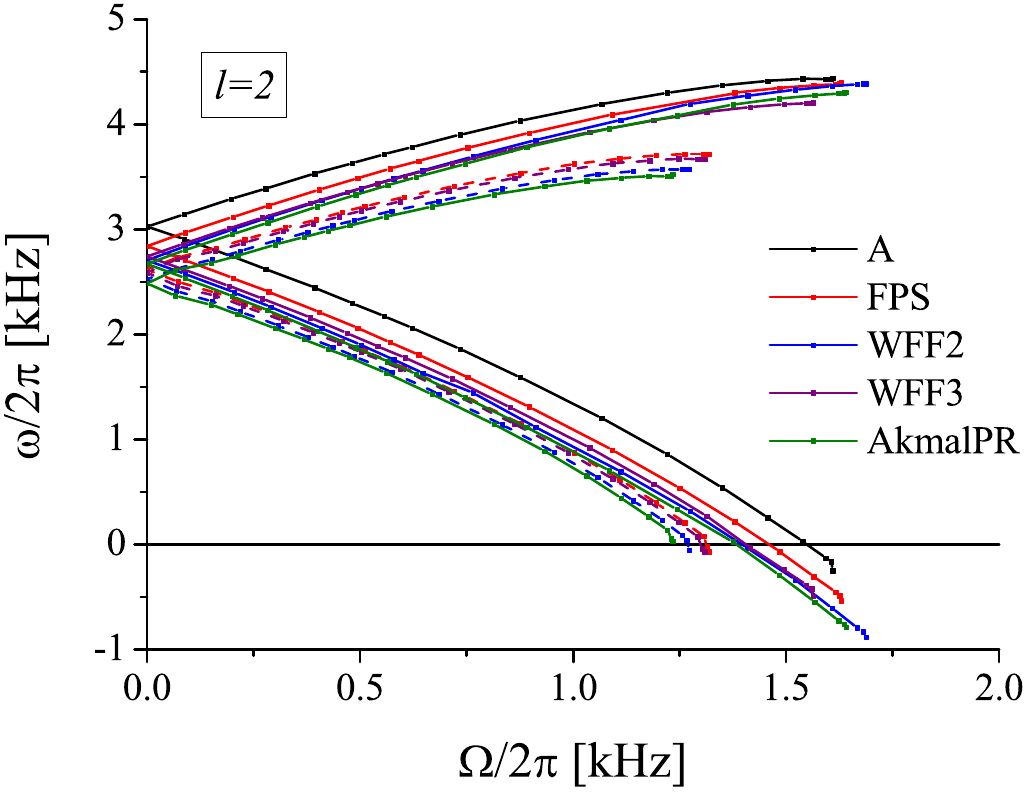}
      \includegraphics[width=0.455\textwidth]{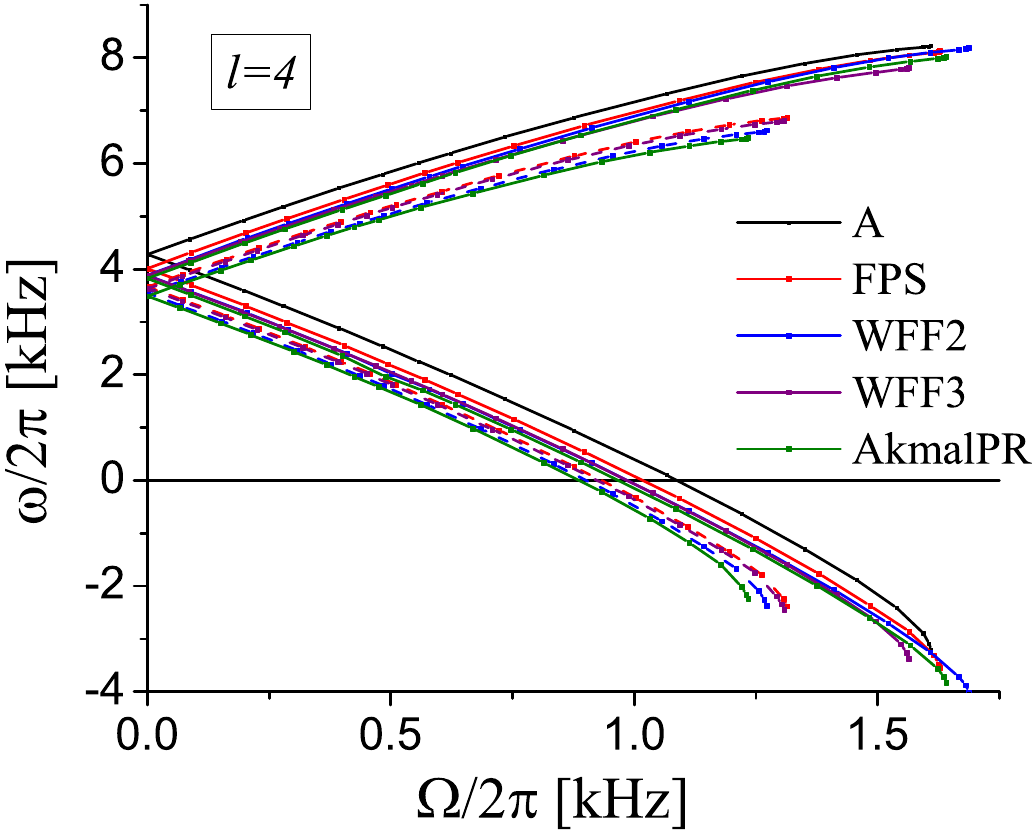}
    \caption{$f$-mode frequencies in the inertial frame as a function
      of the stellar rotation angular frequency for different EOSs
      labeled as A, FPS, WFF2, WFF3 and AkmalPR
      (see~\cite{DonevaGaert2013} for more details. Left: $l = | m | =
      2$. Right: $l = | m | = 4$.  (Image reproduced with permission from
      \cite{DonevaGaert2013}, copyright by APS.)  }
    \label{fig:fmode_GR_real_EOS}
\end{figure*}

A substantial step forward in the field of gravitational wave
asteroseismology is taken by the work of Doneva et
al.~\cite{DonevaGaert2013} who extend the work by Gaertig and
Kokkotas~\cite{Gaertig2008PhRvD..78f4063G} in several different ways:
a) they consider realistic equations of state, b) they treat higher
modes up to $l=m=4$, c) they address the problem of inferring the
properties of a neutron star from observations of observed frequencies
and damping timescales using $f$-modes $l>2$. The authors use the {\tt
  rns} code to build the equilibrium rotating configurations adopting
five realistic EOSs, constructing two constant-central-density
rotational sequences up to the mass-shedding limit for most of them.
For each EOS, the first sequence starts with a non-spinning star with
mass $1.4M_\odot$, and the second with a non-spinning star near the
TOV limit. The characteristic mode splitting of the $f$-modes for
different values of $m$ as the stars are spun-up can be seen in
Fig.~\ref{fig:fmode_GR_real_EOS} with the upper branch being the
stable one and the lower branch being (potentially) unstable.

When normalizing the oscillation frequencies in the corotating frame
as in Eq.~\eqref{eq:asteroseismology}, they find that
Eq.~\eqref{eq:asteroseismology} is still a good approximation with
interpolation parameters for the unstable modes given by
\begin{itemize}

\item $l=m=2$:\ \ \ $C_{22}^{(1)} = 0.402$ and $C_{22}^{(2)} = -0.406$, \\ with $\frac{\sigma_{0,l=2}}{2\pi}[{\rm kHz}]=1.562+1.151\left(\frac{\bar M_0}{\bar R_0^3}\right)^{1/2}$.

\item $l=m=3$:\ \ \ $C_{33}^{(1)} = 0.373$ and $C_{33}^{(2)} = -0.485$, \\ with $\frac{\sigma_{0,l=3}}{2\pi}[{\rm kHz}]=1.764+1.577\left(\frac{\bar M_0}{\bar R_0^3}\right)^{1/2}$.

\item $l=m=4$:\ \ \ $C_{44}^{(1)} = 0.360$ and $C_{44}^{(2)} = -0.543$, \\ with $\frac{\sigma_{0,l=4}}{2\pi}[{\rm kHz}]=1.958+1.898\left(\frac{\bar M_0}{\bar R_0^3}\right)^{1/2}$.

\end{itemize}

In the expressions for $\sigma_{0,l}$ above the mass and radius are
normalized as $\bar M_0= M_0/1.4M_\odot$ and $\bar R_0= R_0/10{\rm
  km}$, and stand for the masses and radii of the nonspining
configurations, respectively. Interestingly, Doneva et al. find a
universal fitting relation for all stable-mode ($l=-m=2$, $l=-m=3$ and
$l=-m=4$) oscillation frequencies, where the coefficients in
Eq.~\eqref{eq:asteroseismology} are given by $C_{lm}^{(1)} = -0.235$
and $C_{lm}^{(2)} = -0.358$. The Kepler limit $\Omega_K$ in
Eq.~\eqref{eq:asteroseismology} is well described (within 2\%
accuracy) by $(1/2\pi)\Omega_K[{\rm kHz}]= 1.716(\bar M_0/\bar
R_0^3)^{1/2}-0.189$. Finally, the authors find that the masses and
radii of the rotating configurations are well-described as functions
of the masses and radii of the non-spinning counterparts and the
angular frequency by the following expressions
\be
\frac{M}{M_0}=0.991+9.36\times 10^{-3}e^{3.28\frac{\Omega}{\Omega_K}}
\ee
and
\be
\frac{R}{R_0}=0.997+2.77\times 10^{-3}e^{4.74\frac{\Omega}{\Omega_K}}
\ee

Doneva et al.~\cite{DonevaGaert2013} also provide approximate
relations for the damping (growth) timescale of the stable (unstable)
modes. These are relations are approximate because the authors adopt
the Cowling approximation and as a result they can only estimate the
gravitational wave damping timescale. The expressions are given in the
form
\be
\left(\frac{\tau_l}{\tau_0}\right)^{1/2l}=\sum_{n=0}^3 c_{ln}\left(\frac{\sigma}{\sigma_0}\right)^n, \ \ l=2,3,4
\ee
which the authors argue will remain valid even if the Cowling
approximation is lifted because these involve properly rescaled
quantities. The authors also provide fits for non-spinning limit
$\tau_0$ for the different l modes which scale as $\tau_0^{-1} =
(1/\bar R_0)(\bar M_0/\bar R_0)^{l+1}(\bar c_{l0}+\bar c_{l1}(\bar
M_0/\bar R_0)$. With these approximately EOS-independent expressions
for the $f$-mode frequencies and damping timescales, one can obtain
the mass and the radius of a rotating neutron star following a
determination of the nonrotating-limit parameters.  However, as the
authors point out, not all combinations of frequencies and damping
times are capable of providing information about the neutron star
mass and radius. For example, measuring two frequencies alone can
provide information for $\Omega$ and $M/R^3$, but not of $M$ and $R$
separately. Therefore, measuring the damping timescale of at least one
$f$-mode is necessary to break the degeneracy and estimate $M$ and $R$
separately. But, this is going to be a rather challenging task because
of the long integration times required in noisy detector data.

\vspace{1 em}\noindent{\bf Going further:} A new approach for
performing asteroseismology for neutron stars was introduced by Doneva
and Kokkotas~\cite{Doneva:2015jba}. The $f$-mode oscillation
frequencies in modified gravity theories have recently been addressed
by Staykov et al.~\cite{StaykovDon2015}. The authors adopt the Cowling
approximation and treat $R^2$ gravity as a first case. The authors
derive the $R^2$-gravity asteroseismology relations which they find
they are approximately EOS-independent as in the case of general
relativity. By varying the $R^2$-gravity coupling constant within the
range allowed by current observations, the authors estimate that the
$R^2$-gravity asteroseismology relations deviate from those in general
relativity by up to 10\%. This implies that it will be difficult to
further constrain $R^2$-gravity via gravitational wave
asteroseismology. A study of $f$-modes of rapidly rotating stars in
a scalar-tensor theory of gravity are studied for the first time by
Yazadjiev, Doneva and Kokkotas~\cite{Yazadjiev:2017vpg}.


\subsection{Nonaxisymmetric instabilities}


\subsubsection{Introduction}

Rotating cold neutron stars, detected as pulsars, have a remarkably
stable rotation period. But, at birth or during accretion, rapidly
rotating neutron stars can be subject to various nonaxisymmetric
instabilities, which will affect the evolution of their rotation rate.

If a proto-neutron star has a sufficiently high rotation rate (so
that, e.g., $T/W > 0.27$ in the case of Maclaurin spheroids), it will
be subject to a dynamical instability driven by hydrodynamics and
gravity, typically referred to as the dynamical bar-mode
instability. Through the $l=2$ mode, the instability will deform the
star into a bar shape. This highly nonaxisymmetric configuration will
emit strong gravitational waves with frequencies in the kHz
regime. The development of the instability and the resulting waveform
have been computed numerically in the context of Newtonian gravity by
Houser et al.~\cite{HCS94} and in full general relativity by
Shibata et al.~\cite{Shibata00c} (see
Section~\ref{s:dynamical}).

At lower rotation rates, the star can become unstable to secular
nonaxisymmetric instabilities, driven by gravitational radiation or
viscosity. Gravitational radiation drives a nonaxisymmetric
instability when a mode that is retrograde in a frame corotating with
the star appears as prograde to a distant inertial observer, via the
Chandrasekhar-Friedman-Schutz (CFS) mechanism~\cite{C70,FS78}: A mode
that is retrograde in the corotating frame has negative angular
momentum, because the perturbed star has less angular momentum than
the unperturbed one. If, for a distant observer, the mode is prograde,
it removes positive angular momentum from the star, and thus the
angular momentum of the mode becomes increasingly negative.

The instability evolves on a secular timescale, during which the
star loses angular momentum via the emitted gravitational waves.
When the star rotates more slowly than a critical value, the mode
becomes stable and the instability proceeds on the longer
timescale of the next unstable mode, unless it is suppressed by
viscosity.

Neglecting viscosity, the CFS instability is generic in rotating
stars for both polar and axial modes. For polar modes, the
instability occurs only above some critical angular velocity,
where the frequency of the mode goes through zero in the inertial
frame. The critical angular velocity is smaller for increasing
mode number $l$. Thus, there will always be a high enough mode
number $l$ for which a slowly rotating star will be unstable.
Many of the hybrid inertial modes (and in particular the
relativistic $r$-mode) are generically unstable in all rotating
stars, since the mode has zero frequency in the inertial frame
when the star is nonrotating~\cite{A97,FM97}.

The shear and bulk viscosity of neutron star matter is able to
suppress the growth of the CFS instability except when the star
passes through a certain temperature window. In Newtonian
gravity, it appears that the polar mode CFS instability can occur
only in nascent neutron stars that rotate close to the mass-shedding
limit~\cite{IL91a,IL91b,IL92,YE95,Lindblom95}, but the
computation of neutral $f$-modes in full relativity~\cite{SPHD,SF97}
shows that relativity enhances the instability, allowing it to occur
in stars with smaller rotation rates than previously thought.

\vspace{1 em}\noindent{\bf Going further:}~~ A numerical method for the
analysis of the ergosphere instability in relativistic stars, which
could be extended to nonaxisymmetric instabilities of fluid modes, is
presented by Yoshida and Eriguchi in~\cite{YE96}.


\subsubsection{CFS instability of polar modes}

The existence of the CFS instability in rotating stars was first
demonstrated by Chandrasekhar~\cite{C70} in the case of the $l=2$ mode
in uniformly rotating, uniform density Maclaurin spheroids. Friedman
and Schutz~\cite{FS78} show that this instability also appears in
compressible stars and that all rotating, self-gravitating perfect
fluid configurations are generically unstable to the emission of
gravitational waves. In addition, they find that a nonaxisymmetric
mode becomes unstable when its frequency vanishes in the inertial
frame. Thus, zero-frequency outgoing modes in rotating stars are
neutral (marginally stable).

\begin{figure*}[htbp]
  \center
  \includegraphics[width=0.7\textwidth]{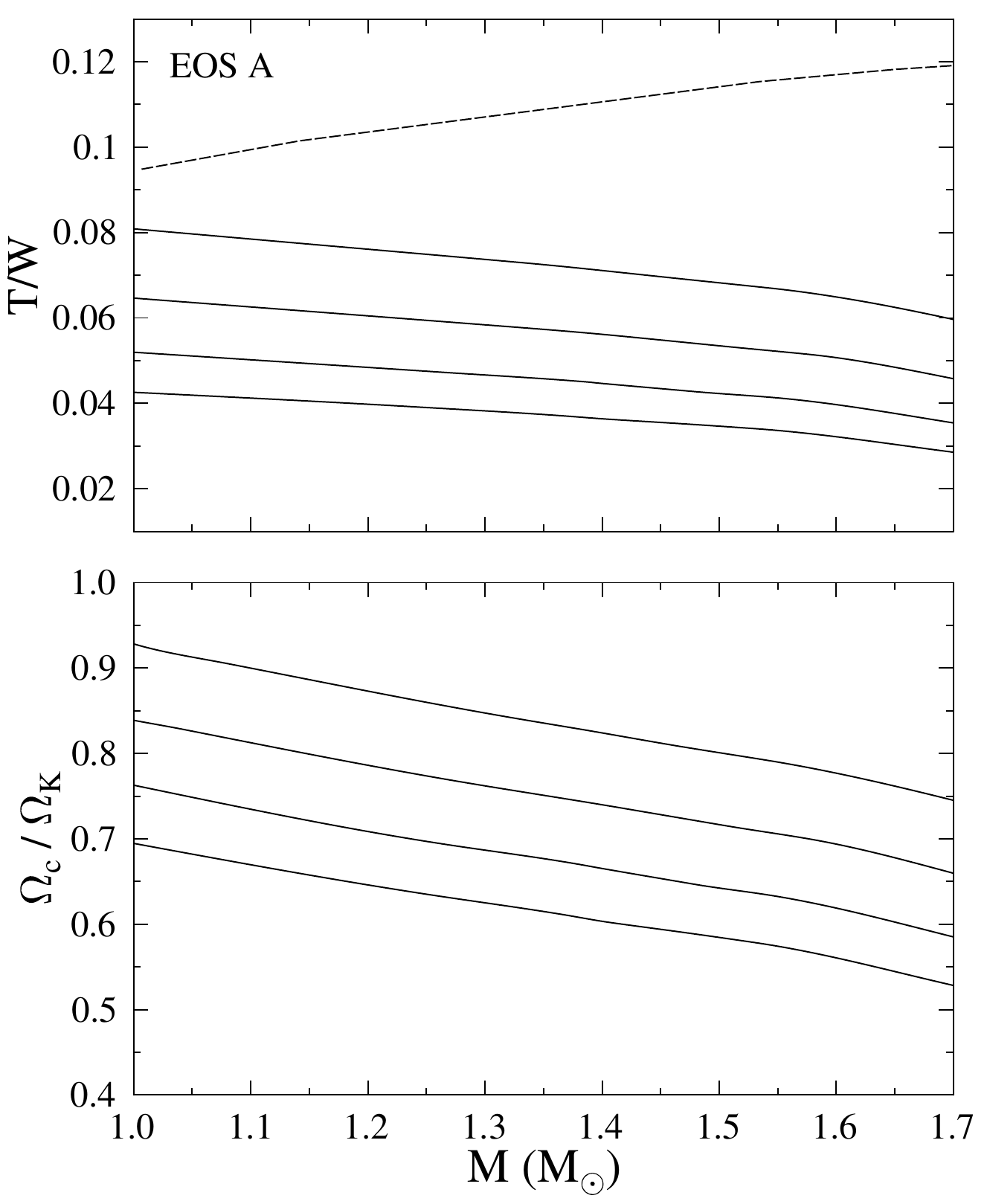}
  \caption{The $l=m$ neutral $f$-mode sequences for EOS A. Shown
    are the ratio of rotational to gravitational energy $T/W$ (upper
    panel) and the ratio of the critical angular velocity
    $\Omega_{\mathrm{c}}$ to the angular velocity at the mass-shedding
    limit for uniform rotation (lower panel) as a function of
    gravitational mass. The solid curves are the neutral mode
    sequences for $l=m=2, 3, 4$, and $5$ (from top to bottom), while
    the dashed curve in the upper panel corresponds to the
    mass-shedding limit for uniform rotation. The $l=m=2$ $f$-mode
    becomes CFS-unstable even at 85\% of the mass-shedding limit,
    for $1.4\,M_{\odot}$ models constructed with this
    EOS. (Image reproduced with permission from~\cite{MSB98}, copyright by Ap. J.)}
  \label{fig:msb}
\end{figure*}

In the Newtonian limit, neutral modes have been determined for several
polytropic EOSs~\cite{IFD85,M85,IL90,YE95}. The instability first
sets in through $l=m$ modes. Modes with larger $l$ become unstable at
lower rotation rates, but viscosity limits the interesting ones to $l
\leq5$.  For an $N=1$ polytrope, the critical values of $T/W$ for the
$l=3,4$, and 5 modes are 0.079, 0.058, and 0.045, respectively, and
these values become smaller for softer polytropes. The $l=m=2$ ``bar''
mode has a critical $T/W$ ratio of 0.14 that is almost independent of
the polytropic index. Since soft EOSs cannot produce models with high
$T/W$ values, the bar mode instability appears only for stiff
Newtonian polytropes of $N \leq 0.808$~\cite{Ja64,SL96}. In addition,
the viscosity-driven bar mode appears at the same critical $T/W$ ratio
as the bar mode driven by gravitational radiation~\cite{IM85} (we will
see later that this is no longer true in general relativity).

The post-Newtonian computation of neutral modes by Cutler and
Lindblom~\cite{CL92,Li95} has shown that general relativity tends to
strengthen the CFS instability. Compared to their Newtonian
counterparts, critical angular velocity ratios
$\Omega_{\mathrm{c}}/\Omega_0$ (where $\Omega_0=(3M_0/4R_0^3)^{1/2}$, and
$M_0$, $R_0$ are the mass and radius of the nonrotating star in the
sequence) are lowered by as much as 10\% for stars obeying the $N=1$
polytropic EOS (for which the instability occurs only for $l=m \geq 3$
modes in the post-Newtonian approximation).

In full general relativity, neutral modes have been determined for
polytropic EOSs of $N \geq 1.0$ by Stergioulas and
Friedman~\cite{SPHD,SF97}, using a new numerical scheme. The scheme
completes the Eulerian formalism developed by Ipser and Lindblom in
the Cowling approximation (where $\delta g_{ab}$ was
neglected)~\cite{IL92}, by finding an appropriate gauge in which the
time independent perturbation equations can be solved numerically for
$\delta g_{ab}$. The computation of neutral modes for polytropes of
$N=1.0$, 1.5, and 2.0 shows that relativity significantly strengthens
the instability. For the $N=1.0$ polytrope, the critical angular
velocity ratio $\Omega_{\mathrm{c}} / \Omega_{\mathrm{K}}$, where
$\Omega_{\mathrm{K}}$ is the angular velocity at the mass-shedding
limit at same central energy density, is reduced by as much as 15\%
for the most relativistic configuration (see Figure~\ref{fig:msb}). A
surprising result (which was not found in computations that used the
post-Newtonian approximation) is that the $l=m=2$ bar mode is unstable
even for relativistic polytropes of index $N=1.0$. The classical
Newtonian result for the onset of the bar mode instability
($N_{\mathrm{crit}} <0.808$) is replaced by
\begin{equation}
  N_{\mathrm{crit}} < 1.3
\end{equation}
in general relativity. For relativistic stars, it is evident that the
onset of the gravitational-radiation-driven bar mode does not coincide
with the onset of the viscosity-driven bar mode, which occurs at
larger $T/W$~\cite{BFG97}. The computation of the onset of the CFS
instability in the relativistic Cowling approximation by Yoshida and
Eriguchi~\cite{YE97} agrees qualitatively with the conclusions
in~\cite{SPHD,SF97}.

Morsink, Stergioulas, and Blattning~\cite{MSB98} extend the method
presented in~\cite{SF97} to a wide range of realistic equations of
state (which usually have a stiff high density region, corresponding
to polytropes of index $N=0.5\mbox{\,--\,}0.7$) and find that the
$l=m=2$ bar mode becomes unstable for stars with gravitational mass as
low as $1.0\mbox{\,--\,}1.2\,M_{\odot}$. For $1.4\,M_{\odot}$ neutron
stars, the mode becomes unstable at 80\,--\,95\% of the maximum
allowed rotation rate. For a wide range of equations of state, the
$l=m=2$ $f$-mode becomes unstable at a ratio of rotational to
gravitational energies $T/W \sim 0.08$ for $1.4\,M_{\odot}$ stars and
$T/W \sim 0.06$ for maximum mass stars. This is to be contrasted with
the Newtonian value of $T/W \sim 0.14$. The empirical formula
\begin{equation}
  \left( T/W \right)_2 = 0.115\mbox{\,--\,}0.048 \frac{M}{M_{\mathrm{max}}^{\mathrm{sph}}},
  \label{emp}
\end{equation}
where $M_{\mathrm{max}}^{\mathrm{sph}}$ is the maximum mass for a
spherical star allowed by a given equation of state, gives the
critical value of $T/W$ for the bar $f$-mode instability, with an
accuracy of 4\,--\,6\%, for a wide range of realistic EOSs.

\begin{figure*}[htbp]
  \center
  \includegraphics[width=0.7\textwidth]{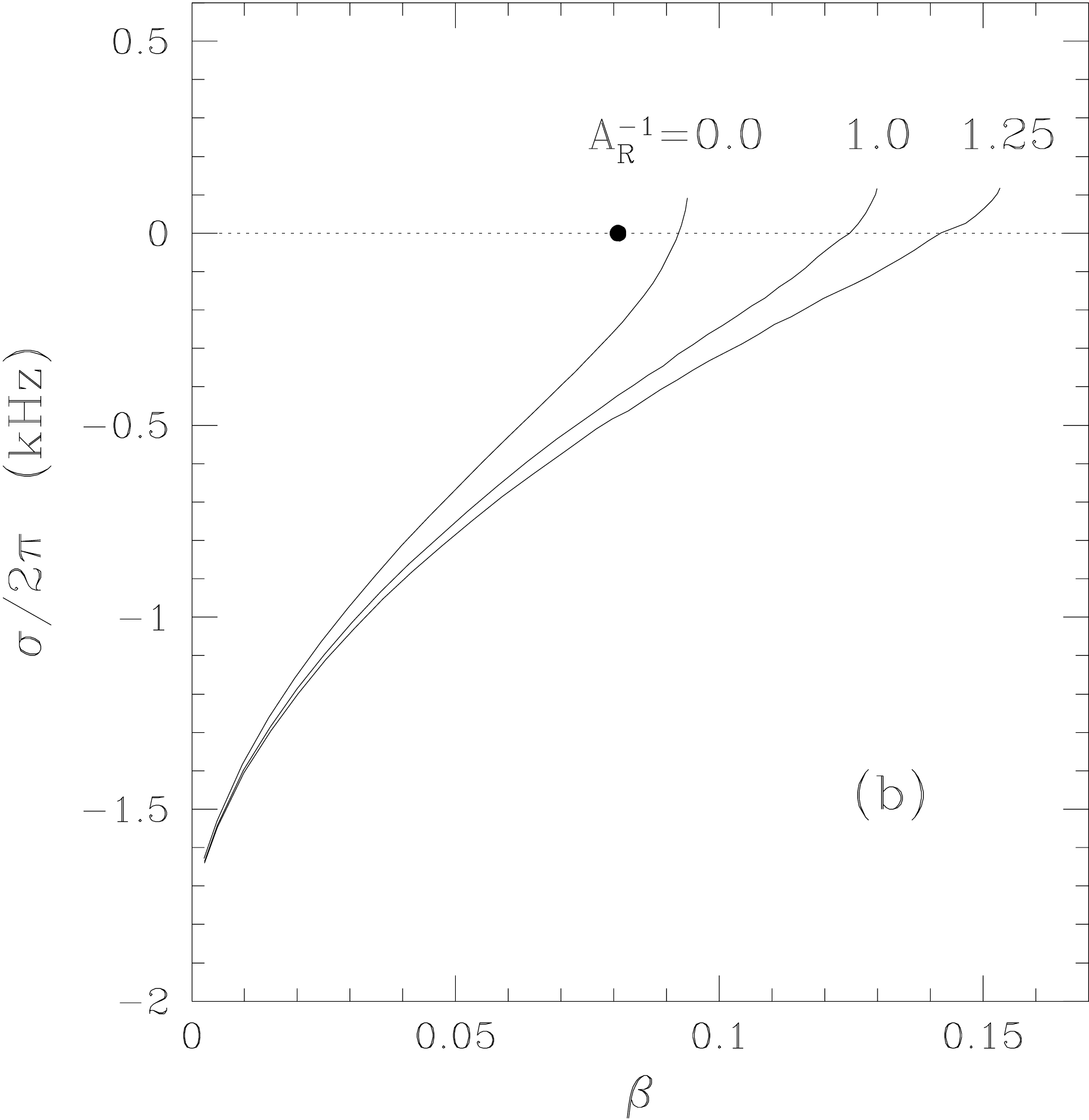}
  \caption{Eigenfrequencies (in the Cowling approximation) of the
    $m=2$ mode as a function of the parameter $\beta=T/|W|$ for three
    different sequences of differentially rotating neutron stars (the
    $A_{\mathrm{r}}^{-1}=0.0$ line corresponding to uniform rotation). The
    filled circle indicates the neutral stability point of a uniformly
    rotating star computed in full general relativity (Stergioulas and
    Friedman~\cite{SF97}). Differential rotation
    shifts the neutral point to higher rotation
    rates. (Image reproduced with permission from~\cite{Yoshida02}, copyright by Ap. J..)}
  \label{fig:diff}
\end{figure*}

In newly-born neutron stars the CFS instability could develop while
the background equilibrium star is still differentially rotating. In
that case, the critical value of $T/W$, required for the instability
in the $f$-mode to set in, is larger than the corresponding value in
the case of uniform rotation~\cite{Yoshida02} (Figure~\ref{fig:diff}).
The mass-shedding limit for differentially rotating stars also appears
at considerably larger $T/W$ than the mass-shedding limit for uniform
rotation. Thus, Yoshida et al.~\cite{Yoshida02} suggest that
differential rotation favours the instability, since the ratio
$(T/W)_{\mathrm{critical}}/(T/W)_{\mathrm{shedding}}$ decreases with
increasing degree of differential rotation.

\begin{figure*}[t]
    \center
      \includegraphics[trim =0cm 1.5cm 1.5cm 0.cm,clip=True,width=0.5\textwidth]{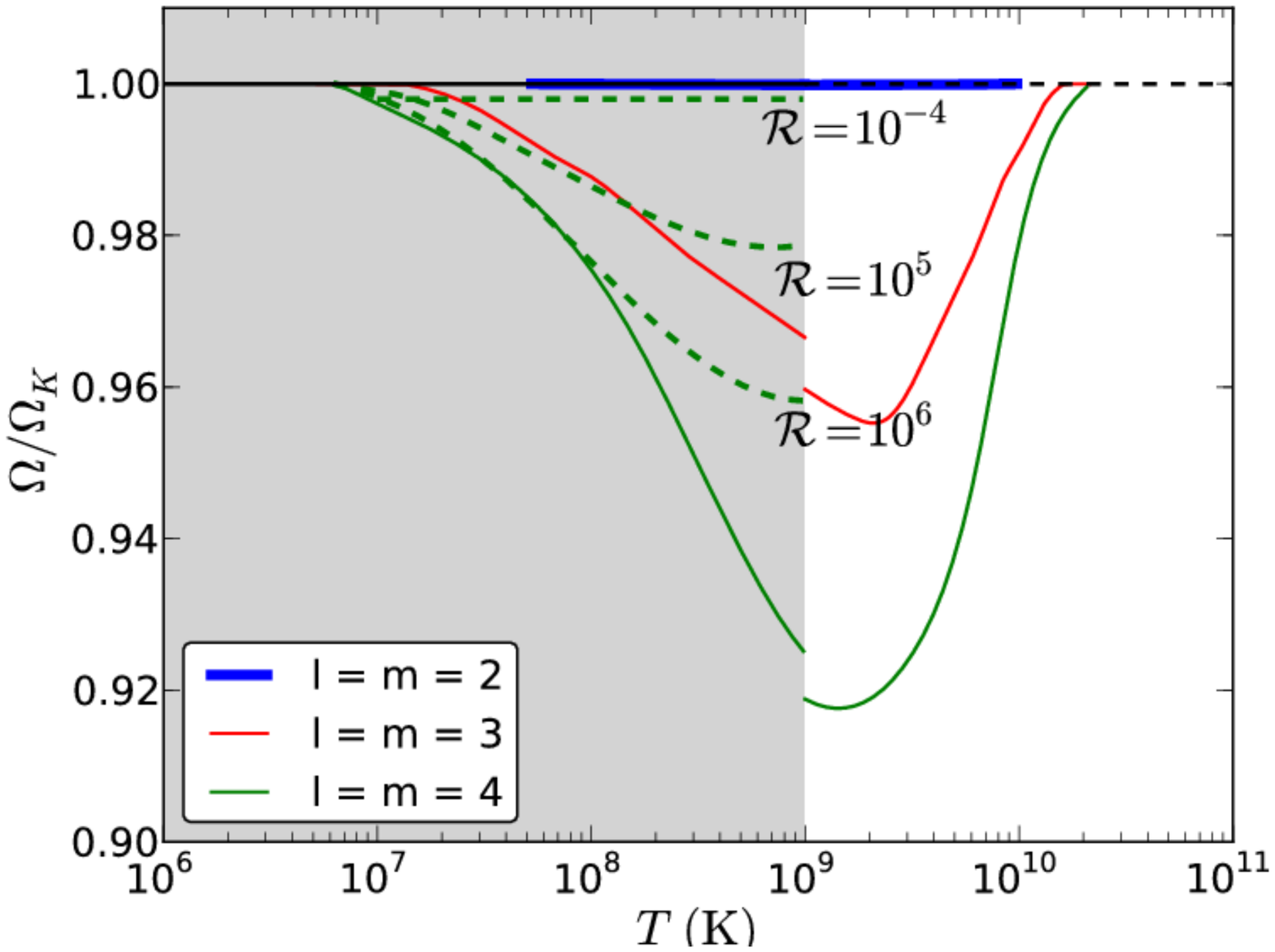}
      \includegraphics[width=0.48\textwidth]{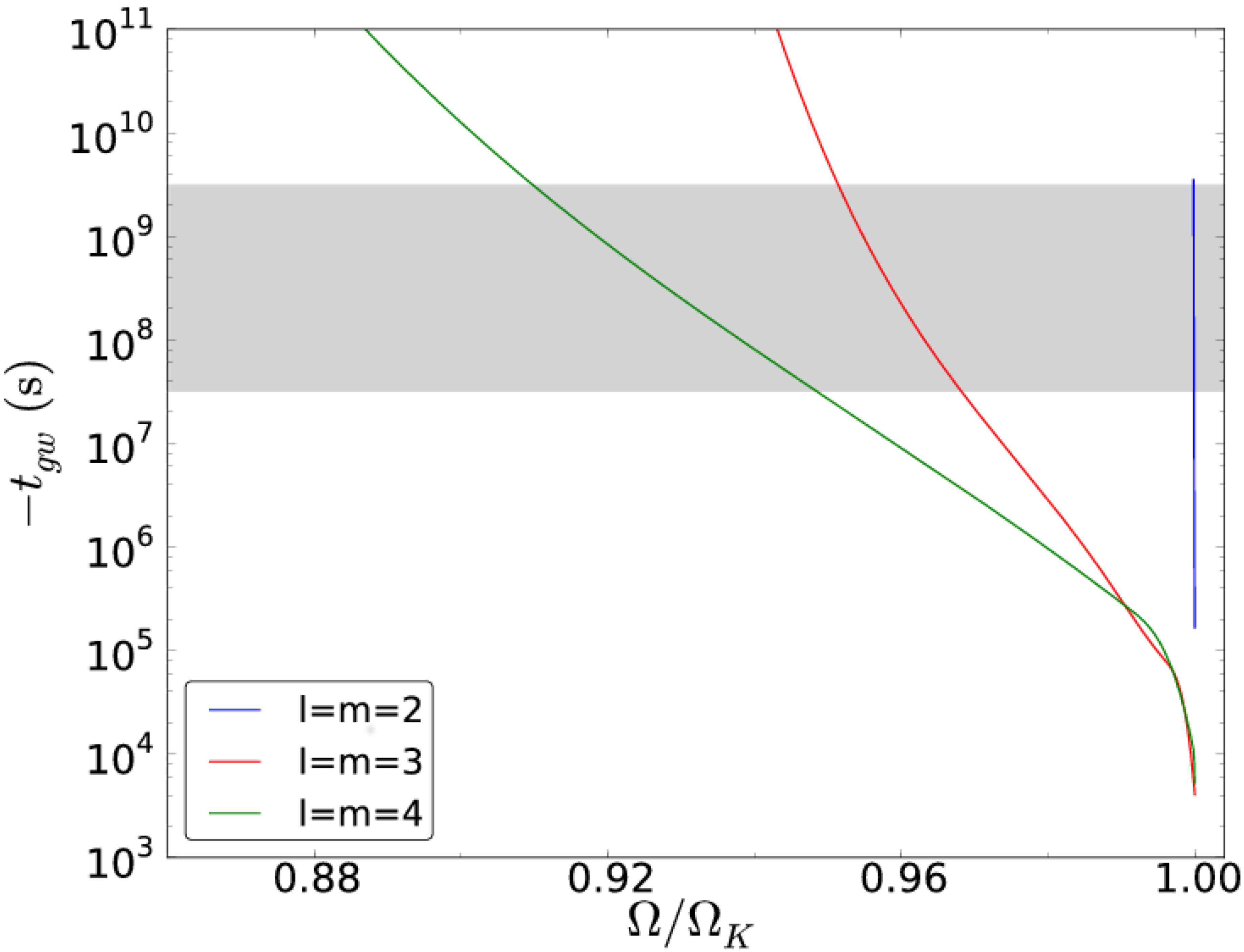}
    \caption{Left: The $f$-mode instability window neutron star rotation
      angular frequency vs temperature. Here $\Omega_K\simeq 6868 \rm
      rad/s$ is the angular frequency at the mass-shedding limit for a
      $n=0.73$, and mass $M=1.48M_\odot$ model with radius $R=10.47$
      km (in the TOV limit). The shaded area shows the region of
      superfluidity, and the dashed curves represent different values
      for the mutual friction drag parameter $\mathcal{R}$ (shown only
      for the $m = 4$ mode).  Right: The $f$-mode instability growth
      time as a function of $\Omega$ for the same model as in the left
      panel. The shaded area show the range of cooling timescales from
      an initial temperature $T = 5 \times 10^{10}$ K to a final $T =
      5 - 9 \times 10^8$ K.  (Image reproduced with permission from
      \cite{Gaertig2011PhRvL}, copyright by APS.)  }
    \label{fig:f_mode_inst_gr_cowling}
\end{figure*}

Gaertig et al.~\cite{Gaertig2011PhRvL} perform a calculation of the
CFS $f$-mode instability in rapidly rotating, relativistic neutron
stars employing the Cowling approximation while treating dissipation
through shear and bulk viscosity as well as superfluid mutual
friction~\cite{Lindblom00} of the nuclear matter. They adopt a
polytropic equation of state and construct, self-consistent equilibria
with the {\tt RNS} code. The focus of the study is primarily on the
$l=4,m=4$ mode, which is the dominant one because it has the largest
instability window. For the full dissipative system Gaertig et
al. compute the $f$-mode instability growth time and instability
window, i.e., the curve $\Omega(T)$ - the angular velocity at which
the growth time of the instability equals the dissipation timescale
due to viscous effects as a function of the temperature $T$. For the
analysis they consider one background solution which is a star with
polytropic index $n=0.73$, mass $M=1.48M_\odot$ and radius $R=10.47$
km (in the TOV limit). The main results are shown in
Fig.~\ref{fig:f_mode_inst_gr_cowling} from which it becomes clear that
the instability window for the $m=4$ mode is the widest with $\Omega
\gtrsim 0.92 \Omega_K$, where $\Omega_K$ is the mass-shedding limit
angular velocity.  The minimum $f$-mode instability growth time found
is on the order of $10^3$ s. The authors conclude that the $f$-mode
instability is more likely to be excited in nascent neutron stars,
spinning with $\Omega \gtrsim 0.9\Omega_K$ and having a temperature $T
\lesssim 2\times 10^{10}$ K.
%

Passamonti et al.~\cite{Passamonti2013PhRvD..87h4010P} adopt linear
perturbation theory and the Cowling approximation to study the
evolution of the $f$-mode instability in rapidly rotating, polytropic
relativistic neutron stars, while treating thermal effects, magnetic
fields, and the impact of unstable $r$-modes. The authors focus on two
polytropic stars with indices $n=1$ and $n=0.62$, and masses
$1.4M_\odot$ and $1.98M_\odot$, respectively. They report that
magnetic fields affect the evolution and the gravitational waves
generated during the instability, if the strength is larger than
$10^{12}$ G. An unstable $r$-mode dominates over the $f$-mode when the
$r$-mode reaches saturation but these conclusions are limited by the
unknown $r$-mode amplitudes at saturation. Finally, the authors find
that the thermal evolution suggests that heat generated by shear
viscosity during the saturation phase balances exactly cooling by
neutrinos, and prevents mutual friction from ever becoming important.

Doneva et al.~\cite{DonevaGaert2013} adopt the Cowling approximation
to study the $f$-mode instability in rapidly rotating stars with
realistic equations of state focusing on a constant mass sequence with
$M=2.0M_\odot$. The authors confirm the earlier result that $l=m=4$
modes have a larger instability window~\cite{Gaertig2011PhRvL}. In
addition, the authors report that realistic EOSs have a larger
instability window than polytropic EOSs, thus, favouring the $f$-mode
CFS instability.


\subsubsection{CFS instability of axial modes}
\label{s_axial}
     
In nonrotating stars, axial fluid modes are degenerate at zero
frequency, but in rotating stars they have nonzero frequency and are
called $r$-modes in the Newtonian limit~\cite{PP78,Sa82}. To order
${\cal O}(\Omega)$, their frequency in the inertial frame is
\begin{equation}
  \omega_{\mathrm{i}} = -m\Omega \Bigl( 1-\frac{2}{l(l+1)} \Bigr),
  \label{e:om}
\end{equation}
while the radial eigenfunction of the perturbation in the velocity can
be determined at order $\Omega^2$~\cite{Koj97}. According to
Equation~(\ref{e:om}), $r$-modes with $m>0$ are prograde
($\omega_{\mathrm{i}}<0$) with respect to a distant observer but
retrograde ($\omega_{\mathrm{r}} = \omega_{\mathrm{i}}+m\Omega >0$) in
the comoving frame for all values of the angular velocity. Thus,
$r$-modes in relativistic stars are generically unstable to the
emission of gravitational waves via the CFS instability, as was first
discovered by Andersson~\cite{A97} for the case of slowly rotating,
relativistic stars. This result was proved rigorously by Friedman and
Morsink~\cite{FM97}, who showed that the canonical energy of the modes
is negative.

\begin{figure*}[htbp]
  \center
  \includegraphics[trim =0.0cm 0.0cm 0.0cm 0.025cm,clip=true,width=0.7\textwidth]{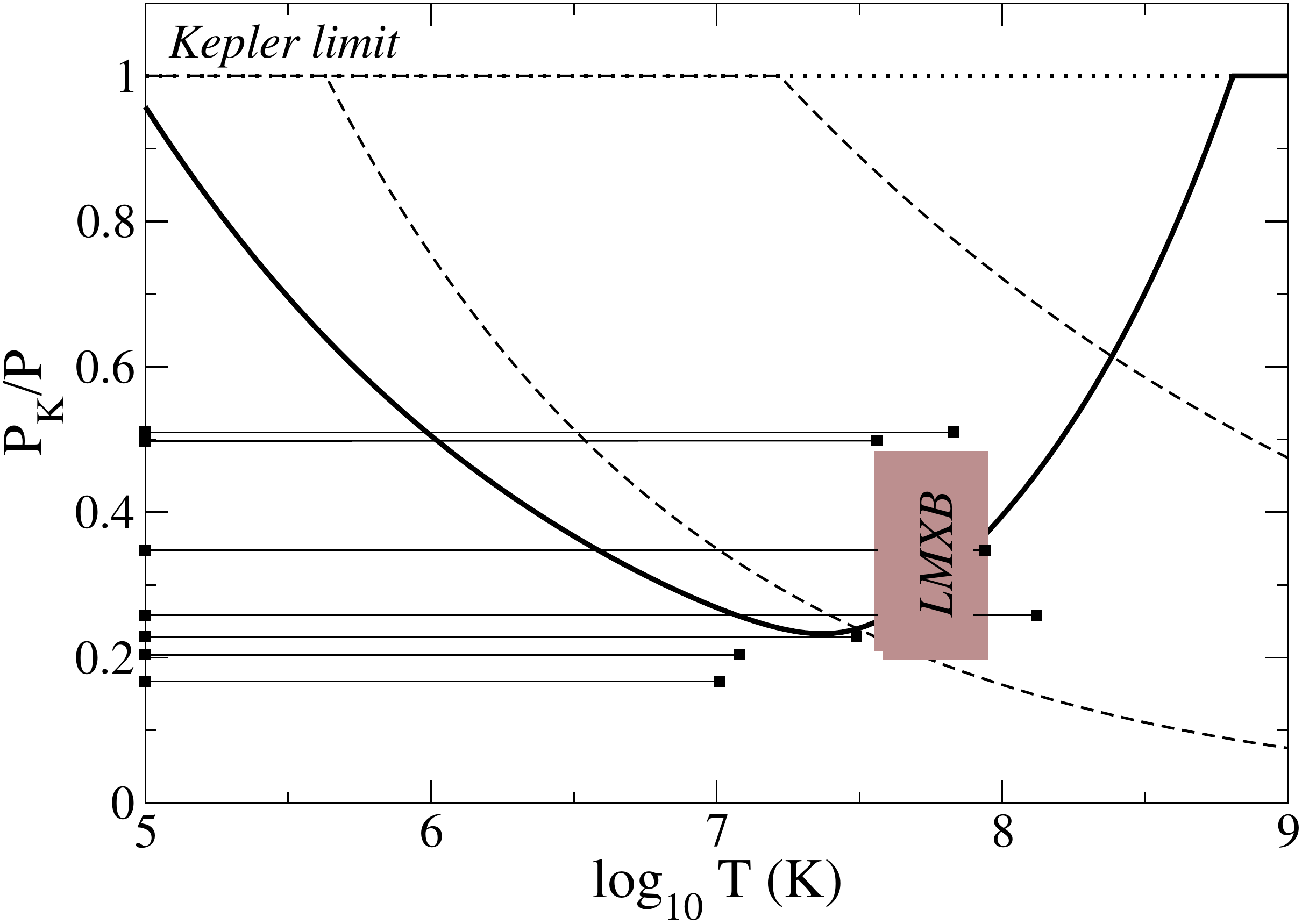}
  \caption{The $r$-mode instability window for a strange star of
    $M=1.4\,M_{\odot}$ and $R$~=10~km (solid line). Dashed curves show
    the corresponding instability windows for normal npe fluid and
    neutron stars with a crust. The instability window is compared to
    i) the inferred spin-periods for accreting stars in LMBXs [shaded
      box], and ii) the fastest known millisecond pulsars (for which
    observational upper limits on the temperature are available)
    [horizontal lines]. (Image reproduced with permission
    from~\cite{Andersson02c}, copyright by MNRAS.)}
  \label{fig:rstrange}
\end{figure*}

Two independent computations in the Newtonian Cowling approximation
\cite{LOM98,AKS98} showed that the usual shear and bulk viscosity
assumed to exist for neutron star matter is not able to damp the
$r$-mode instability, even in slowly rotating stars. In a temperature
window of $10^5 \mathrm{\ K} < T < 10^{10} \mathrm{\ K}$, the growth
time of the $l=m=2$ mode becomes shorter than the shear or bulk
viscosity damping time at a critical rotation rate that is roughly one
tenth the maximum allowed angular velocity of uniformly rotating
stars. Gravitational radiation is dominated by the mass quadrupole
term. These results suggested that a rapidly rotating proto-neutron
star will spin down to Crab-like rotation rates within one year of its
birth, because of the $r$-mode instability. Due to uncertainties in
the actual viscous damping times and because of other dissipative
mechanisms, this scenario is also consistent with somewhat higher
initial spins, such as the suggested initial spin period of several
milliseconds for the X-ray pulsar in the supernova remnant
N157B~\cite{Ma98}. Millisecond pulsars with periods less than a few
milliseconds can then only form after the accretion-induced spin-up of
old pulsars and not in the accretion-induced collapse of a white
dwarf.

\begin{figure*}[htbp]
  \center
  \includegraphics[width=0.7\textwidth]{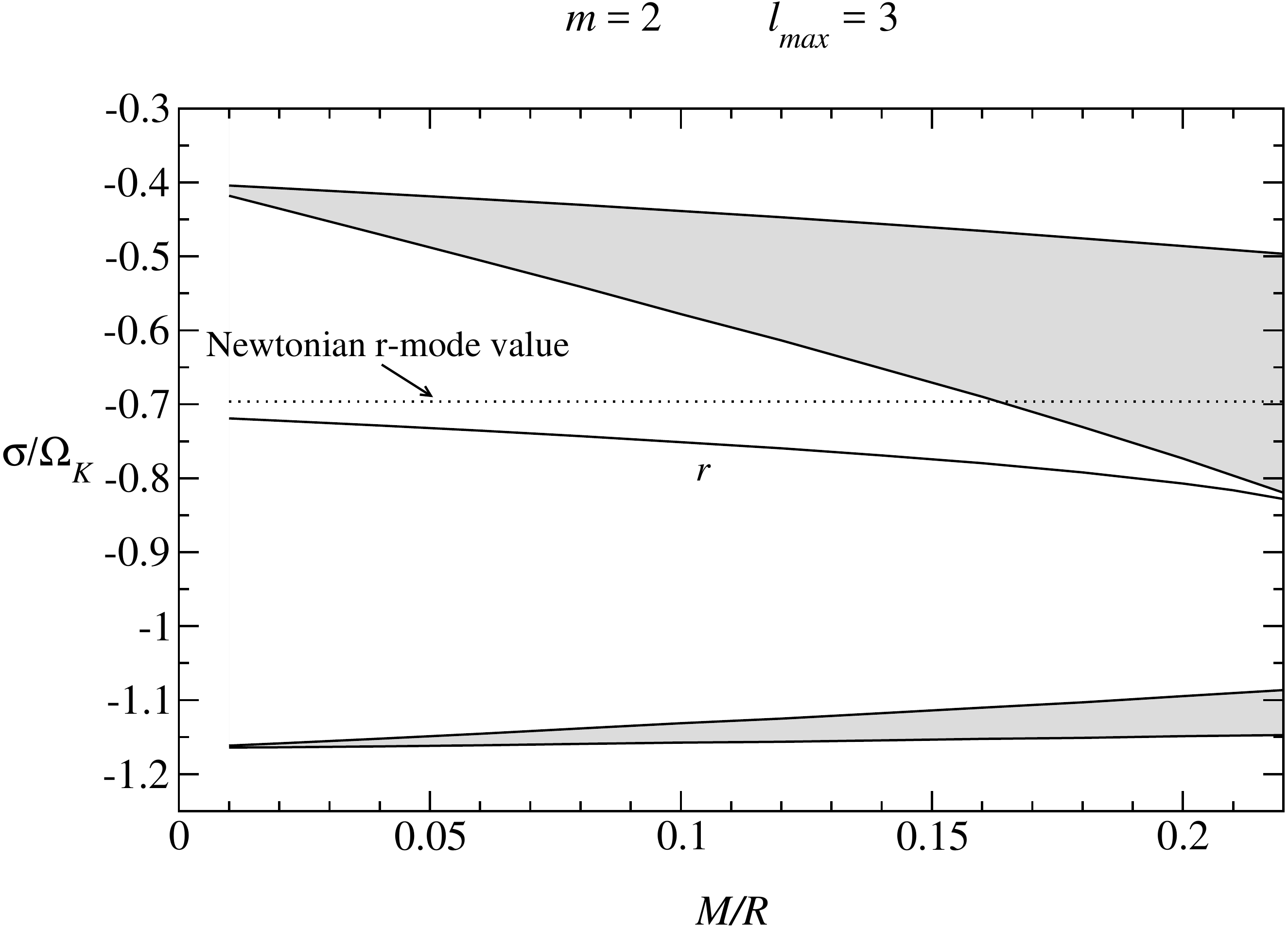}
  \caption{Relativistic $r$-mode frequencies for a range of the
    compactness ratio $M/R$. The coupling of polar and axial terms,
    even in the order ${\cal O}(\Omega)$ slow rotation approximation
    has a dramatic impact on the continuous frequency bands (shaded
    areas), allowing the $r$-mode to exist even in highly compact
    stars. The Newtonian value of the $r$-mode frequency is plotted as
    a dashed-dotted line. (Image reproduced with permission
    from~\cite{Ruoff02}, copyright by MNRAS.)}
  \label{fig:rspect}
\end{figure*}

The precise limit on the angular velocity of newly-born neutron stars
will depend on several factors, such as the strength of the bulk
viscosity, the cooling process, superfluidity, the presence of
hyperons, and the influence of a solid crust. In the uniform density
approximation, the $r$-mode instability can be studied analytically to
${\cal O}(\Omega^2)$ in the angular velocity of the
star~\cite{KS98}. A study on the issue of detectability of
gravitational waves from the $r$-mode instability was presented
in~\cite{OW98} (see Section~\ref{grw}), while Andersson, Kokkotas, and
Stergioulas~\cite{AKSt98} and Bildsten~\cite{Bildsten98} proposed
that the $r$-mode instability is limiting the spin of millisecond
pulsars spun-up in LMXBs and it could even set the minimum observed
spin period of $\sim$~1.5~ms (see~\cite{Andersson00}). This
scenario is also compatible with observational data, if one considers
strange stars instead of neutron stars~\cite{Andersson02c} (see
Figure~\ref{fig:rstrange}).

\begin{figure*}[htbp]
  \center
  \includegraphics[width=0.7\textwidth]{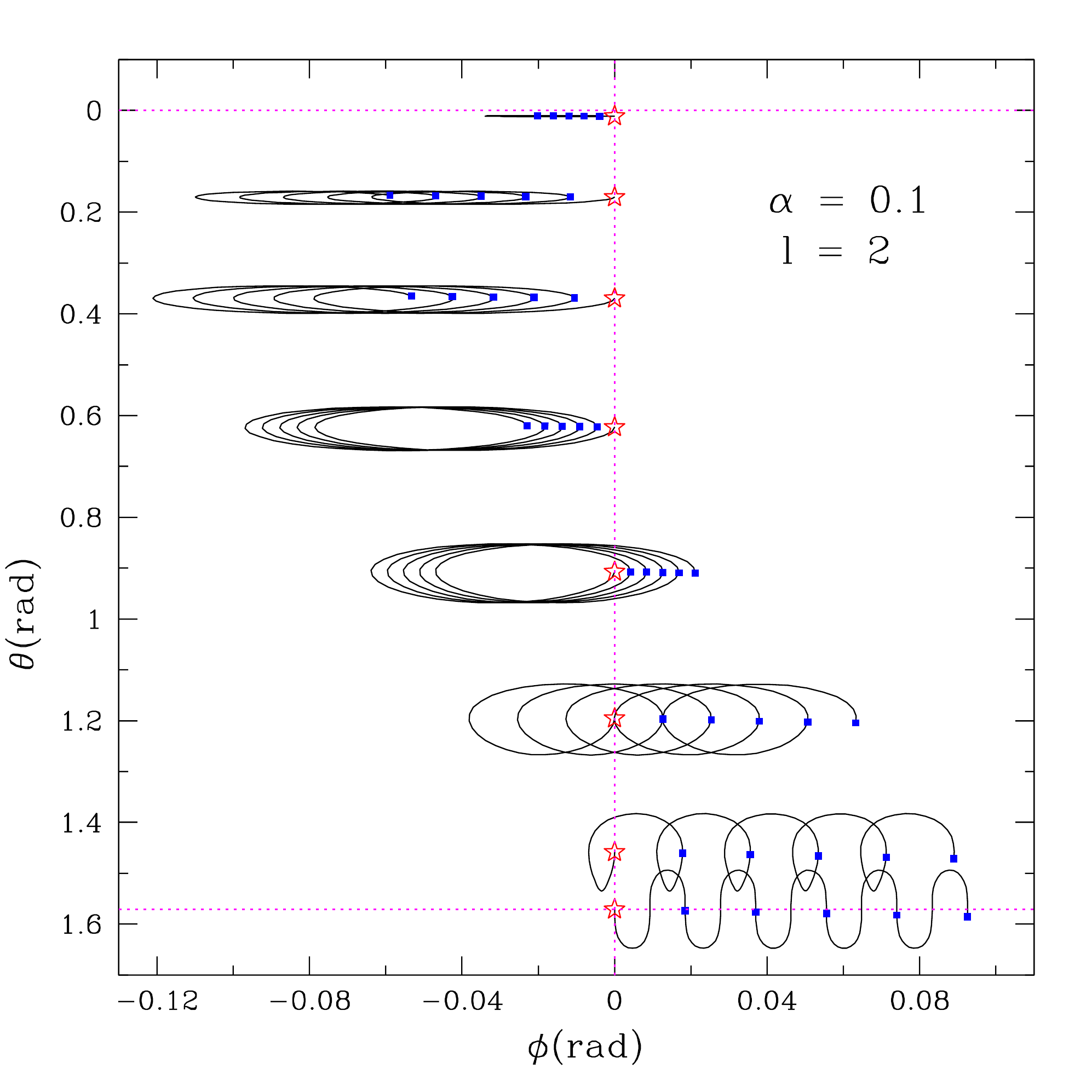}
  \caption{Projected trajectories of several fiducial fluid elements
    (as seen in the corotating frame) for an $l=m=2$ Newtonian
    $r$-mode. All of the fluid elements are initially positioned on
    the $\phi_0=0$ meridian at different latitudes (indicated with
    stars). Blue dots indicate the position of the fluid elements
    after each full oscillation period. The $r$-mode induces a
    kinematical, differential drift. (Image reproduced with permission
    from~\cite{Rezzolla01b}, copyright by APS.)}
  \label{fig:rdiff}
\end{figure*}

Since the discovery of the $r$-mode instability, a large number of
authors have studied the development of the instability and its
astrophysical consequences in more detail. Unlike in the case of the
$f$-mode instability, many different aspects and interactions have
been considered. This intense focus on the detailed physics has been
very fruitful and we now have a much more complete understanding of
the various physical processes that are associated with pulsations in
rapidly rotating relativistic stars. The latest understanding of the
$r$-mode instability is that it may not be a very promising
gravitational wave source (as originally thought), but the important
astrophysical consequences, such as the limits of the spin of young
and of recycled neutron stars are still considered plausible. The most
crucial factors affecting the instability are magnetic
fields~\cite{Spruit99,Rezzolla00,Rezzolla01b,Rezzolla01c}, possible
hyperon bulk viscosity~\cite{JonesPB01,Lindblom01,Haensel02} and
nonlinear
saturation~\cite{Stergioulas01,Lindblom01b,Lindblom02,Arras02}. The
question of the possible existence of a continuous spectrum has also
been discussed by several authors, but the most recent analysis
suggests that higher order rotational effects still allow for discrete
$r$-modes in relativistic stars~\cite{Yoshida02b,Ruoff02} (see
Figure~\ref{fig:rspect}).

Haskell and Andersson~\cite{Haskell2010} study the effects of
superfluid hyperon bulk viscosity on the $r$-mode instability window
using a multifluid formalism. They find that although the extra bulk
viscosity does not alter the instability window qualitatively, it
could become substantial and even suppress the $r$-mode instability
altogether in a range of temperatures and neutron star radii.
However, hyperons are predicted only by certain equations of state and
the relativistic mean field theory is not universally accepted. Thus,
our ignorance of the true equation of state still leaves a lot of room
for the $r$-mode instability to be considered a viable mechanism for
the generation of detectable gravitational radiation.

In a subsequent paper Andersson et al.~\cite{AnderssonGlampe2013}
study the superfluid $r$-mode instability which arises in rotating
stars in which there is ``differential'' rotation between the crust
and the underlying superfluid, and which was first discovered by
Glampedakis and Andersson in~\cite{Glampedakis2009PhRvL} as a new
mechanism for explaining the unpinning of vortices in pulsar
glitches. In~\cite{AnderssonGlampe2013} the analysis goes beyond the
strong-drag limit adopted in~\cite{Glampedakis2009PhRvL} and it is
shown that there exist dynamically unstable modes (growth time
comparable to the stellar rotation period), and that the $r$-modes
undergo a secular instability.

Magnetic fields can affect the $r$-mode instability, as the $r$-mode
velocity field creates differential rotation, which is both
kinematical and due to gravitational radiation reaction (see
Figure~\ref{fig:rdiff}). Under differential rotation, an initially
weak poloidal magnetic field is wound-up, creating a strong toroidal
field, which causes the $r$-mode amplitude to saturate. On the other
hand a more recent study of the effects of a dipole magnetic field on
$r$-modes of slowly rotating, relativistic neutron stars by Chirenti
and Sk{\'a}kala~\cite{Chirenti2013} reports that magnetic fields
affect the $r$-mode oscillation frequencies and the $r$-mode
instability growth time very little even for strengths as large as $B
\sim 10^{15}$ G.

\begin{figure*}[htbp]
  \center
  \includegraphics[width=0.7\textwidth]{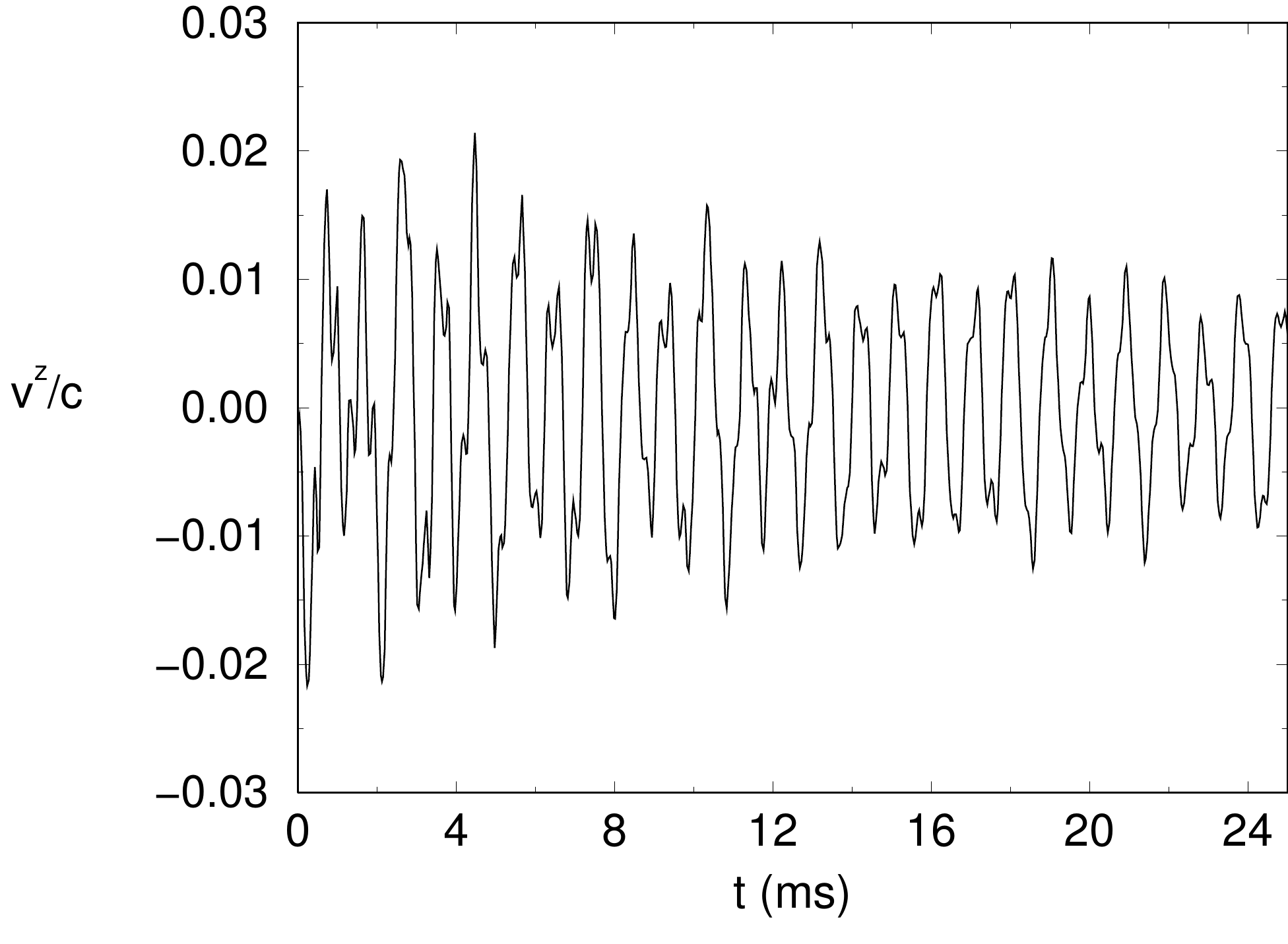}
  \caption{Evolution of the axial velocity in the equatorial plane for
    a relativistic $r$-mode in a rapidly rotating $N=1.0$ polytrope
    (in the Cowling approximation). Since the initial data used to
    excite the mode are not exact, the evolution is a superposition of
    (mainly) the $l=m=2$ $r$-mode and several inertial modes. The
    amplitude of the oscillation decreases due to numerical
    (finite-differencing) viscosity of the code. A beating between the
    $l=m=2$ $r$-mode and another inertial mode can also be
    seen. (Image reproduced with permission from~\cite{Stergioulas01},
    copyright by APS.)}
  \label{fig:rmode}
\end{figure*}

The detection of gravitational waves from $r$-modes depends crucially
on the nonlinear saturation amplitude. A first study by Stergioulas
and Font~\cite{Stergioulas01} suggests that $r$-modes can exist at
large amplitudes of order unity for dozens of rotational periods in
rapidly rotating relativistic stars (Figure~\ref{fig:rmode}). The
study used 3D relativistic hydrodynamical evolutions in the Cowling
approximation. This result was confirmed by Newtonian 3D simulations
of nonlinear $r$-modes by Lindblom, Tohline, and
Vallisneri~\cite{Lindblom01,Lindblom01b}. Lindblom et al.\ went
further, using an accelerated radiation reaction force to artificially
grow the $r$-mode amplitude on a hydrodynamical (instead of the
secular) timescale. At the end of the simulations, the $r$-mode grew
so large that large shock waves appeared on the surface of the star,
while the amplitude of the mode subsequently collapsed. Lindblom {\it
  et al.}\ suggested that shock heating may be the mechanism that
saturates the $r$-modes at a dimensionless amplitude of $\alpha \sim
3$. 

Other studies of nonlinear couplings between the $r$-mode and higher
order inertial modes~\cite{Arras02} and new 3D nonlinear Newtonian
simulations~\cite{Gressman02} seem to suggest a different picture. The
$r$-mode could be saturated due to mode couplings or due to a
hydrodynamical instability at amplitudes much smaller than the
amplitude at which shock waves appeared in the simulations by Lindblom
et al.\ Such a low amplitude, on the other hand, modifies the
properties of the $r$-mode instability as a gravitational wave source,
but is not necessarily bad news for gravitational wave detection, as a
lower spin-down rate also implies a higher event rate for the $r$-mode
instability in LMXBs in our own Galaxy~\cite{Andersson02c,Heyl02}.
The 3D simulations need to achieve significantly higher resolutions
before definite conclusions can be reached, while the Arras {\it et
  al.}\ work could be extended to rapidly rotating relativistic stars
(in which case the mode frequencies and eigenfunctions could change
significantly, compared to the slowly rotating Newtonian case, which
could affect the nonlinear coupling coefficients). Spectral methods
can be used for achieving high accuracy in mode calculations; first
results have been obtained by Villain and Bonazzolla~\cite{Villain02}
for inertial modes of slowly rotating stars in the relativistic
Cowling approximation.

More recently Bondarescu, Teukolsky and
Wasserman~\cite{Bondarescu:2008qx} perform a study of the non-linear
development of the $r$-mode instability, including three-mode
couplings, neutrino cooling and viscous heating effects on rotating
stars near the mass-shedding limit.  In their most optimistic
scenarios, the authors conclude that gravitational waves from the
r-mode instability in young, rapidly spinning neutron stars may be
detectable by advanced LIGO out to 1 Mpc for years, and perhaps
decades, after formation. In follow up work, Bondarescu and
Wasserman~\cite{Bondarescu:2013xwa} include interactions of the
$\ell=m=2$ $r$-mode with ``pairs of daugther modes'' close to
resonance.  They find that if dissipation occurs at the crust–core
boundary layer, the $r$-mode saturation amplitude is too large for the
star to be spun up by accretion to even 300 Hz because of angular
momentum loss to gravitational radiation. Spin up to higher
frequencies seems to require that the core-crust transition occur over
a lengthscale much longer than 1 cm.

The idea of utilizing X-ray and UV observations of low-mass X-ray
binaries to constrain the physics of the $r$-mode instability is
discussed by Haskell et al. in~\cite{Haskell2012}.

For a more extensive coverage of the numerous articles on the $r$-mode
instability that appeared in recent years, the reader is referred to
several review and recent
articles~\cite{Andersson01c,Friedman01,Lindblom01c,Kokkotas02,Andersson02b,Alford:2014jha,Kokkotas2016EPJA...52...38K,Jasiulek:2016epr}.
 
 \vspace{1 em}\noindent{\bf Going further:}~~ If rotating stars with
very high compactness exist, then $w$-modes can also become unstable,
as was found by Kokkotas, Ruoff, and
Andersson~\cite{Kokkotas03}. The possible astrophysical implications
are still under investigation.


\subsubsection{Effect of viscosity on the CFS instability} 

In the previous sections, we have discussed the growth of the 
CFS instability driven by gravitational radiation in an otherwise
non-dissipative star. The effect of neutron star matter viscosity on
the dynamical evolution of nonaxisymmetric perturbations can be
considered separately, when the timescale of the viscosity is much
longer than the oscillation timescale. If $\tau_{\mathrm{gr}}$ is the
computed growth rate of the instability in the absence of viscosity,
and $\tau_{\mathrm{s}}$, $\tau_{\mathrm{b}}$ are the timescales of shear and
bulk viscosity, then the total timescale of the perturbation is
\begin{equation}
  \frac{1}{\tau} =
  \frac{1}{\tau_{\mathrm{gr}}} + \frac{1}{\tau_{\mathrm{s}}} + \frac{1}{\tau_{\mathrm{b}}}.
\end{equation}
Since $\tau_{\mathrm{gr}} < 0$ and $\tau_{\mathrm{b}}$, $\tau_{\mathrm{s}}>0$, a mode
will grow only if $\tau_{\mathrm{gr}}$ is shorter than the viscous
timescales, so that $1/\tau<0$.

In normal neutron star matter, shear viscosity is dominated by
neutron--neutron scattering with a temperature dependence of
$T^{-2}$~\cite{FI76}, and computations in the Newtonian limit and
post-Newtonian approximation show that the CFS instability is
suppressed for $T <10^6 \mathrm{\ K}$\,--\,$10^7
\mathrm{\ K}$~\cite{IL91a,IL91b,YE95,Li95}. If neutrons become a
superfluid below a transition temperature $T_{\mathrm{s}}$, then
mutual friction, which is caused by the scattering of electrons off
the cores of neutron vortices could significantly suppress the
$f$-mode instability for $T<T_{\mathrm{s}}$~\cite{Lindblom95}, but the
$r$-mode instability remains unaffected~\cite{Lindblom00}. The
superfluid transition temperature depends on the theoretical model for
superfluidity and lies in the range $10^8 \mathrm{\ K}$\,--\,$6 \times
10^9 \mathrm{\ K}$~\cite{Pa94}.
   
In a pulsating fluid that undergoes compression and expansion, the
weak interaction requires a relatively long time to re-establish
equilibrium. This creates a phase lag between density and pressure
perturbations, which results in a large bulk viscosity~\cite{Sa89}.
The bulk viscosity due to this effect can suppress the CFS instability
only for temperatures for which matter has become transparent to
neutrinos~\cite{LS95,BFG96}. It has been proposed that for $T>5
\times 10^9 \mathrm{\ K}$, matter will be opaque to neutrinos and the
neutrino phase space could be blocked (\cite{LS95}; see
also~\cite{BFG96}). In this case, bulk viscosity will be too weak to
suppress the instability, but a more detailed study is needed.

In the neutrino transparent regime, the effect of bulk viscosity on
the instability depends crucially on the proton fraction
$x_{\mathrm{p}}$. If $x_{\mathrm{p}}$ is lower than a critical value
($\sim 1/9$), only modified URCA processes are allowed. In this case
bulk viscosity limits, but does not completely suppress, the
instability~\cite{IL91a,IL91b,YE95}. For most modern EOSs, however,
the proton fraction is larger than $\sim 1/9$ at sufficiently high
densities~\cite{Lat91}, allowing direct URCA processes to take
place. In this case, depending on the EOS and the central density of
the star, the bulk viscosity could almost completely suppress the CFS
instability in the neutrino transparent regime~\cite{Zd95}. At high
temperatures, $T>5 \times 10^9 \mathrm{\ K}$, even if the star is
opaque to neutrinos, the direct URCA cooling timescale to $T \sim 5
\times 10^9 \mathrm{\ K}$ could be shorter than the growth timescale
of the CFS instability.

Bildsten and Ushomirsky~\cite{Bildsten2000ApJ...529L..33B} considered the dissipation of $r$-modes due to the presence of a viscous boundary
layer between the oscillating fluid and the crust and found it to be several orders of magnitude higher than the dissipation due to shear in the
neutron star interior, if the crust is rigid. Subsequently, Lindblom, Owen, and
Ushomirsky~\cite{Lindblom2000PhRvD..62h4030L} included the effects of the Coriolis force
in more realistic neutron-star models, finding that an $r$-mode amplitude value of
$\sim 5\times 10^{-3}$ for maximally rotating stars would result in sufficient heating at the crust-core boundary layer for the crust to melt.  These initial computations used a rigid crust that did not participate in the $r$-mode oscillation, but the magnitude of the effect strongly depends on a slippage
parameter $\cal S$ that measures the fractional difference in velocity of the
normal fluid between the crust and the core and the fractional pinning of
vortices in the crust \cite{LU01,Glampedakis2006,Glampedakis2006b}.
The dependence of the crust-core slippage on the spin frequency is complicated, and is very sensitive to the physical thickness of the crust.
 \vspace{1 em}\noindent{\bf \\ Going further:}~~ For more recent work on
viscous damping of $r$-modes see Alford et al.~\cite{Alford:2011pi}
(and references therein) who treat non-linear viscous effects in the
large-amplitude regime, as well as consider hadronic stars, strange
quark stars, and hybrid stars. See also Kolomeitsev and
Voskresensky~\cite{Kolomeitsev:2014gfa} (and references therein) for a
microphysical computation of the shear and bulk viscosities from
various processes and applications to viscous damping of $r$-modes.
For recent work on the $r$-mode instability window and applications on
pulsar recycling see~\cite{Gusakov:2013jwa,Gusakov:2016drh}, and
references therein.


\subsubsection{Gravitational radiation from CFS instability}
\label{grw}
 
Conservation of angular momentum and the inferred initial period
(assuming magnetic braking) of a few milliseconds for the X-ray pulsar
in the supernova remnant N157B~\cite{Ma98} suggests that a fraction
of neutron stars may be born with very large rotational energies. The
$f$-mode bar CFS instability thus appears as a promising source for
the planned gravitational wave detectors~\cite{LS95}. It could also
play a role in the rotational evolution of merged binary neutron
stars, if the post-merger angular momentum exceeds the maximum allowed
to form a Kerr black hole~\cite{BaS98} or if differential rotation
temporarily stabilizes the merged object.

Lai and Shapiro~\cite{LS95} have studied the development of the
$f$-mode instability using Newtonian ellipsoidal
models~\cite{LRS93,LRS94}. They consider the case when a rapidly
rotating neutron star is created in a core collapse. After a brief
dynamical phase, the proto-neutron star becomes secularly
unstable. The instability deforms the star into a nonaxisymmetric
configuration via the $l=2$ bar mode. Since the star loses angular
momentum via the emission of gravitational waves, it spins down until
it becomes secularly stable. The frequency of the waves sweeps
downward from a few hundred Hz to zero, passing through LIGO's ideal
sensitivity band. A rough estimate of the wave amplitude shows that,
at $\sim$~100~Hz, the gravitational waves from the CFS instability
could be detected out to the distance of 140~Mpc by the advanced LIGO
detector. This result is very promising, especially since for
relativistic stars the instability will be stronger than the Newtonian
estimate~\cite{SF97}. More recent work by Passamonti et
al.~\cite{Passamonti2013PhRvD..87h4010P} suggests that the
gravitational wave signal generated during the $f$-mode instability,
could potentially be detectable by Advanced LIGO/Virgo from a source
located in the Virgo cluster, as long as the star was massive enough.

Pnigouras and
Kokkotas~\cite{Pnigouras2015PhRvD..92h4018P,Pnigouras2016PhRvD..94b4053P}
develop a formalism to study the saturation of the $f$-mode
instability as a result of nonlinear coupling of modes. They find that
parent (unstable) modes couple resonantly to daughter modes which
drain energy from the parent modes leading to saturation of the
instability. These results can be applicable to neutron stars formed
in core collapse and following neutron star mergers. Doneva, Kokkotas
and Pnigouras~\cite{Doneva2015PhRvD..92j4040D} report that
gravitational waves generated by the $f$-mode instability in
supramassive neutron stars (that could form following binary neutron
star mergers) could be detectable by advanced LIGO at 20 Mpc (where,
however, the event rate is very low, so that a more sensitive
instrument is needed for realistic detection rates). The stochastic
gravitational wave background due to the $f$-mode instability in
neutron stars is estimated by Surace, Kokkotas and Pnigouras
in~\cite{Surace2016A&A...586A..86S}. They find that for the $l=m=2$
$f$-mode $\Omega_{\rm GW} \sim 10^{-9}$ which could be detectable
through cross correlating data from pairs of grounds based detectors.

  \begin{figure*}[t]
    \center
      \includegraphics[width=0.49\textwidth]{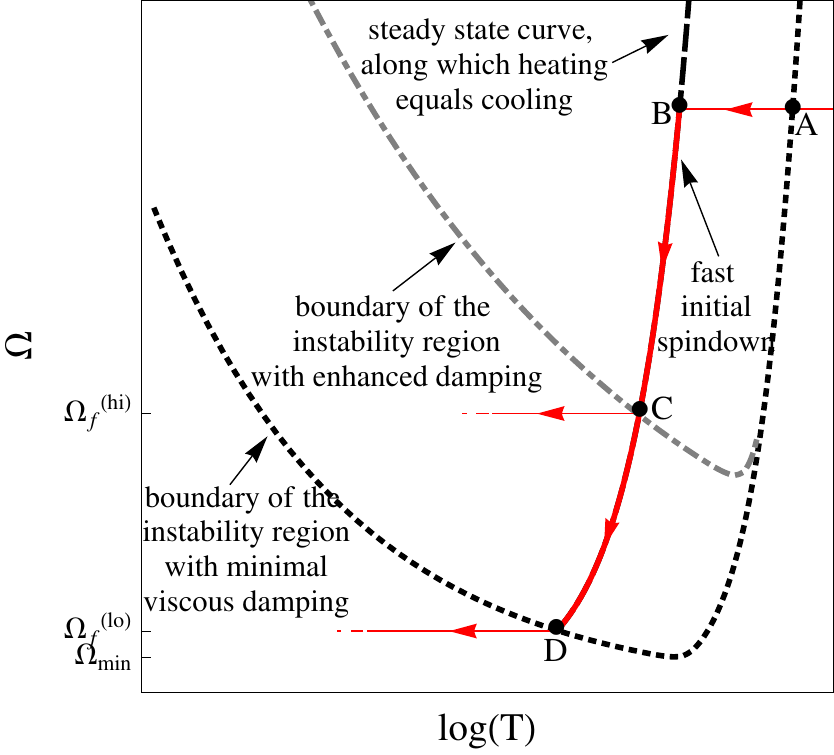}
      \includegraphics[width=0.49\textwidth]{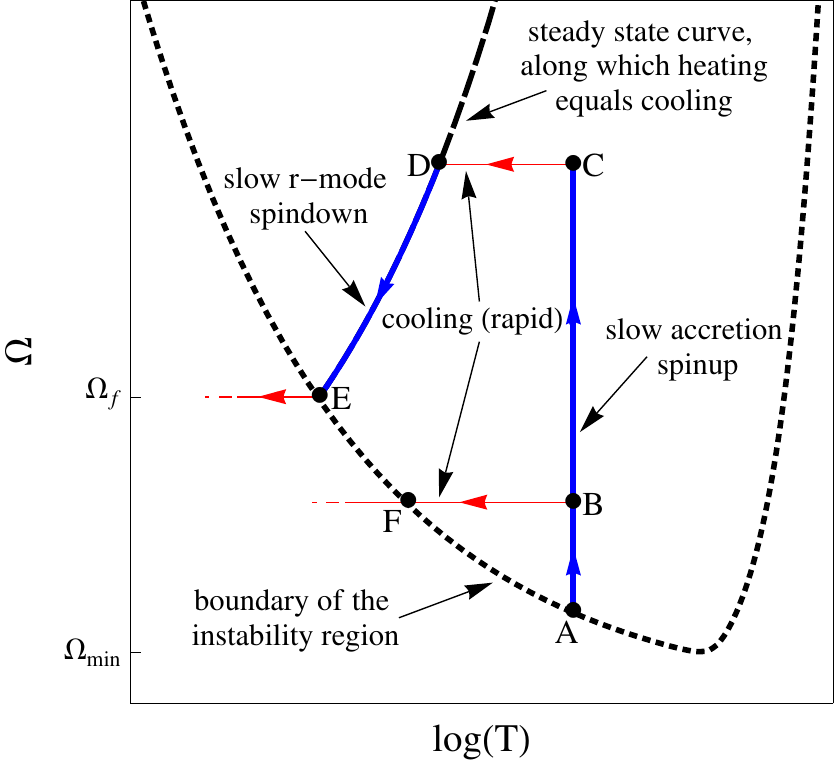} 
      \includegraphics[width=0.49\textwidth]{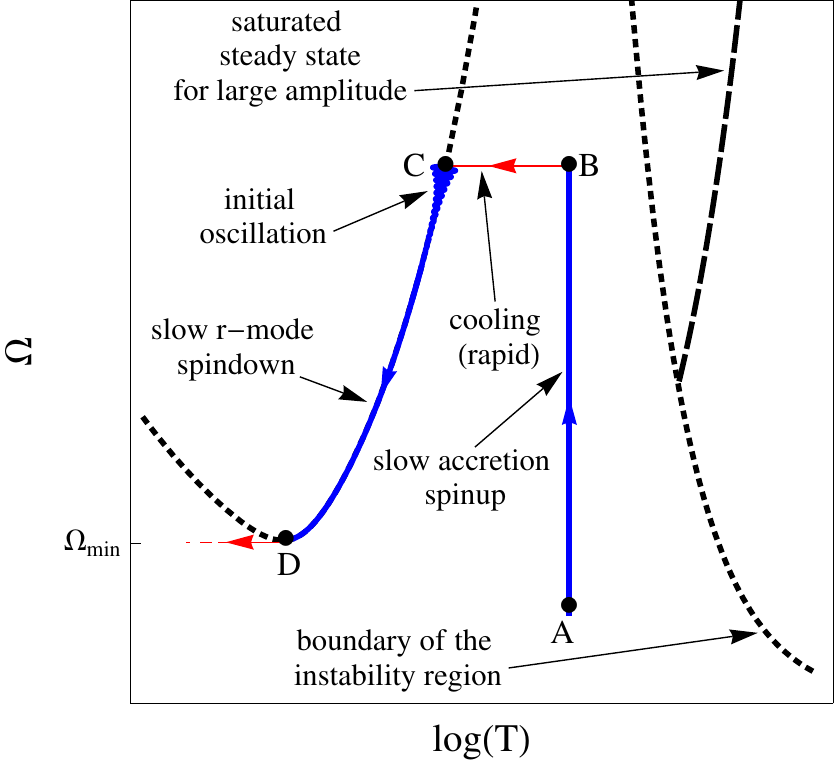} 
    \caption{Different scenarios in which the r-mode instability can
      generate gravitational waves in the angular velocity ($\Omega$) -
      temperature ($T$) plane. Top left panel: Spindown of nascent sources;
      Top right panel: Pulsar recycling in LMXBs and spindown of
      millisecond pulsars; Lower panel: Recycling and spindown in
      sources with increased damping. Each time the evolution goes
      through or lies at the boundary of the instability region
      (region within the dotted lines) gravitational wave emission is
      switched on. (Image reproduced with permission from
      \cite{Kokkotas2016EPJA...52...38K}, copyright by EPJ.)  }
    \label{rmodescenarios}
\end{figure*}

Whether $r$-modes should also be considered a promising gravitational
wave source depends crucially on their nonlinear saturation amplitude
(see Section~\ref{s_axial}). Nevertheless, the issues of detectability
and interpretation of gravitational waves generated by the $r$-mode
instability is discussed by Owen in~\cite{Owen2010}, and the effects
of realistic equations of state and the potential for gravitational
waves from the $r$-mode instability to constrain the nuclear equation
of state are studied by Idrisy et
al. in~\cite{Idrisy2015}. Applications of the different $r$-mode
instability scenarios (see Fig.~\ref{rmodescenarios}) in
gravitational-wave astronomy were recently presented by Kokkotas and
Schwenzer~\cite{Kokkotas2016EPJA...52...38K}.

\vspace{1 em}\noindent{\bf Going further:}~~ The possible ways for
neutron stars to emit gravitational waves and their detectability are
reviewed in~\cite{BG96,BGou,GBG96,FL98,Th96,Sc98,Cutler02}.


\subsubsection{Viscosity-driven instability}

A different type of nonaxisymmetric instability in rotating stars is
the instability driven by viscosity, which breaks the circulation of
the fluid~\cite{RS63,Ja64}. The instability is suppressed by
gravitational radiation, so it cannot act in the temperature window in
which the CFS instability is active. The instability
sets in when the frequency of an $l=-m$ mode goes through zero in the
rotating frame. In contrast to the CFS instability, the
viscosity-driven instability is not generic in rotating stars. The
$m=2$ mode becomes unstable at a high rotation rate for very stiff
stars, and higher $m$-modes become unstable at larger rotation rates.

In Newtonian polytropes, the instability occurs only for stiff
polytropes of index $N<0.808$~\cite{Ja64,SL96}. For relativistic
models, the situation for the instability becomes worse, since
relativistic effects tend to suppress the viscosity-driven instability
(while the CFS instability becomes stronger). According to recent
results by Bonazzola et al.~\cite{BFG97}, for the most
relativistic stars, the viscosity-driven bar mode can become unstable
only if $N<0.55$. For $1.4\,M_{\odot}$ stars, the instability is
present for $N<0.67$.

These results are based on an approximate computation of the
instability in which one perturbs an axisymmetric and stationary
configuration, and studies its evolution by constructing a series of
triaxial quasi-equilibrium configurations. During the evolution only
the dominant nonaxisymmetric terms are taken into account. The method
presented in~\cite{BFG97} is an improvement (taking into account
nonaxisymmetric terms of higher order) of an earlier method by the
same authors~\cite{BFG96}. Although the method is approximate, its
results indicate that the viscosity-driven instability is likely to be
absent in most relativistic stars, unless the EOS turns out to be
unexpectedly stiff. 

An investigation by Shapiro and Zane~\cite{SZ97} of the
viscosity-driven bar mode instability, using incompressible, uniformly
rotating triaxial ellipsoids in the post-Newtonian approximation,
finds that the relativistic effects increase the critical $T/W$ ratio
for the onset of the instability significantly. More recently, new
post-Newtonian~\cite{DiGirolamo02} and fully relativistic calculations
for uniform density stars~\cite{Gondek02} show that the
viscosity-driven instability is not as strongly suppressed by
relativistic effects as suggested in~\cite{SZ97}. The most promising
case for the onset of the viscosity-driven instability (in terms of
the critical rotation rate) would be rapidly rotating strange
stars~\cite{Gondek03}, but the instability can only appear if its
growth rate is larger than the damping rate due to the emission of
gravitational radiation -- a corresponding detailed comparison is
still missing. 

The non-linear evolution of the bar mode instability has been studied
via post-Newtonian hydrodynamic simulations by Ou et al.~\cite{Ou04},
and Shibata and Karino~\cite{shibatakarino04}.  Ou et al. find that
the instability goes through a ``Dedekind-like"
configuration~\cite{Chandrasekhar69} before becoming unstable due to a
hydrodynamical shearing instability. Shibata and Karino find that the
end state of the instability is an ellipsoidal star of ellipticity
$e\gtrsim 0.7$.

\subsubsection{One-arm (spiral) instability}
\label{s:one-arm}

A remarkable feature about highly differentially rotating neutron
stars is that they can also become unstable to a dynamical one-arm
($m=1$) ``spiral'' instability.

The one-arm instability in differentially rotating stars was
discovered in Newtonian hydrodynamic simulations with soft polytropic
equations of state and a high degree of differential rotation by
Centrella et al.~\cite{Centrella2001}. The instability growth occurs
on a dynamical (rotational period) timescale and saturates within a
few tens of rotational periods. During the development of the
instability a perturbation displaces the stellar core from the center
of mass resulting in the core orbiting around the center of mass at
roughly constant angular frequency. The $m=1$ deformation leads to a
time-changing quadrupole moment which results in the emission of
gravitational waves which may be detectable. This, in part, motivates
the study of this instability and the conditions under which it
develops.

Shortly after the discovery of the $m=1$ instability, Saijo et
al.~\cite{Saijo2003} confirmed its existence with further Newtonian
hydrodynamic simulations and suggested that a toroidal configuration
may be necessary to trigger the one-arm instability but not
sufficient. Guided by observations reported by Watts et
al.~\cite{Watts2005} that the low-$T/|W|$ dynamical (bar-mode)
instability (discovered by Shibata et
al. in~\cite{ShibataKarino2002,ShibataKarino2003} for highly
differentially rotating stars, see
also~\cite{Saijo2006,CerdaD2007,PassamontiAnderson2015} ) develops
near the corotation radius, i.e., the radius where the angular
frequency of the unstable mode matches the local angular velocity of
the fluid, Saijo et al.~\cite{Saijo2006} argue that the one-arm spiral
instability is also excited near the corotation radius. Hydrodynamic
simulations of differentially rotating stars by Ou and Tohline in
Newtonian gravity~\cite{Ou2006}, and by Corvino et
al.~\cite{Corvino2010} in general relativity seem to confirm this
picture, although Ou and Tohline also point to the significance of the
existence of a minimum of the vortensity within the star.  These
studies seem to point to a type of resonant excitation of the unstable
mode. Ou and Tohline~\cite{Ou2006} further find that the one-arm
spiral instability can develop even for stiff equations of state
($\Gamma=2$), as well as for non-toroidal configurations, as long as
the radial vortensity profile exhibits a local minimum. A recent
simplified Newtonian perturbative analysis by Saijo and
Yoshida~\cite{Saijo2016} (see also~\cite{Yoshida:2016kol}) solving an
eigenvalue problem on the equatorial plane of a star with $j=\rm
const.$ differential rotation law, suggests that when a corotation
radius is present $f$-modes become unstable giving rise to the class
of ``low-$T/|W|$'', shearing instabilities. In Yoshida and
Saijo~\cite{Yoshida:2016kol} the role of the corrotation radius is
further explored and the authors suggest that low-$T/|W|$
instabilities may arise due because of ``over-reflection'' of sound
waves between the stellar surface and the corotation band. More
recently, Muhlberger et al.~\cite{Muhlberger2014}, find that $m=1$
modes were excited in general-relativistic magnetohydrodynamic
simulations of the low-$T/|W|$ instability in isolated neutron stars
(see also Fu and Lai~\cite{FuLai2011}). Despite multiple studies of
the $m=1$ instability, a clear interpretation of how and under what
conditions the instability arises is still absent.


\section{Dynamical Simulations of Rotating Stars in Numerical Relativity}
\label{rotating-stars}

 In the framework of the
3+1 split of the Einstein equations~\cite{Smarr78}, the spacetime
metric obtains the Arnowitt-Deser-Misner (ADM) form \cite{ADM2008}
\begin{equation}
  ds^2=-(\alpha^2-\beta_i\beta^i) dt^2+2\beta_idx^idt+\gamma_{ij}dx^idx^j,
\end{equation}
where $\alpha$ is the lapse function, $\beta^i$ is the shift
three-vector, and $\gamma_{ij}$ is the spatial three-metric, with
$i=1\ldots 3$. Casting the spacetime metric of a stationary,
axisymmetric rotating star [see Eq.~(\ref{e:metric})] in the ADM form,
the metric has the following properties:

\begin{itemize}
\item The metric function $\omega$ describing the dragging of inertial
  frames by rotation is related to the shift vector through
  $\beta^\phi =-\omega $. This shift vector satisfies the
  \emph{minimal distortion shift} condition.
\item The metric satisfies the \emph{maximal slicing} condition, while
  the lapse function is related to the metric function $\nu$
  in~(\ref{e:metric}) through $\alpha =e^\nu$.
\item The quasi-isotropic coordinates are suitable for numerical
  evolution, while the radial-gauge coordinates~\cite{Bardeen83} are
  not suitable for nonspherical sources (see~\cite{BGSM93} for
  details).
\item The ZAMOs are the Eulerian observers, whose worldlines are
  normal to the $t = \mathrm{\ const.}$ hypersurfaces.
\item Uniformly rotating stars have $\Omega = \mathrm{\ const.}$ in the
  \emph{coordinate frame}. This can be shown by requiring a vanishing
  rate of shear.
\item Normal modes of pulsation are discrete in
  the coordinate frame and their frequencies can be obtained by
  Fourier transforms (with respect to coordinate time $t$) of evolved
  variables at a fixed coordinate location~\cite{Font00}.
\end{itemize}

Crucial ingredients for the successful long-term and accurate
evolution of rotating stars in numerical relativity are the
Baumgarte-Shapiro-Shibata-Nakamura (BSSN)
(see~\cite{Nakamura87,Shibata95,Baumgarte99,Alcubierre00}) or
Generalized-Harmonic \cite{Pretorius04,Lindblom06} formulations for
the spacetime evolution, and high-order, finite volume
(magneto)hydrodynamical schemes that have been shown to preserve the
sharp features at the surface of the star (see
e.g.~\cite{Font00,Stergioulas01,Font02,Shibata03,Baiotti05,Duez05,Anderson06,Giacomazzo07,Yamamoto08,East12,ET12}).


\subsection{Numerical evolution of equilibrium models}


\subsubsection{Stable equilibrium}

\begin{figure*}[htbp]
  \center
  \includegraphics[width=0.7\textwidth]{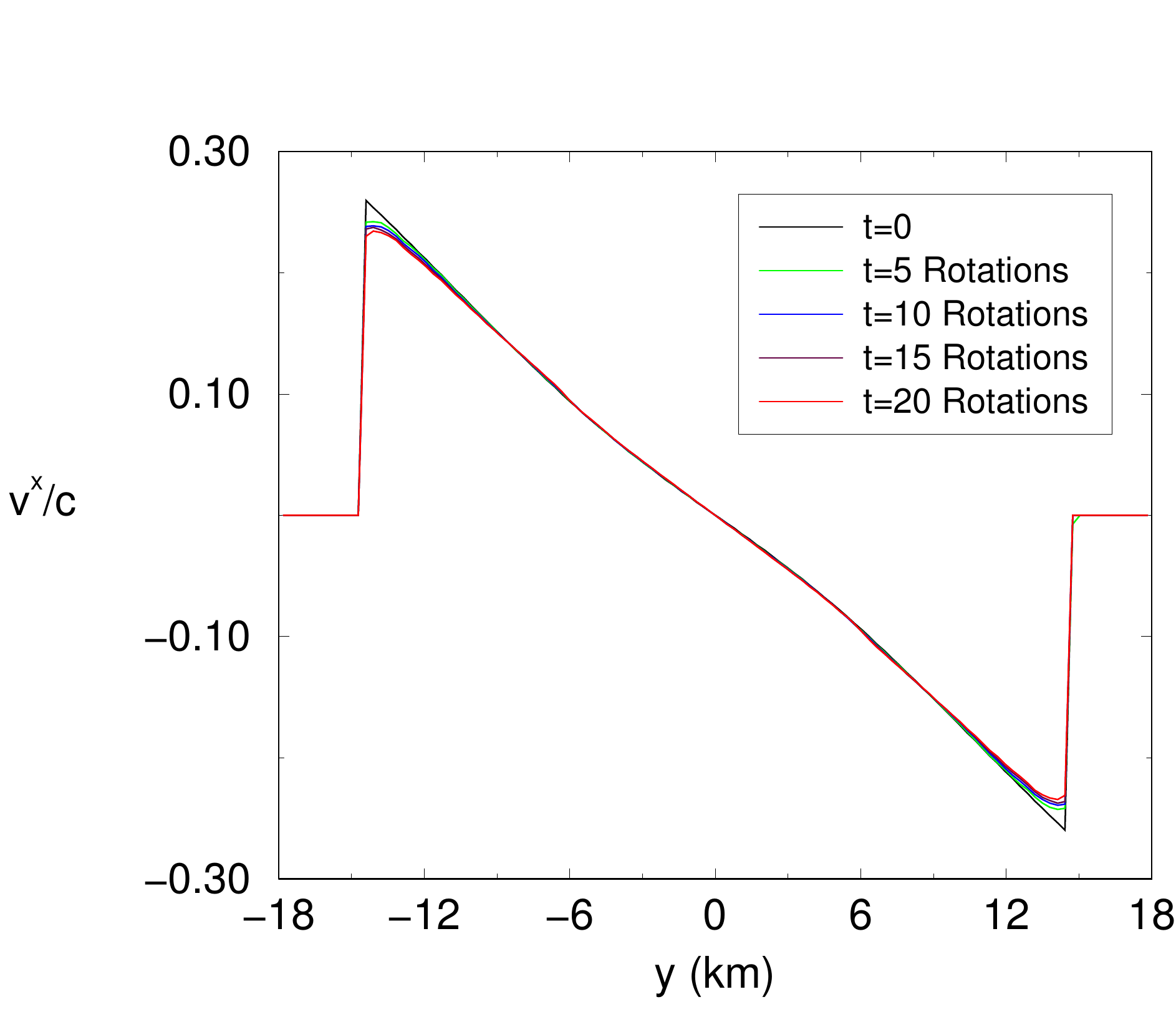}
  \caption{Time evolution of the rotational velocity profile for a
    stationary, rapidly rotating relativistic star (in the Cowling
    approximation), using the 3rd order PPM scheme and a $116^3$
    grid. The initial rotational profile is preserved to a high degree
    of accuracy, even after 20 rotational periods. (Image reproduced
    with permission from~\cite{Stergioulas01}, copyright by APS.)}
  \label{fig:rotprof}
\end{figure*}

Preserving the equilibrium of a stable rotating neutron star has now
become a standard test for numerical relativity codes. The long-term
stable evolution of rotating relativistic stars in 3D simulations has
become possible through the use of High-Resolution Shock-Capturing
(HRSC) methods (see~\cite{Font00b} for a review). Stergioulas and
Font~\cite{Stergioulas01} evolve rotating relativistic stars near the
mass-shedding limit for dozens of rotational periods (evolving only
the equations of hydrodynamics) (see Figure~\ref{fig:rotprof}), while
accurately preserving the rotational profile, using the 3rd order PPM
reconstruction~\cite{Collela84}. This method was shown to be superior to
other, commonly used methods, in 2D evolutions of rotating
relativistic stars~\cite{Font00}.

Fully coupled hydrodynamical and spacetime evolutions in 3D have been
obtained by Shibata~\cite{Shibata99b} and by Font {\it et
  al.}~\cite{Font02}. In~\cite{Shibata99b}, the evolution of
approximate (conformally flat) initial data is presented for about two
rotational periods, and in~\cite{Font02} the simulations extend to
several full rotational periods, using
numerically exact initial data and a monotonized central difference
(MC) slope limiter~\cite{vanLeer77}. The MC slope limiter is somewhat
less accurate in preserving the rotational profile of equilibrium
stars than the 3rd order PPM method, but, on the other hand, it is
easier to implement in a numerical code. 

Other evolutions of uniformly and differentially rotating stars in 3D,
using different gauges and coordinate systems, are presented
in~\cite{Duez03}, while 2D evolutions are presented
in~\cite{Shibata03}. In~\cite{Duez05}, the axisymmetric dynamical
evolution of a rapidly, uniformly rotating neutron star for 10
rotation periods shows that the PPM reconstruction preservers the
maximum value of the rest-mass density better than either the MC or
the Convex Essentially Non-oscillatory (CENO) \cite{CENO}
reconstruction method. It is reported that PPM achieves similar
performance also for full 3D evolution of the same rotating neutron
star models. The initial data for these simulations are equilibrium
numerical solutions of the Einstein equations generated using the Cook
et al.  code~\cite{CST96}. Evolutions of stable, uniformly rotating
neutron stars are also performed in \cite{Liebling2010} with initial
data generated in \cite{BBGN95}.


\subsubsection{Instability to collapse}

\paragraph{Hydrodynamic Simulations:}
Shibata, Baumgarte, and Shapiro~\cite{Shibata00b} study the stability
of supramassive neutron stars rotating at the mass-shedding limit, for
a $\Gamma=2$ polytropic EOS. Their 3D simulations in full general
relativity show that stars on the mass-shedding sequence, with central
energy density somewhat larger than that of the maximum mass model,
are dynamically unstable to collapse. Thus, the dynamical instability
of rotating neutron stars to axisymmetric perturbations is close to
the corresponding secular instability. The initial data for these
simulations are approximate, conformally flat axisymmetric solutions,
but their properties are not very different from exact axisymmetric
solutions even near the mass-shedding limit~\cite{Cook96}. It should
be noted that the approximate minimal distortion (AMD) shift condition
does not prove useful in the numerical evolution, once a horizon
forms. Instead, modified shift conditions are used
in~\cite{Shibata00b}. In the above simulations, no massive disk around
the black hole is formed, because the equatorial radius of the initial
model is inside the radius which becomes the ISCO of the final black
hole, a result also confirmed in \cite{Baiotti05} via 3D hydrodynamic
evolution in full general relativity, using HRSC methods and excision
technique to follow the evolution past the black hole formation.

To study the effects of the stiffness of the equation of state,
Shibata~\cite{Shibata2003} performs axisymmetric hydrodynamic
simulations in full GR of polytropic supramassive neutron stars with
polytropic index between $2/3$ and $2$. The initial data are
marginally stable and to induce gravitational collapse, Shibata
initially reduced the pressure uniformly by 0.5\% subsequently solving
the Hamiltonian and momentum constraints. Independently of the
polytropic index he finds the final state to be a Kerr black hole, and
the disk mass to be $<10^{-3}$ of the initial stellar mass.

  \begin{figure*}[t]
    \center
      \includegraphics[width=0.24\textwidth]{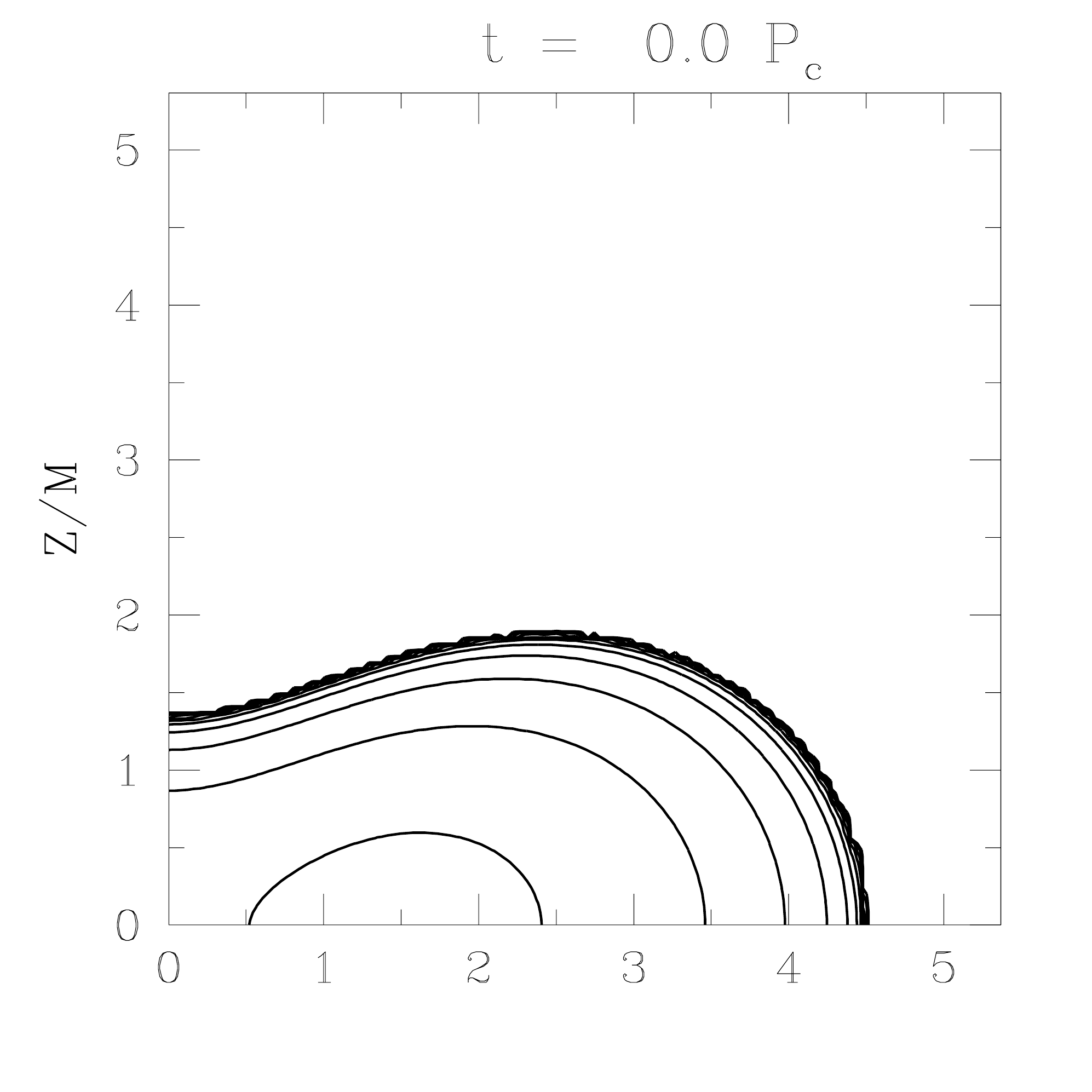}
      \includegraphics[width=0.24\textwidth]{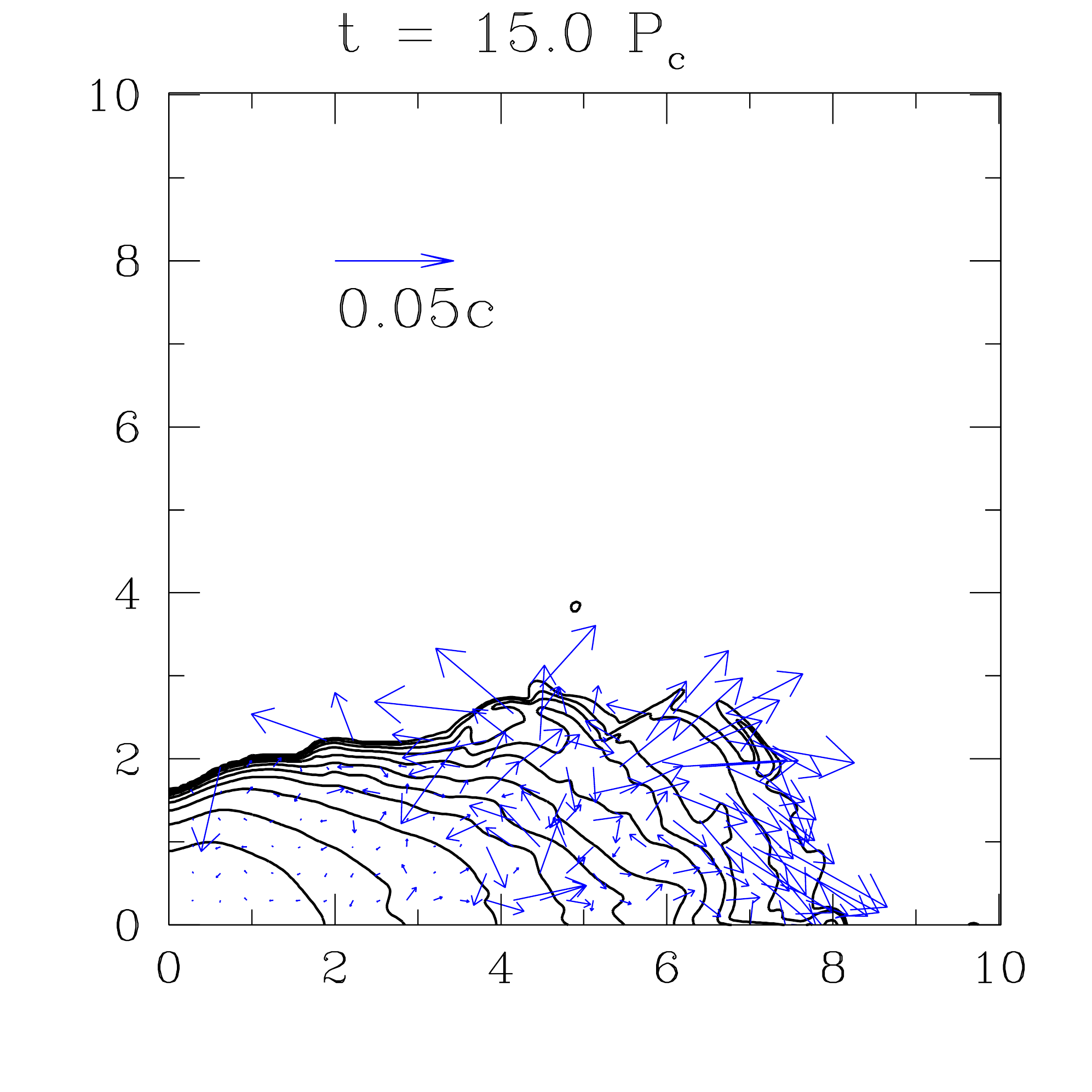}
      \includegraphics[width=0.24\textwidth]{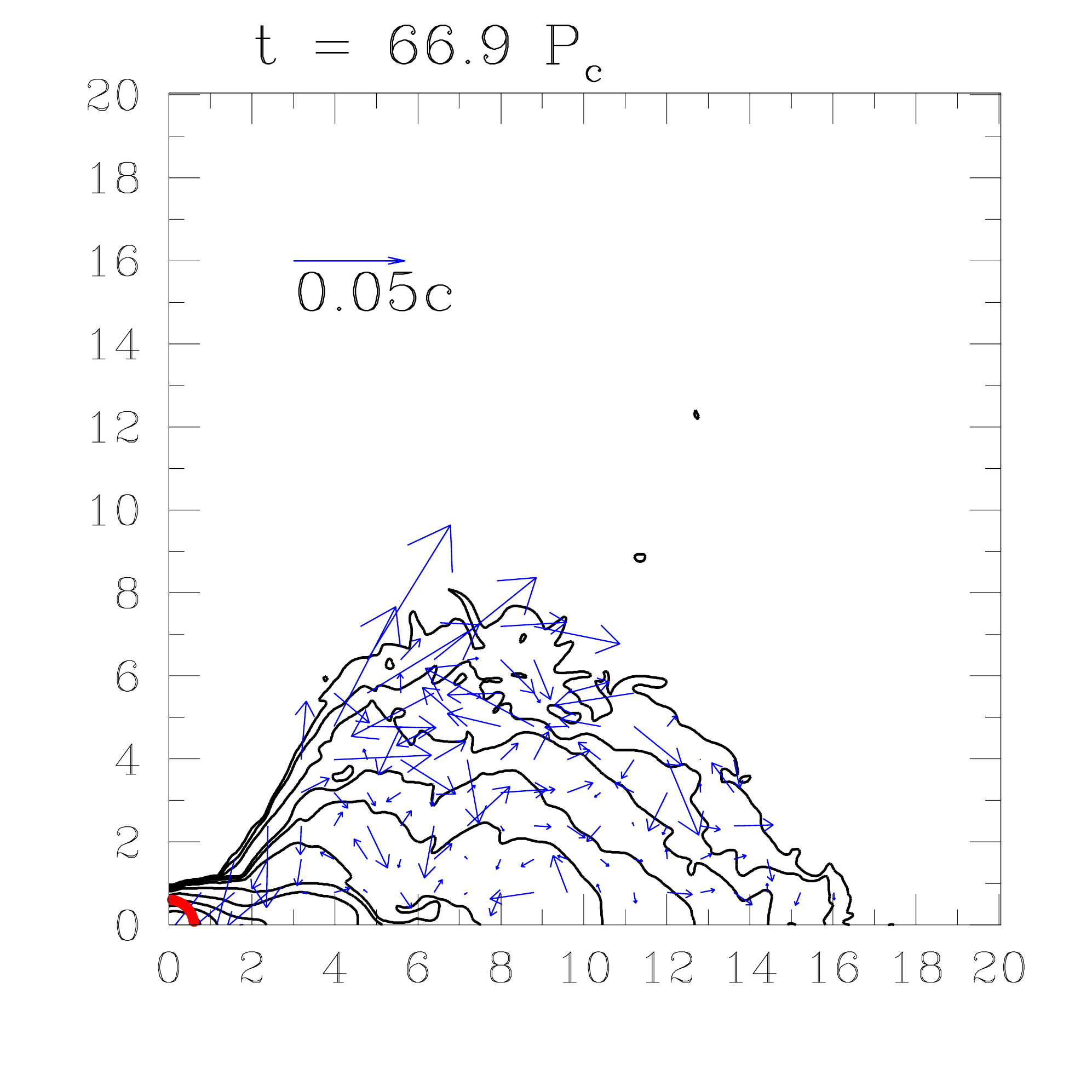}
      \includegraphics[width=0.24\textwidth]{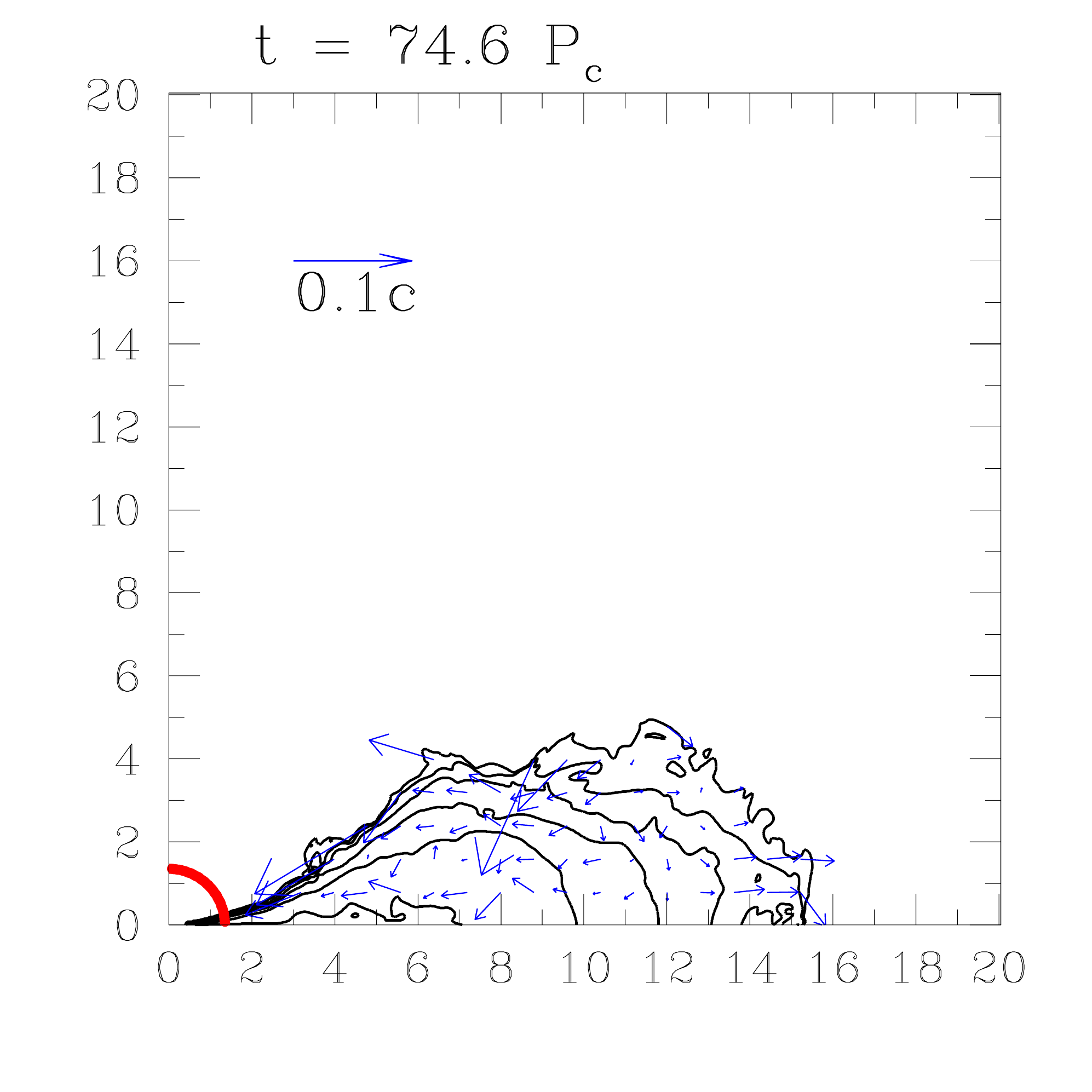}
      \includegraphics[width=0.24\textwidth]{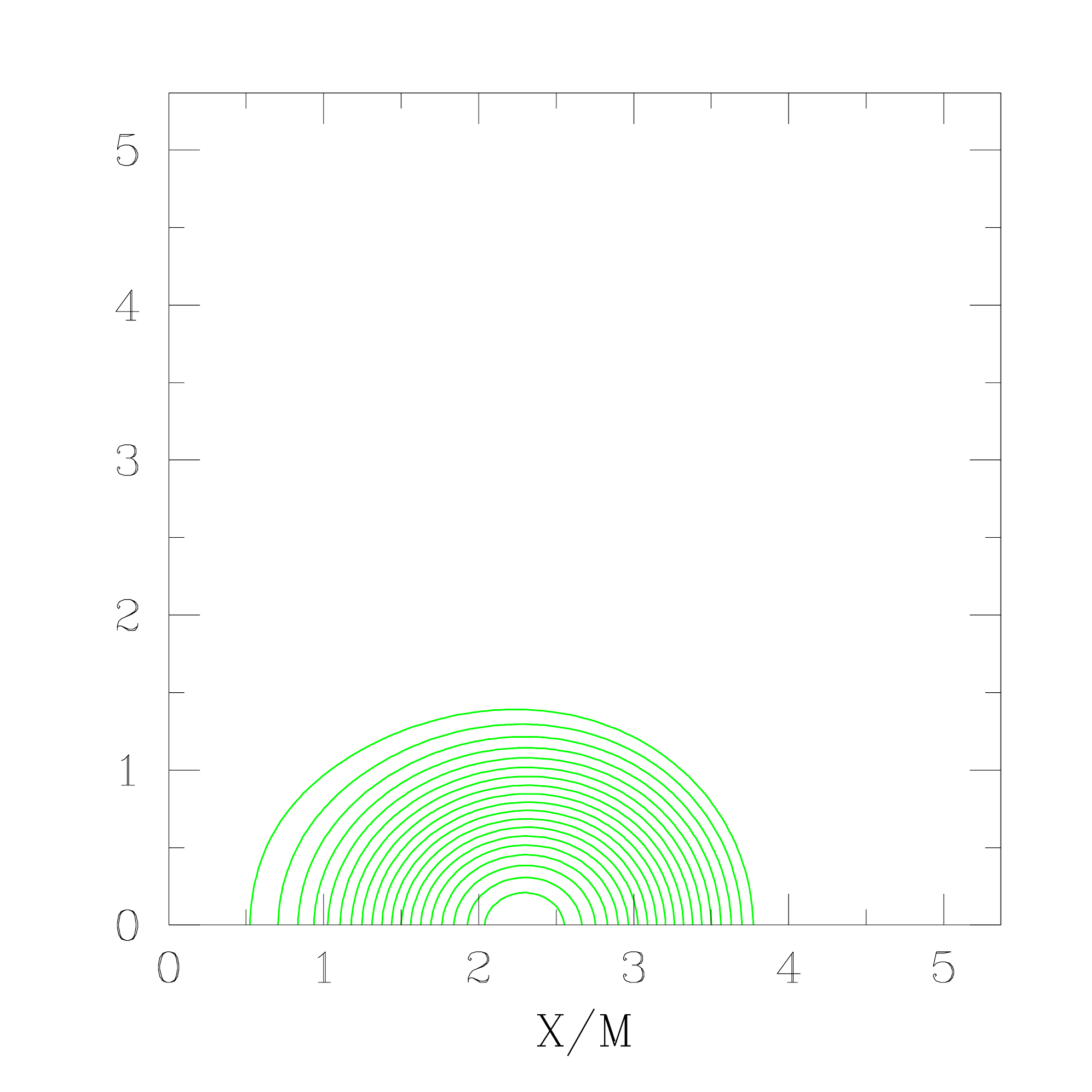}
      \includegraphics[width=0.24\textwidth]{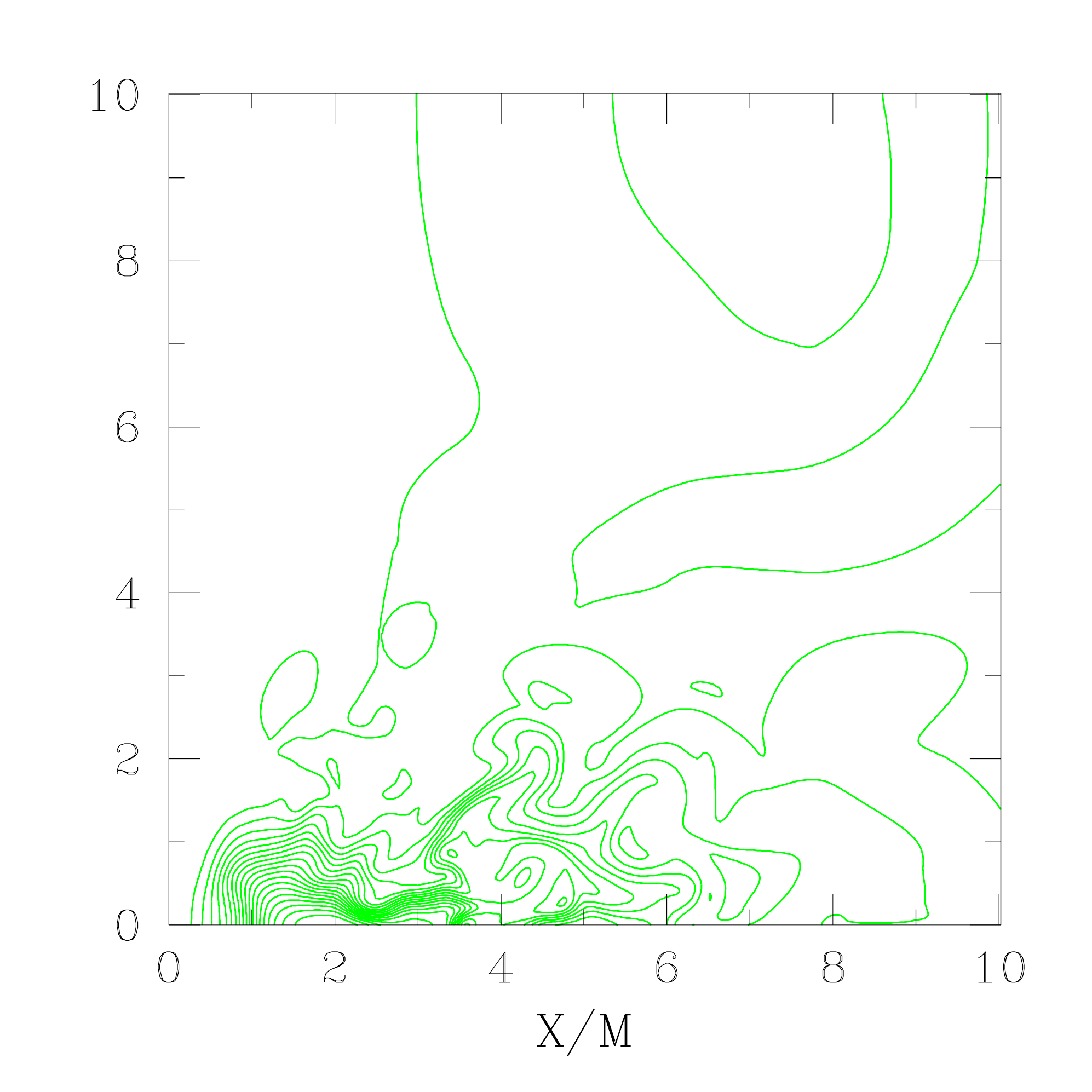}
      \includegraphics[width=0.24\textwidth]{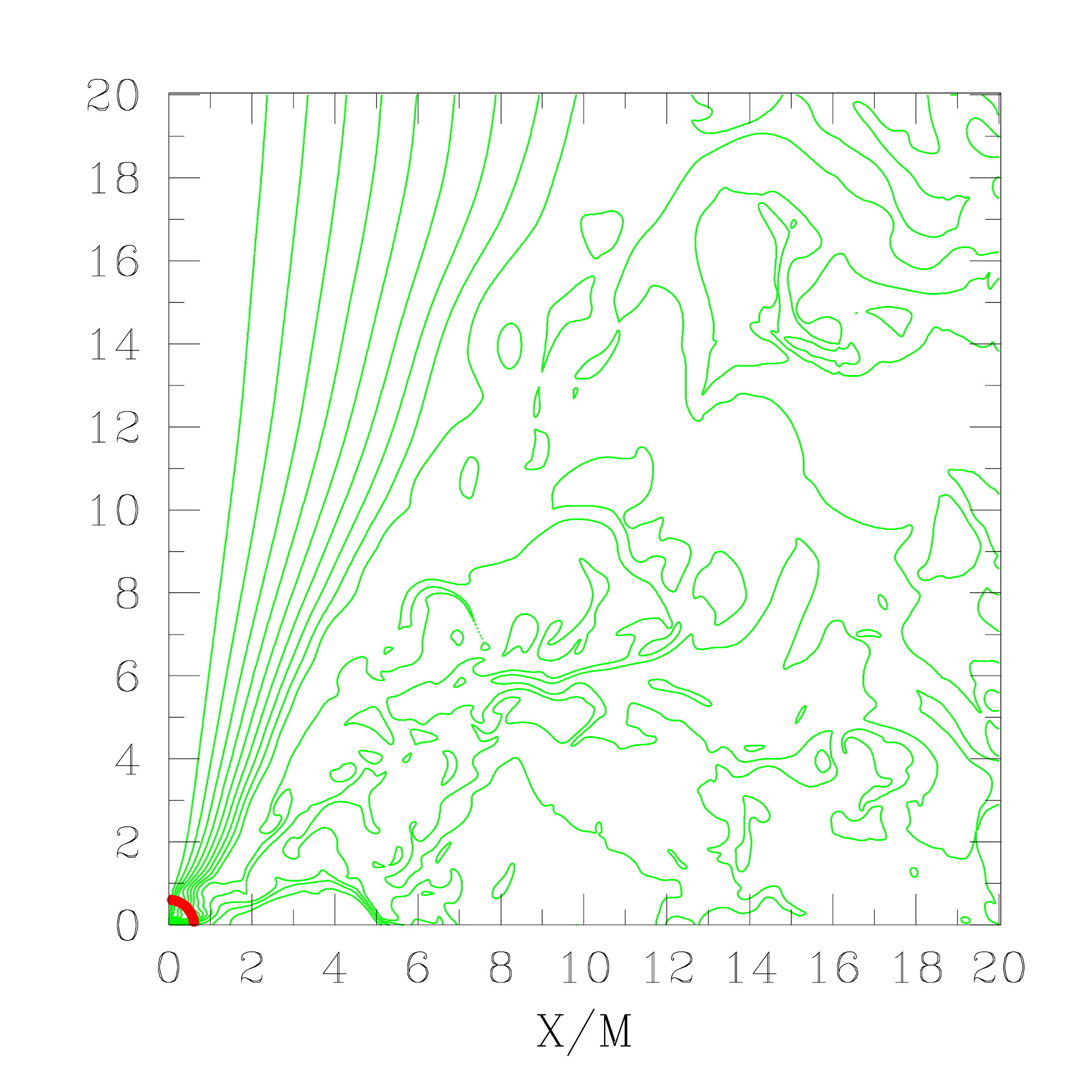}
      \includegraphics[width=0.24\textwidth]{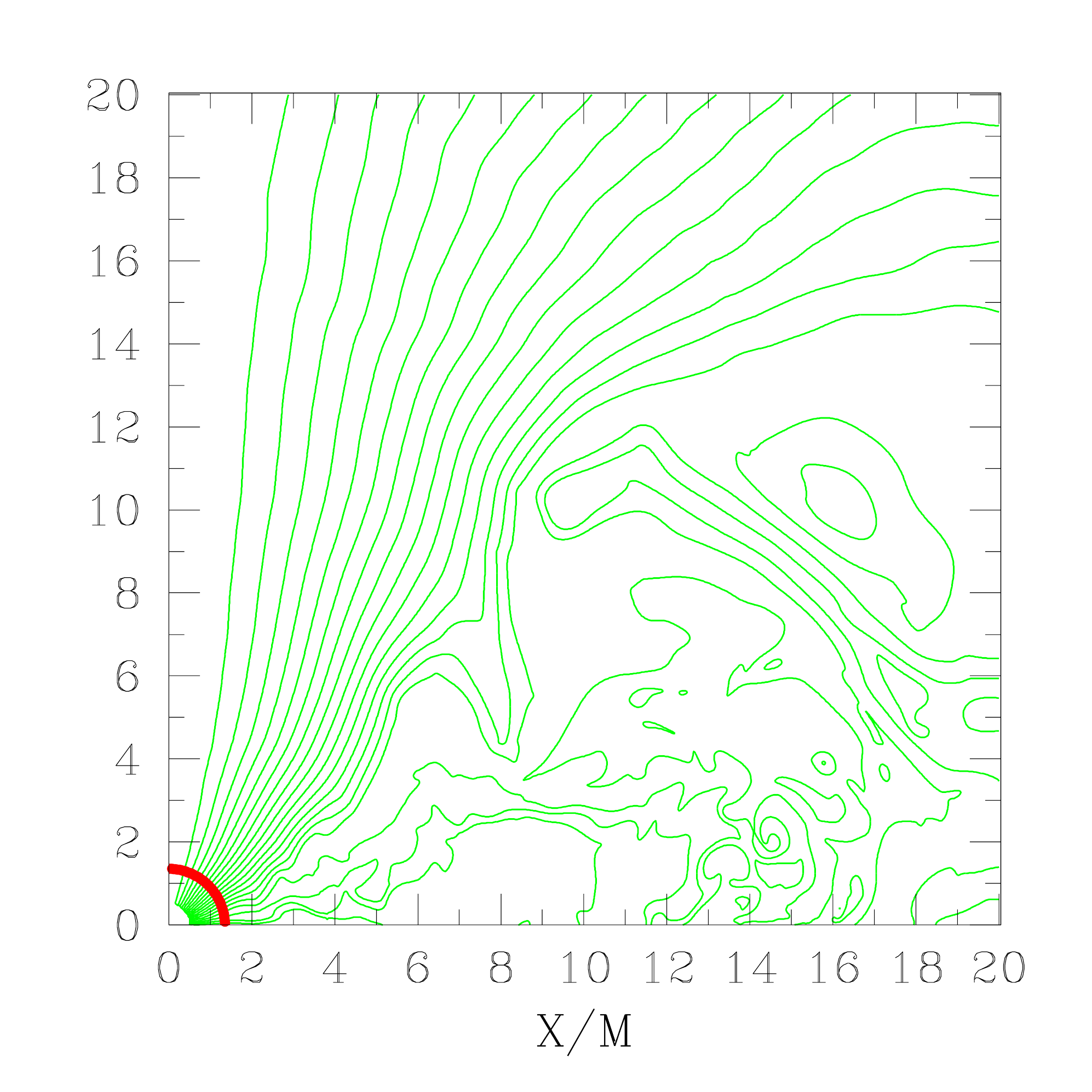}
    \caption{Magnetohydrodynamic evolution in full GR of a
      hypermassive, $n=1$ polytropic neutron star which is initially
      $70\%$ more massive than the $n=1$ supramassive-limit
      mass. Upper row: rest-mass density contours corresponding to
      $\rho/\rho_{\rm max,0}= 10^{-0.3 i-0.09}~(i=0$--12) and velocity
      arrows. Lower row: magnetic field lines are drawn for $A_{\phi}
      = A_{\phi,\rm min} + (A_{\phi,\rm max} - A_{\phi,\rm min})
      i/20~(i=1$--19), where $A_{\phi,\rm max}$ and $A_{\phi,\rm min}$
      are the maximum and minimum value of $A_{\phi}$ (the toroidal
      component of the vector potential) respectively at the given
      time.  The thick solid (red) curves denote the black hole
      apparent horizon.  (Image reproduced with permission from
      \cite{Duez2006a}, copyright by APS.)  }
    \label{mag_HMNS}
\end{figure*}

Adopting the BSSN formulation, Duez et al.~\cite{Duez2004} perform
evolutions of rapidly (differentially) rotating, hypermassive, $n=1$
polytropic neutron stars with strong shear viscosity in full general
relativity both in axisymmetry and in 3D, as a means for predicting
the outcome after the loss of differential rotation. Like magnetic
fields, shear viscosity redistributes angular momentum, braking the
differential rotation until the star is eventually uniformly
rotating. During this process the outer layers of the star gain
angular momentum and the star expands. The loss of the differential
rotation support may lead to collapse to a black hole.  The initial
data used are self-consistent general relativistic equilibria
generated with the Cook et al. code \cite{CST94a,CST94b}.  It is found that
without rapid cooling, if the hypermassive NS is sufficiently massive
($38\%$ more massive than the $n=1$ supramassive-limit mass) the star
collapses and forms a black hole after about 28 rotational periods
surrounded by a massive disk. The evolution is continued through black
hole formation by using excision methods. Hypermassive neutron stars
whose mass is only 10\% larger than the supramassive limit, do not
promptly collapse to a black hole due to the additional support provided by
thermal pressure which is generated through viscous heating. However,
rapid cooling (e.g. due to neutrinos) can remove the excess heat and
eventually these hypermassive neutron star models collapse to a black
hole, too. In all cases where a black hole forms, a massive disk
surrounds the black hole with rest-mass 10-20\% of the initial stellar
rest mass.

A different aspect of the neutron star collapse to a black hole is
investigated in Giacomazzo, Rezzola, and Stergioulas
\cite{Giacomazzo2011}, where the focus is on cosmic censorship. They
performed hydrodynamic simulations in full general relativity of
differentially rotating, polytropic neutron star models generated by
the {\tt RNS} code~\cite{RNS}. The evolutions are performed using the
BSSN formulation, and the {\tt Whisky} MHD code \cite{Giacomazzo07}.
They consider 5 values for the polytropic index
$(0.5,0.75,1.0,1.25,1.5)$, and initial models that are both sub-Kerr
$J/M^2<1$ and supra-Kerr $J/M^2>1$ to answer the following two
questions: (1) Do dynamically unstable stellar models exist with
$J/M^2 > 1$? (2) If a stable stellar model with $J/M^2 > 1$ is artificially induced to
collapse to a black hole, does it violate cosmic censorship? The
answer to question (1) is that finding supra-Kerr models which are
dynamically unstable to gravitational collapse will be difficult as at least a parameter survey for different polytropes and different strengths of (one-parameter) differential rotation did not produce such models. The
answer to (2) is that a supra-Kerr model can be induced to collapse
only if a severe pressure depletion is performed. However, even in
this case, prompt formation of a rotating black hole does not take
place. This result does not exclude the possibility that a naked
singularity can be produced by the collapse of a supra-Kerr,
differentially rotating star. However, the authors argue that a
generic supra-Kerr progenitor does not form a naked singularity,
thereby indicating that cosmic censorship still holds in the collapse
of differentially rotating neutron stars.

\paragraph{Magnetohydrodynamic Simulations:}
Duez et al. \cite{Duez2006a} perform axisymmetric ideal
magnetohydrodynamic (MHD) evolutions in full GR of a $n=1$ polytropic,
equilibrium model of hypermassive neutron star which is initially
seeded with dynamically unimportant purely poloidal magnetic fields
(plasma $\beta$ parameter $\sim 10^3$), but sufficiently strong so
that the fastest growing mode of the MRI can be resolved. The mass of
the neutron star model is $70\%$ larger than the corresponding TOV
limit. The spacetime evolution is performed using the BSSN formulation
and a non-staggered, flux-CT constrained transport method is employed
to enforce the ${\bf \nabla \cdot B}=0$ constraint to machine
precision (see \cite{Duez05} and references therein).  They find that
magnetic winding and the MRI amplify the magnetic field and
magnetically brake the differential rotation through redistribution of
angular momentum. Following 74.6 rotational periods of evolution, the
star eventually collapses to form a black hole surrounded by a
turbulent, magnetized, hot accretion torus with large scale collimated
magnetic fields (see Figure~\ref{mag_HMNS}). Due to the chosen gauge
conditions the evolution could not be continued sufficiently long
after the black hole formation to observe collimated outflows, but the
remnant system provides a promising engine for a short-hard Gamma-ray
Burst (sGRB). In a companion paper Shibata et al.~\cite{Shibata2006}
perform axisymmetric magnetohydrodynamic simulations in full GR of the
magnetorotational, catastrophic collapse of initially piecewise
polytropic hypermassive neutron star models seeded with weak, purely
poloidal magnetic fields, and find that the remnant torus has a
temperature $\geq 10^{12}$~K and can hence lead to copious
($\nu\bar\nu$) thermal radiation. In follow up work \cite{Duez2006b},
the authors use both polytropic and piecewise polytropic
differentially rotating neutron star models and find that catastrophic
collapse to a BH and a plausible sGRB engine forms only for
sufficiently massive hypermassive neutron stars (as low as 14\% more
massive than the supramassive limit mass). The end state of initially
differentially rotating neutron stars whose mass is smaller than the
supramassive limit mass is a uniformly rotating neutron star core,
surrounded by a differentially rotating torus-like envelope. The
remnant black hole-tori systems are evolved in~\cite{Stephens2008}
using a combination of black hole excision and the Cowling
approximation, finding that these systems launch mildly relativistic
outflows (Lorentz factors $\sim1.2\ -\ 1.5$), but that a stiff
equation of state is likely to suppress these outflows. We note that
GW searches triggered by sGRBs have tremendous potential to constrain
sGRB progenitors such as collapsing hypermassive neutron stars that
are formed following binary neutron star
mergers~\cite{Abbott:2016cjt}.

Magnetohydrodynamic evolutions of a magnetized and uniformly rotating,
unstable neutron star in full general relativity and 3 spatial
dimensions are performed in~\cite{Liebling2010}. Instead of the BSSN
formulation, the generalized harmonic formulation with excision is
adopted. The ${\bf \nabla \cdot B}=0$ constraint is controlled by means
of a hyperbolic divergence cleaning method (see \cite{Anderson06} and
references therein). A $\Gamma$-law ($\Gamma=2$) equation of state is
used and the initial neutron star models are self-consistent, $n=1$
polytropic, uniformly rotating, magnetized (polar magnetic field
strength $10^{16}$G), equilibrium, self-consistent solutions of the
Einstein equations generated with the {\tt Magstar} code
\cite{BBGN95}. Without any initial perturbation the unstable star
collapses and forms a black hole. In agreement with earlier studies,
the calculations demonstrate no evidence of a significant remnant
disk. However, evidence for critical phenomena is found when the
initial star is perturbed: a slight increase of the initial pressure
leads to collapse and black hole formation, however if the initial
perturbation increases the pressure above a threshold value, the star
expands and oscillates around a new, potentially stable solution.

\paragraph{Gravitational radiation from rotating neutron star collapse:}

Baiotti et al.~\cite{Baiotti05b} study gravitational wave emission
from rotating collapse of a neutron star. They perform 3D hydrodynamic
simulations in full general relativity using the {\tt Whisky}
hydrodynamics code~\cite{Baiotti05}. Singularity excision is used to
follow the black hole formation. The equilibrium initial data
correspond to a uniformly rotating neutron star near its mass-shedding
limit (of dimensionless spin parameter $J/M^2=0.54$), and are
solutions to the Einstein equations. The matter is modeled as an $n=1$
polytrope, and the collapse is triggered by a $2\%$ pressure
depletion. The constraints are re-solved after the pressure reduction
to start the evolution with valid, constraint-satisfying initial
data. A $\Gamma$-law equation of state is used for the evolution to
allow for shock heating. The gravitational waves are extracted on
spheres of large radius using the gauge invariant Moncrief
method~\cite{Moncrief1974}. The authors find that a characteristic
amplitude of the gravitational wave burst is $h_c = 5.77\times
10^{-22}(M/M_\odot)(r/10\rm kpc)^{-1}$ at a characteristic frequency
$f_c = 931\rm Hz$. The total energy emitted in gravitational waves is
found to be $E/M = 1.45\times 10^{-6}$, and they report that these
waves could be detectable by ground based gravitational wave laser
interferometers, but only for nearby sources. In a follow-up paper,
Baiotti and Rezzolla~\cite{Baiotti06} use the {\tt Whisky}
hydrodynamics code to perform simulations of collapsing, slowly and
rapidly, uniformly rotating neutron stars. The rapidly rotating model
is the same as the one in~\cite{Baiotti05b}, while the slowly rotating
model has a dimensionless spin parameter ($J/M^2=0.21$).  For these
simulations they adopted the puncture gauge
conditions~\cite{Campanelli2006,Baker2006}, instead of singularity
excision, allowing them to continue the integration of the Einstein
equations for much longer times and even study the black hole
ring-down phase. With the complete gravitational wave train the
authors report that the energy lost to gravitational wave emission
becomes $E/M = 3.7\times 10^{-6}$ ($E/M = 3.3\times 10^{-6}M$) for the
rapidly (slowly) rotating progenitor. The role of the puncture gauge
conditions in collapse simulations has been investigated
in~\cite{Thierfelder:2010dv,Dietrich:2014wja}.

  \begin{figure*}[t]
    \center
      \includegraphics[width=0.5\textwidth]{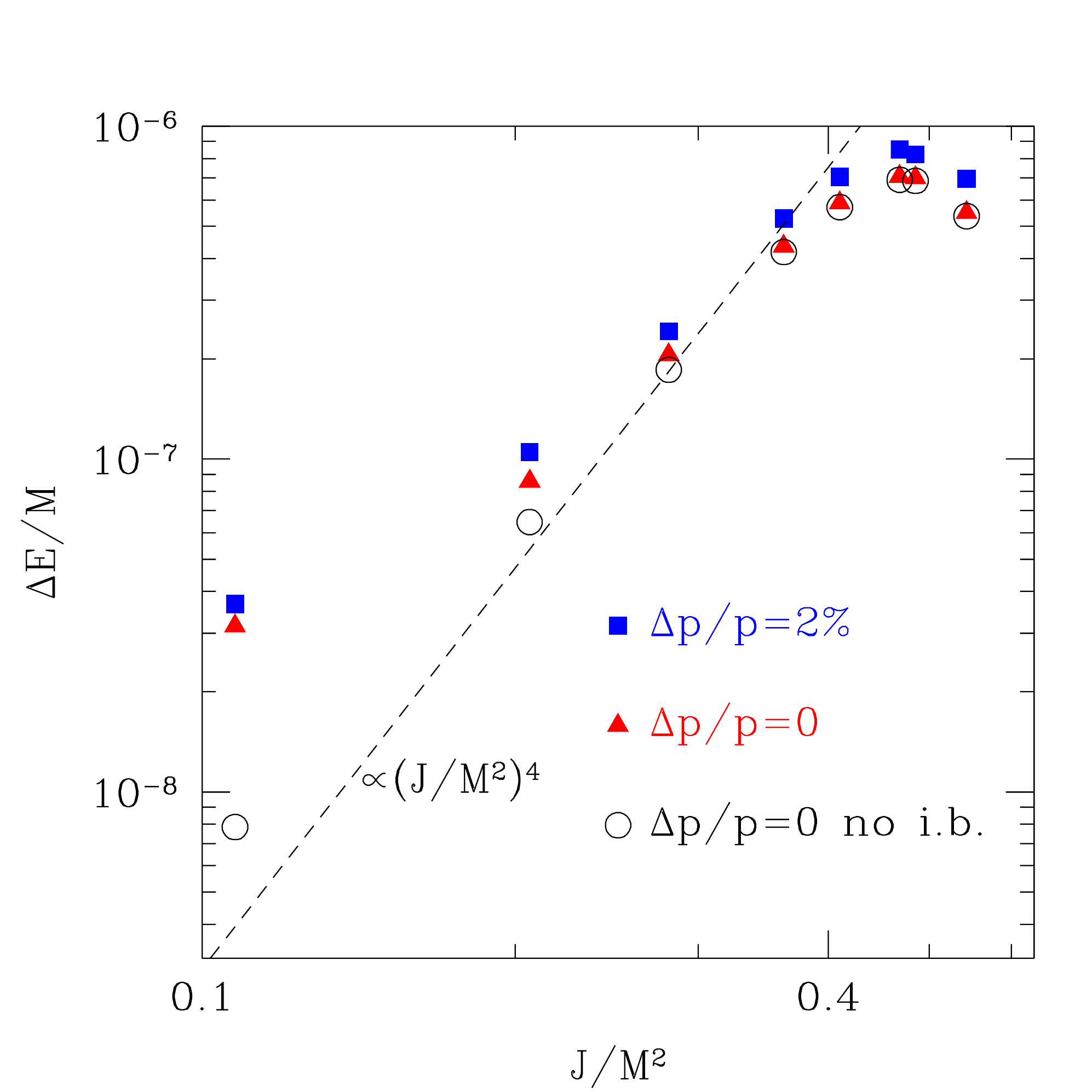}
      \includegraphics[width=0.5\textwidth]{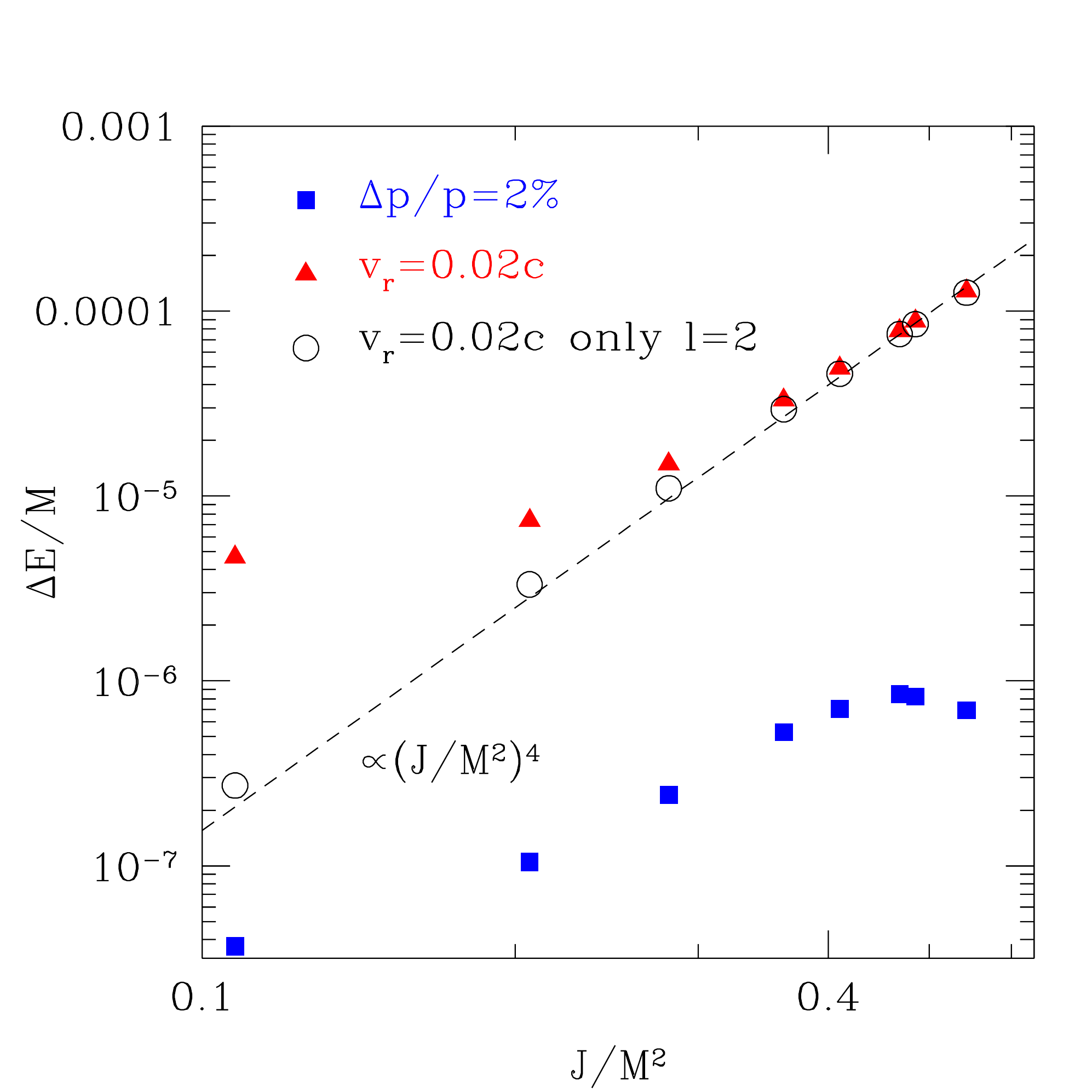}
    \caption{ Energy carried off by gravitational waves during the
      collapse of uniformly rotating neutron stars for different
      values of dimensionless spin parameter $J/M^2$ and initial
      perturbations. Left panel: Filled squares and triangles denote
      models with a $2\%$ pressure depletion and unperturbed models,
      respectively. Open triangles, refer to the same models as the
      filled ones, but exclude the initial (potentially spurious)
      burst in the waveforms. Right panel: Filled triangles denote
      models with an inward radial velocity perturbation of magnitude
      $0.02c$, and the open circles the same models but considering
      only the $l =2$ contribution to the emitted energy; for
      comparison the pressure-depleted models are plotted (filled
      squares). In both plots the dashed lines indicate a scaling
      $\sim (J/M^2)^4$. Image reproduced with permission
      from~\cite{Baiotti2007}, copyright by IOP.}
  \label{Baiotti07GWrotcollapse}
\end{figure*}


To study the effects of rotation and different perturbations on the
gravitational waves arising from the collapse of uniformly rotating
neutron stars, Baiotti, Hawke and Rezzola~\cite{Baiotti2007} perform
hydrodynamic simulations in full GR using the {\tt Whisky} code and
similar evolution techniques as in~\cite{Baiotti06}. For initial data
they consider 9 equilibrium neutron star models along a sequence of
dynamically unstable stars, all modeled as $n=1$ polytropes, whose
dimensionless spin parameter ranges from $0$ to $0.54$, and their
ratio of kinetic to gravitational binding energy ranges from $0$ to
$7.67$. For all cases the collapse is induced either by a $2\%$
pressure depletion or the addition of an inward, radial velocity
perturbation of magnitude $0.02c$, but this time the constraints are
not resolved leading to a small initial violation of the
constraints. The gravitational waves were extracted on a sphere of
radius $50M$, and the results of their study are summarized in
Fig.~\ref{Baiotti07GWrotcollapse}, where the total energy carried off
by gravitational waves $E$ is plotted vs $J/M^2$ for various
perturbations.  It is clear that over the range of dimensionless spin
parameters, rotation influences the gravitational wave amplitude by
about 2 orders of magnitude. When excluding the initial (potentially
spurious) burst of gravitational waves, $E$ scales roughly as $(J/M^2)^4$
and the models with velocity perturbations emit gravitational waves
more strongly (total energy emitted is about 2 orders of magnitude
larger than $2\%$ pressure perturbations). 

  \begin{figure*}[t]
    \center
      \includegraphics[width=0.55\textwidth]{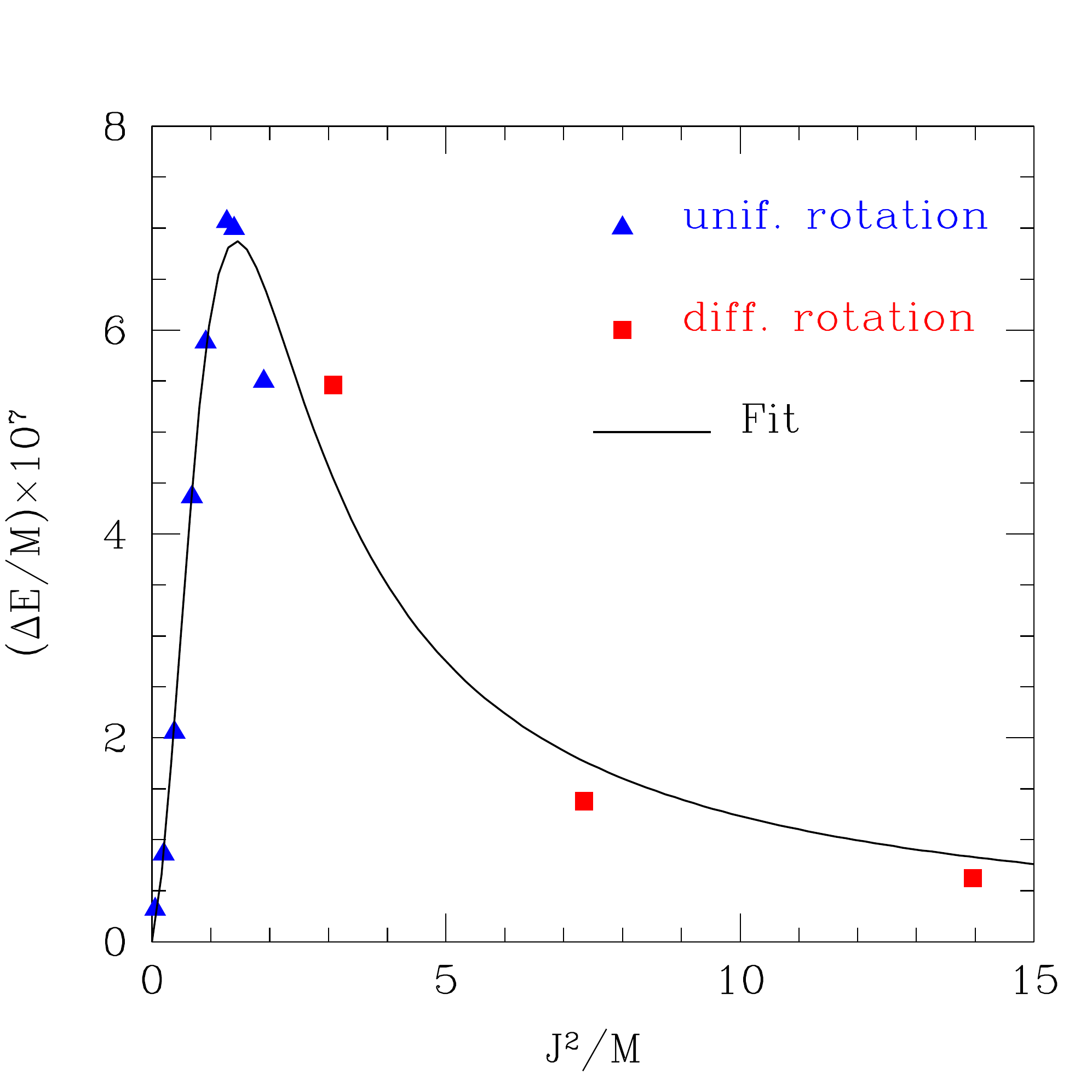}
  \caption{Energy carried off by gravitational waves, normalized to
    the total initial mass $M$, as a function of $J^2/M$ for
    collapsing, rapidly rotating neutron stars initially modeled as
    $n=1$ polytropes. Triangles represent uniformly rotating models,
    whereas the squares refer to the differentially rotating models
    discussed here. The solid line is the best fit (see
    Eq.~\eqref{EvsM_NScollapse}). (Image reproduced with permission
    from \cite{Giacomazzo2011}, copyright by APS.)}
  \label{fig:EvsJ2oM}
\end{figure*}

In \cite{Giacomazzo2011} Giacomazzo, Rezzolla and Stergioulas, in
addition to addressing cosmic censorship, they also extended these
results by considering the gravitational wave emission arising from
differentially rotating, collapsing neutron stars with initially $J/M^2
> 0.54$. They consider the energy lost in gravitational waves as a
function of $J^2/M$ (which is proportional to the initial quadrupole
moment). They find that as $J^2/M$ increases past $J^2/M \approx 1$,
$E$ decreases (see Fig.~\ref{fig:EvsJ2oM}) and suggest the following
fitting formula for the energy carried off by gravitational waves
\be
\label{EvsM_NScollapse} 
\frac{E}{M}=\frac{(J^2/M)^{n_1}}{a_1(J^2/M)^{n_2}+a_2}, 
\ee
where $n_1=1.43\pm 0.74$, $n_2=2.63\pm 0.53$, $a_1=(5.17\pm
4.37)\times 10^{5}$, $a_2=(1.11\pm 0.57)\times 10^{5}$.
The authors also analyze the signal-to-noise ratio for these
sources assuming a fiducial distance of $10$kpc, finding that
ratios of order 50 are possible for advanced LIGO and VIRGO, 
around 1700 for the Einstein telescope, thus these objects 
could be detectable by third-generation, ground-based laser
interferometers at distances of $\sim 1$ Mpc.

Collapse of the NS can potentially occur not only to a black hole but
also to a hybrid quark star, i.e., a compact star with a deconfined
quark matter core and outer layers made of neutrons. The collapse can
proceed through a first-order phase transition in the core. Following
up on the Lin et al. study~\cite{LinCheng2006}, this scenario (known
as phase-transition-induced collapse) is studied in Abdikamalov et
al.~\cite{Abdikamalov09} via axisymmetric, general relativistic
hydrodynamic simulations adopting the conformal flatness approximation
using the {\tt CoCoNut} code~\cite{Dimmelmeier2005}. The initial
neutron star models are both non-rotating and rotating, $\Gamma=2$
polytropes. The evolution adopts an approximate, phenomenological
hybrid equation of state: a $\Gamma$-law ($\Gamma=2$) is adopted for
hadronic matter -- ``hm'' for short -- (rest-mass density ($\rho_0$)
such that $\rho_0 < \rho_{\rm hm}=6.97\times 10^{14}\rm g\ cm^{-3}$),
the EOS of the MIT bag model for massless and non-interacting quarks
at zero temperature is adopted for the quark matter -- ``qm'' for
short -- ($\rho_0 > \rho_{\rm qm}=9\rho_{\rm nuc}$, where $\rho_{\rm
  nuc}=2.7\times 10^{14}\rm g\ cm^{-3}$ is the nuclear saturation
density), and a linear combination of the two for densities $\rho_{\rm
  hm} \le \rho_0 \le\rho_{\rm qm}$. The gravitational waves emitted by
the collapsing NS are computed using the quadrupole formula integrated
in time as in~\cite{Dimmelmeier02}.  Abdikamalov et al. find that the
emitted gravitational-wave spectrum is dominated by the fundamental
quasi-radial and quadrupolar pulsation modes, but that the strain
amplitudes are much smaller than suggested previously by Newtonian
simulations~\cite{LinCheng2006}. Therefore, it will be challenging to
detect gravitational waves from phase-transition-induced collapse.

The gravitational waveforms from collapse have also been  studied
through a new code adopting multi-patch methods in Reisswig et
al.~\cite{Reisswig:2012nc}, by Tietrich and Bernuzzi
in~\cite{Dietrich:2014wja} and through simulations of the
accretion-induced collapse of neutron stars by Giacomazzo and
Perna~\cite{Giacomazzo:2012bw}.


\subsubsection{Dynamical bar-mode instability}
\label{s:dynamical}

\begin{figure*}[htbp]
  \center
  \includegraphics[width=0.7\textwidth]{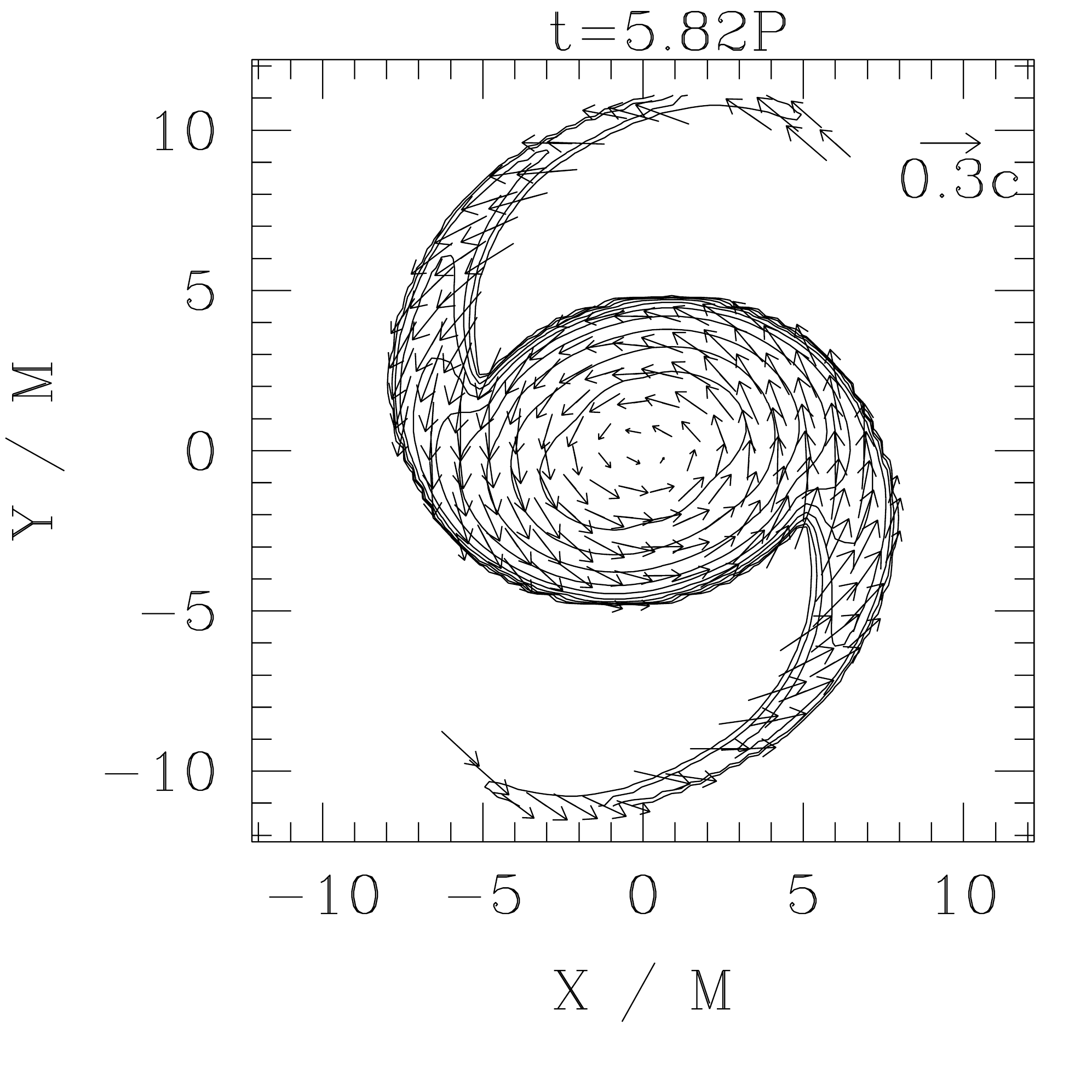}
  \caption{Density contours and velocity flow for a neutron star model
    that has developed spiral arms, due to the dynamical bar-mode
    instability. The computation was done in full General
    Relativity. (Image reproduced with permission
    from~\cite{Shibata00c}, copyright by Ap. J.).}
  \label{fig-spiral}
\end{figure*}

Shibata, Baumgarte, and Shapiro~\cite{Shibata00c} study the dynamical
bar-mode instability in differentially rotating neutron stars, in
fully relativistic 3D simulations. They find that stars become
unstable when rotating faster than a critical value of $\beta \equiv
T/|W| \sim 0.24 \mbox{\,--\,} 0.25$. This is only somewhat smaller
than the Newtonian value of $\beta\sim 0.27$. Models with rotation
only somewhat above critical become differentially rotating
ellipsoids, while models with $\beta$ much larger than critical also
form spiral arms, leading to mass ejection, see
for example Figure~\ref{fig-spiral}. In any case, the differentially rotating
ellipsoids formed during the bar-mode instability have $\beta >0.2$,
indicating that they will be secularly unstable to bar-mode formation
(driven by gravitational radiation or viscosity). The decrease of the
critical value of $\beta$ for dynamical bar formation due to
relativistic effects has been confirmed by post-Newtonian
simulations~\cite{Saijo01}.

In \cite{Shibata2005} Shibata and Sekiguchi study nonaxisymmetric
dynamical instabilities in the context of (differentially) rotating
stellar core collapse. The initial data corresponding to rotating,
$\Gamma=4/3$ polytropic models with maximum rest-mass density
$10^{10}\rm g/cm^3$, various degrees of differential rotation and
values for the rotational parameter $\beta=T/|W|$ ranging from
$0.00232$ to $0.0263$. The adopted differential rotation law is given
by
\be
u^tu_\phi = \varpi^2 (\Omega_a-\Omega), 
\ee 
where $\Omega=u^\phi/u^t$, $\Omega_a$ is the angular velocity at 
the location of the rotation axis, and $\varpi_d$ is a constant.

For the evolution a hybrid $\Gamma$-law equation of state is adopted
that has a cold part and a thermal part. The cold part has polytropic
exponent $\Gamma_1$ for densities less than the nuclear density and
$\Gamma_2$ otherwise. Most of the models in this study are evolved
using $\Gamma_1=4/3$ and $\Gamma_2=2$, but other values are
considered, too. The thermal part $\Gamma_{\rm th}=\Gamma_1$. The
early stage of the collapse is followed by an axisymmetric
hydrodynamic code in full GR and when the core becomes sufficiently
compact -- the minimum value of the lapse function becomes 0.8-0.85 --
they add a bar-mode density perturbation and after resolving the
Hamiltonian and momentum constraints assuming conformal flatness and
maximal slicing the evolution is followed using a 3D code. They find
that a dynamical bar more instability can occur in newly formed
neutron stars following the collapse of a stellar core, and that
$\beta$ can be amplified even beyond the value critical value $\sim
0.27$ when: i) the initial stellar model is highly differentially
rotating $\varpi_d/R_e \lesssim 0.1$, where $R_e$ is the equatorial
radius of the star; ii) the initial value of the rotational parameter
is in the range $0.01 \lesssim \beta_{\rm init} \lesssim 0.02$; iii)
the initial star is massive enough to become sufficiently compact so
that it is rapidly spinning, but less massive than the critical mass
value that leads to catastrophic collapse to a black hole. They find
that the maximum $\beta$ value reached by a proto-neutron star is
$0.36$ for a stiff equation of state with $\Gamma_2=2.75$.

  \begin{figure*}[t]
    \center
      \includegraphics[width=0.55\textwidth]{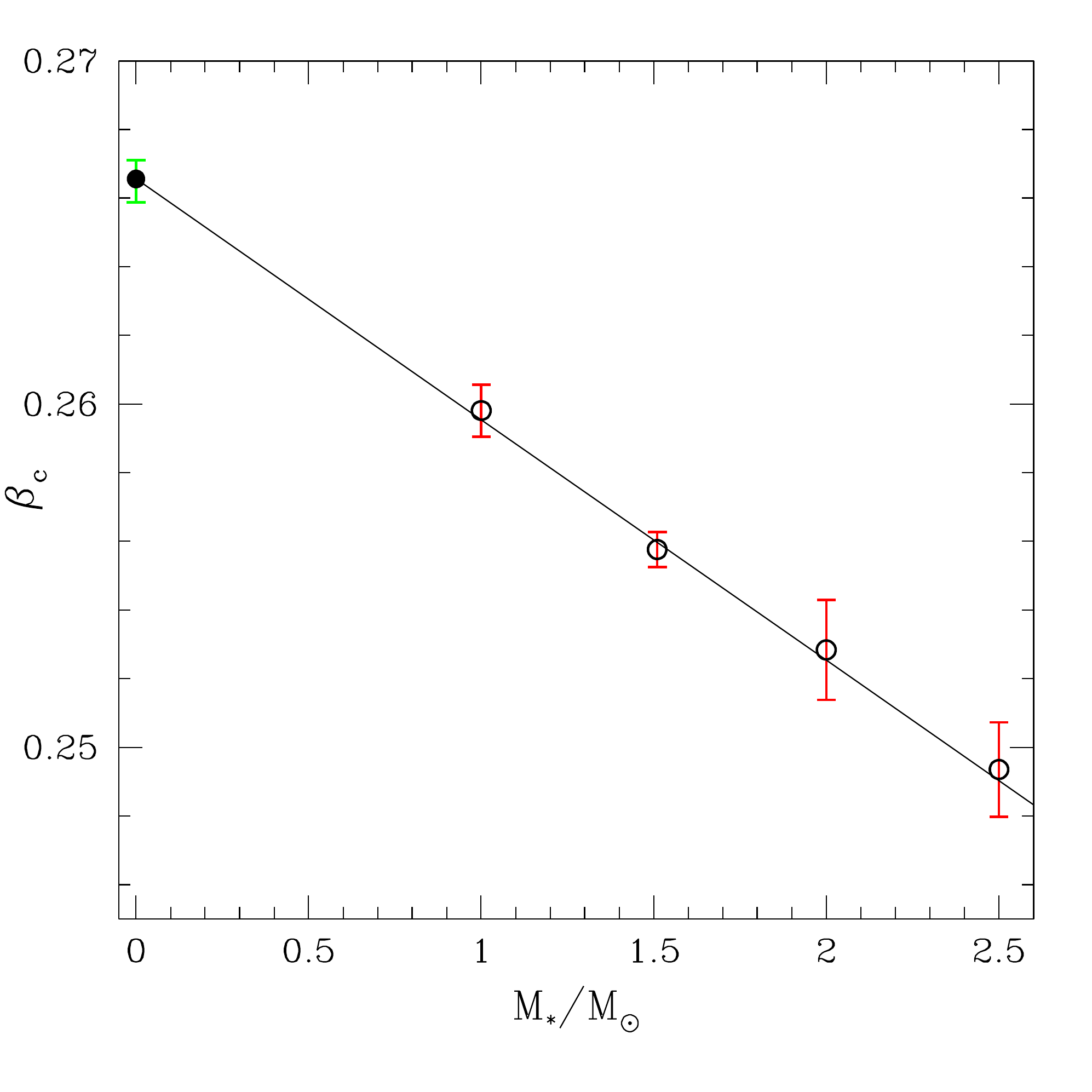}
  \caption{The open circles represent the extrapolated $\beta_c$ as a
    function of the stellar rest-mass (for the masses considered the
    compactness also increases along the positive horizontal axis). The
    filled circle represents the Newtonian limit for $\beta_c$. The
    plot corresponds to $\Gamma=2$ polytropes. (Image reproduced with
    permission from \cite{Manca07}, copyright by IOP.)}
  \label{fig:beta_vs_M}
\end{figure*}

Baiotti et al. \cite{Baiotti2007b} use the {\tt Whisky} code to
perform 3D hydrodynamic studies of the dynamical bar mode instability
in full GR. The initial data used correspond to equilibrium,
differentially rotating, polytropic neutron star models with
$\Gamma=2$, $K=100$ and have a constant rest mass of $M_0\approx
1.51M_\odot$ and various initial $\beta$ parameters. In contrast to
earlier studies that added an $m=2$ perturbation to trigger the bar
mode instability, in this work the instability is triggered primarily
by truncation error and additional simulations are performed to
investigate the effects of initial $m=1$ and $m=2$ perturbations. The
evolutions adopt of $\Gamma$-law equation of state.  They find that:
i) An initial $m=1$ or $m=2$ mode perturbation affects the lifetime of
the bar, but not the growth timescale of the instability, unless the
initial $\beta$ is near the threshold value for instability; ii) For
models with $\beta\sim \beta_{c}$ imposing $\pi$ symmetry can
radically change the dynamics and extend the lifetime of the
bar. However, this does not hold for models with initial $\beta\gg
\beta_c$, in which case even symmetries cannot produce long-lived bar;
iii) The bar lifetime depends strongly on the ratio $\beta/\beta_c$
and is generally of the order of the dynamical timescale ranging from
$\sim 6$ms to $\sim 24$ms (for comparison the initial equatorial
period considered range from $2$ms to $3.9$ms); iv) Nonlinear
mode-coupling takes place during the instability development, i.e, an
$m=2$ mode also excites an $m=1$ mode and vice versa. This mixing can
severely limit the bar lifetime and even suppress the bar.

Manca et al. \cite{Manca07} use the same methods as in
\cite{Baiotti2007b} to analyze the effects of initial stellar
compactness on $\beta_c$ for the onset of the dynamical bar-mode
instability through hydrodynamic simulations in full GR. They evolve
four sequences of models of constant baryonic mass ($1.0M_\odot$,
$1.51M_\odot$, $2.0M_\odot$, and $2.5M_\odot$,) for a total of 59,
$\Gamma=2$-polytropic stellar models. Using an extrapolation technique
for these models they estimate $\beta_c$ for each constant-mass
sequence, finding that the higher the initial compactness the smaller
$\beta_c$ becomes. Their results are summarized in
Fig.~\ref{fig:beta_vs_M}. In addition to the dependence of $\beta_c$
on the compactness, it is also found that for stars with sufficiently
large mass (rest-mass greater than $2.0M_\odot$) and compactness
(greater than $\sim 0.1$), the fastest growing mode corresponds to
$m=3$, and that for all 59 models the nonaxisymmetric instability
occurs on a dynamical timescale with the $m=1$ mode being dominant
toward the final stages of the instability.

Recently De Pietri et al.~\cite{DePietri2014} and L\"{o}ffler et
al.~\cite{Loffler:2014jma} use the {\tt Einstein Toolkit} \cite{ET12}
to perform a study very similar to the one in~\cite{Manca07} but
changing the polytropic exponent to $\Gamma=2.25,2.5,2.75,3.0$ and
considering five constant-mass sequences of differentially rotating
equilibrium neutron stars with masses $0.5M_\odot$, $1.0M_\odot$,
$1.5M_\odot$, $2.0M_\odot$, and $2.5M_\odot$. Using a similar
extrapolation method as in~\cite{Manca07} they find that the threshold
value for $\beta$ is reduced by $\sim 5\%$ when compared to the
$\Gamma=2$ case, concluding that a stiffer, realistic equation of
state is expected to have smaller $\beta_c$ for the onset of the
dynamical bar-mode instability.

Franci et al.~\cite{Franci2013} adopt the {\tt Whisky} code to perform
magnetohydrodynamic simulations in full GR in order to study the
effects of magnetic fields on the development of the bar-mode
instability. The initial $\Gamma=2$ polytropic, differentially
rotating, equilibrium stars are seeded with an initially purely
poloidal magnetic field confined in the neutron star interior.  The
magnitude of the magnetic field at the center of the star is chosen in
the range $10^{14}-10^{16}$G. Their magnetohydrodynamic calculations
show that strong initial magnetic fields, $B\gtrsim 10^{16}$G, can
suppress the instability completely, while smaller magnetic fields
have negligible impact on the instability.

We note here that some preliminary studies in full GR of the
low-$T/|W|$, bar-mode ($m=2$) instability have been carried out in
\cite{CerdaD2007,Corvino2010,DePietri2014} via hydrodynamic
simulations and in \cite{Muhlberger14} via magnetohydrodynamic
simulations, where it was shown that magnetic fields can suppress the
development of the instability, but only for a narrow range of the
magnetic field strength.



\subsection{Pulsations of rotating stars}
\label{pulsrot}

\begin{figure*}[htbp]
  \center
  \includegraphics[width=0.7\textwidth]{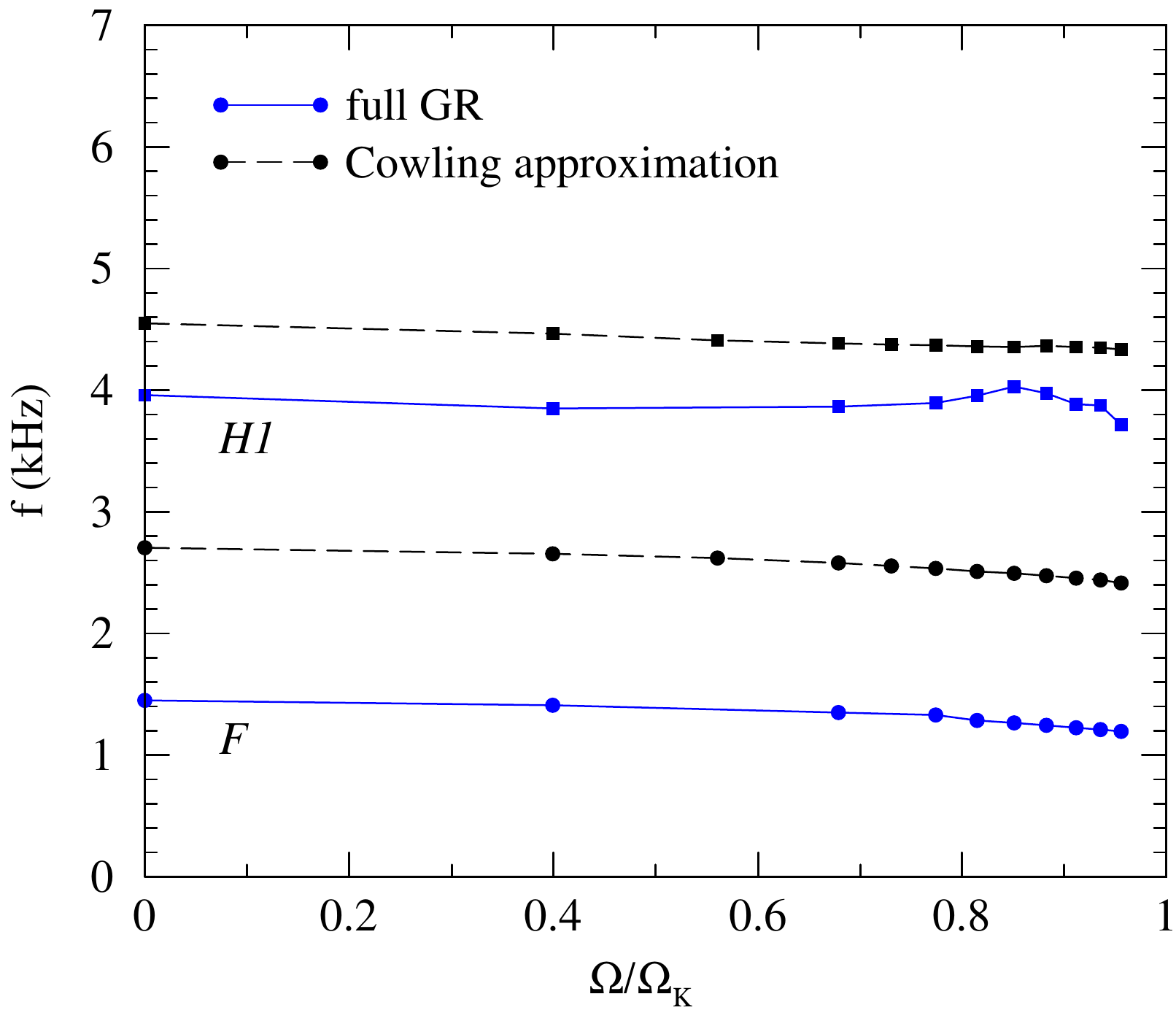}
  \caption{The first fully relativistic,
    quasi-radial pulsation frequencies for a sequence of rapidly
    rotating $N=1$ polytropes, up to the mass-shedding limit (with fixed central density along the sequence).  The frequencies of the fundamental
    mode $F$ (filled circles) and of the first overtone $H_1$ (filled
    squares) are obtained through \emph{coupled} hydrodynamical and
    spacetime evolutions (blue solid lines). The corresponding frequencies obtained from
    computations in the relativistic Cowling approximation (fixed spacetime)~\cite{Font01} are shown as black dashed lines. (Image
    reproduced with permission from~\cite{Font02}, copyright by APS.)}
  \label{fig:last}
\end{figure*}

Pulsations of rotating relativistic stars are traditionally studied
(when possible) as a time independent, linear eigenvalue problem, but
recent advances in numerical relativity also allow the study of such
pulsations via numerical time evolutions. Quasi-radial mode
frequencies of rapidly rotating stars in full general relativity have
been obtained in~\cite{Font02}, something that has not been achieved
yet with linear perturbation theory. The fundamental quasi-radial mode
in full general relativity has a similar rotational dependence as in
the relativistic Cowling approximation, and an empirical relation
between the full GR computation and the Cowling approximation can be
constructed (Figure~\ref{fig:last}). For higher order modes, apparent
intersections of mode sequences near the mass-shedding limit do not
allow for such empirical relations to be constructed. 

For a comparison
study of linear and non-linear evolution methods, in the case of nonrotating polytropic models, as well as for a
comparison of different gravitational wave extraction techniques see,
e.g., Baiotti et al.~\cite{Baiotti:2008nf}

In the relativistic Cowling approximation, 2D time evolutions have
yielded frequencies for the $l=0$ to $l=3$ axisymmetric modes of
rapidly rotating relativistic polytropes with
$N=1.0$~\cite{Font01}. The higher order overtones of these modes show
characteristic apparent crossings near mass-shedding (as was observed
for the quasi-radial modes in~\cite{Yoshida00}). For a recent code
aimed at computing oscillation frequencies adopting the conformal
flatness approximation see Yoshida~\cite{Yoshida2012}, who finds good
agreement between $f$ and $p$-mode frequencies when compared to the
full theory in the case of slowly rotating stars.

Numerical relativity has also enabled the first study of nonlinear
$r$-modes in rapidly rotating relativistic stars (in the Cowling
approximation) by Stergioulas and Font~\cite{Stergioulas01}. For
several dozen dynamical timescales, the study shows that nonlinear
$r$-modes with amplitudes of order unity can exist in a star rotating
near mass-shedding. However, on longer timescales, nonlinear effects
may limit the $r$-mode amplitude to smaller values (see
Section~\ref{s_axial}).

In another study, Siebel, Font and Papadopoulos
\cite{SiebelFontPapa2002} perform spherically symmetric simulations in
full GR of the Einstein-Klein-Gordon-perfect fluid system to study the
interaction of a scalar field with a spherical neutron star using a
characteristic approach to solve the dynamical equations. The initial
data correspond to a TOV, $\Gamma=2$ polytropic neutron star for the
fluid and metric variables, and a Gaussian pulse is chosen for the
initial scalar field. For a small amplitude scalar field pulse, radial
oscillations are excited on the NS, and a Fourier analysis shows that
the oscillation frequencies corresponding to the fundamental, first
and second overtone modes computed through the full non-linear
evolution are very close to those of a linearized analysis, but
generally smaller by $\lesssim 1-2\%$.

The gravitational waveforms from oscillating spherical and rigidly
rotating (near the mass shedding limit), $\Gamma=2$ polytropic neutron
stars have been computed by Shibata and Sekiguchi in
\cite{Shibata2004}. The equilibrium polytropic stars were perturbed
using a velocity perturbation with magnitude $0.1c$ at the neutron
star surface and evolved using a fully general relativistic
hydrodynamics code in axisymmetry. Gauge invariant methods and a
quadrupole formula were used to compute the gravitational wave
signature and it is found that the wave phase and modulation of the
amplitude can be computed accurately using a quadrupole formula but
not the amplitude. It is also found that for both spherical and
rotating stars the gravitational wave frequency is associated with the
fundamental $l=2$ mode, and that for rotating stars another frequency
in the gravitational wave signal is detected, which is likely
associated with the quasiradial oscillation $p_1$ mode.

Stergioulas, Apostolatos and Font~\cite{SAF} perform 2D hydrodynamic
simulations of differentially rotating neutron stars to study
non-linear pulsations in the Cowling approximation. It is found that
for $\Gamma=2$ polytropic stars near the mass shedding limit, shocks
forming near the stellar surface damp the oscillations and this
mechanism may set a small saturation amplitude for modes that are
unstable to the emission of gravitational waves.

  \begin{figure*}[t]
    \center
      \includegraphics[width=0.65\textwidth]{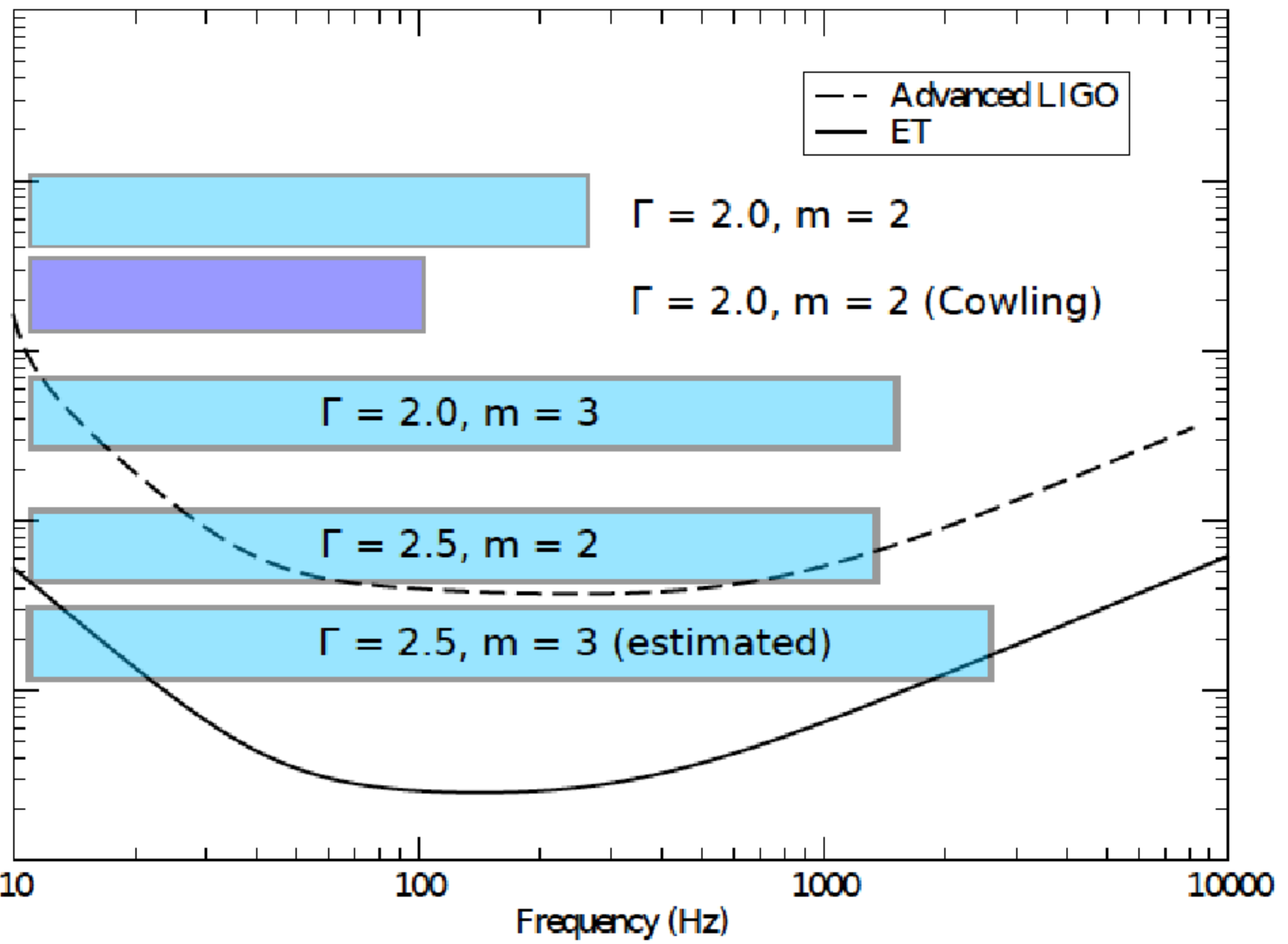}
  \caption{Frequency band of the CFS instability based on $m = 2$ and
    $m = 3$ f -mode oscillations for two sequences of rigidly rotating
    polytropic neutron stars with $\Gamma=2$ and $\Gamma=2.5$. Each
    band is limited on the left by the neutral point and on the right
    by the frequency of the counter-rotating mode in the most rapidly
    rotating model. The sensitivity noise curves of aLIGO and the
    future Einstein Telescope are also plotted. (Image reproduced with
    permission from \cite{Zink10}, copyright by APS.)}
  \label{frequency_bands}
\end{figure*}

Zink et al.~\cite{Zink10} study the location of the neutral-points for
the $l=|m| = 2$ and $l=|m| = 3$ $f$-mode oscillations of uniformly
rotating polytropes using 2D, hydrodynamic simulations both in the
Cowling approximation and in full general relativity. These
frequencies are important because rapidly rotating neutron stars can
become unstable to the gravitational-wave-driven CFS instability. A
polytropic equation of state is adopted both for the construction of
initial data (generated with the {\tt RNS} code) and for the
evolution. To assess the effects of the stiffness of the equation of
state, two sequences of models are considered: one with $\Gamma=2$ and
one with $\Gamma=2.5$.  The rotational parameter $\beta$ ranges from
$0$ to $0.08$ for the $\Gamma=2$ sequence, and from $0$ to $0.12$ for
the $\Gamma=2.5$ sequence. All models are chosen to have an initial
central rest-mass density close to that of the maximum mass TOV star
with the same equation of state, i.e., $\rho_c=2.6823\times 10^{-3}
(100/K)$ ($\rho_c=5.0\times 10^{-3} (1000/K)^{3/2}$) for $\Gamma=2$
($\Gamma=2.5$). To excite the modes the authors add a small-amplitude,
$l=|m|$ density perturbation and evolve the initial data. The authors
find that the Cowling approximation in some cases underestimates
the lower limit of the $f$-mode CFS instability window by $10\%$ and
the upper limit by $60\%$. The authors conclude that general
relativity enhances the detectability of a CFS-unstable neutron star
substantially and derive limits on the observable gravitational-wave
frequency band available to the instability, finding.  These results
are summarized in Fig.~\ref{frequency_bands}.

Kastaun, Willburger and Kokkotas~\cite{kastaun2010} perform
axisymmetric and 3D general relativistic hydrodynamic simulation in
the Cowling approximation using the {\tt PIZZA} code \cite{Kastaun06}
to qualitatively understand what mechanisms set the saturation
amplitude of the $f$-mode instability. Using the {\tt RNS} code they
build initial data corresponding to uniformly rotating neutron star
models with various masses and degree of rotation and polytropic
indices $n=2,1,0.6849$. For one of the models studied (MA65) the
$l=m=2$ $f$-mode is excited by the CFS instability in full GR, and
hence is a candidate for detectable gravitational waves. To excite
high amplitude oscillations, they linearly scale the eigenfunctions of
specific energy and 3-velocity, and add them to the background
solution. The primary focus is on the $l=|m|=2$ and $l=2,\ m=0$
modes. It is found that the saturation amplitude of high-amplitude
axisymmetric $f$-mode oscillations is rapidly determined by shock
formation near the surface of the star. It is also found that stiffer
EOSs allow higher amplitudes before shocks begin to dissipate the
$f$-modes, and rotation affects the damping of axisymmetric modes only
weakly, until the Kepler limit is reached, at which point damping
occurs via mass shedding. For nonaxisymmetric $f$-mode oscillations,
the saturation amplitude is not determined by shocks, but is primarily
determined by damping due to wave breaking and non-linear mode
coupling.

In follow up work Kastaun \cite{Kastaun2011} performs relativistic
hydrodynamical simulations in the Cowling approximation to study
non-linear, $r$-mode oscillations using the {\tt PIZZA} code.  The
initial data correspond to two rigidly rotating, $\Gamma=2$ polytropic
models of neutron stars with rest-masses $1.6194M_\odot$ and
$1.7555M_\odot$, and ratio of polar to equatorial radius $0.85$ and
$0.7$, respectively. The initial data are seeded with an $l=m=2$
perturbation based on the eigenvectors from linearized studies of
$r$-mode oscillations, and then scaled to larger amplitudes such that
the energy in the $r$-mode divided by the stellar binding energy is of
order $10^{-3}$. The initial data are then evolved adopting a
polytropic equation of state for the hydrodynamics. It is found that
the frequencies of the $r$-modes in the inertial frame agree to better
than $0.1\%$ with those found by linearized studies in
\cite{Krueger09}. As found in earlier studies, Kastaun finds that
differential rotation develops during the evolution of the $r$-mode
with high initial amplitude. Related with the onset of differential
rotation is also the decay of the $r$-modes, which is why it is
hypothesized that the saturation of the differential rotation should
occur when the $r$-mode decay due to differential rotation is balanced by
the differential rotation due to the presence of the $r$-mode. Kastaun
also argues that for the models under study the $r$-mode decay is not
due to shock formation near the stellar surface. Finally, it is
pointed out that while mode-mode coupling is probably not the main
cause of the $r$-mode decay for these models, as found for other
models in \cite{Gressman02,LinSuen2006} by Newtonian simulations,
mode-mode coupling cannot be ruled out as a possible cause for the
$r$-mode decay.

Note that the perturbative work by
Chugunov~\cite{Chugunov2015MNRAS.451.2772C} finds that, for the stable
$r$-mode, differential rotation is pure gauge, reflecting only the
differential rotation in the initial conditions. For the unstable
$r$-mode, Friedman, Lindblom and Lockitch~\cite{Friedman:2015iqa} find
that the 2nd-order differential rotation is unique.

It should be noted that the work of Kastaun should be considered
preliminary, because the presence of magnetic fields could brake the
differential rotation~\cite{Rezzolla00,Rezzolla01b,Rezzolla01c} and
also give rise to a turbulent environment. For example, for large
enough amplitudes Rezzolla, Lamb and Shapiro~\cite{Rezzolla00} argue
that the magnetic fields could grow large enough to completely
suppress the $r$-mode instability. Hence, a magnetohydronamic study is
required to fully understand this feedback mechanism. In a recent
work, Friedman et al.~\cite{Friedman:2017wfi} explore how much
differential rotation, that is induced by the $r$-mode instability can
boost the magnetic fields. The authors argue that magnetic-field
amplification is restricted in strength by the saturation amplitude of the
unstable mode, and if the saturation amplitude is weak enough
$\lesssim 10^{-4}$, then the magnetic field cannot grow to levels that
suppress or modify that $r$-mode in neutron stars with ``normal'' and
type II superconducting interiors.


\subsection{Rotating core collapse}


  \begin{figure*}[th!]
    \center
    \includegraphics[width=0.69\textwidth]{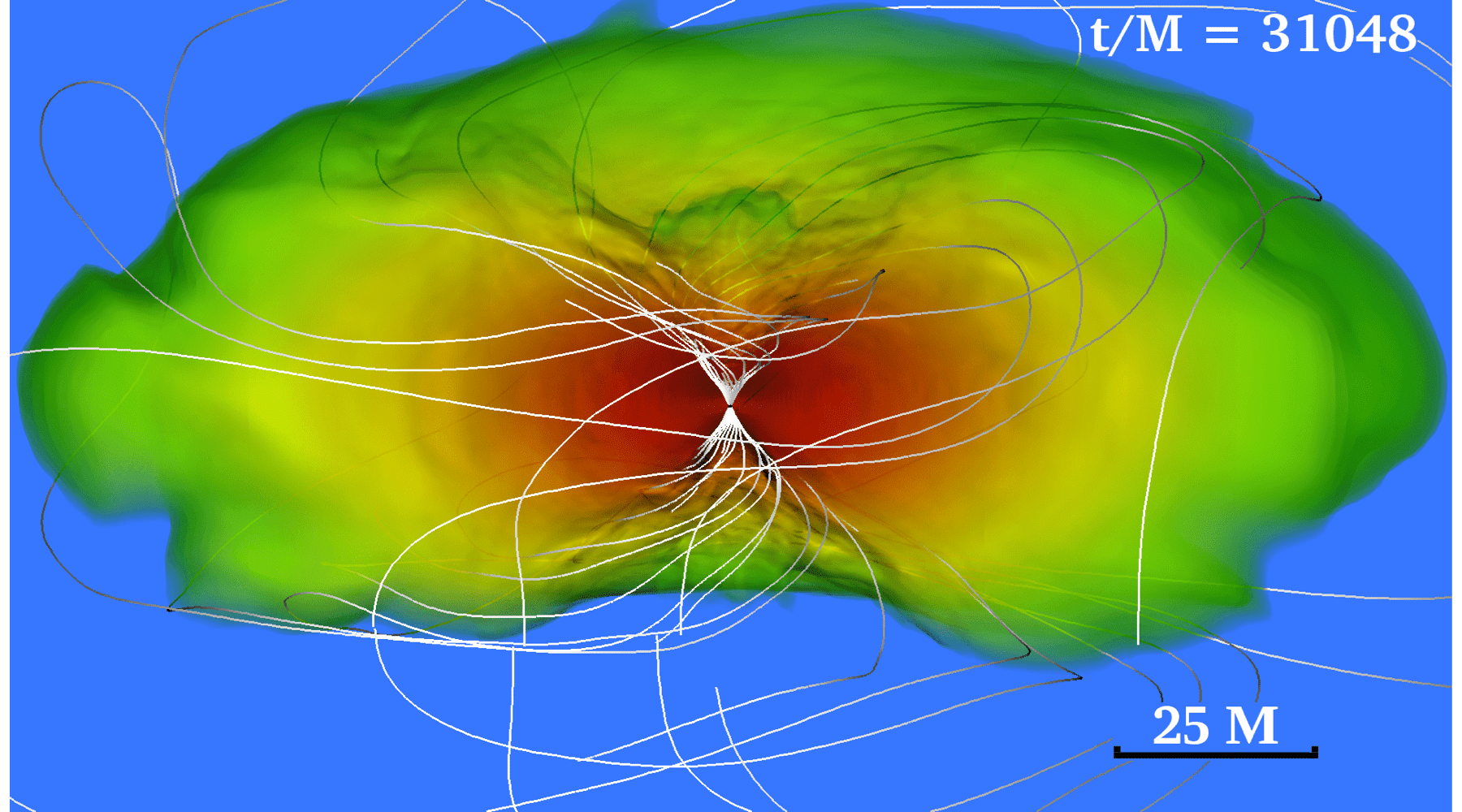}
    \includegraphics[width=0.69\textwidth]{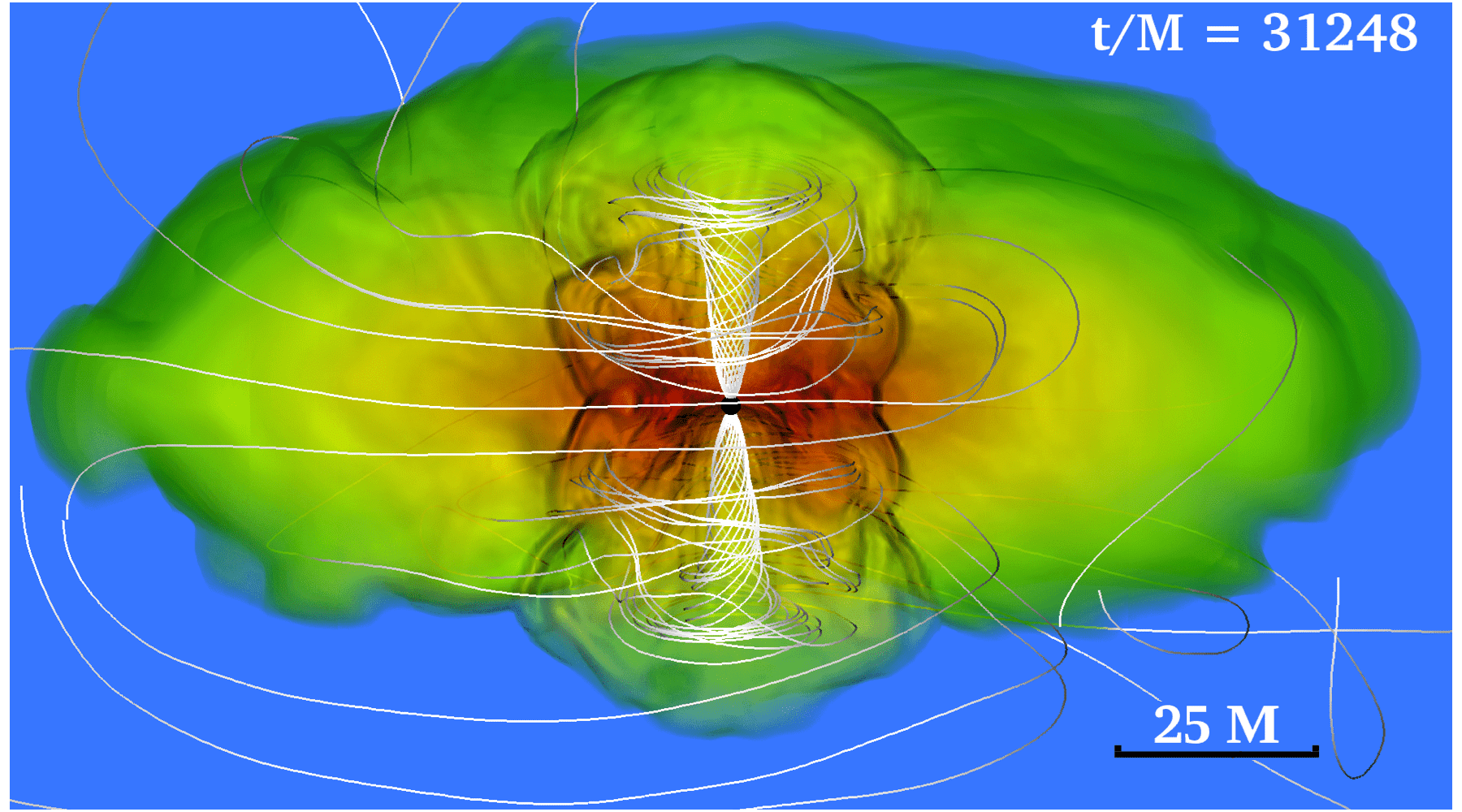}
    \includegraphics[width=0.69\textwidth]{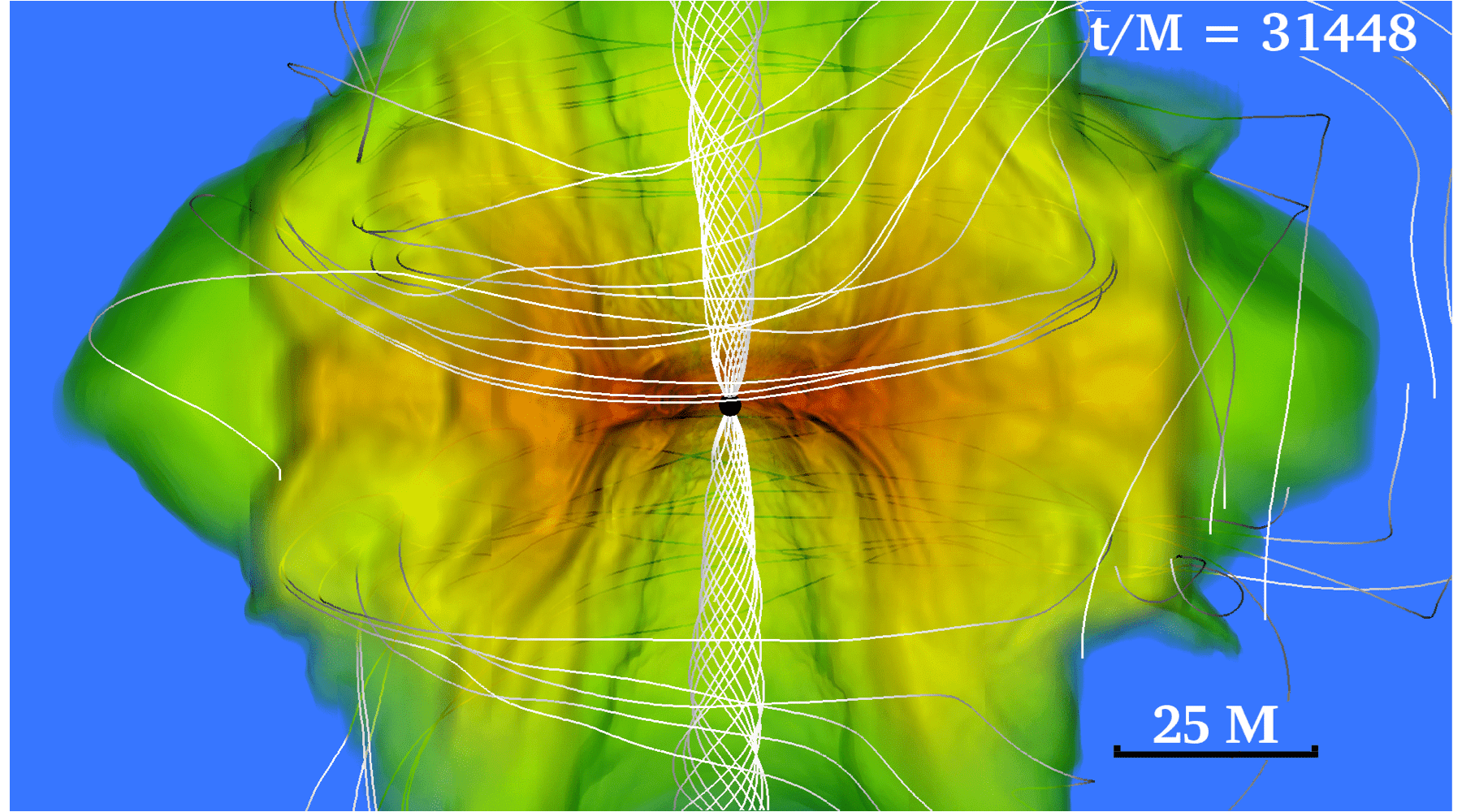}
    \caption{Cut of 3D rest-mass density profile (colored volume
      rendering) of a magnetized rotating, collapsing supermassive
      star with magnetic-field lines indicated by white curves.  The
      top panel corresponds to the time near BH formation, the middle
      panel is during the development of the incipient jet, and the
      bottom panel shows the fully developed incipient jet. (Image
      reproduced with permission from~\cite{Sun:2017voo}.)}
  \label{SMSjets}
\end{figure*}

  \begin{figure*}[th!]
    \center
      \includegraphics[width=0.69\textwidth]{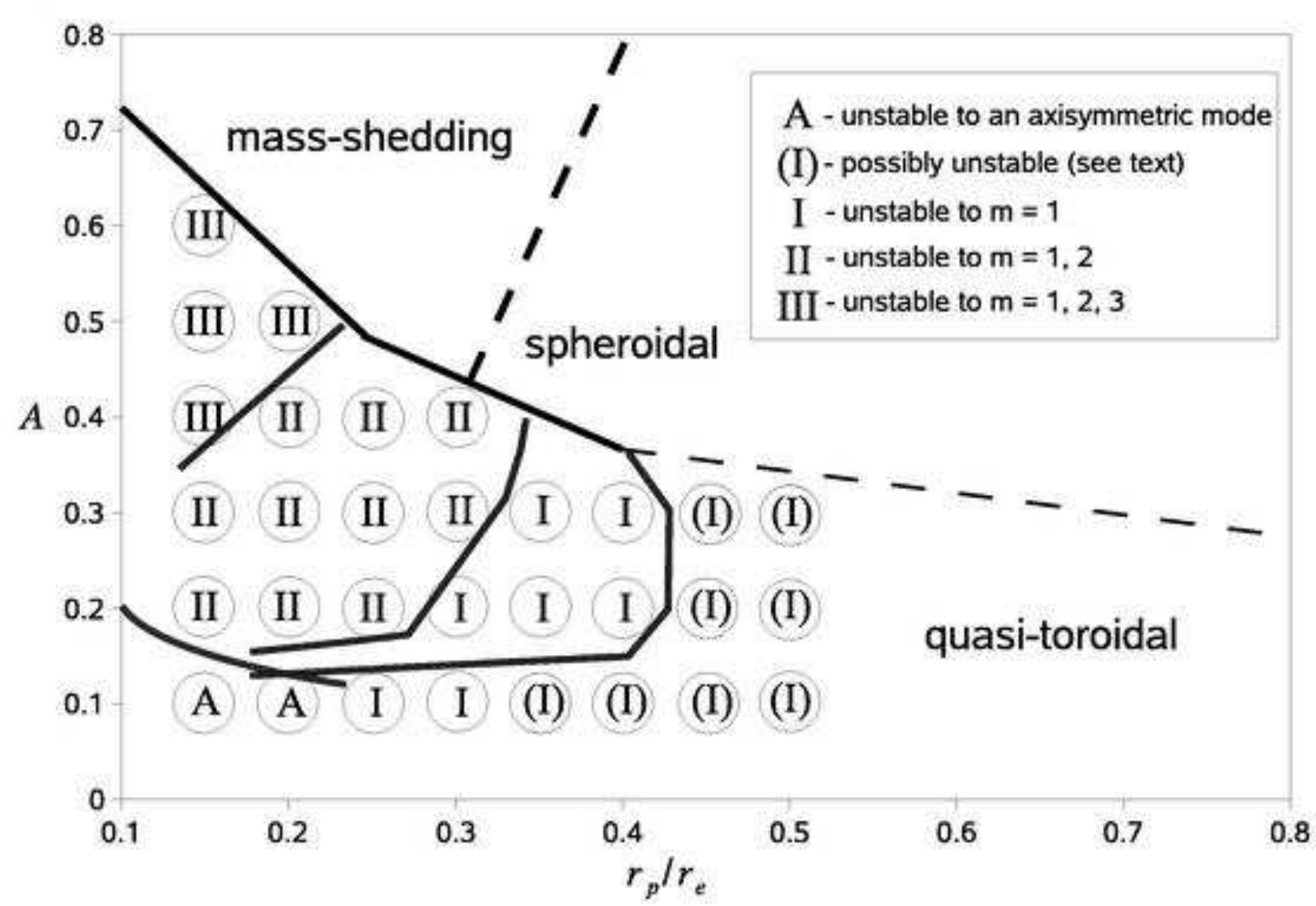}
      \includegraphics[width=0.69\textwidth]{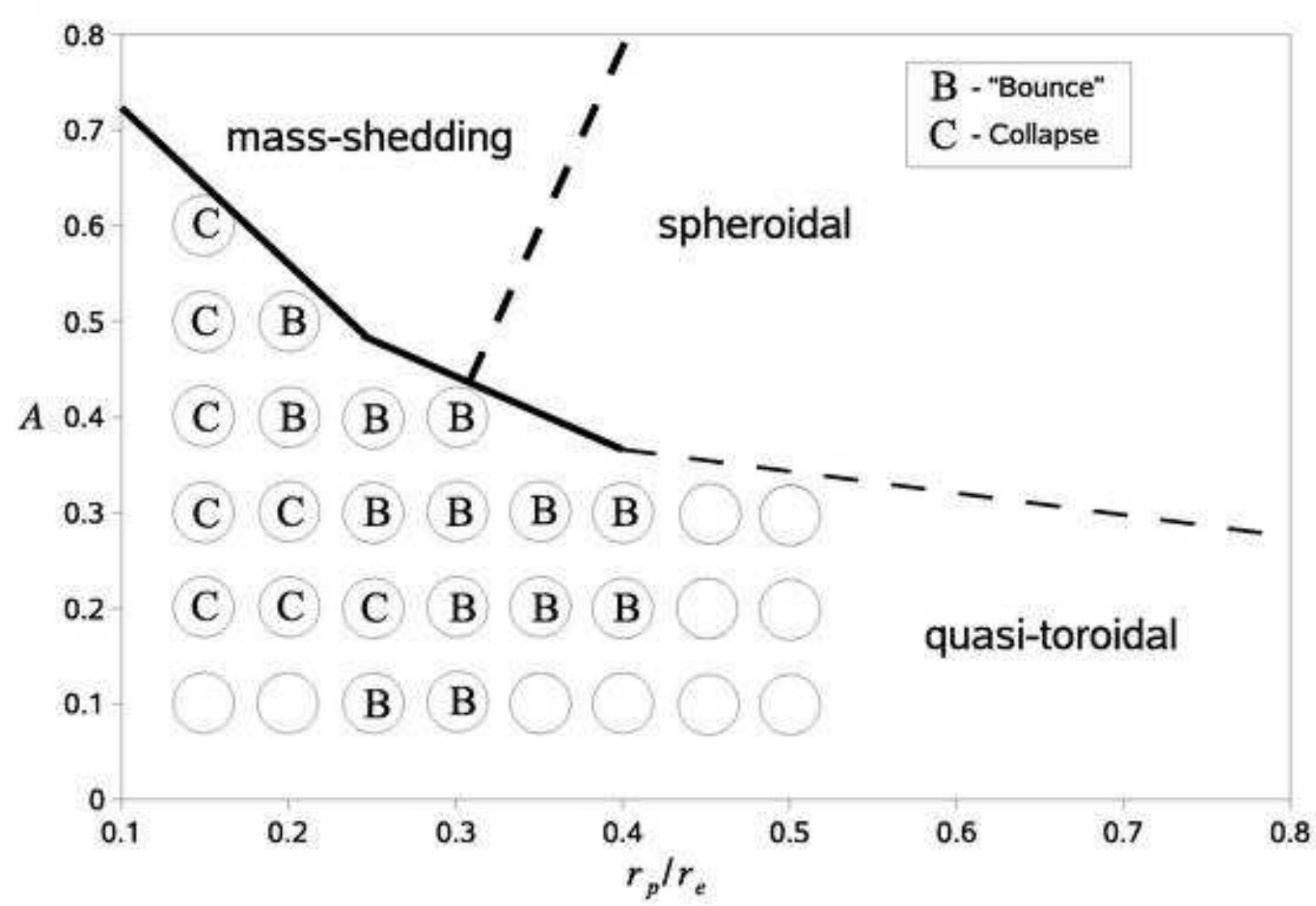}
      \caption{Top panel: Stability of quasi-toroidal models with
        $\rho_c =10^{-7}$ on the $A$ -- $r_p/r_e$ plane where $A$ is a
        constant controlling the degree of differential rotation. (We
        point out that, here, $A$ does not coincide with $A$ in        Eq.~\eqref{eq:diffrot}, but it equals the $A$ appearing in
        that equation divided by the equatorial coordinate radius $r_e$). A Latin number denotes the
        highest azimuthal order more that becomes unstable, i.e. I
        implies that only $m = 1$ is unstable, II implies $m = 1, 2$
        are unstable, and III implies $ m = 1, 2, 3$ are
        unstable. Models denoted by (I) are either secularly unstable
        -- growth times $\tau > t_{dyn}$ (where $t_{dyn}$ is the
        dynamical timescale), or stable (see \cite{Zink2007}).  Models
        denoted by A exhibit an axisymmetric instability. The line in
        the lower left indicates the location of the sequence $J/M^2 =
        1$, and the three lines inside the quasi-toroidal region
        indicate the locations of sequences with $T/|W| = 0.14$
        (right), $T/|W| = 0.18$ (middle) and $T/|W| = 0.26$
        (left). Lower panel: Remnants of the models from left panel,
        which are unstable with respect to nonaxisymmetric modes. The
        nonlinear behaviour has been analyzed by observing the
        evolution of the function minimum value of the lapse
        $\alpha_{min}$ Models which show a minimum in this function
        are marked by B for bounce, while models exhibiting an
        exponential collapse of the lapse are marked by C for
        collapse. (Image reproduced with permission
        from~\cite{Zink2007}, copyright by APS.)}
  \label{StabZink07}
\end{figure*}

\subsubsection{Collapse to a rotating black hole}

Black hole formation in relativistic core collapse was first studied
in axisymmetry by Nakamura~\cite{Nakamura81,Nakamura83}, using the
(2+1)+1 formalism~\cite{Maeda80}. The outcome of the simulation
depends on the rotational parameter
\begin{equation}
  q \equiv J/M^2.
\end{equation}
A rotating black hole is formed only if $q<1$, indicating that cosmic
censorship holds. Stark and Piran~\cite{Stark85,Piran86} use the 3+1
formalism and the radial gauge of Bardeen--Piran~\cite{Bardeen83} to
study black hole formation and gravitational wave emission in
axisymmetry. In this gauge, two metric functions used in determining
$g_{\theta\theta}$ and $g_{\phi\phi}$ can be chosen such that at large
radii they asymptotically approach $h_+$ and $h_\times$ (the even and
odd transverse traceless amplitudes of the gravitational waves, with
$1/r$ fall-off at large radii; note that $h_+$ defined
in~\cite{Stark85} has the opposite sign as that commonly used, e.g.,
in~\cite{Thorne83}). In this way, the gravitational waveform is
obtained at large radii directly in the numerical evolution. It is
also easy to compute the gravitational energy emitted, as a simple
integral over a sphere far from the source: $\Delta E \sim r^2\int
dt(h_{+,r}^2+h_{\times,r}^2)$. Using polar slicing, black hole
formation appears as a region of exponentially small lapse, when
$q<{\cal O}(1)$. The initial data consists of a nonrotating, pressure
deficient TOV solution, to which angular momentum is added by
hand. The obtained waveform is nearly independent of the details of
the collapse: It consists of a broad initial peak (since the star
adjusts its initial spherical shape to a flattened shape, more
consistent with the prescribed angular momentum), the main emission
(during the formation of the black hole), and an oscillatory tail,
corresponding to oscillations of the formed black hole spacetime. The
energy of the emitted gravitational waves during the axisymmetric core
collapse is found not to exceed $7\times 10^{-4}\,M_{\odot} c^2$ (to
which the broad initial peak has a negligible contribution). The
emitted energy scales as $q^4$, while the energy in the even mode
exceeds that in the odd mode by at least an order of magnitude.  The
qualitative morphology of the gravitational wave signal from collapse
has been studied through perturbative approaches by Seidel and
Moore~\cite{Seidel1987PhRvD..35.2287S}, Seidel, Myra and
Moore~\cite{Seidel1988PhRvD..38.2349S} and
Seidel~\cite{Seidel1990PhRvD..42.1884S}.

Shibata~\cite{Shibata00} carried out axisymmetric simulations of
rotating stellar collapse in full general relativity, using a
Cartesian grid, in which axisymmetry is imposed by suitable boundary
conditions. The details of the formalism (numerical evolution scheme
and gauge) are given in~\cite{Shibata99}. It is found that rapid
rotation can prevent prompt black hole formation. When $q={\cal
  O}(1)$, a prompt collapse to a black hole is prevented even for a
rest mass that is 70\,--\,80\% larger than the maximum allowed mass of
spherical stars, and this depends weakly on the rotational profile of
the initial configuration. The final configuration is supported
against collapse by the induced differential rotation. In these
axisymmetric simulations, shock formation for $q<0.5$ does not result
in a significant heating of the core; shocks are formed at a
spheroidal shell around the high density core. In contrast, when the
initial configuration is rapidly rotating ($q={\cal O}(1)$), shocks
are formed in a highly nonspherical manner near high density regions,
and the resultant shock heating contributes in preventing prompt
collapse to a black hole. A qualitative analysis in~\cite{Shibata00}
suggests that a disk can form around a black hole during core
collapse, provided the progenitor is nearly rigidly rotating and
$q={\cal O}(1)$ for a stiff progenitor EOS. On the other hand, $q \ll
1$ still allows for a disk formation if the progenitor EOS is soft. At
present, it is not clear how much the above conclusions depend on the
restriction to axisymmetry or on other assumptions -- 3-dimensional
simulations of the core collapse of such initially axisymmetric
configurations have still to be performed.

Shibata and Shapiro perform axisymmetric, hydrodynamic simulations in
full general relativity to follow the collapse of a rigidly rotating,
supermassive star (SMS) to a supermassive black hole (SMBH) in
\cite{Shibata2002}. The initial, equilibrium $\Gamma=4/3$ polytropic,
SMS of arbitrary mass $M$ rotates at the mass-shedding limit, is
marginally unstable to collapse, and has $T/|W| \simeq 0.009$ and
$J/M^2\simeq 0.97$. The collapse is induced via a $1\%$ pressure
depletion and proceeds homologously early on, until eventually an
apparent horizon forms at the center. Shibata and Shapiro estimate
that the final black hole will contain $\sim 90\%$ of the total mass
of the system and have a spin parameter $J/M^2 \sim 0.75$, with the
remaining gas forming a disk around the black hole. In follow up work,
Liu, Shapiro and Stephens~\cite{Liu:2007cf} study the
magnetorotational collapse of supermassive stars in axisymmetry and
find that following black hole formation the magnetic-field lines
partially collimate along the hole's spin axis speculating that these
systems may be able to launch jets.

In a more recent studies in full general relativity, Shibata et
al.~\cite{Shibata:2016vzw} and Sun, Paschalidis, Ruiz and
Shapiro~\cite{Sun:2017voo} study the gravitational wave emission from
such collapsing, rotating supermassive stars. Both studies find that
for stellar masses $M\sim 10^6M_\odot$, the gravitational waves peak
in the LISA band and could be detectable out to cosmological redshift
$z\sim 3$. Sun, Paschalidis, Ruiz and Shapiro also point out that for
supermassive stars with mass $M\sim 10^4$, the gravitational waves
from collapse could be detectable by DECIGO/BBO out to redshift $z\sim
11$. This scenario is very interesting because future space-based
gravitational wave observatories have the potential to probe whether
supermassive stars exist and can form seed black holes that later on
could grow through accretion to form the supermassive black holes we
observe at the centers of quasars as early as $z=7$.

Sun, Paschalidis, Ruiz and Shapiro~\cite{Sun:2017voo} also investigate
the effects of magnetic fields and find that shortly after black hole
formation the black hole - accretion disk engine that forms can launch
jets (see Fig.~\ref{SMSjets}) with characteristic jet luminosity
$L_{\rm jet} \sim 10^{51} \rm erg\ s^{-1}$ which could be observable
as a very long gamma-ray burst by current satellites such as
Swift. Thus, collapsing supermassive stars could be multimessenger
sources.

In a recent work Uchida et al.~\cite{Uchida:2017qwn} perform
axisymmetric calculations in full general relativity to investigate
the effects of nuclear burning in the collapse of supermassive stars,
finding that the collapse proceeds nearly unaffected. In addition,
they find that if a supermassive star core is sufficiently rapidly
rotating about 1\% of the initial rest-mass becomes unbound with
characteristic velocity and kinetic energy $0.2c$ and $10^{54-56}$
erg.

%
  \begin{figure*}[h!]
    \center
      \includegraphics[width=0.65\textwidth]{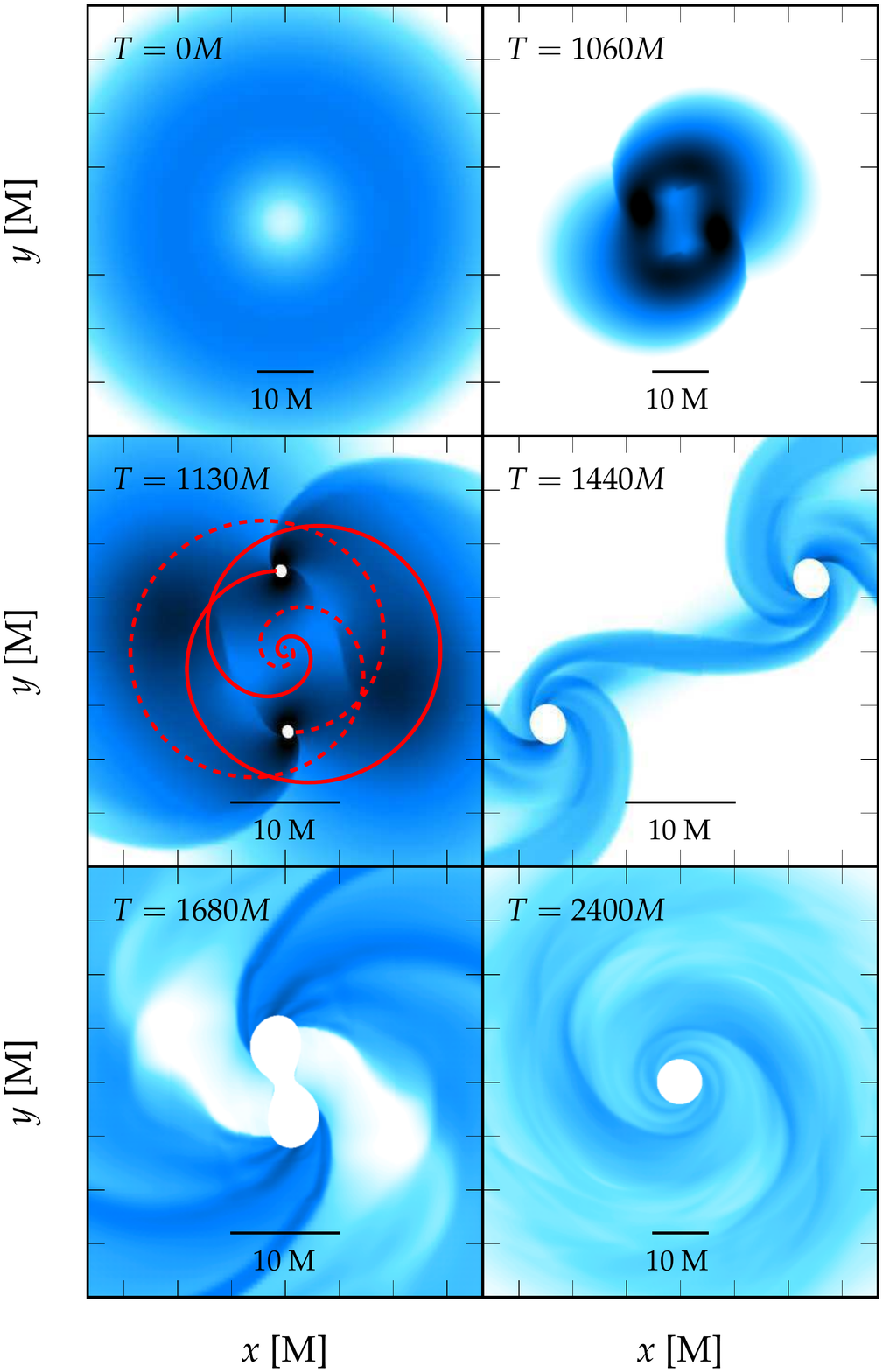}
      \caption{Equatorial density contours at select times of a
    fragmenting, collapsing supermassive star, forming a supermassive
    black hole binary. Dark colors indicate high density, light colors
    indicate low density. The logarithmic density colormap ranges from
    $10^{-7}M^{-2}$ (white) to $10^{-3}M^{-2}$ (black). In the bottom
    two panels, the colormap is rescaled to the range $[10^{-8}M^{-2},
    10^{-4}M^{-2}]$.  The upper two and the bottom right panels show
    physical dimensions of $\pm40M$, while the remaining panels show
    physical dimensions of $\pm20M$.  The white disks roughly indicate
    the black hole apparent horizons. (Image reproduced with permission
    from~\cite{Reisswig2013}, copyright by APS.)}
  \label{Reisswig2013BBH}
\end{figure*}
%

A 3D dimensional hydrodynamics code capable of following the collapse
of a massive relativistic star in full general relativity is presented
in \cite{Font02}.  A different numerical code for axisymmetric
gravitational collapse in the (2+1)+1 formalism is described
in~\cite{Choptuik03}.

Zink et al. \cite{Zink05} perform hydrodynamic simulations of
supermassive stars in full general relativity to study for the first
time the off-center formation of a black hole through fragmentation of
a general relativistic polytrope. They adopt the {\tt Cactus} code and
the {\tt Whisky} module for the spacetime and hydrodynamics, in
conjunction with a $\Gamma$ law equation of state. The initial data
correspond to $n=3$ equilibrium polytrope that is differentially
rotating with rotation law~\eqref{eq:diffrot}
$u^tu_\phi=A^2(\Omega_c-\Omega)$, where $A$ is a constant that
regulates the degree of differential rotation, $\Omega_c$ is the
angular velocity at the center of the star, and $\Omega$ the angular
velocity at a given (cylindrical) radius.  The particular initial
model they consider is generated using the {\tt RNS} code, and has
$A=r_e/3$, where $r_e$ is the equatorial coordinate radius of the
star, a central density $\rho_c = 3.38\times 10^{-6}$ (in geometrized,
polytropic units where $G=1=c=K$ and $K$ is the polytropic constant),
the ratio of the polar $r_p$ to equatorial radius is $r_p/r_3=0.24$,
and $T/|W|=0.227$. After the initial data are generated small
nonaxisymmetric density perturbations are added of the form
\be
\rho \rightarrow \rho(1+\frac{1}{\lambda r_e}\sum_{m=1}^{4}\lambda_mBr\sin(m\phi)),
\ee
where $\lambda_m=0,1$ and $\lambda=\sum_{m}\lambda_m$, and the
collapse is induced by a $0.1\%$ pressure depletion. After the initial
perturbation the constraints are not solved again because the
amplitude $B$ is chosen sufficiently small that the truncation-error
induced constraint violation dominates. They find that the star is
unstable to $m = 1$ and $m = 2$, and these modes grow from the linear
to the nonlinear regime with time, and eventually, depending on the
initial perturbation, lead to the formation of one or more off-center
fragments. Using an adaptive-mesh refinement type of method, they
follow the behavior in the case where one off-center fragment forms,
and find that it collapses to form a black hole. Based on these
results the authors argue that the fragmentation could turn a massive
star into a binary black hole with a massive accretion disk around it.
In a follow-up paper \cite{Zink2007} Zink et al. use the same codes to
study many more cases including different values for compactness, equation of
state stiffness, and different axes ratios (corresponding to even low
$T/|W|$ models). It is found that: 1) the growth time of the $m = 1$
and $m = 2$ modes increases with lower $r_p/r_e$, 2) the $m = 1$ and
$m = 2$ modes are stabilized with increasing $\Gamma$ -- stiffness of
the equation of state -- and $T/|W|$ decreasing from 0.227 to 0.159,
3) the instability growth time is approximately similar for stars of
different compactness (with $T/|W|$ approximately constant), but the
outcome of the fragmentation can differ drastically -- the fragments
of more compact stars $M/R \gtrsim 0.044$ seem to collapse and form
black holes, but stars with low compactness $M/R \lesssim 0.022$ seem
to prevent black hole formation.  The results summarizing whether the
instability develops and whether the fragments collapse to a black
hole or not are presented in Fig.~\ref{StabZink07}. Zink et al. also
conclude that along a sequence of increasing $T/|W|$ and restricted to
a few dynamical timescales, the $m = 1$ perturbation is dominant
before higher-order modes become unstable, suggesting the (off-center)
formation of a single black hole with a massive accretion
disk. However, the authors note that these results do not exclude the
possibility that on a longer timescale a higher-order mode will be
activated before the $m=1$ mode becomes unstable so that multiple
black holes could form. The authors also find that in the cases where
two fragments form and collapse, a runaway instability takes over,
leading eventually to a central collapse.

Montero et al.~\cite{Montero2012} perform axisymmetric, HRSC
hydrodynamic simulations in full general relativity (adopting the BSSN
formulation) to study the collapse and explosion of rotating
supermassive stars while accounting for thermonuclear effects. Their
simulations adopt an equation of state that accounts for the gas
pressure, and the pressure associated with radiation and
electron-positron pairs. In addition, they include the effects of
thermonuclear energy released by hydrogen and helium burning and
neutrino cooling through thermal processes. The initial models are
$n=3$ polytropic, rigidly rotating equilibrium configurations
constructed with the {\tt LORENE} library. They find that non-rotating
supermassive stars with a mass of $\sim 5 \times 10^5M_\odot$ and an
initial metallicity less than $Z_{CNO} \sim 0.007$ collapse to a black
hole, while the threshold metallicity is reduced to $Z_{CNO} \sim
0.001$ for uniformly rotating supermassive stars.  The critical
initial metallicity is increased for $10^6 M_\odot$ stars. It is noted
that, for some models, collapse to a black hole does not occur unless
the effects of $e^{\pm}$ pairs are accounted for, which render the
star unstable to gravitational collapse by reducing the effective
adiabatic index. For the stars that collapse the evolution is
continued past black hole formation, and the computed peak neutrino
and antineutrino luminosities for all flavors is $L \sim 10^{55}\rm
erg/s$.

Reisswig et al. \cite{Reisswig2013} inspired by the results
of~\cite{Montero2012} revisit the fragmentation instabilities in
differentially rotating supermassive stars studied
in~\cite{Zink05,Zink2007}, who, in turn, extended the configurations
studied by Saijo in~\cite{Saijo2004}. However, instead of a
$\Gamma=4/3$, they adopt a slightly softer $\Gamma=1.33$ equation of
state and consider nonaxisymmetric perturbations of the form $\rho
\rightarrow \rho (1 + A_m r \sin(m\phi))$, on initially equilibrium
$n=3$ polytropes with rotational parameter $J/M^2=1.0643$ which are
generated with the {\tt RNS} code. The hydrodynamic evolutions in full
general relativity are performed using the {\tt Einstein Toolkit}. For
the case where only an $m=2$ perturbation is considered, the authors
find that the initial star gives rise to two fragments which
subsequently collapse and form a bound supermassive black hole binary,
which in turn inspirals and merges in the gaseous environment of the
star (see Fig.~\ref{Reisswig2013BBH}). The authors compute the
associated gravitational wave signature and conclude that if $m = 2$,
fragmentation and formation of a supermassive black hole binary occurs
in supermassive stellar collapse, the coalescence of the binary will
result in a unique gravitational wave signal that can be detected at
redshifts $z \gtrsim 10$ with DECIGO and the Big Bang Observer, if the
supermassive star's mass is in the range $10^4 - 10^6M_\odot$.

Ott et al. \cite{Ott2011} use the Einstein Toolkit \cite{ET12} to
perform 3D hydrodynamic simulations of rotating core collapse in full
general relativity to study gravitational wave emission from collapsar
model for long gamma-ray bursts. The initial data correspond to the
inner $\sim 5700$km profile of the realistic 75-$M_\odot$,
$10^{-4}$-solar metallicity model u75 of~\cite{Woosley2002}, which
corresponds to the inner $\sim 4.5M_\odot$ of the star and imposing
different rotational profiles.  A hybrid, piecewise polytropic,
$\Gamma$-law equation of state is adopted for the evolution (as in
e.g.~\cite{Shibata03}), such that $\Gamma_1= 1.31$ at subnuclear
densities, and $\Gamma_2=2.4$ at supernuclear densities, and
$\Gamma_{\rm th}=4/3$. Octant symmetry is imposed throughout the
evolution and crude neutrino cooling is accounted for. Gravitational
waves are extracted using Cauchy-characteristic matching as described
in~\cite{Reisswig2011}. The initial data are evolved through collapse,
core bounce, proto-neutron star (PNS) formation through collapse of
the PNS to a BH. Gravitational waves are computed for all rotational
profiles and it is found that the peak amplitude at bounce is
approximately proportional to the model spin, and following bounce the
signal is dominated by quadrupole motion due to turbulence behind the
post-bounce shock. Following PNS collapse, a second burst in the
waveform appears which corresponds to the BH formation, which
subsequently rings down. As expected, characteristic gravitational
wave frequencies are $1-3$kHz, and it is estimated that for such an
event taking place $10$kpc away the signal-to-noise ratio for aLIGO
will be $\sim 50$.

The collapsar scenario is also studied in Debrye et
al. \cite{Debrye2013} through hydrodynamic simulations in general
relativity adopting the conformal flatness approximation and performed
using the {\tt CoCoNuT} code. More detailed microphysics and a
neutrino leakage scheme is implemented to account for deleptonization
and neutrino cooling, and the initial stellar mass, metallicity, and
rotational profile of the stellar progenitor are varied to determine
their influence on the outcome. It is shown that shown that
sufficiently fast rotating cores collapse due to the fall-back of
matter surrounding the compact remnant and due to neutrino cooling,
eventually forming spinning BHs.


\subsection{Formation of rotating neutron stars}

Rotating neutron stars can be formed following the collapse of a
massive star to a neutron star. Moreover, rapidly differentially
rotating neutron stars are a natural outcome of the merger of a binary
neutron star system which is not sufficiently massive for the merger remnant
to promptly collapse to a black hole.

\subsubsection{Stellar collapse to a rotating neutron star}
First attempts to study the formation of rotating neutron stars in
axisymmetric collapse were initiated by Evans~\cite{Evans84,Evans86}.
Dimmelmeier, Font and M{\"u}ller~\cite{Dimmelmeier01,Dimmelmeier01b}
have performed general relativistic simulations of neutron star
formation in rotating collapse. In the numerical scheme, HRSC methods
are employed for the hydrodynamical evolution, while for the spacetime
evolution the \emph{conformal flatness approximation}~\cite{Wilson95}
is used. Surprisingly, the gravitational waves obtained during the
neutron star formation in rotating core collapse are weaker in general
relativity than in Newtonian simulations. The reason for this result
is that relativistic rotating cores bounce at larger central densities
than in the Newtonian limit (for the same initial conditions). The
gravitational waves are computed from the time derivatives of the
quadrupole moment, which involves the volume integration of $\rho
r^4$. As the density profile of the formed neutron star is more
centrally condensed than in the Newtonian case, the corresponding
gravitational waves turn out to be weaker. Details of the numerical
methods and of the gravitational wave extraction used in the above
studies can be found in~\cite{Dimmelmeier02,Dimmelmeier02b}. In
addition, the rotational core collapse to proto-neutron star
simulations performed in~\cite{Dimmelmeier02b} suggest that types of
rotational supernova core collapse and gravitational waveforms
identified in earlier Newtonian simulations~\cite{Zwerger1997}
(regular collapse, multiple bounce collapse, and rapid collapse) are
also present in conformal gravity.

Fully relativistic axisymmetric simulations with coupled
hydrodynamical and spacetime evolution in the light-cone approach,
have been obtained by Siebel et al.~\cite{Siebel02,Siebel03}. One of
the advantages of the light-cone approach is that gravitational waves
can be extracted accurately at null infinity, without spurious
contamination by boundary conditions. The code by Siebel et
al. combines the light-cone approach for the spacetime evolution with
HRSC methods for the hydrodynamical evolution. In~\cite{Siebel03} it
is found that gravitational waves are extracted more accurately using
the Bondi news function than by a quadrupole formula on the null cone.

Shibata~\cite{Shibata03} presents an axisymmetric hydrodynamics code
based on HRSC methods and considers rotating stellar core collapse of
a realistic, uniformly rotating, equilibrium star near the mass
shedding limit with central density $\sim 10^{10} \rm g/cm^3$, and
$\Gamma =4/3$ yielding a mass $M=1.491M_\odot$ and radius $R=1910\rm
km$.  The evolution adopts a $\Gamma$-law-type equation of state
consisting of cold component and a thermal component (allowing for
shock heating). The cold part is piecewise polytropic with exponents
$\Gamma_1$ (for rest-mass densities $\rho_0\le \rho_{nuc}=2\times
10^{14} \rm g cm^{-3}$) and $\Gamma_2$ for ($\rho_0 >
\rho_{nuc}$). The collapse is triggered by a small reduction of
$\Gamma_1$ from the $4/3$ value, i.e., $\Gamma_1 = 1.325$. Due to the
absence of centrifugal force, after the shock formation, shock fronts
of {\it prolate} shape spread outward. As the collapse proceeds, the
central density monotonically increases until it exceeds $\rho_{\rm
nuc}$. When the central density is $\sim 3.5\rho_{nuc}$, the collapse
is halted and a proto neutron star is formed, which demonstrates
approximate quasi-periodic oscillations. In a follow-up
paper~\cite{Shibata2004} Shibata and Sekiguchi perform hydrodynamic
simulations in full GR of neutron star formation from stellar
collapse adopting similar methods as in~\cite{Shibata03} and adopting
the same parametric equation of state as in~\cite{Dimmelmeier02b}, but
focusing on the gravitational wave signatures of such events. As
in~\cite{Dimmelmeier02b} gravitational waves are computed based on a
quadrupole formula and it is found that waveforms computed based on
their fully general relativistic simulations are only qualitatively in
good agreement with the ones in~\cite{Dimmelmeier02b} which were
computed based on the conformal flatness
approximation. Quantitatively, the quadrupole formula used in the
conformal flatness calculations~\cite{Dimmelmeier02b}, yields
different results and Shibata and Sekiguchi suggest the use of a
quadrupole formula which is calibrated based on fully general
relativistic calculations.

Cerd{\'a}-Dur{\'a}n et al.~\cite{Cerda2005} introduce a new formalism
based on the conformal flatness approximation that extends the
original formulation~\cite{Wilson95} by adding to the conformally flat
3-metric, second-order post-Newtonian terms that lead to deviations
from isotropy. This new approximation is termed by the authors the
CFC+ formulation and a numerical implementation is described. After
testing the code using oscillating stars, the authors find that the
resulting oscillation frequencies using the CFC+ formalism are
practically the same as those using the original conformal flatness
formalism. The authors conclude that even for stars near the
mass-shedding limit the CFC+ formalism accounts for corrections at the
level of $1\%$. The first application of the code is axisymmetric
rotational core collapse to a proto-neutron star. It is shown that the
gravitational waves extracted using the quadrupole formula are not
substantially different between CFC+ and the original conformal
flatness approach.

Obergaulinger et al.~\cite{Obergaulinger2006} perform axisymmetric,
magnetohydrodynamic simulations of magnetorotational core collapse
accounting for relativistic effects with a modified TOV potential.
The initial data are Newtonian, equilibrium (differentially) rotating
polytropes which are seeded with a dynamically weak, dipolar magnetic
field (central field strength of $10^{10}-10^{13}$G). The evolution
adopts HRSC schemes for the magnetohydrodynamic equations and the
constrained transport method of Evans and Hawley~\cite{Evans1988} for
the ${\bf \nabla \cdot B}=0$ constraint, and a hybrid $\Gamma$-law
equation of state as in ~\cite{Dimmelmeier02b}. The main effects of
magnetic fields are to trigger the MRI, and brake the differential
rotation of the initial star. It is found that both of these effects
operate in their simulations and that the saturated magnetic field
reaches magnitudes of order $10^{16}$. However, it is not reported
whether a proto-magnetar forms in these simulations. It is generally
found that only stronger initial magnetic fields can affect the
gravitational wave signatures significantly, and that the
gravitational waves should be detectable by aLIGO, if the source is
about 10kpc away.

The gravitational-wave signal from rotating core collapse has been
investigated via hydrodynamic evolutions in 2 and 3 spatial
dimensions, both in Newtonian gravity (see M\"uller and Hillebrandt
\cite{Mueller1981}, M\"uller \cite{Mueller1982}, M\"onchmeyer et
al. \cite{Moenchmeyer1991}, Zwerger and M\"uller \cite{Zwerger1997},
Kotake et al.~\cite{Kotake2003}, Ott et al. \cite{Ott2004}) and in
general relativity (Dimmelmeier et
al.~\cite{Dimmelmeier02,Dimmelmeier02b}, Shibata and Sekiguchi
\cite{Shibata2004}, Obergaulinger et al.~\cite{Obergaulinger2006}, Ott
et al.~\cite{Ott2007}, Dimmelmeier et al. \cite{Dimmelmeier2008}) and
four different types of gravitational wave signals have been
identified so far (see also Ott \cite{Ott2009} for a comprehensive
review on core collapse supernovae):

\begin{enumerate}

\item[I.] In this type the stellar core bounces due to the stiffening
of the equation of state at nuclear densities and subsequently rings
down into equilibrium. The gravitational wave train possesses
one large peak corresponding to core bounce, and then undergoes a
damped ring-down phase.

\item[II.] In this type the stellar core bounce is driven by
  centrifugal forces occurring at sub-nuclear densities and unlike
  type I the post bounce phase consists of multiple bounces that are
  gradually damped. As a result the gravitational waveform is
  characterized by multiple distinct peaks corresponding to each
  bounce.

\item[III.] In this type the stellar core undergoes rapid collapse
  following bounce. The gravitational waveforms are low-amplitude and
  possess a subdominant peak.

\item[IV.] For magnetized progenitors with $B\gtrsim 10^{12}$G, the
  magnetic fields can affect the bounce dynamics. The gravitational
  wave signature, which has been referred to as magnetic-type,
  initially resembles the multiple-bounce signal. However, after the
  first shock launching the gravitational wave signal shows
  high-amplitude oscillations whose frequencies increase as the
  collapse proceeds.

\end{enumerate}

  \begin{figure*}[t]
    \center
      \includegraphics[width=0.69\textwidth]{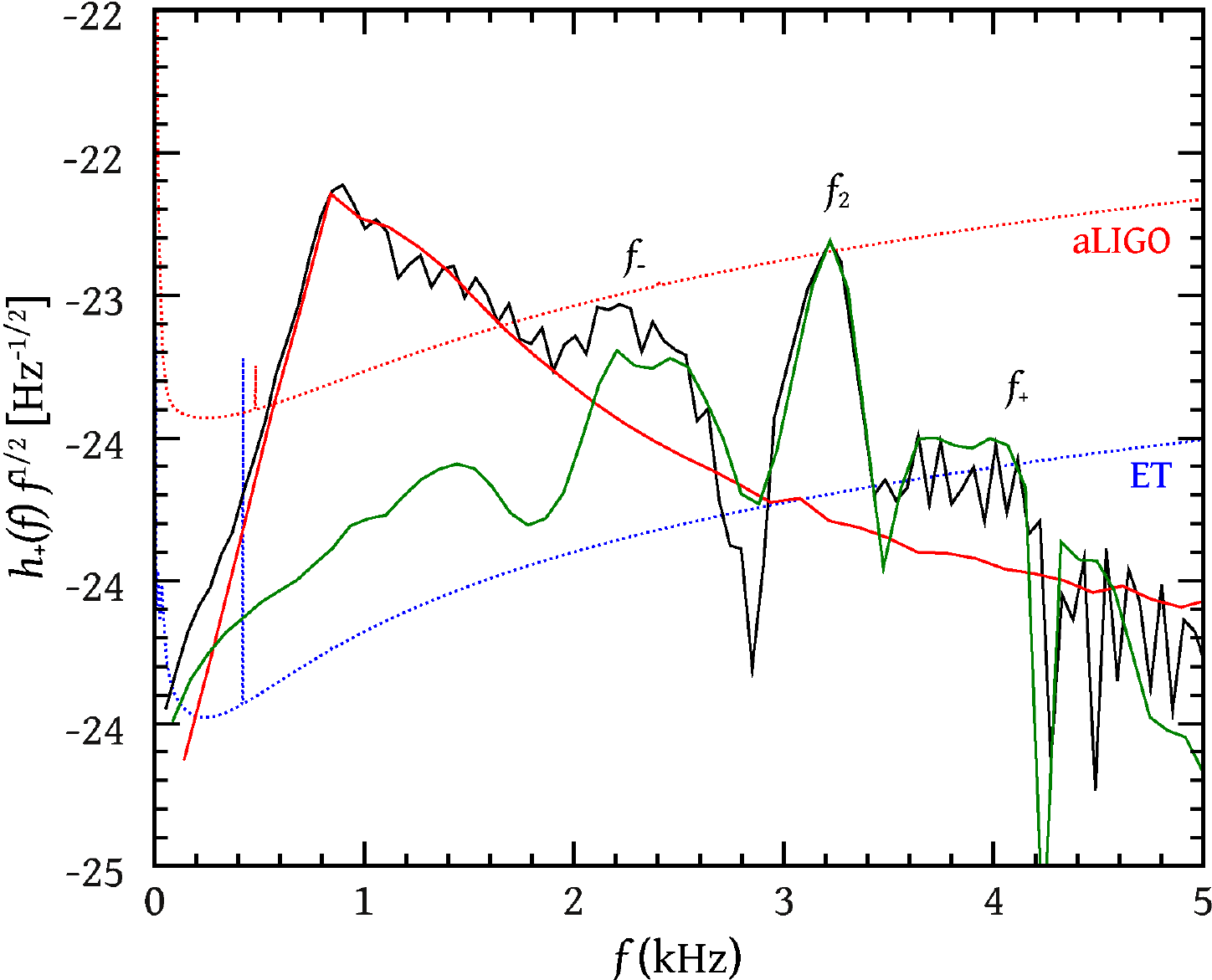}
       \includegraphics[width=0.69\textwidth]{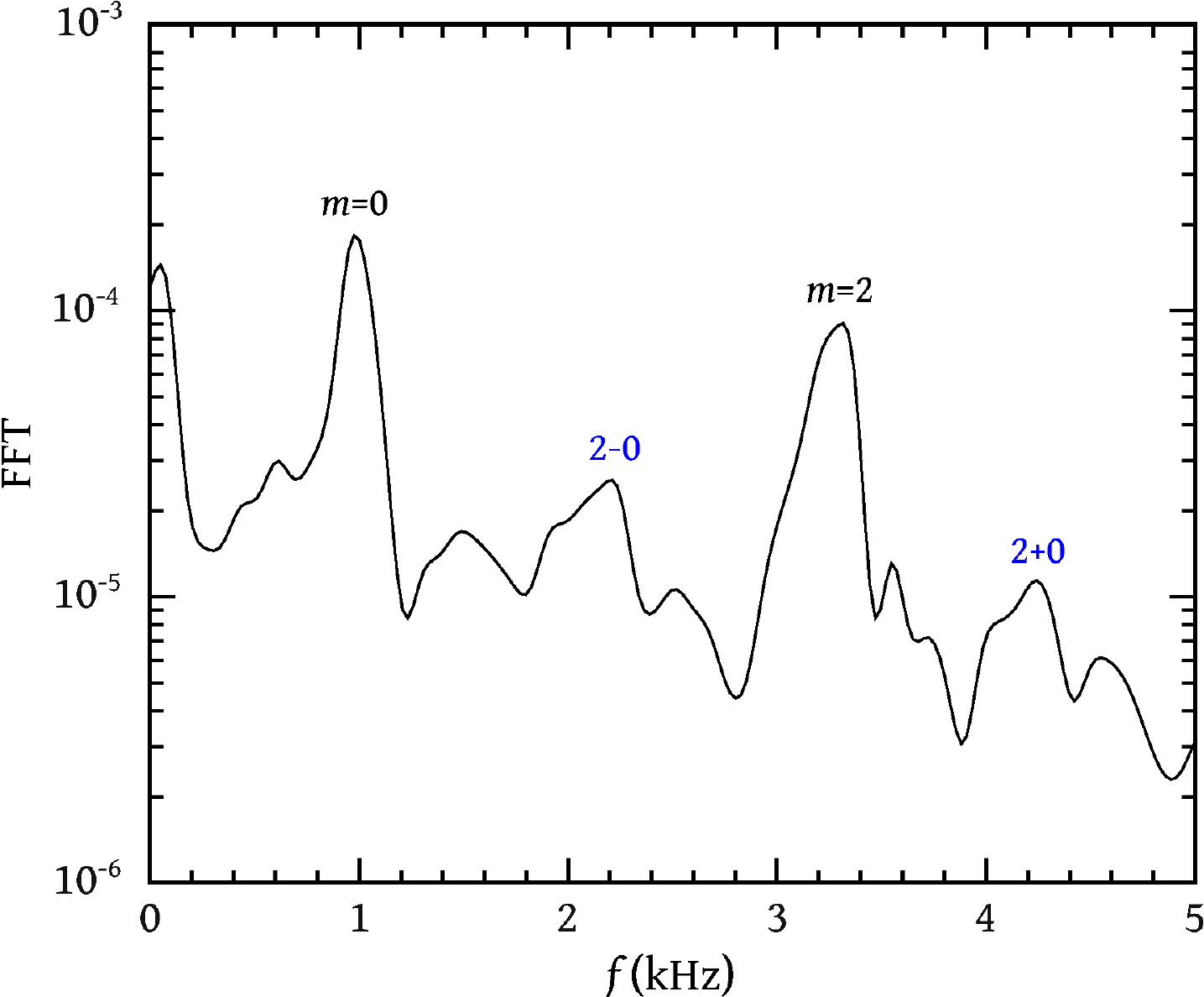}
    \caption{{\it Top panel}: Total (black), pre-merger (red) and post-merger (green)
      scaled power spectral density, compared to the aLIGO and ET
      unity SNR sensitivity curves for a HMNS formed in the merger of
      an equal mass NSNS system using the Lattimer-Swesty (LS). Each  neutron star has a mass of      $1.35M_\odot$ and the distance to the source is set at a nominal value of
      $100$Mpc. {\it Lower panel}: Corresponding FFT of the evolution of pressure in the equatorial plane, where discrete oscillation frequencies and their quasi-linear combinations can be seen. (Image reproduced with permission
      from~\cite{Stergioulas2011}, copyright by MNRAS.) }
  \label{fig:LS135135GW}
\end{figure*}

More detailed studies in ~\cite{Ott2007,Dimmelmeier2008} that account
for both general relativistic and microphysics effects suggest that the
generic core-collapse gravitation signal is of type I. Some work on
understanding the different oscillation modes of proto-neutron stars
was performed by Fuller et al.~\cite{Fuller2015}, where the effects of
relativity were largely ignored.

Recently, magnetorotational core collapse has been studied by M\"osta
et al. \cite{Moesta2014} via ideal magnetohydrodynamic simulations in
full GR that accounts for microphysics by adopting a finite-temperature
nuclear equation of state and a neutrino leakage scheme. The
simulations are performed using the {\tt Einstein Toolkit} and the
focus is on outflows and magnetic instabilities, rather than the
gravitational wave signal. Fundamental differences are reported
between axisymmetric and full 3D simulations in which a kink develops
breaking the axisymmetry of the expanding lobes. 

In a more recent work Andresen~\cite{Andresen2016} perform 3D
multi-group neutrino hydrodynamic simulations of core-collapse
supernovae focusing on the gravitational wave signatures generated
during the first few hundreds of milliseconds from the post-bounce
phase. Approximate general relativistic effects are accounted for by
use of a pseudorelativistic effective potential. The authors find that
gravitational waves from models dominated by the
standing-accretion-shock instability (SASI) are clearly distinct from
models that are convection-dominated. The main difference arises in
the low-frequency band around 100-200 Hz. The authors also find that
the gravitational wave strain above 250 Hz in 3D is considerably lower
than in 2D simulations. The authors' results suggest that
second-generation detectors will be able to detect only very nearby
events, but that third-generation detectors could distinguish SASI-
and convection-dominated models at distances of 10 kpc.

\subsubsection{Binary neutron star mergers}

Binary neutron stars have been simulated using numerical relativity
techniques for over a decade. There are recent technical reviews of
the topic focusing primarily on the history of relevant studies,
numerical techniques and the final fate of the merger remnant, see
e.g., \cite{Duez2010} and
\cite{FaberRasio2012,BaiottiRezz2016review,Paschalidis:2016agf}. Here,
we focus on the remnant NS properties highlighting the most recent
results related to the remnant hypermassive neutron star (HMNS)
oscillations, and how these can help to constrain the nuclear equation
of state.

Observational determination of masses in the known binary neutron star
(NSNS)
systems~\cite{Chamel2013IJMPE..2230018C,MillerMiller2015PhR,OzelFreire2016ARA&A..54..401O,MillerLamb2016EPJA...52...63M,Oertel2016arXiv161003361O}
indicates that a likely range of the total binary mass (sum of
individual TOV masses) is $2.4 M_\odot < M_{\rm tot} < 3.0 M_\odot$
with a peak around $2.7 M_\odot$.  Since observations
\cite{Demorest2010,Antoniadis2013} require a TOV limit mass of
$\gtrsim 2.0M_\odot$, a likely outcome of a NSNS merger is a
long-lived ($\gtrsim 10$ms) HMNS (see,
e.g. \cite{hotokezaka2011,Bauswein2012}).

The rotational profile of hypermassive neutron stars formed following
binary neutron star mergers have been studied by a number of
authors~\cite{Shibata:1999wm,ShibataTani2006PhRvD..73f4027S,2008PhRvD..78h4033B,2008PhRvD..77b4006A,2008PhRvD..78b4012L,Bernuzzi:2011aq,Kastaun:2014fna,DePietri:2015lya,Kastaun:2016yaf}. The
common outcome in these studies is that the actual differential
rotation profile does not seem to match the $j-$constant rotation law
that is usually adopted in models of isolated differentially rotating
neutron stars. Instead, the post-merger remnants almost universally
exhibit a rotation profile that is approximately uniform in the core
that smoothly turns into quasi-Keplerian in the outer layers.  An
extended study in Hanauske et al.~\cite{Hanauske:2016gia} argues that
this profile seems to also be EOS-independent.

The gravitational-wave spectrum in the post-merger phase comprises
several distinct peaks that could be used for characterizing the
hypermassive compact object (see
e.g. \cite{Zhuge1994,Oechslin2002PhRvD..65j3005O,ShibataUryu2002,Shibata2005PhRvD..71h4021S,ShibataTani2006PhRvD..73f4027S,Kiuchi2009PhRvD..80f4037K,Dietrich:2016lyp}).
That several post-merger GW peaks do in fact originate from specific
oscillation modes of the remnants was established in Stergioulas et
al. \cite{Stergioulas2011}, by extracting eigenfunctions in the
equatorial plane for the dominant oscillation
frequencies. Gravitational wave spectra were split into pre- and
post-merger parts and it was shown that several peaks in the
post-merger GW spectrum have discrete counterparts in the evolution of
the fluid that correspond to specific normal modes of oscillation. The
dominant peak was identified as being the co-rotating $m=2$ $f$-mode
(denoted as $f_2$ or $f_{\rm peak}$), while additional frequencies
($f_{-}$ and $f_{+}$ ) were shown to originate from the quasi-linear
combination between $f_2$ and the quasi-radial oscillation frequency
$f_0$. The quasi-radial frequencies satisfy $f_- =f_2 - f_0$ and $f_+
=f_2 + f_0$, forming an equidistant triplet with $f_2$ (see
Fig.~\ref{fig:LS135135GW}). Since the amplitude of $f_+$ is much
smaller than other frequency peaks, the quasi-linear combination
frequency $f_-$ is the more important one from the observational point
of view (after $f_2$) and it has been renamed to $f_{2-0}$ in
subsequent studies, in order to emphasize its origin. Since $f_{2-0}$
is a quasi-linear feature, its amplitude quickly decays (it is the
product of the amplitudes of $f_2$ and $f_0$).

\begin{figure*}[t]
    \center
      \includegraphics[width=0.69\textwidth]{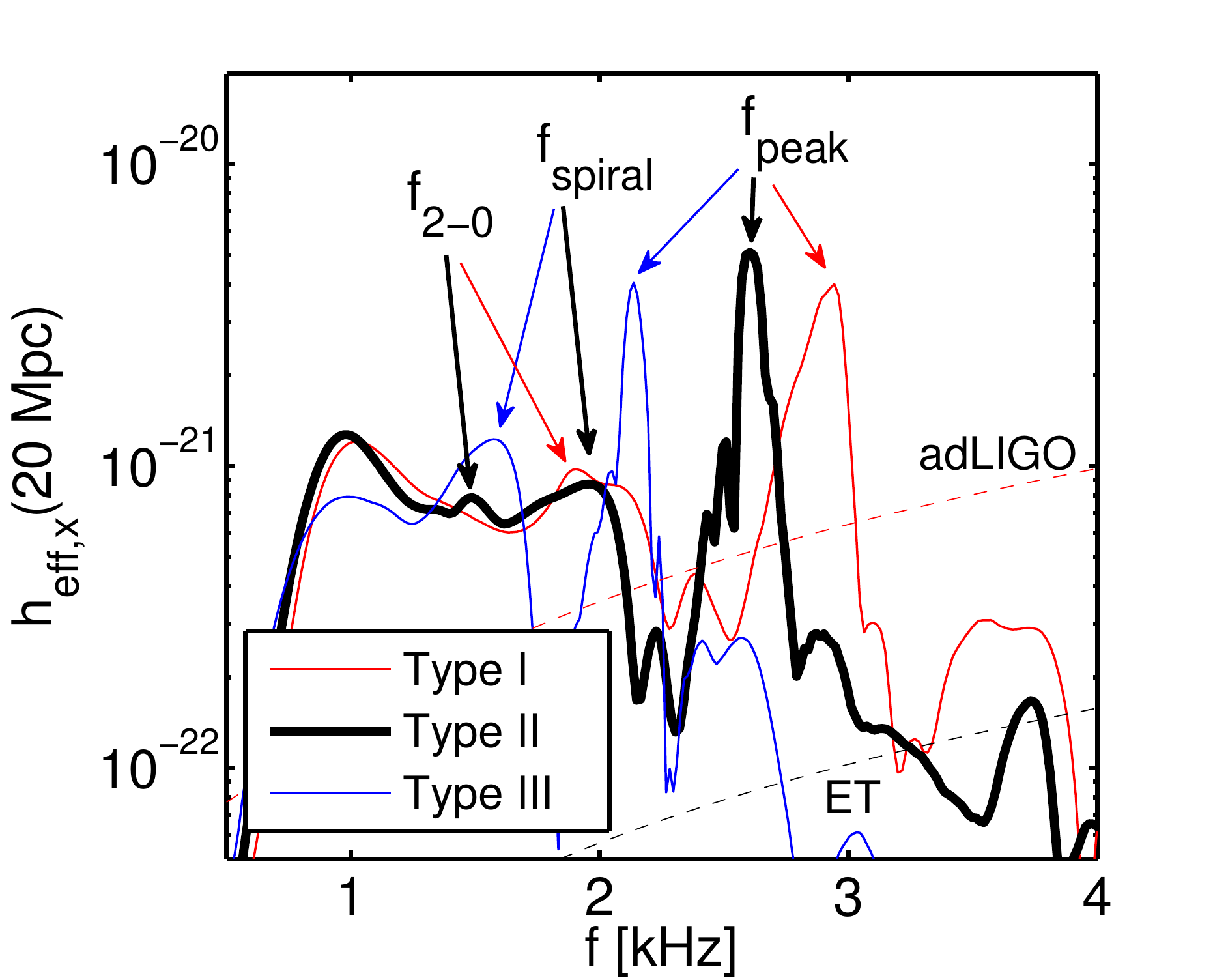}
       \includegraphics[width=0.69\textwidth]{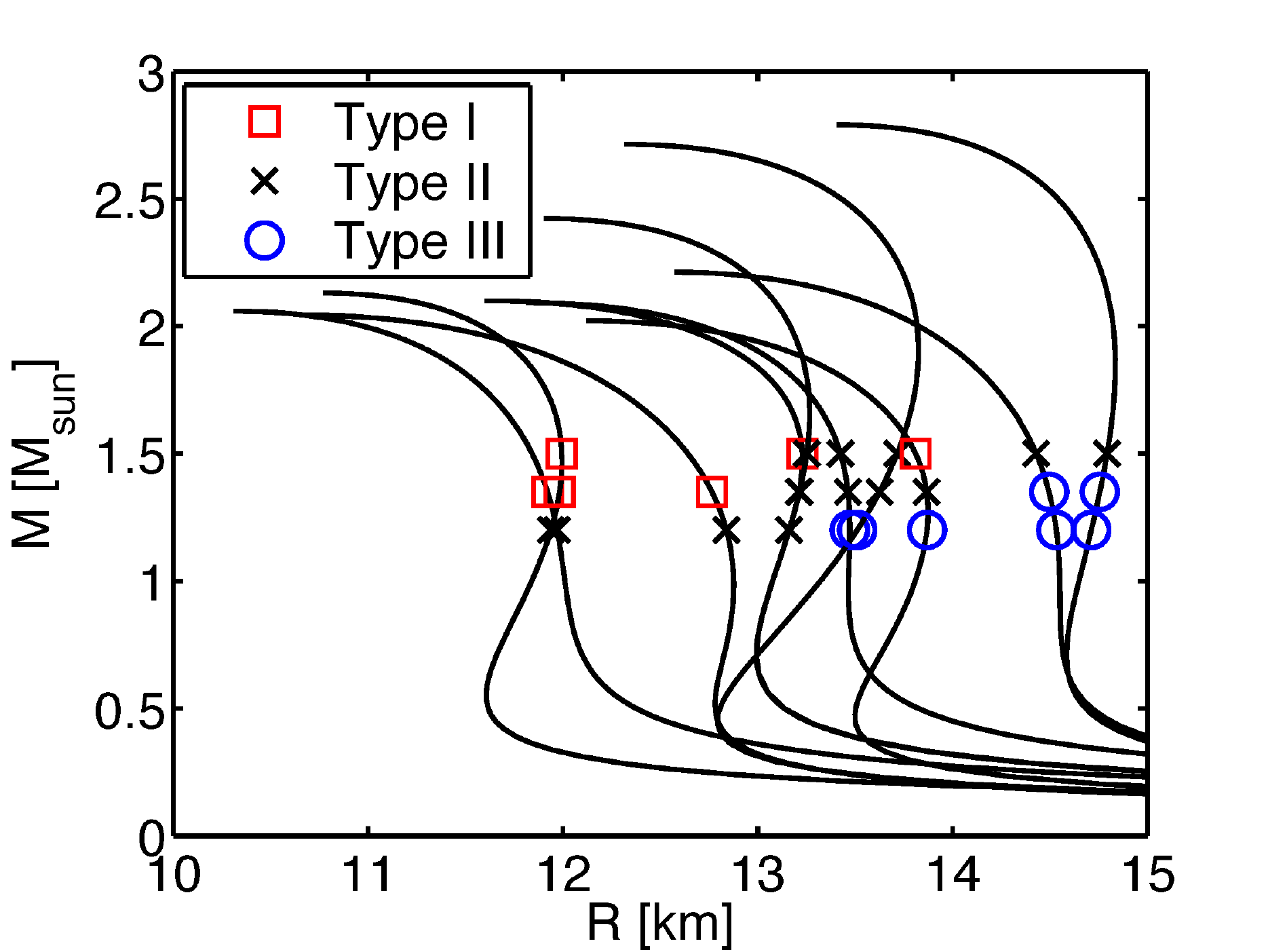}
    \caption{{\it Top panel}: GW spectra of 1.35-1.35~$M_\odot$
      mergers with a soft (red), intermediate (black) and stiff EOS
      (blue), at a reference distance of 20~Mpc. {\it Lower panel}:
      Different types of merger dynamics are indicated for several
      EOSs and different masses (in each case, half of the sum of
      individual pre-merger masses is shown). (Image reproduced with
      permission from~\cite{Bauswein2015PhRvD..91l4056B}, copyright by
      PRD.) }
  \label{fig:BS2015}
\end{figure*}

A more extensive parameter search by Bauswein \&
Stergioulas~\cite{Bauswein2015PhRvD..91l4056B} revealed that apart
from the $f_{2-0}$ quasi-linear peak, a fully nonlinear peak (denoted
as $f_{\rm spiral}$) exists in most cases, originating from the 
  transient appearance of a spiral deformation with two antipodal bulges at the time of
  merging. Investigating a large number of EOSs and different masses,
Bauswein \& Stergioulas (2015)~\cite{Bauswein2015PhRvD..91l4056B}
found that the post-merger phase can be broadly classified as
belonging to one of three different types:
\begin{enumerate}
\item {\it Type} I (soft EOS/high mass): $f_{2-0}$ is the strongest secondary peak.
\item {\it Type} II (intermediate EOS/intermediate mass): $f_{2-0}$ and $f_{\rm spiral}$ have roughly comparable amplitudes.
\item {\it Type} III (stiff EOS/low mass): $f_{\rm spiral}$ is the
  strongest secondary peak. 
\end{enumerate} 
For a broad sample of EOSs and for initial masses of $2.4M_\odot\leq
M_{\rm tot} \leq 3.0 M_\odot$, the frequency of $f_{\rm spiral}$ is in
the range $f_{\rm peak} - 0.5{\rm kHz} < f_{\rm spiral}< f_{\rm peak}
- 0.9{\rm kHz}$, while $f_{2-0}$ is in the range $f_{\rm peak} -
0.9{\rm kHz} < f_{\rm 2-0}< f_{\rm peak} - 1.3{\rm kHz}$. The fact
that the two ranges do not overlap can be used in search strategies
and in identifying the type of the merger dynamics.
Fig. \ref{fig:BS2015} (left panel) displays GW spectra for three
representative cases (corresponding to the three types described
above), while the right panel shows the dependence of the different
types on the initial mass ($M_{\rm tot}/2$ is shown) in a mass
vs. radius plot for nonrotating models.

  \begin{figure*}[th!]
    \center
      \includegraphics[width=0.69\textwidth]{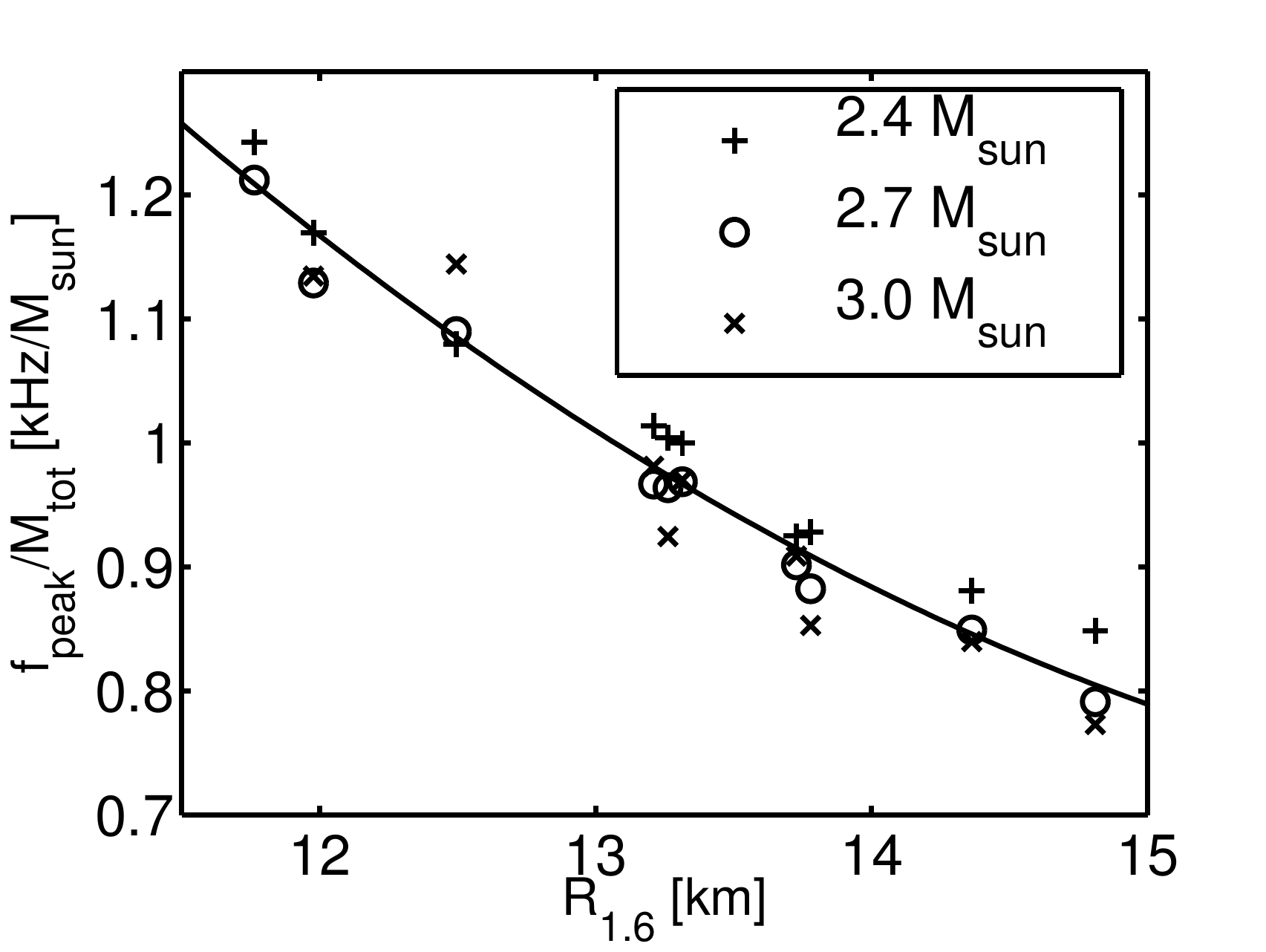}
    \caption{Peak frequency $f_{\rm peak}$ scaled by the total mass
      $M_{\rm tot}$ versus the radius of a nonrotating NS of mass
      $1.6M_\odot$ for different EOSs. The symbols correspond to
      different values of the total mass. The solid line shows the
      quadratic fit in Eq.~\eqref{eq:fvsR2}, which can be used to
      determine $R_{1.6}$ with a maximum uncertainty of a few percent
      for a given $f_{\rm peak}$\ measurement in a system where the
      total mass $M_{\rm tot}$ is determined from the inspiral
      gravitational waveform. (Image reproduced with permission
      from~\cite{Bauswein2016EPJA...52...56B}, copyright by EPJ.)}
  \label{fig:f2vsR}
\end{figure*}

Bauswein \& Janka \cite{Bauswein2012PhRvL.108a1101B} found that the
peak frequency $f_{\rm peak}$ is directly related to the radius of
nonrotating neutron stars through an EOS-independent empirical
relation, which can be used to observationally determine neutron star
radii with high accuracy, when the total mass of the system is
known. Because the remnants for mergers in the $2.4M_\odot\leq M_{\rm
  tot} \leq 3.0 M_\odot $ have a central density comparable to that of
a $\sim1.6M_\odot$ nonrotating neutron star, the uncertainty in the above
empirical relation is reduced  when it is cast in terms of
the radius $R_{1.6}$ of a $\sim1.6M_\odot$ nonrotating star \cite{Bauswein2012}. In
\cite{Bauswein2012PhRvL.108a1101B} representative examples of three initial binary setups (focusing on the
$1.35+1.35 M_\odot$ case) were discussed. Relations for different binary masses and mass ratios (using also a larger
set of EOSs) were discussed and presented  in~\cite{Bauswein2012}. For specific binary masses such an empirical relation can have an uncertainty
of only a few percent. In the case of a
$1.35+1.35M_\odot$\ merger, the relation yielding the radius of a $\sim 1.6M_\odot$ nonrotating star is
\cite{Bauswein2014PhysRevD.90.023002}%
\begin{equation}
R_{1.6}=1.099\cdot f_{\rm peak}^2-8.574\cdot f_{\rm peak}+28.07.
\end{equation}The $f_{\rm peak }$
vs. radius relation can be scaled by the total mass, to become a
universal relation, which is (to high accuracy) quadratic in the
radius:

\begin{equation} \label{eq:fvsR2}
f_{\mathrm{peak}}[{\rm kHz}]/M_{\rm tot}[M_\odot]= 0.0157\cdot R_{1.6}^2-0.5495\cdot
R_{1.6} +5.5030,
\end{equation}
see~\cite{Bauswein2016EPJA...52...56B} and Fig. \ref{fig:f2vsR}.  This
relation depends only weakly on the mass ratio. Similar relations can
easily be constructed for the radius of nonrotating stars at lower or
higher masses than $1.6 M_\odot$, but then the accuracy of radius
determinations deteriorates for $1.35M_\odot+1.35M_\odot$ mergers. For
other total binary masses, other TOV radii are obtained with minimal
uncertainty \cite{Bauswein2012}.

A single event in the most likely range of $2.4M_\odot\leq M_{\rm tot}
\leq 3.0 M_\odot $ will thus suffice to significantly constrain the
EOS in the density range that corresponds to a TOV mass of $1.6
M_\odot$. At significantly higher densities (close to $M_{\rm
  max}>2M_\odot$), it is unlikely that direct constraints can be
obtained. On the one hand, the expected merger rate may diminish above
$M_{\rm tot} > 3.0 M_\odot$, since all known double neutron star
systems have masses smaller than this (notice that measuring neutron
star radii from inspiral waveforms is similarly restricted to low
masses). On the other hand, even in the rare case of a merger with an
unusually high total mass it is quite possible that the remnant will
promptly collapse to a black hole, before the radius can be measured
through the detection of post-merger gravitational waves. However,
Bauswein, Stergioulas \& Janka \cite{Bauswein2014PhysRevD.90.023002}
devised a method to extrapolate the mass and radius of the
maximum-mass TOV\ model from at least two well-separated low-mass
$f_{\rm peak}$ measurements. The method is based on the observation
that for a given EOS $f_{\rm peak}$ is almost a linear function of
$M_{\rm tot}$, while the slope of this relation can be used to
determine empirically the threshold mass of binary systems to black
hole collapse, $M_{\rm thres}$. From an empirical relation between
$M_{\rm thres}$ and the maximum TOV mass $M_{\rm max}$, found in
\cite{2013PhRvL.111m1101B}, one thus arrives at a determination of
$M_{\rm max}$ with an uncertainty of order $0.1 M_\odot$. Furthermore,
an empirical relation between the peak frequency $f_{\rm peak}^{\rm
  thres}$ for a binary system with mass equal to the threshold mass
$M_{\rm thres}$ and the radius $R_{\rm max}$ of the maximum-mass TOV
model then permits a determination of $R_{\rm max}$ with an
uncertainty of order 5\%. Similar considerations allow the
determination of the central density of the maximum-mass TOV star,
$\rho_{c,{\rm max}}$ with an uncertainty of order 10\%.

  \begin{figure*}[th!]
    \center
      \includegraphics[width=0.69\textwidth]{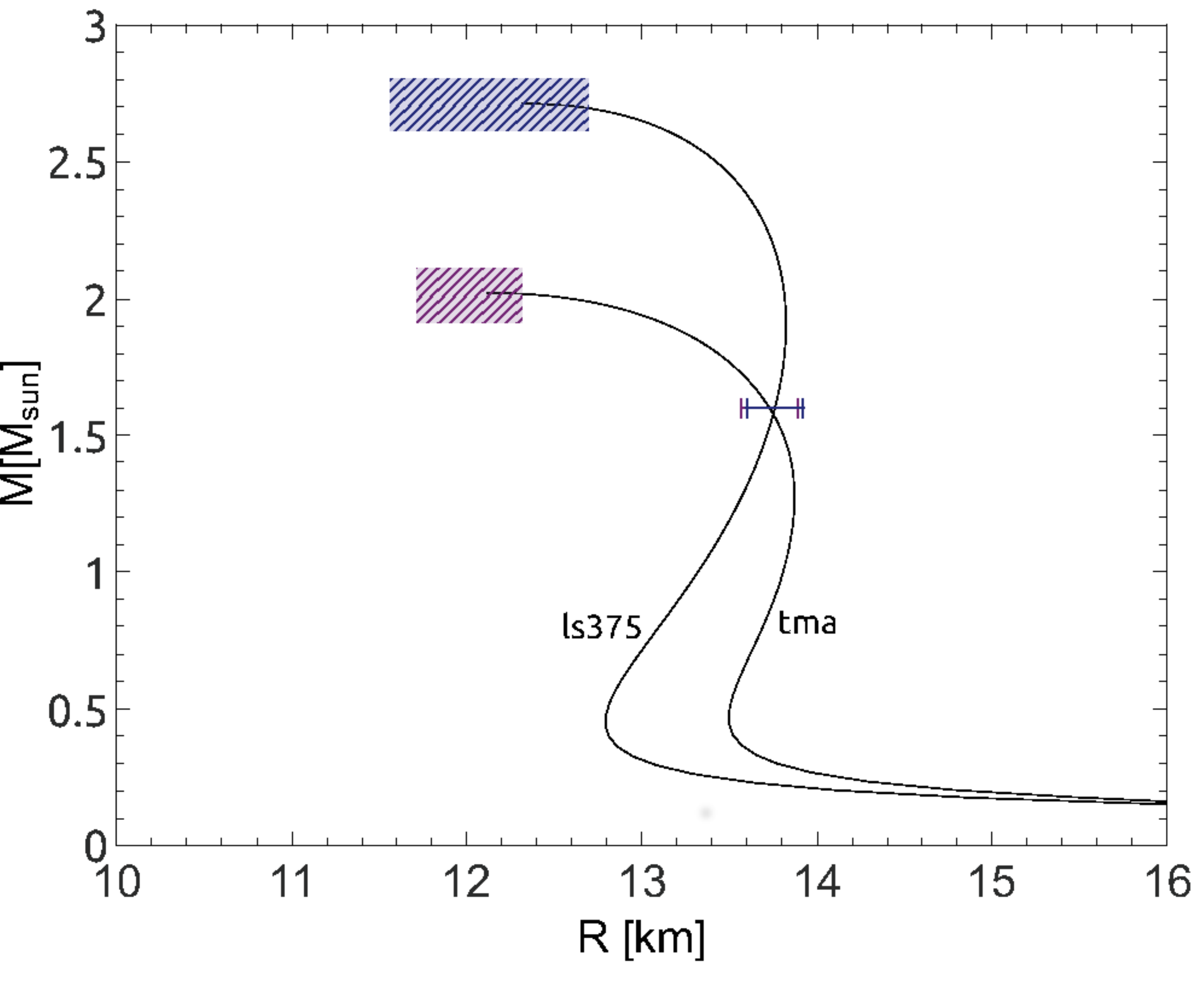}
    \caption{Mass-radius relations for two EOSs with similar stellar
      properties in the intermediate mass range around 1.6~$M_{\odot}$
      where the two mass-radius relations cross. Bars at
      1.6~$M_{\odot}$ indicate the maximum deviation of the estimated
      radius inferred from a single GW detection of a low-mass binary
      NS merger.  Using the extrapolation procedure described in
      \cite{Bauswein2014PhysRevD.90.023002} the two EOSs can clearly
      be distinguished. Boxes illustrate the maximum deviation of the
      estimated properties of the maximum-mass configuration. (Image
      reproduced with permission
      from~\cite{Bauswein2014PhysRevD.90.023002}, copyright by PRD.)}
  \label{fig:extrap}
\end{figure*}

Two representative cases of the determination of the radius and mass
$(R_{\rm max},M_{\rm max}$) of the maximum-mass TOV model are shown in
Fig. \ref{fig:extrap}.  These two EOSs cross at about $1.6 M_\odot$,
so that they cannot be distinguished by a single low-mass merger
event. However, extrapolating two well-separated low-mass $f_{\rm
  peak}$ measurements (using the procedure described in
\cite{Bauswein2014PhysRevD.90.023002}) allows for a clear distinction
of the EOS.

Lehner et al.~\cite{Lehner2016arXiv160300501L} study hypermassive
neutron stars formed in equal and unequal-mass NSNS mergers with
realistic, hot nuclear equations of state while employing an
approximate neutrino cooling scheme. The authors find agreement with
earlier findings in~\cite{Bauswein2015PhRvD..91l4056B}, as well as
that for a given total mass, the mass ratio has only a small effect on
$f_{\rm peak}$ \cite{Bauswein2012}. They also discuss an interesting
empirical relation between $f_{\rm peak}$ and the GW frequency at
contact. A different correlation between the post-merger oscillation
frequency and the tidal coupling constant $\kappa_2^T$ has been
discussed by Bernuzzi, Dietrich and Nagar~\cite{Bernuzzi:2015rla}. The
authors report that their proposed correlation exhibits small scatter
with the binary total mass, mass-ratio, EOS, and thermal
effects. However, only total masses in the range of $2.5-2.7M_\odot$
where considered and thermal effects where not based on realistic
finite temperature EOSs.

The above results suggest that a postmerger gravitational wave
detection can potentially determine neutron star radii to high
accuracy and thus constrain the EOS. The model can be further refined
by taking into account additional effects. For example, although
preliminary MHD studies suggest that realistic magnetic fields do not
have a significant direct impact on the $f_{\rm peak}$ frequency (see
e.g. \cite{Endrizzi2016,Kawamura2016PhRvD..94f4012K}), the timescale
on which MRI could modify the background requires further studies.

Takami, Rezzolla and Baiotti~\cite{Takami2014} perform binary neutron
star merger simulations for different EOSs (which are fitted by
piecewise polytropes) using the {\tt Whisky} code, and suggest a
universal (EOS and mass-independent) empirical relation between a
secondary peak in the GW spectrum and the compactness $M/R$ of the
progenitor neutron stars, although their analysis was for a restricted
set of EOSs and for varying mass ranges, without distinguishing
between $f_{2-0}$ and $f_{\rm spiral}$. In~\cite{Takami2016}, a
somewhat more extensive set of models is considered. While their GW
spectra appear to be broadly consistent with the unified picture
presented in~\cite{Bauswein2015PhRvD..91l4056B}, a different
interpretation of the secondary peaks (not consistent with
~\cite{Bauswein2015PhRvD..91l4056B}) is presented. 

Maione et al.~\cite{Maione:2017aux} also perform a large number of
binary neutron star merger simulations (using the {\tt Einstein
  Toolkit}~\cite{ET12}) surveying the effects of total mass, the EOS
stiffness and the mass ratio. They test their results against the two
competing interpretations of the sub-dominant frequencies in the
post-merger spectrum that have been presented in
~\cite{Bauswein2015PhRvD..91l4056B} and \cite{Takami2014}. The authors
conclude that they agree with~\cite{Bauswein2015PhRvD..91l4056B}
(which includes two different mechanisms for producing
\textit{mass-dependent} sub-dominant frequencies, $f_{2-0}$ and
$f_{\rm spiral}$) in that at least two different mechanisms should be
considered for the interpretation of these sub-dominant frequencies.
In several models, Maione et al. were able to confirm another
prediction of the unified model of \cite{Bauswein2015PhRvD..91l4056B},
that the presence of $f_{\rm spiral}$ would leave an observable
imprint also in the maximum density evolution, as a modulation with
frequency $f_{2-0}-f_{\rm spiral}$, due to the relative instantaneous
orientation of the external spiral structure with respect to the
internal double core structure.

Shibata and Kiuchi~\cite{Shibata:2017xht} perform viscous hydrodynamic
simulations of binary neutron star mergers in full GR and argue that
for large values of the shear viscosity any post-merger oscillations
could be damped within 5 ms. Nevertheless, the values of the shear
viscosity the authors adopted may be large compared to realistic
values anticipated due to magnetic fields. More recently, Alford et
al.~\cite{Alford:2017rxf} explored various dissipative mechanisms that
may operate in a binary neutron star merger remnant and argue that
bulk viscosity may be sufficiently strong to dampen any post-merger
oscillations. A careful investigation of the effects of bulk viscosity
in future relativistic calculations of mergers will show whether bulk
viscosisty plays an important role.  For another review summarizing
binary neutron star post-merger oscillation properties
see~\cite{BaiottiRezz2016review}.  Properties of the post-merger
remnants have also been investigated
in~\cite{Kaplan2014,KastaunGaleazzi2015,KastaunCiolfiGiacomazzo2016}.

The detectability of such post-merger oscillations has recently also
been considered. Clark et al. \cite{Clark2016CQGra} developed a
post-merger GW detection template, based on the method of principal
component analysis (PCA) and evaluated the prospects for detectability
when using present and planned gravitational wave
interferometers. Adopting a signal-to-noise ($SNR$) ratio detection
threshold of 5, an optimally oriented source and the galactic merger
rate of~\cite{Abadie:2010cf} they calculated that post-merger
oscillations would not be detectable by advanced LIGO at design
sensitivity, but could become detectable out to $\sim 110-180$ Mpc
(depending on the EOS) with the proposed LIGO\ Voyager upgrade
(LV)~\cite{LIGOVoyager}, out to $\sim 200-340$ Mpc with the planned
third-generation Einstein Telescope (ET)~\cite{amaro2009einstein} and
out to $\sim 330-530$ Mpc with the planned Cosmic Explorer
(CE)~\cite{abbott2017exploring} detector. These results translate to
EOS-dependent detection rates of $\sim 0.2-0.9\ \rm yr^{-1}$ for LV,
$\sim 1-6 \ \rm yr^{-1}$ for ET and $\sim 5-23 \ \rm yr^{-1}$ for CE.

Yang et al.~\cite{Yang:2017xlf} focus on the detectability of the
dominant component of the $l=2,m=2$ mode, exploring different EOSs and
considering the ET and CE detectors. They adopt the same SNR detection
threshold of 5 as Clark et al. \cite{Clark2016CQGra}. Instead of the
galactic merger rate of~\cite{Abadie:2010cf} adopted by Clark et al.,
they adopt the galactic merger rate
of~\cite{Belczynski:2015tba,Dominik:2014yma,deMink:2015yea} which is
consistent with the latest pulsar beaming corrections and improved
modeling of PSR J0737-3039B~\cite{Kim2015MNRAS.448..928K}, and roughly
10 times lower than the one adopted by Clark et al.. In addition, Yang
et al. consider sources that are not optimally oriented, and instead
have random sky orientations. Moreover, they adopt the Gaussian
mass-distribution for binary neutron stars
of~\cite{OzelFreire2016ARA&A..54..401O} and reject models with total
mass larger than the threshold mass for prompt collapse. Under the
above assumptions, Yang et al. perform 100 Monte-Carlo (MC)
realizations assuming a one-year of observations, and (for the EOSs
treated) they find it is not likely that a detection of the dominant
component of the post-merger $l=2,m=2$ mode from individual sources
will be made. However, this conclusion depends on the EOS. For
example, for the TM1 EOS the rate of individual detections exceeds $1
\ \rm yr^{-1}$ only in 15\% of the MC realizations, but rises to $\sim
60$\% of MC realizations for the Shen EOS.

To increase the prospects for detection, Yang et al. develop a method
that takes advantage of the empirical relation between the peak
frequency (when scaled by the total binary mass) and the radius of a
nonrotating neutron star of mass $1.6 M_{\odot}$ (as given
in~\cite{Bauswein2016EPJA...52...56B}; see Eq.~\eqref{eq:fvsR2}) to
coherently stack an ensemble of post-merger signals. The authors find
that after coherently stacking the post-merger oscillations from
different sources, the percentage of MC realizations with stacked SNR
above the detection threshold rises considerably to 91\% for the TM1
EOS and CE. In agreement with Clark et al. \cite{Clark2016CQGra}, Yang
et al. find that systematic errors (e.g. scatter in the $f_{\rm
  peak}$-EOS universal relationship) at this time dominate the
statistical errors in inferring the NS radius and hence constrain the
EOS. Another work exploring stacking of post-merger signals was
recently presented by Bose et al.~\cite{Bose:2017jvk} (but without
considering the detection probability). As pointed in Yang et
al.~\cite{Yang:2017xlf} the results from all of these studies should
be considered preliminary, because they depend on the mass
distribution and the merger rates of binary neutron stars as well as
imprecise models of the post-merger gravitational waveforms which
should all improve with time.

\subsubsection{One-arm instability}
\label{s:one-arm-NSNS}

Apart from the dynamical bar-mode (m=2), highly differentially
rotating stars can also become unstable to a dynamical one-arm ($m=1$)
``spiral'' instability. Such, highly differentially rotating neutron
stars can form either during core-collapse or binary neutron star
mergers. Hence, one might expect that the one-arm instability could
arise in these dynamical scenarios. Indeed, the instability has been
found to operate in the differentially rotating neutron star cores
formed in general relativistic hydrodynamic core-collapse simulations
by Ott et al.~\cite{Ott2005,Ott2007PhRvL} and Kuroda et
al.~\cite{Kuroda2014}.  Shibata and Sekiguchi~\cite{Shibata2005} also
report the emergence of $m=1$ modes in core-collapse simulations, but
the $m=1$ perturbations are not reported to grow
significantly. However, until recently the $m=1$ instability has never
been found to operate in hypermassive neutron stars formed in a binary
neutron star merger.  Nevertheless, we note that Anderson et
al.~\cite{Anderson2008PhRvL} performed magnetized neutron star mergers
in full general relativity and reported the emergence of $m=1$ modes
following merger, which were attributed to magnetic Tayler
instabilities~\cite{tayler73,Tayler1973,Tayler1973MNRAS.162...17T,Markey1973}. In
addition, $m=1$ density modes in hypermassive neutron stars formed
following binary neutron star mergers were reported by Bernuzzi et
al. in~\cite{Bernuzzi2014PhRvD..89j4021B}, where they were explained
to arise due to mode couplings.

\begin{figure*}[th!]
    \center
    \includegraphics[width=0.49\textwidth]{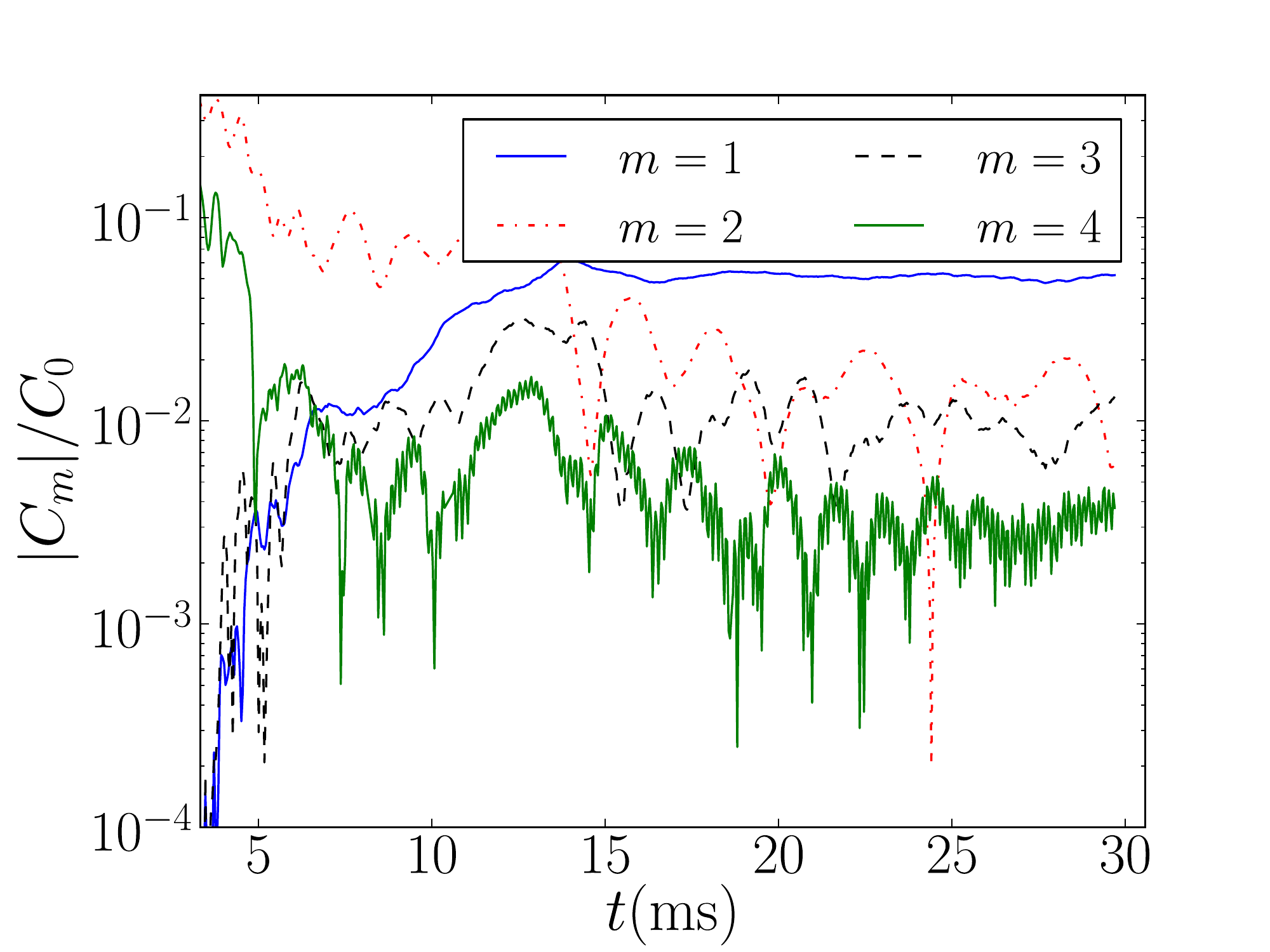}
       \includegraphics[width=0.49\textwidth]{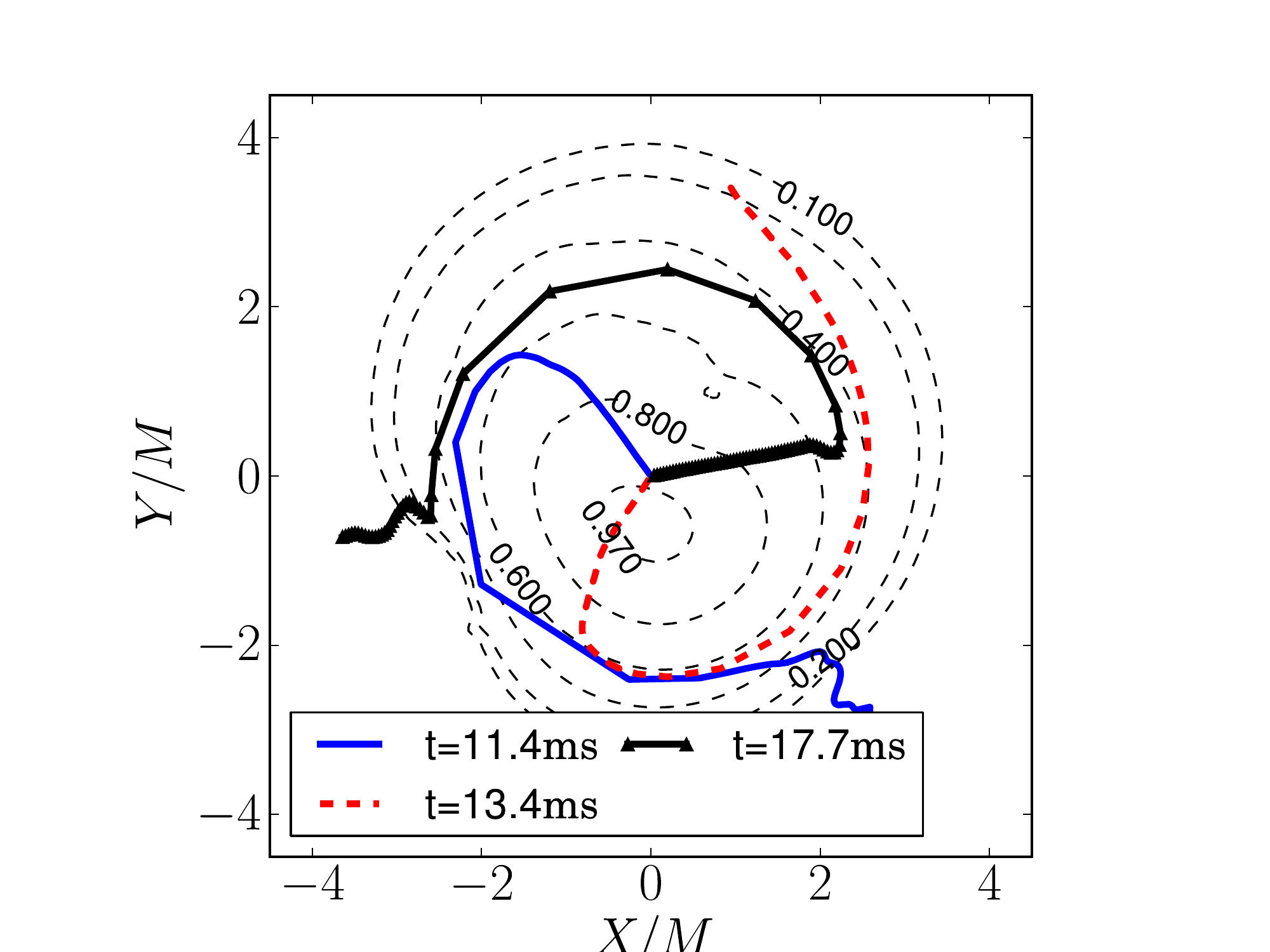}
        \includegraphics[width=0.49\textwidth]{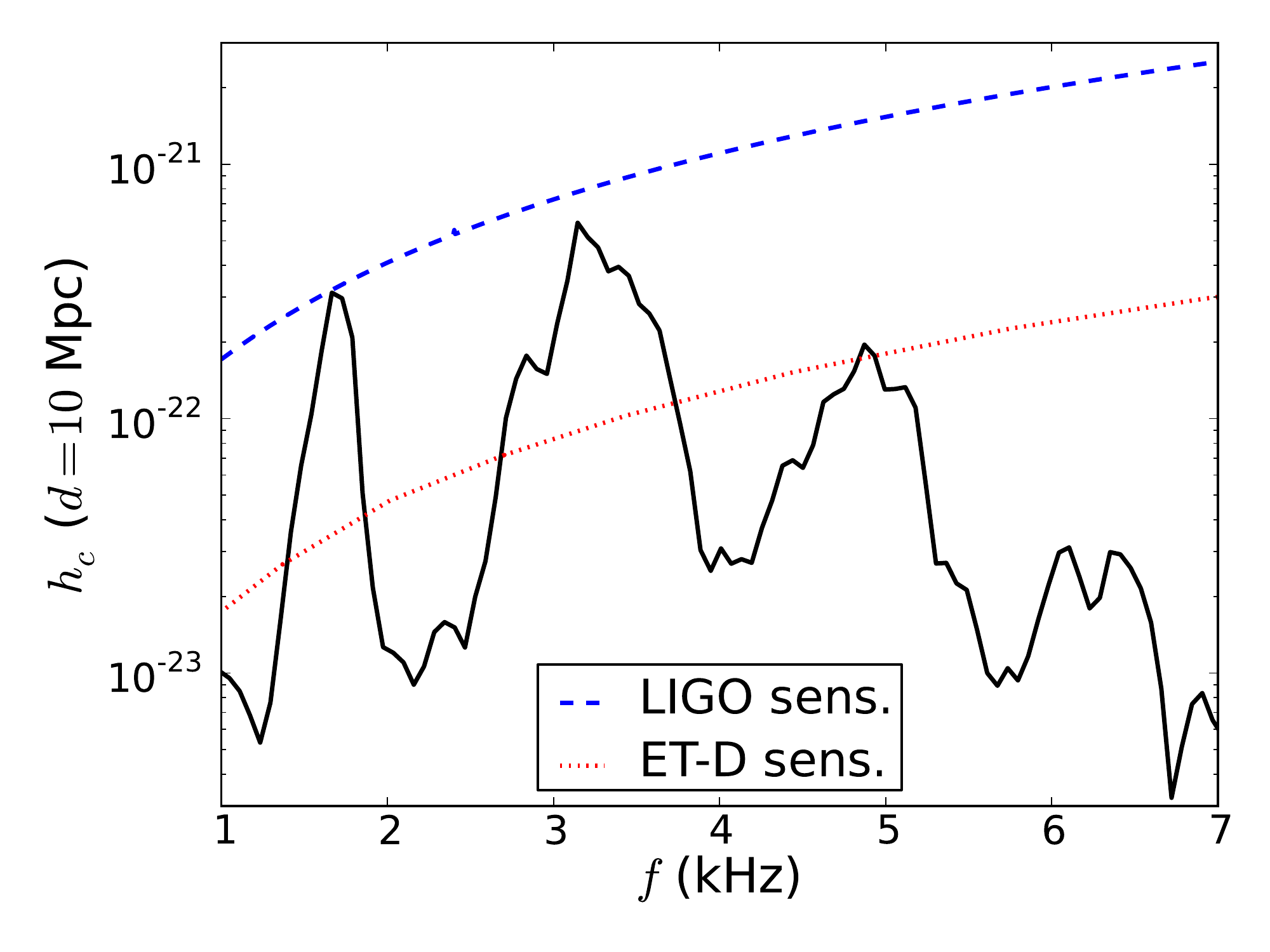}
    \caption{\textit{Top left panel}: Magnitude of $C_m$ normalized to $C_0$ as a
      function of coordinate time.  \textit{Top right panel}: Thick lines indicate the
      phase of the mode $C_1$ as a function of cylindrical radius
      $\varpi$ at select times. Dashed thin lines are contours of the
      rest-mass density normalized to its maximum value at $t=13.4$
      ms. The numbers inlined are the values of the level
      surfaces. The small contour at $X/M\approx Y/M \approx 1$ has a
      value of 0.6 and is at the cite of a strong vortex. \textit{Lower panel}:
      gravitational wave characteristic strain (solid black curve) vs
      gravitational wave frequency for the last $\sim 15$ ms (after
      the $m=1$ instability has saturated) seen by an observer located
      edge-on at $r=10$ Mpc. The first peak on the left corresponds to
      the $m=1$ mode. Notice that the frequencies of the peaks satisfy
      $f_m\simeq m f_1, \ \ m=1,\ldots 4$, with $f_1 \simeq 1.7$
      kHz. Dashed (blue) curve: the aLIGO sensitivity curve. Dotted
      red curve: the proposed Einstein Telescope (ET-D) sensitivity
      curve~\cite{ET_D} (Image reproduced with permission
      from~\cite{PEPS2015}, copyright by APS.)}
  \label{fig:onearminst}
\end{figure*}

But, recent hydrodynamic simulations in full general relativity
adopting a piecewise polytropic equation of state of moderate
stiffness by Paschalidis et al. in~\cite{PEPS2015}, report the
development of the one-arm instability in the highly differentially
rotating hypermassive neutron star remnant for the first time. The
emergence of the instability in the merger remnants of eccentric
binary neutron star mergers was subsequently studied in East et
al.~\cite{EPPS2016} with a larger survey of hydrodynamic simulations
in full general relativity. Paschalidis et al. and East et al. argued
that the trigger of the instability is the post-merger vortex dynamics
during the merger of the two stars. The growth time of the instability
in the cases studied was $\sim 1-2$ ms and the instability saturates
within $\sim 10-20{\rm ms}$ from merger. The $m=1$ instability is most
easily observed in an azimuthal density decomposition through the
quantities
\be
C_m(\varpi,z) = \frac{1}{2\pi}\int_0^{2\pi}\rho_0 u^0\sqrt{-g}e^{im\phi} d\phi
\ee
and
\be
C_m = \frac{1}{2\pi}\int_0^{2\pi}\rho_0 u^0\sqrt{-g}e^{im\phi} d^3x,
\ee
where $g$ is the determinant of the metric, $\rho_0$ the rest-mass
density, $u^\mu$ the fluid 4-velocity, and $\phi$ an azimuthal angle
defined in a center-of-mass frame. Here, $\varpi$ is the cylindrical
radius from the center of mass.  These quantities are shown in the
left and middle panels of Fig.~\ref{fig:onearminst} for a binary
neutron star merger remnant that underwent the one-arm instability
in~\cite{PEPS2015}, where it is clear (left panel) that $\sim 10$ms
following merger the $m=1$ density azimuthal mode amplitude is larger
than all other non-zero $m$ modes, signaling the saturation and
dominance of the instability. Notice the almost constant amplitude of
the $m=1$ mode throughout the evolution, which acts as a
quasistationary source of gravitational waves. The middle panel also
plots the rest-mass density contours and the phase of the $C_1$ mode
in the center of mass and on the equatorial plane at select
times. Notice how the high-density hypermassive neutron star core is
displaced from the center of mass -- a signature of the $m=1$
instability. Observe also the spiral pattern of the phase of the $m=1$
mode as it becomes sheared toward the surface of the remnant (although
it is unclear at the moment whether the spiralling of the phase of
the mode has any physical significance).

Paschalidis et al. also investigated whether previous criteria for the
development of the instability hold in these cases, too. They find
that there exists a corotation radius within the star prior to the
development of the instability. This extends earlier criteria for the
development of shear instabilities from isolated cold stars to hot
hypermassive neutron stars formed by binary neutron star mergers.

%

Both~\cite{PEPS2015} and~\cite{EPPS2016} demonstrated that the
instability is imprinted on the gravitational waves generated during
the post-merger phase. In particular, the m=1 instability gives rise
to an $l=2,m=1$ mode of gravitational waves that is quasi-periodic and
almost constant in amplitude, and with the gravitational wave
fundamental frequencies being consistent with the dominant rotational
frequencies of azimuthal density modes. Moreover, the $l=2,m=1$
gravitational wave signature occurs at roughly half the frequency of
the $l=2,m=2$ mode (higher modes have frequencies $f_m\simeq m f_1$ --
another signature of the one-arm instability) and hence lies in a
regime where the LIGO detector is more sensitive (see right panel of
Fig.~\ref{fig:onearminst}). If the $m=1$ mode persists during the
hypermassive neutron star lifetime $t_{\rm HMNS} \sim O(1)$
s~\cite{PEPS2015,EPPS2016}, the peak power at the $m=1$ mode frequency
can be amplified by a factor $\sqrt{t_{\rm HMNS}/(15 \ {\rm ms})}\sim
O(10)$. Thus, for long-lived (1-2s) hypermassive neutron stars for
which the one-arm instability persists Paschalidis et al. and East et
al. predicted that the GWs could be detectable by aLIGO at $\sim 10$
Mpc and by the Einstein Telescope at $\sim 100$ Mpc, and speculated
that the GWs from the instability may help to constrain the EOS of the
matter above nuclear saturation density.

While these simulations had high eccentricity at merger, Paschalidis
et al. and East et al. found that the instability arises for cases
where the total angular momentum at merger $J/M^2 \sim 0.9-1.0$, where
$J$ is the ADM angular momentum and $M$ the ADM mass. Since this part
of the parameter space is also relevant for quasicircular mergers,
Paschalidis et al. and East et al. predicted that the $m=1$
instability should arise in quasicircular binary neutron star mergers,
too.  These predictions were subsequently confirmed by Radice et
al.~\cite{Radice2016arXiv160305726R} and Lehner et
al.~\cite{Lehner2016arXiv160300501L} who performed hydrodynamic
simulations in full GR for equal-mass, and equal- and unequal-mass
binary neutron star mergers, respectively and confirmed that the
one-arm instability develops in quasicircular mergers. Radice et
al. employed piecewise polytropic equations of state, while Lehner et
al. adopted realistic equations of state. The Radice et al. study
concluded that aLIGO is unlikely to detect the $l=2,m=1$ modes arising
from the one-arm instability, but, as Lehner et al. pointed out, it is
worth searching for such quasimonochromatic signatures, because the
gravitational wave signal from the inspiral will reduce the
signal-to-noise ratio required for detection of the post-merger
gravitational wave signal. Both studies found that for hypermassive
neutron stars that undergo delayed collapse to black hole $\gtrsim
20$ms following merger, the instability develops for all equations of
state they considered. Finally, Lehner et
al.~\cite{Lehner2016arXiv160300501L} confirm the hypothesis
of~\cite{PEPS2015,EPPS2016} that the frequencies of the peaks in the
gravitational wave power spectrum should correlate with the nuclear
equation of state. The same conclusion was also reached in the more
recent relativistic studies of East, Paschalidis and
Pretorius~\cite{EPP2016}, where it was shown that softer equations of
state result in higher frequency $l=2,m=1$ gravitational wave
modes. East, Paschalidis and Pretorius~\cite{EPP2016} find that the
one-arm instability can be triggerred almost independently of the
background configuration that forms following merger, i.e.,
independently of whether the hypermassive neutron star is toroidal,
ellipsoidal or a double core configuration. They also estimate that
typical signal-to-noise ratios of the $l=2,m=1$ GW modes generated by
the one-arm instability would be $\sim 3$ for aLIGO at 10 Mpc and
$\sim 3$ at 100 Mpc for the Einstein Telescope. These signal-to-noise
estimates are more optimistic that those presented
in~\cite{Radice2016arXiv160305726R}, but less optimistic than the ones
in~\cite{Lehner2016arXiv160300501L}.  Thus, gravitational waves from
the $m=1$ instability could potentially be used to probe the nuclear
equation of state, although more work is needed to solidify this idea.

\subsection{Evolution of magnetized, rotating neutron stars}

Recent advances in the field of numerical relativity that combine HRSC
methods with the BSSN or the Generalized harmonic formulation as well
as approaches that control the no-magnetic-monopole constraint ${\bf
  \nabla}\cdot {\bf B} = 0$ (see e.g.~\cite{Etienne10,Etienne11} for a
summary of such methods), have allowed the evolution of
magnetohydrodynamic models of neutron stars that enabled the study
of magnetic effects such as magnetic instabilities as well as
magnetically driven outflows and magnetospheric phenomena.

  \begin{figure*}[th!]
    \center
      \includegraphics[width=0.45\textwidth]{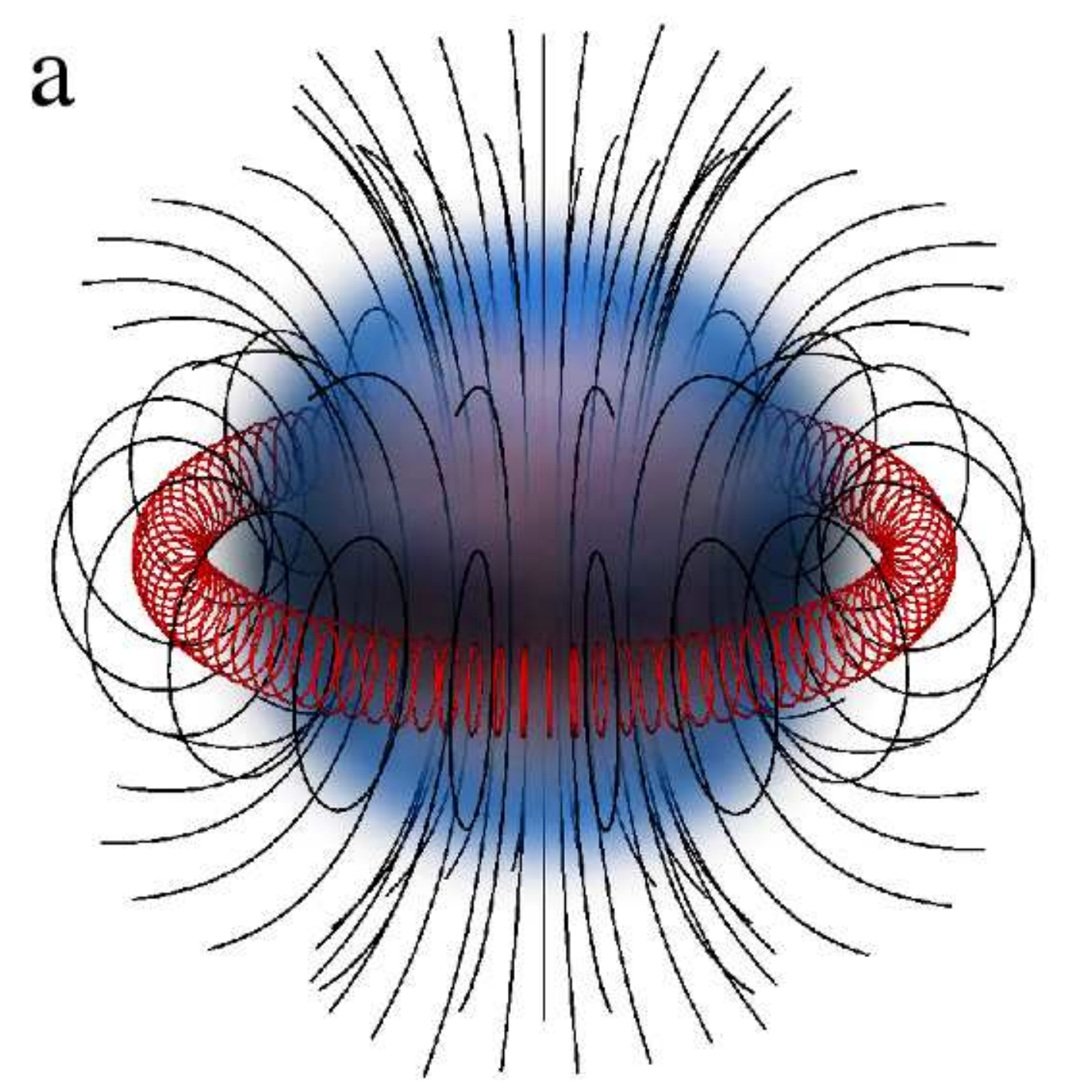}
      \includegraphics[width=0.45\textwidth]{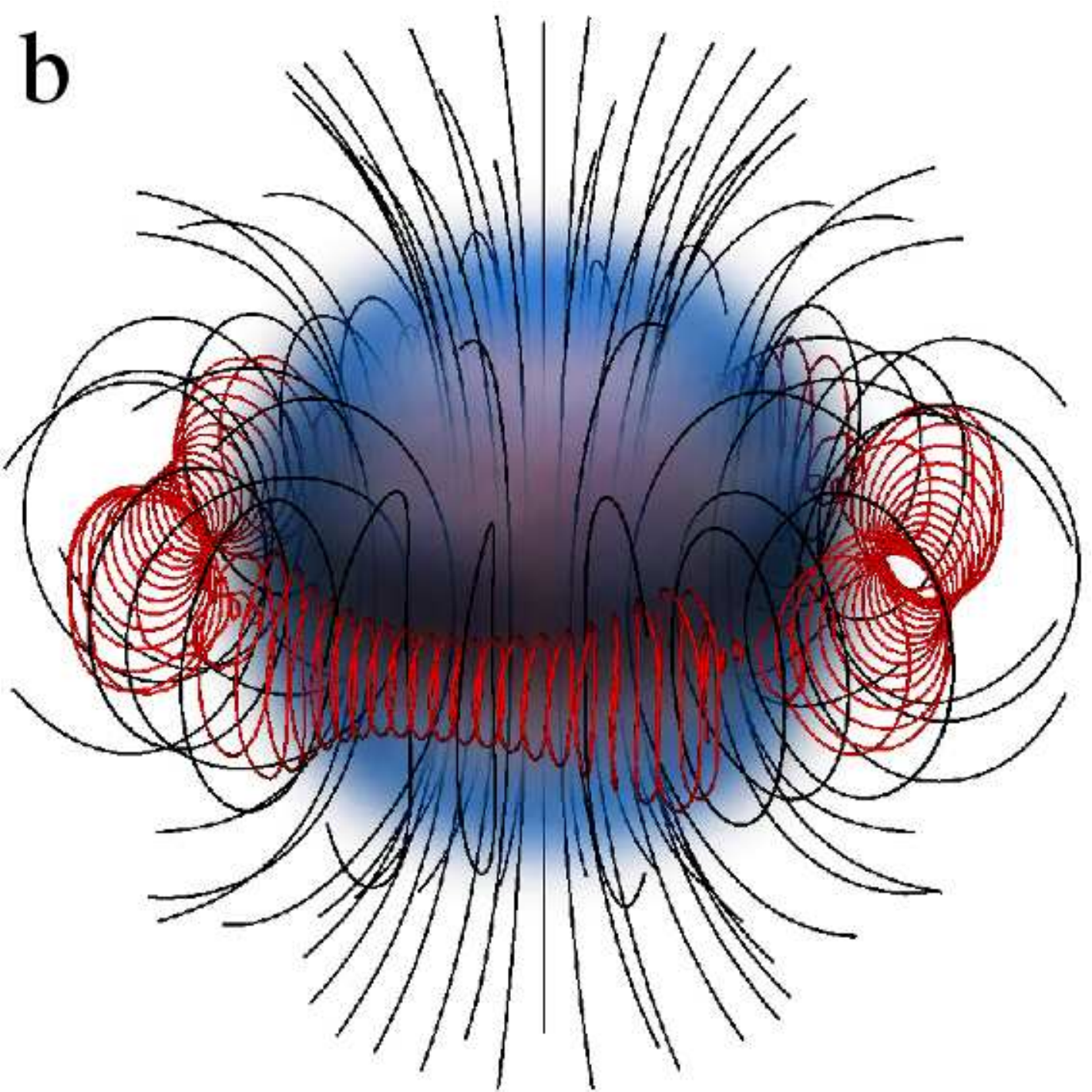} 
      \includegraphics[width=0.45\textwidth]{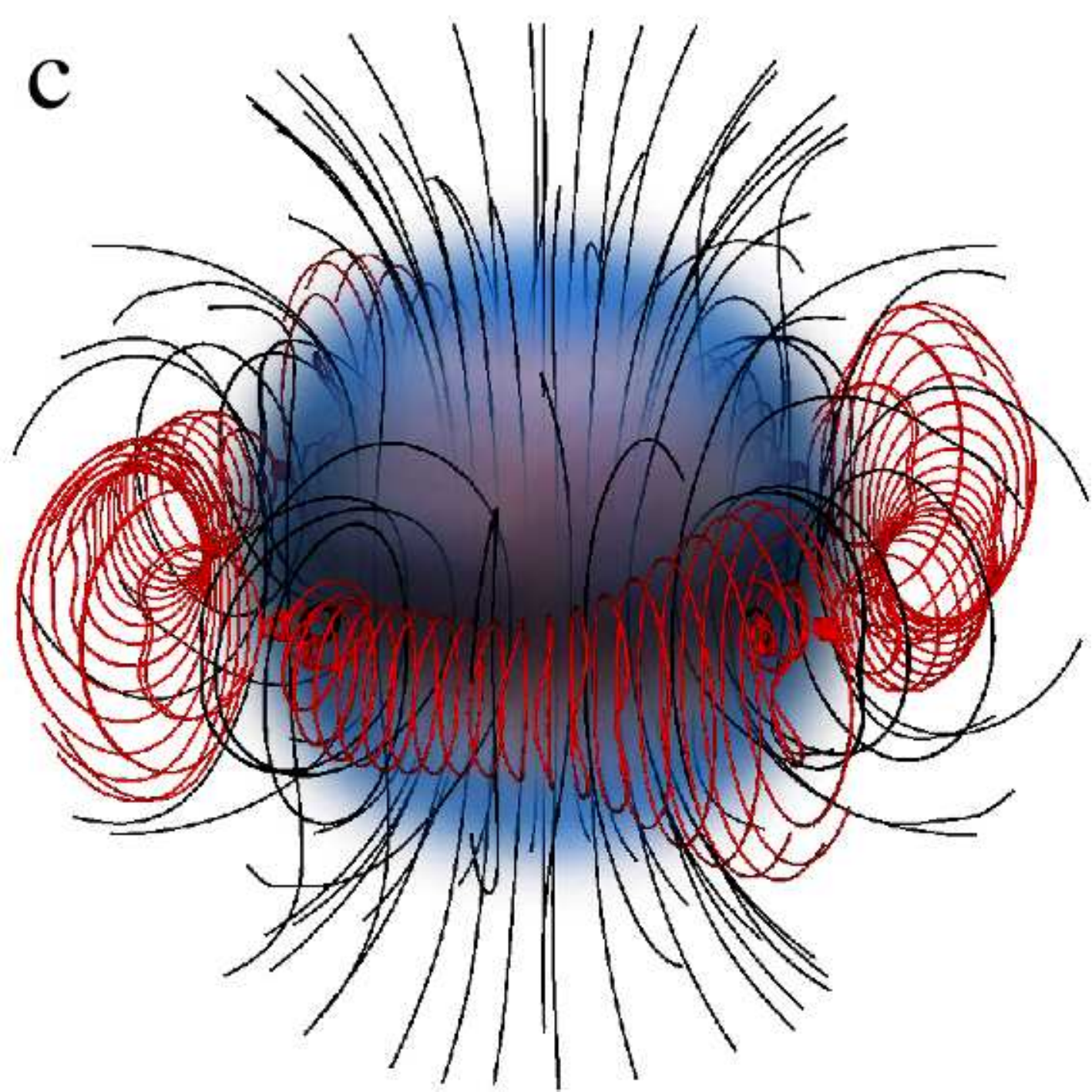}
      \includegraphics[width=0.45\textwidth]{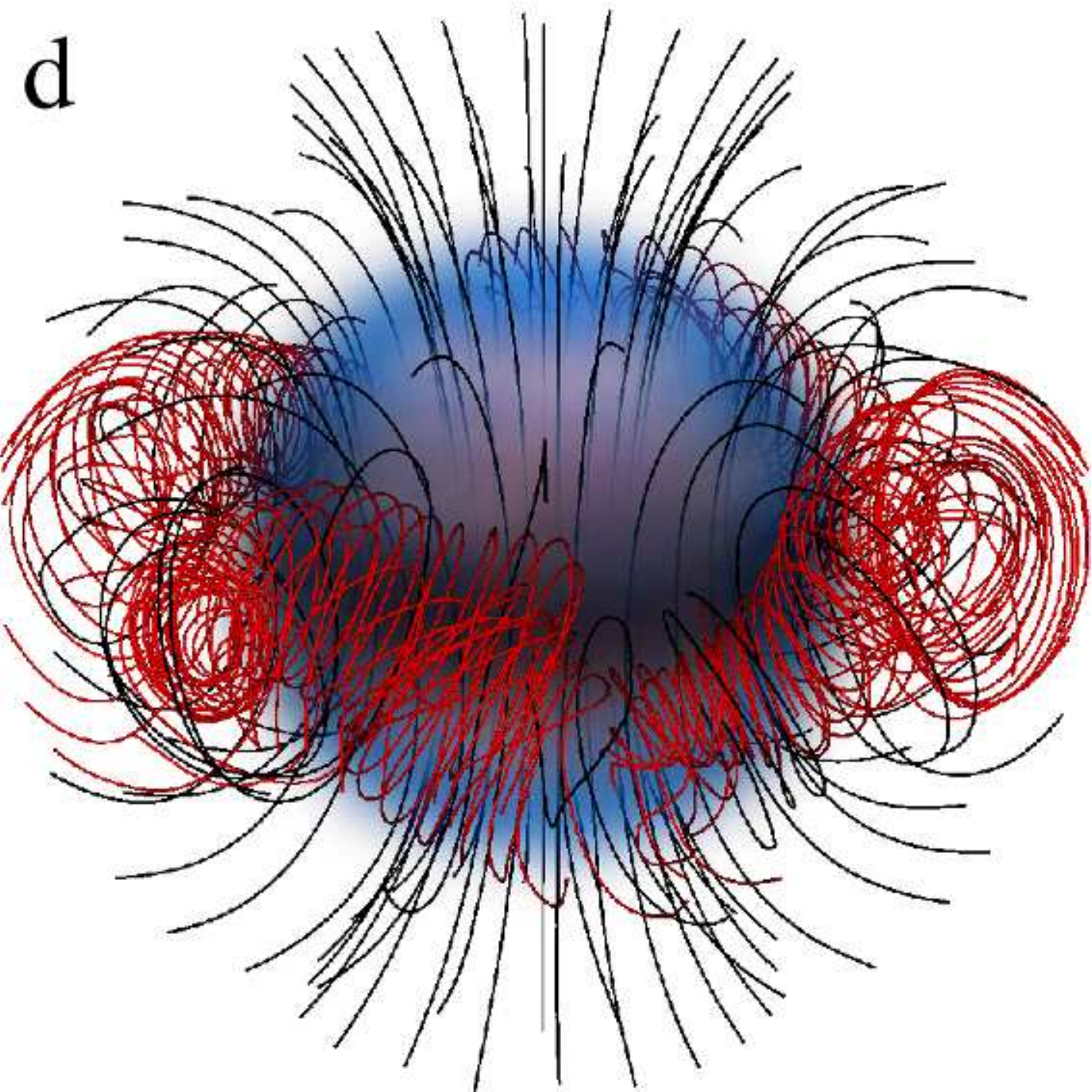} 
    \caption{Time evolution of a magnetized, rotating neutron star
      model with central magnetic field $10^{17}$G, initially rotating
      at 100Hz with a polar to equatorial radius ratio of
      $0.99$. Lines represent the magnetic field lines and the volume
      rendering the rest-mas density. The red field lines are seeded
      on the equatorial plane close to the neutral line, and the black
      field lines are seeded on the equatorial plane interior to the
      neutral line. The volume rendering is an contour surface at
      $37\%$ of the central rest-mass density. The plots correspond to
      times a) $t = 0$ms, b) $t = 17$ms, c) $t = 27$ms and d) $t =
      42$ms. (Image reproduced with permission
      from~\cite{Lasky2012}.)}
  \label{fig:MagRotstar}
\end{figure*}

\subsubsection{Magnetic Instabilities}

Axisymmetric, magnetohydrodynamic simulations of neutron stars endowed
with purely toroidal magnetic fields are performed in full GR by
Kiuchi, Shibata and Yoshida in \cite{Kiuchi2008}. Both rotating and
non-rotating, magnetized, equilibrium $\Gamma=2$ polytropes are
evolved and the stability of such configurations is investigated
varying the compaction, profile and strength of magnetic fields and
degree of rotation. The equilibrium initial data are constructed as
described in \cite{Kiuchi2008b}, and the toroidal magnetic field is
set such that 
\be 
b_\phi=B_0 u^t(\rho h\alpha^2\gamma_{\phi\phi})^k,\label{btoroidal}
\ee 
where $B_0$ and $k$ are constants that determine the $B$-field
strength and profile, $u^\mu$ is the fluid four-velocity, $h$ the
relativistic enthalpy, $\alpha$ the lapse function and $\gamma_{ij}$
is the 3-metric. Note that Eq.~\eqref{btoroidal} assumes geometrized
polytropic units. As is clear from Eq.~\eqref{btoroidal} the magnetic
field is confined in the NS interior. For the evolution, the GR
magnetohydrodynamics code described in \cite{Shibata2005b} is adopted
in conjunction with a $\Gamma$-law equation of state. It is found that
for $k=1$ the stars are stable, but for $k\geq 2$ a dynamical
instability sets in, which occurs on an Alfv\'en timescale until a new
state is reached which is dynamically stable against axisymmetric
perturbations. It is also found that rotation tends to stabilize the
stars, and overall these results are in agreement with earlier studies
by Tayler \cite{Tayler1973}, Acheson \cite{Acheson1978}, and Pitts
\cite{Pitts1985}.  In a follow-up study Kiuchi, Yoshida and Shibata
\cite{Kiuchi2011} use similar methods to study the stability of
toroidal magnetic fields in non-rotating and rapidly, rigidly
rotating, $\Gamma=2$ polytropes in full general relativity and 3+1
dimensions, focusing on the $k=1$ case and nonaxisymmetric
perturbations. Very strong initial magnetic fields of
$10^{16}-10^{17}$G are chosen, which may not be realistic but are
useful to accelerate the development of instabilities. It is found
that the Parker~\cite{Parker1966,Parker1967} and/or Tayler
instabilities operate in both nonrotating and rotating stars
triggering long-term turbulence. In contrast to the axisymmetric
simulations, it is found that the magnetic fields never reach a
dynamically stable state after the onset of turbulence. It is
concluded that unlike linearized studies even rotation cannot
stabilize the $k=1$ case against nonaxisymmetric perturbations.

Lasky et al. \cite{Lasky2011,Lasky2012} study the stability of
initially purely poloidal magnetic fields threading non-rotating and
uniformly rotating, polytropic, equilibrium neutron star models which
are generated with the {\tt magstar} module of the {\tt LORENE}
libraries. The studies vary several quantities: i) the degree of
rotation, ii) the strength of the magnetic field, iii) the stiffness
of the equation of state, while the mass of the star is fixed at
$1.31M_\odot$. For the evolutions the {\tt THOR} and {\tt HORIZON}
ideal magnetohydrodynamic codes are used \cite{Zink10,Zink2011}
keeping the spacetime fixed. Consistent with perturbation studies
\cite{Markey1973,Wright1973}, it is found that on an Alfv\'en
timescale a magnetohydrodynamic instability develops (``kink''
instability) leading to violent re-arrangement of the magnetic
fields. Such a re-arrangement may be the engine behind magnetar
flares.  The simulations demonstrate that the re-arrangement leads to
$f$-mode oscillations, but that gravitational waves from $f$-modes are
not likely to be detected by current or near-future gravitational wave
observatories. On the other hand, gravitational waves from Alfv\'en
waves propagating inside the neutron star are more promising candidates.
It is found that rotation separates the timescales of different
instabilities, varicose vs kink instability, but both modes are always
present regardless of the degree of rotation.  The end-state magnetic
field geometries derived from the simulations are nonaxisymmetric,
with approximately 65\% of the magnetic energy in the poloidal field
and the authors conclude that these resemble twisted torus
configurations, i.e., the toroidal magnetic field component is
confined within the closed poloidal field lines. The development of
the instability and the final magnetic field configuration in one of
their rotating star cases is shown in Fig.~\ref{fig:MagRotstar}.

Lasky and Melatos \cite{Lasky2013} study tilted torus magnetic fields,
which are defined as a superposition of a poloidal component extending
from the stellar interior to its exterior, with symmetry axis tilted
with respect to the spin axis, and an interior toroidal component,
with symmetry axis aligned with the spin axis. Using the {\tt HORIZON}
code they perform a general relativistic magnetohydrodynamics
evolution of a magnetized, differentially rotating, polytropic neutron
star model (prepared with the {\tt RNS} code), to argue that such
tilted torus magnetic fields arise naturally. The significance of the
result is that tilted torus magnetic fields, if they are of magnetar
strength, lead to triaxial deformations on the star, and hence the
star becomes a quasi-periodic emitter of gravitational waves. The
authors argue that these configurations have a distinguishable
gravitational wave signature and could be discerned from other
magnetic field configurations, if detected by gravitational wave
observatories.

Ciolfi et al.~\cite{Ciolfi2011,Ciolfi2012} also study the stability of
initially purely poloidal magnetic fields threading non-rotating
$\Gamma=2$ polytropic equilibrium neutron star models which are
generated with the {\tt magstar} module of the {\tt LORENE}
libraries. The mass of the neutron star is chosen to be $1.4M_\odot$
and the strength of the dipole magnetic field at the pole in the range
$1-9.5\times 10^{16}$G.  The magnetic field extends from the NS
interior to its exterior and to handle the exterior magnetic field the
authors add a non-gauge invariant ``resistive'' term to the
right-hand-side of the evolution equation for the vector potential of
the form $\eta\nabla^2 {\bf A}$, where $\eta$ is a constant. The
evolutions are performed with the {\tt Whisky} code and the spacetime
is held fixed (Cowling approximation). To shorten the time for the
development of the instability, a small, $m=2$ perturbation is added
to the initial $\theta$ component of the fluid velocity. In agreement
with perturbation theory \cite{Markey1973,Wright1973}, it is found
that the instability is triggered and accompanied by the production of
toroidal magnetic field. As in Lasky et
al. \cite{Lasky2011,Lasky2012}, the instability occurs on an Alfv\'en
timescale and saturates when the strength of the toroidal field is
comparable to that of the poloidal one. Major rearrangements of the
magnetic field take place that could lead to electromagnetic emission,
and excite $f$-mode stellar oscillations. The evolutions settle to a
solution that is stable on a dynamical/Alfv\'en timescale and the
authors argue that a stable neutron star magnetic field configuration
should comprise  both toroidal and poloidal components.

\subsubsection{Magnetically driven outflows}

\begin{figure*}[t]
  \center
      \includegraphics[width=0.95\textwidth]{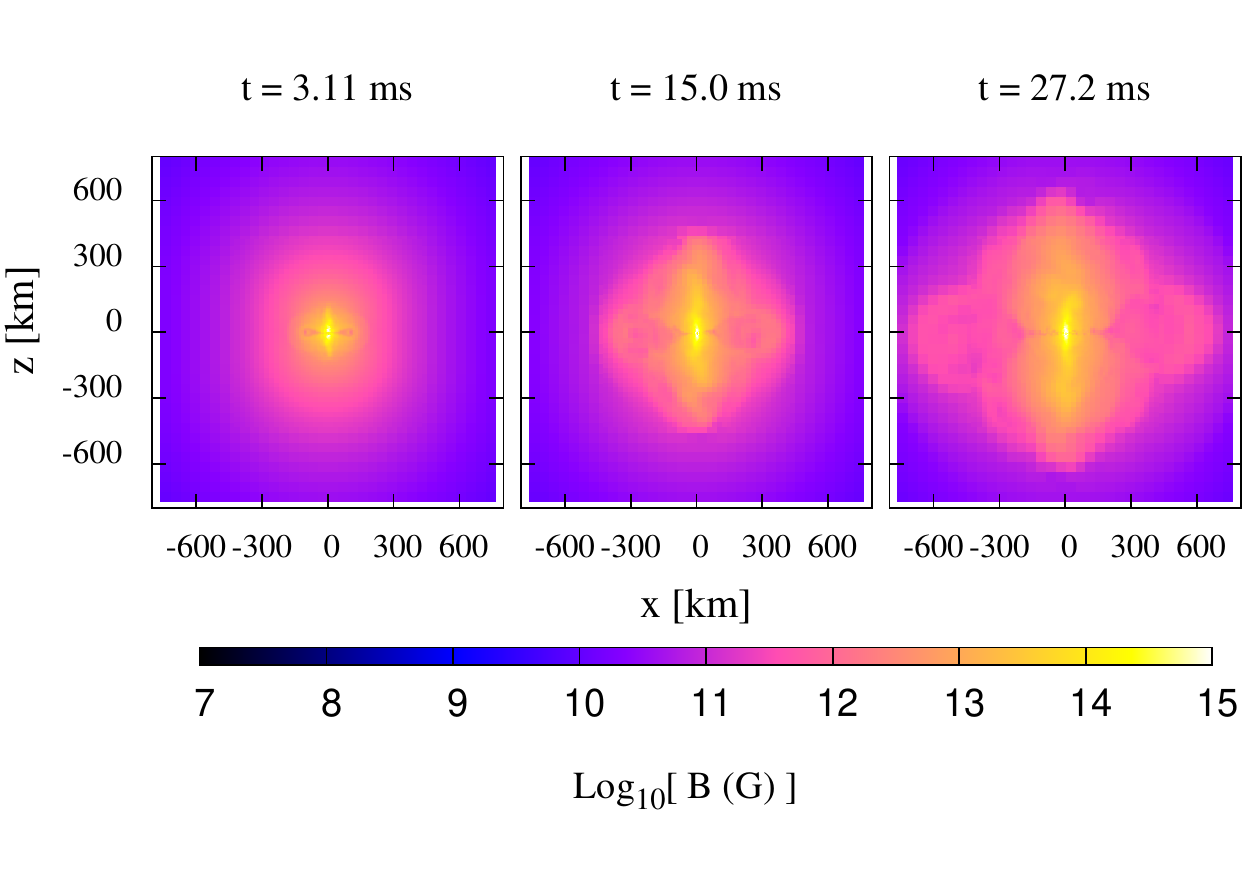}
  \caption{Snapshots, at select times, of the magnetic field strength
    on a the x-z (meridional) plane for the case with magnetic field
    strength $1.7\times 10^{14}\rm G$ evolved in full 3D
    dimensions. Image reproduced with permission
    from~\cite{Kiuchi2012}, copyright by APS.)}
  \label{fig:HMNSoutflow}
\end{figure*}

Shibata et al. \cite{Shibata2011} and Kiuchi et al.~\cite{Kiuchi2012}
perform axisymmetric and 3D, magnetohydrodynamic simulations of a
differentially rotating star in full general relativity adopting the
BSSN formulation. Their 3D evolutions are performed using a
fixed-mesh-refinement hierarchy with the Balsara divergence-free
interpolation scheme~\cite{Balsara01,Balsara09} coupled to the flux-CT
constrained transport method for the magnetic field to remain
divergence-free to machine precision even across refinement levels.
They adopt a piecewise polytropic equation of state for initial
equilibrium rotating neutron star model, and a hybrid $\Gamma$-law
equation of state which consists of a cold part and a thermal part.
The ADM mass of the differentially rotating neutron star
($2.02M_\odot$) is slightly larger than the TOV limit with the adopted
equation of state ($2.01M_\odot$) but smaller than the corresponding
supramassive limit ($2.27M_\odot$). They seed the star with a weak
purely poloidal, dipolar magnetic field such that the maximum B-field
strength (as measured in a frame comoving with the fluid) is $4.2\times
10^{13}\rm G$ or $1.7\times 10^{14}\rm G$, and such that the magnetic
dipole moment is aligned with the angular momentum of the star. It is
found that after the evolution begins strong outflows are launched with
the Poynting ($L_B$) and matter ejection ($L_M$) luminosities scaling as
\begin{equation}
L_B \sim 10^{47}\left(\frac{B_0}{10^{13}\rm
G}\right)^2\left(\frac{R_e}{10^{6}\rm
cm}\right)^3\left(\frac{\Omega}{10^{4}\rm rad/s}\right) \rm erg/s,
\end{equation}
\begin{equation}
L_M \sim 10^{48}\left(\frac{B_0}{10^{13}\rm
G}\right)^2\left(\frac{R_e}{10^{6}\rm
cm}\right)^3\left(\frac{\Omega}{10^{4}\rm rad/s}\right) \rm erg/s.
\end{equation}

These results hold both in axisymmetry and in 3 spatial dimensions.
While the authors report that in 3 dimensions a kink instability
\cite{Goebloed2004} develops, they find that the instability does not
affect the outgoing luminosities because it saturates at a small
amplitude.  However, it does affect the geometry of the outflow (see
Fig.~\ref{fig:HMNSoutflow}) leading to nonaxisymmetric features. As
noted by the authors these outflows could shine electromagnetically,
but it is not very likely that the signals will be detectable.

Motivated by these earlier results, Siegel, Ciolfi and
Rezzolla~\cite{Siegel2014} perform 3D ideal magnetohydrodynamics
simulations of a differentially rotating $\Gamma=2$ polytropic,
hypermassive neutron star (with a mass of  $2.43M_\odot$) which is initially
seeded with either a dipolar magnetic field or a random magnetic
field. The evolutions are performed using the {\tt Whisky} code and
adopting a vector potential formulation for maintaining the ${\bf
\nabla}\cdot {\bf B}=0$ constraint, coupled to the generalized Lorenz
gauge~\cite{Farris2012}. As in Shibata et al. \cite{Shibata2011} and
Kiuchi et al.~\cite{Kiuchi2012} the authors also find outflows soon
after the evolutions start. In the cases where a dipole magnetic field
is initially seeded in the star, a collimated outflow along the
stellar rotation axis is also launched in addition to a magnetized
wind, whereas in the random magnetic field case, the outflow is more
in a form of a wind and isotropic. The typical Poynting luminosity
associated with these outflows is found to be
\begin{equation}
L_B \sim 10^{48}\left(\frac{B_0}{10^{14}\rm
G}\right)^2\left(\frac{R_e}{10^{6}\rm
cm}\right)^3\left(\frac{P}{10^{-4}\rm rad/s}\right)^{-1} \rm erg/s,
\end{equation}
where $P$ is the rotation period at the location of the spin axis.

\subsubsection{Magnetospheric studies}

Lehner et al. \cite{Lehner2011} develop a novel scheme for matching
the equations of ideal general relativistic magnetohydrodynamic
stellar interiors to Maxwell's equations for force-free electrodynamic
or vacuum exteriors. The spacetime is dynamical and evolved using the
generalized harmonic formulation in conjunction with black hole
excision. Validating their method and code using the force-free
aligned rotator test (see
e.g.~\cite{GJ1969,Contopoulos1999,Spitkovsky2006,Contopoylos2006,Komissarov2006,Mckinney2006}),
and Michel monopole solution~\cite{Michel1973}, they study the
electromagnetic emission arising from both non-rotating and rotating,
collapsing, polytropic neutron star models. The initial data are
generated using the {\tt magstar} module of the {\tt LORENE}
libraries, and correspond to self-consistent rotating $\Gamma=2$
polytropes in unstable equilibrium, threaded by dipole magnetic
fields. It is found that in the non-rotating stellar collapse
approximately $1\%$ ($10\%$) of the stored energy in the initial
magnetosphere is radiated away during the collapse in the force-free
(electrovacuum) cases. The average outgoing Poynting luminosity for a
non-rotating collapsing neutron star with a force-free exterior scales
as
\be
L_{\rm EM}\approx 10^{48} \bigg(\frac{B_{\rm pole}}{10^{15}\rm G}\bigg)^2\rm erg/s,
\ee
and has a predominantly dipolar distribution. Here, $B_{\rm pole}$ is
the strength of the initial magnetic field at the stellar pole. On the
other hand, for rotating stellar collapse approximately $20\%$ of the
stored energy in the initial magnetosphere is radiated away during the
collapse both in the force-free and electrovacuum cases. The average
outgoing Poynting luminosity for rotating collapsing neutron stars
with a force-free exterior scales as
\be
L_{\rm EM}\approx 1.3\times 10^{48} \bigg(\frac{B_{\rm pole}}{10^{15}\rm G}\bigg)^2\rm erg/s,
\ee
and has a predominantly quadrupolar distribution. 

The electromagnetic emission from non-rotating collapsing neutron
stars is also studied in Dionysopoulou et al.~\cite{Dionysopoulou2013}
using a general relativistic resistive magnetohydrodynamic scheme in
full general relativity assuming electrovacuum for the stellar
exterior. It is found that up to $5\%$ of the initial energy in the
magnetosphere is radiated away and following the black hole formation
the evolution of the magnetic field follows an exponential decay, with
complex frequency matching the quasinormal mode ringing of a
Schwarzschild black hole~\cite{KScha98} to within a few percent. This
result seems to be in disagreement with the calculations of Baumgarte
and Shapiro~\cite{Baumgarte:2002vu} who studied Oppenheimer-Snyder of
ideal magnetohydrodynamic matter matched onto an exterior
electrovacuum and recovered the power-law decay anticipated from
Price's theorem~\cite{Price1972}.

The aforementioned studies focused on dynamical spacetime
magnetospheric effects.  However, special relativistic studies of
stationary pulsar magnetospheres were performed well before these
general relativistic studies. In particular, many flat-spacetime works
attempted to compute the pulsar spin-down due to dipole emission in
the limit of force-free electrodynamics. The first successful
numerical solution of the pulsar equation was presented by
Contopoulos, Kazanas and Fendt~\cite{Contopoulos1999}, which was later
followed by numerous studies of aligned and oblique rotators (see
e.g.~\cite{Spitkovsky2006,Contopoylos2006,Komissarov2006,Mckinney2006,Gruzinov:2007qa,Gruzinov:2008gb,Kalapotharakos2009A&A...496..495K,Tchekhovskoy:2012hm,Kalapotharakos2012MNRAS.420.2793K,Kalapotharakos2012ApJ...749....2K,Gruzinov:2013pva,Uzdensky:2012tf,Contopoulos2014ApJ...781...46C}
and references therein) that studied global features of the
magnetosphere. These studies did not include the magnetized NS
interior, and modeled the effects of rotation through a boundary
condition on the spherical stellar surface, which is modelled as a
perfect conductor. Simulations of pulsar magnetospheres in flat
spacetime have produced important results, such as a proof of
existence of a stationary force-free magnetospheric configuration, the
calculation of the spin-down luminosity of force-free aligned and
oblique rotators, and the evolution of the obliquity angle
\cite{Philippov:2013aha}, all in flat spacetime. There have also been
some analytic efforts to understand the emission from an accelerated
isolated pulsar in flat spacetime (see
e.g. \cite{Brennan:2013ppa,Gralla:2014yja}).

Recently, a general relativistic resistive magnetohydrodynamics scheme
in full general relativity was introduced by Palenzuela
\cite{Palenzuela2012}.  The scheme is presented and tested using the
force-free aligned rotator solution. Using a rotating relativistic
$\Gamma=2$ polytrope endowed with a dipole magnetic field, Palenzuela
reports that the outgoing Poynting luminosity -- the spin-down
luminosity -- differs by $20\%$ from its flat spacetime value and
several potential sources for this difference are listed, including
resistive, general relativistic effects and the way the flat spacetime
formula is applied to a general relativistic case.

\begin{figure*}[ht]
  \center
      \includegraphics[width=0.97\textwidth]{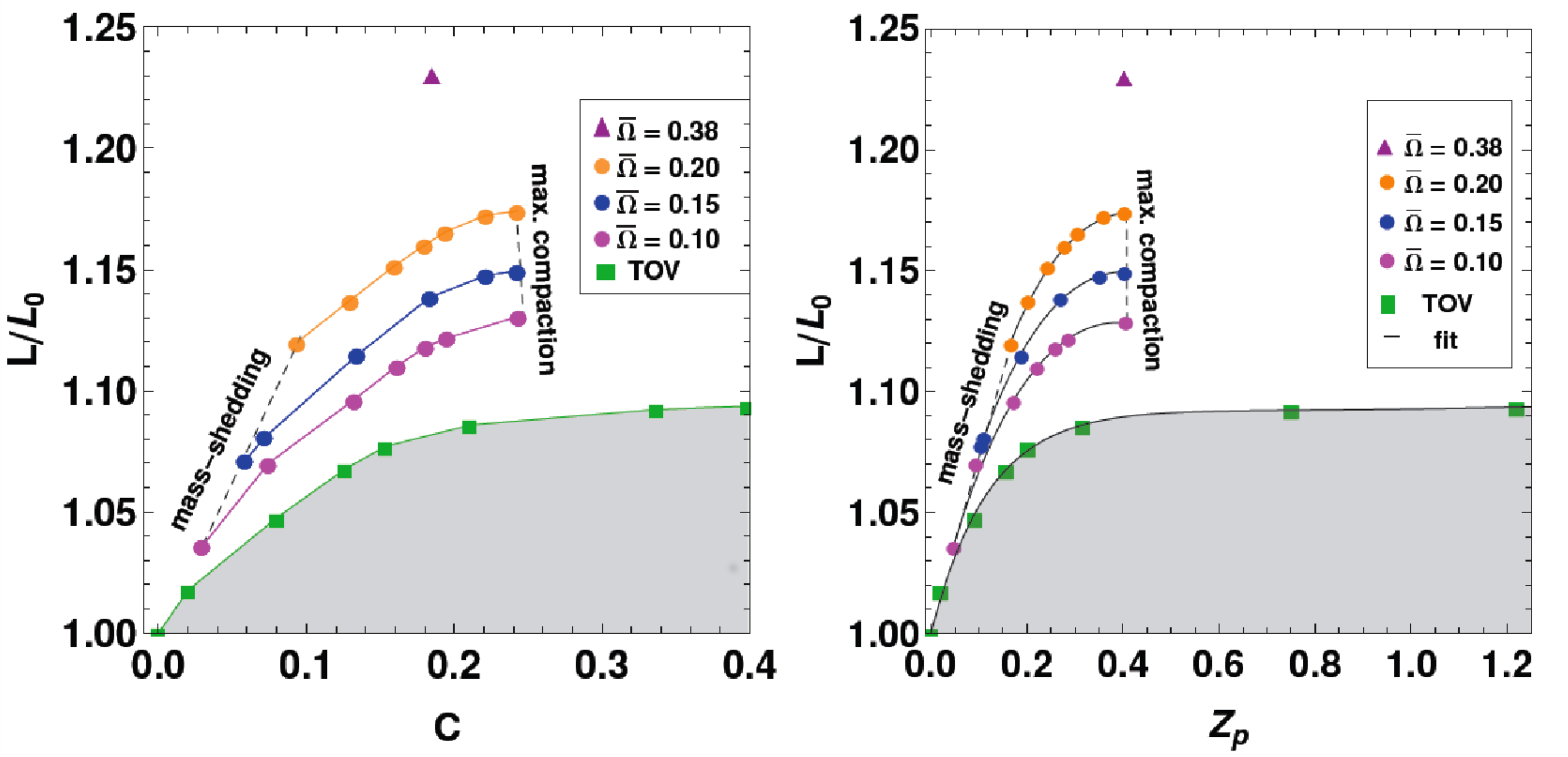}
  \caption{ Pulsar spin-down luminosity $L$ normalized by $L_0 =
    1.02\mu^2 \Omega^4$, the flat-spacetime result. Left panel:
    $L/L_0$ vs stellar compactness $C=M/R_e$, where $R_e$ is the
    equatorial radius. Right panel: $L/L0$ vs polar redshift
    $Z_p$. The parameter space for rotating stars is contained between
    the left dashed line (the mass-shedding limit) and the right
    dashed line (maximum compactness). The top point (triangle)
    represents the value for the supramassive neutron star limit for
    $n = 1$. For these rapidly rotating stars, the lower shaded zone
    is the area of the parameter space that cannot be reached, unless
    flat spacetime is assumed. In the constant angular velocity
    sequences $\bar \Omega=\Omega\cdot K^{n/2}$ is a dimensionless
    angular velocity, with $K$ standing for the polytropic constant.
    (Image reproduced with permission from~\cite{Ruiz2014}, copyright
    by APS.)}
  \label{fig:LMoR}
\end{figure*}

More recently Ruiz, Paschalidis and Shapiro~\cite{Ruiz2014} study the
pulsar spin-down luminosity via time-dependent simulations in general
relativity.  The evolutions are performed using the technique
developed by Paschalidis and Shapiro
\cite{Paschalidis2013a,Paschalidis2013b} for matching general
relativistic ideal magnetohydrodynamics to its force-free
limit. Equilibrium rotating, polytropic neutron star models of
different compactnesses are considered using 3 constant-angular velocity
sequences ranging from the mass-shedding limit to the maximum
compactness configuration. The initial data are prepared using the Cook
et al. code~\cite{CST92,CST94a,CST94b}. Both slowly rotating and
rapidly rotating stars are considered. The stars are endowed with a
general relativistic magnetic dipole~\cite{Wasserman1983} and the
electromagnetic fields are evolved keeping fixed both the spacetime
and the fluid (a valid assumption for weak magnetic fields). The
structure of the final, steady-state magnetosphere reached for all
evolved stars resembles the structure of the magnetosphere found in
flat spacetime studies. However, it is found that general relativity
gives rise to a modest enhancement of the spin-down luminosity when
compared to its flat spacetime value: the maximum enhancement found
for $n=1$ polytropes is $23\%$, and for a rapidly rotating $n=0.5$
polytrope an even greater enhancement of $35\%$ is found. The
spin-down luminosity for all cases studied is shown in
Fig.~\ref{fig:LMoR}. This enhancement in the spin-down luminosity due
to general relativistic effects has been confirmed in more recent
simulations by Philippov et al.~\cite{Philippov2015ApJ} and
P{\'e}tri~\cite{Petri2016MNRAS} who used approximate metrics for the
spacetime around a rotating neutron star. However, semi-analytic work
presented by Gralla et al. in~\cite{Gralla2016} suggests that the
general relativistic corrections in the slow-rotation limit should
disappear (a result also mentioned in~\cite{Philippov2015ApJ}).


\begin{acknowledgements}
We are indebted to all our long-term collaborators for their
contributions in joint projects and for useful discussions. NS is
indebted to John L. Friedman for his work on a recent joint monograph
which also contributed to the substantial revision this review
article. Both authors are grateful to Thomas Baumgarte, Andreas
Bauswein, Daniela Doneva, Georgios Pappas, Pantelis Pnigouras, and
Stuart Shapiro for reading the manuscript and providing valuable
comments. VP acknowledges support from NSF grant PHY-1607449, NASA
Grant NNX16AR67G (Fermi) and the Simons Foundation. This work
benefited from support by the National Science Foundation under Grant
No. PHY-1430152 (JINA Center for the Evolution of the Elements).  This
work benefited from discussions at the Binary Neutron Star Merger
Workshop supported by the National Science Foundation under Grant
No. PHY-1430152 (JINA Center for the Evolution of the Elements).
\end{acknowledgements}

\newpage

\bibliographystyle{spmpsci}
\bibliography{refs}

\end{document}